\pdfoutput=1
\documentclass[12pt]{thesis}
\usepackage{amssymb,amsmath,amsfonts}
\usepackage{bm}
\usepackage{graphicx}
\usepackage{lscape}
\usepackage{indentfirst}
\usepackage{latexsym}
\usepackage{multirow}
\usepackage{tabls}
\usepackage{wrapfig}
\usepackage{slashbox}
\usepackage{color}
\newcommand{\vth}[1][s]{\ensuremath{v_{\mathrm{th}_#1}}}

\newcommand{\gyroR}[2][s]{\ensuremath{{\left< #2 \right>}_\mathbf{R_\mathrm{#1}}}}
\renewcommand{\eqref}[1]{Eq.\ (\ref{#1})}
\newcommand{\eqsref}[2]{Eqs.\ (\ref{#1}) and (\ref{#2})}
\newcommand{\eqsdash}[2]{Eqs.\ (\ref{#1})--(\ref{#2})}
\newcommand{\Secref}[1]{Section \ref{#1}}
\newcommand{\secref}[1]{Sec.\ \ref{#1}}
\newcommand{\apref}[1]{Appendix \ref{#1}}
\renewcommand{\vth}{\ensuremath{ v_{\mathrm{th}} } }
\renewcommand{\gyroR}[1]{\ensuremath{{\left< #1 \right>}_{\bm{R}}}}
\newcommand{\Cgk}{C_{\text{GK}}}
\newcommand{\hk}{h_{\bm{k}}}

\newcommand{\mbf}[1]{\mathbf{#1}}
\newcommand{\lbb}{\left<\left<}
\newcommand{\rbb}{\right>\right>}
\newcommand{\la}{\left|}
\newcommand{\ra}{\right|}
\newcommand{\parl}{\parallel}
\newcommand{\pd}[2]{\frac{\partial #1}{\partial #2}}
\newcommand{\unit}[1]{\mathbf{\hat{#1}}}

\newcommand{\tld}[1]{\tilde{#1}}

\newcommand{\vvol}{d^{3}\mbf{v}}
\newcommand{\rln}{\frac{R}{L_{n}}}

\newcommand{\rlps}{\frac{R}{L_{p_{s}}}}
\newcommand{\ar}{\frac{a}{R}}
\newcommand{\rhoavg}{\left<\la\nabla\rho\ra\right>}
\newcommand{\dn}{\left(n_{k}^{m+1}-n_{k}\right)}
\newcommand{\dpi}{\left(p_{i_{k}}^{m+1}-p_{i_{k}}\right)}
\newcommand{\dpe}{\left(p_{e_{k}}^{m+1}-p_{e_{k}}\right)}
\newcommand{\vpa}{v_{\parallel}}
\newcommand{\vpe}{v_{\perp}}
\newcommand{\lsavg}{\bigg<\bigg<}
\newcommand{\rsavg}{\bigg>\bigg>}
\newcommand{\erf}{\textnormal{Erf}}
\newcommand{\erfi}{\textnormal{Erfi}}
\newcommand{\erfc}{\textnormal{Erfc}}

\newcommand{\beq}{\begin{equation}}
\newcommand{\eeq}{\end{equation}}
\newcommand{\grad}{\nabla}
\newcommand{\bhat}{\unit{b}}

\begin{document}


\hbox{\ }

\renewcommand{\baselinestretch}{1}
\small \normalsize

\begin{center}
\large{{ABSTRACT}} 

\vspace{3em} 

\end{center}
\hspace{-.15in}
\begin{tabular}{ll}
Title of dissertation:    & {\large  \verb#Trinity#: A Unified Treatment of}\\
&      {\large Turbulence, Transport, and Heating} \\
&      {\large  in Magnetized Plasmas} \\
\ \\
&                          {\large  Michael Alexander Barnes, }\\
&                          {\large Doctor of Philosophy, 2008} \\
\ \\
Dissertation directed by: & {\large  Professor William Dorland} \\
&  {\large  Department of Physics } \\
\end{tabular}

\vspace{3em}

\renewcommand{\baselinestretch}{2}
\large \normalsize

To faithfully simulate ITER and other modern fusion devices, one must resolve electron and 
ion fluctuation scales in a five-dimensional phase space and time.  Simultaneously, one must 
account for the interaction of this turbulence with the slow evolution of the large-scale 
plasma profiles.  Because of the enormous range of scales involved and the high 
dimensionality of the problem, resolved first-principles global simulations are very 
challenging using conventional (brute force) techniques.  In this thesis, the problem of
resolving turbulence is addressed by developing velocity space resolution diagnostics and an
adaptive collisionality that allow for the confident simulation of velocity space dynamics
using the approximate minimal necessary dissipation.  With regard to
the wide range of scales, a new approach has been developed
in which turbulence calculations from multiple gyrokinetic flux tube simulations 
are coupled together using transport equations to obtain self-consistent, 
steady-state background profiles and corresponding turbulent fluxes and heating.  This approach
is embodied in a new code, \verb#Trinity#, which is capable of evolving equilibrium profiles for multiple species,
including electromagnetic effects and realistic magnetic geometry, at a fraction of the cost
of conventional global simulations.  Furthermore, an advanced model physical collision 
operator for gyrokinetics has been derived and implemented, allowing for the study
of collisional turbulent heating, which has not been extensively studied.  To demonstrate the 
utility of the coupled flux tube approach, preliminary results from 
\verb#Trinity# simulations of the core of an ITER plasma are presented.

\thispagestyle{empty}
\hbox{\ }
\vspace{1in}
\renewcommand{\baselinestretch}{1}
\small\normalsize
\begin{center}

\large{{\verb#Trinity#: A Unified Treatment of Turbulence, Transport, and Heating in Magnetized Plasmas}}\\
\ \\
\ \\
\large{by} \\
\ \\
\large{Michael Alexander Barnes}
\ \\
\ \\
\ \\
\ \\
\normalsize
Dissertation submitted to the Faculty of the Graduate School of the \\
University of Maryland, College Park in partial fulfillment \\
of the requirements for the degree of \\
Doctor of Philosophy \\
2008
\end{center}

\vspace{7.5em}

\noindent Advisory Committee: \\
Professor William Dorland, Chair/Advisor \\
Professor James Drake \\
Professor Ramani Duraiswami \\
Professor Adil Hassam \\
Professor Edward Ott \\

\thispagestyle{empty}
\hbox{\ }

\vfill
\renewcommand{\baselinestretch}{1}
\small\normalsize

\vspace{-.65in}

\begin{center}
\large{\copyright \hbox{ }Copyright by\\
Michael Alexander Barnes  
\\
2008}
\end{center}

\vfill

\pagestyle{plain}
\pagenumbering{roman}
\setcounter{page}{2}


\renewcommand{\baselinestretch}{2}
\small\normalsize
\hbox{\ }
 
\vspace{-.65in}

\begin{center}
\large{Dedication}
\vspace{+140pt}
\\
\large{To my family, with love}
\end{center} 


\renewcommand{\baselinestretch}{2}
\small\normalsize
\hbox{\ }
 
\vspace{-.65in}

\begin{center}
\large{Acknowledgements} 
\end{center} 

\vspace{1ex}

I owe a deep debt of gratitude to a great many people for helping me along my way
in research and in life over these last several years.
Above all, I want to thank my advisor, Bill Dorland.  He has been an endless source
of new and exciting ideas throughout this research.  His enthusiasm and energy have
provided me with constant encouragement, and his confidence in my abilities has motivated
me to work hard to be a better scientist.  On both a professional and personal level, he has
been an inspiration to myself and others.  I am honored to have had the opportunity to
work with him.

I would also like to thank Kyle Gustafson and Ingmar Broemstrup for accompanying me along
the graduate research path.  Life as a graduate student would have been much emptier without
our shared experiences.

The work presented here has benefited from numerous collaborations with very
clever people.  The Z-pinch calculation in
Chapter 2 arose from a collaboration with Paolo Ricci and Barrett Rogers.  The 
derivation of the transport equations in Chapter 3 is based heavily on the work of
Steve Cowley, Gabe Plunk, Eric Wang, and Greg Howes, who were kind enough to
provide me with many useful discussions, in addition to their calculations.  My understanding
of the development of velocity space structure in Chapter 4 benefited from converstations
with Steve Cowley, Alex Schekochihin, and Tomo Tatsuno.  The derivation of the
collision operator in Chapter 5 is due in large part to Ian Abel, Alex Schekochihin, and
Steve Cowley.  The development of the implicit, conservative treatment of the collision
operator described in Chapter 6 owes a great deal to the insight of Greg Hammett, Alex
Schekochihin, and Tomo Tatsuno.  Greg Hammett also provided numerous useful conversations
that helped guide the development of the coupled flux tube scheme and transport algorithm
given in Chapter 7.

Last, but certainly not least, I'd like to thank my family.  My parents have been there for me 
through all of life's ups and downs (not a few of which occurred over the last several 
years).  Their unwavering support has been a great comfort to me when I've most needed it.
Of course, this research would not have been possible without the inspiration
of my wife, Trinity.  Thanks for sharing the joys and sorrows of life with me.

\renewcommand{\baselinestretch}{1}
\small\normalsize
\tableofcontents 
\newpage
\setlength{\parskip}{1em}
\listoffigures 
\newpage

\newpage
\setlength{\parskip}{0em}
\renewcommand{\baselinestretch}{1}
\small\normalsize

\setcounter{page}{1}
\pagenumbering{arabic}

\renewcommand{\thechapter}{1}

\chapter{\textbf{Introduction}}
\label{chap:intro}
\vspace{+90pt}

\section{Motivation}

%
We are now approaching a significant milestone in the fusion program.  Over the next eight years, a multi-billion
dollar magnetic confinement device (the International Thermonuclear Experimental Reactor, or ``ITER'') 
will be built to demonstrate the feasibility of fusion as an
alternative energy source.  The design for this experiment reflects a myriad of advances made through 
experimental, theoretical, and numerical studies in our understanding of fundamental plasma processes.
However, there are still many issues critical to the success of ITER and to the economic and scientific 
feasibility of future fusion devices that are not well understood.

The main goal of this thesis is to present a set of numerical tools and a sound numerical 
framework within which we can study one of the key fundamental physics issues for magnetic 
confinement fusion devices: the presence of anomalously high levels of particle, momentum, 
and energy transport observed in hot, magnetized plasmas.  This anomalous transport, which 
is due to small-scale turbulence driven by localized instabilities (or ``microinstabilities''), has 
been the subject of intense study within the fusion program for decades.  
Without this turbulence the performance 
of magnetic fusion devices would be considerably improved.  For 
example, a turbulence-free Joint European Torus (JET) would reach fusion ignition.  

The presence of turbulence is certainly not inevitable.  Indeed, JET, TFTR, DIII-D and other 
fusion devices have demonstrated operation with regions of the plasma essentially turbulence-free.  
Understanding, controlling, and ultimately reducing turbulence in magnetic fusion 
experiments is thus a formidable but achievable challenge for the fusion program. 
Much progress has 
been made in our qualitative understanding of turbulent transport, and in some cases 
quantitative agreement between numerical simulations and experiment is remarkably good.
However, plasma turbulence and equilibrium profile evolution are both complex problems, 
and first-principles simulations with experimentally relevant plasma parameters have by 
necessity only addressed either the effect of turbulence on the equilibrium or vice versa.

In this thesis, we present a rigorous theoretical and numerical framework that allows for the 
efficient simulation and routine study of the self-consistent interaction between plasma turbulence and 
equilibrium profiles.  While our approach provides a significant savings over direct
global simulations, it is still very challenging numerically.  We have therefore 
implemented velocity space resolution diagnostics and an adaptive collisionality
that allow us to resolve simulations with an approximately minimal number
of grid points in velocity space.  Furthermore, our collision operator is an improvement
over previous operators: it possesses a number of desirable properties, including local 
conservation of particle number, momentum, and energy, and satisfaction of 
Boltzmann's $H$-Theorem.  The latter property is of particular importance when
considering equilibrium evolution, since it is necessary to ensure that system entropy
is increased and equilibrium profiles are heated by collisions (instead of cooled).
Our use of a theoretically sound collision operator also allows us to conduct 
{\it quantitative} studies of the effect of collisional heating on equilibrium profile evolution --
a topic that has received little attention from the plasma physics community.

We do not claim to have developed a numerical fusion device.  There are a 
number of important processes currently neglected in our model (most notably the development
of equilibrium shear flows and the physics of the edge pedestal, which critically affect the 
power output of fusion devices).  However, the approach 
presented here provides a platform for studying novel effects that may arise from
the self-consistent interaction between turbulence, transport, and heating.  Furthermore,
the code we have developed (named \verb#Trinity#) is capable,
within broad parameter ranges, of providing quantitative predictions of microstability thresholds, turbulent fluctuations and 
tokamak performance from first principles. 

\section{Multiple scales}

The hot, magnetized plasmas present in magnetic confinement fusion devices are rich
and complicated physical systems.  They support an enormous spectrum of processes whose 
time and space scales span many orders of magnitude:  heated to millions of degrees,
charged particles spiral tightly around curved magnetic field lines at a significant fraction of the
speed of light; the same particles drift slowly across magnetic field lines, transporting 
particles, momentum, and heat across the length of the device; a multitude of waves 
propagate through the plasma, from light waves to Alfv\'en waves to drift waves;
kinetic instabilities give rise to a sea of small-scale, rapidly fluctuating turbulence,
and fluid instabilities can lead to bulk motion of the plasma and catastrophic disruptions.
The time and space scales for some of the important processes that affect the performance of 
magnetic confinement fusion devices (in particular, ITER) are presented in 
Tables~\ref{multimak1} and~\ref{multimak2}.

Each of these processes requires often complex modeling.
Consequently, it is neither analytically nor numerically feasible to
work with a single model that simultaneously describes all of the physical
processes present.  Instead, we must determine which processes are of
greatest interest and identify reasonable approximations that will
allow us to develop simplified models of their behavior.  Occasionally,
we may gain insights from these simplified numerical models that allow further 
reductions of the problem, but this cannot always be achieved because of the large 
number of parameters and interactions that are known to be important experimentally.

There are many important issues that must be addressed in order to develop a
scientifically and economically viable fusion reactor.  Fundamentally, however, we are 
interested in achieving high core pressures with minimal power input. 
This requires minimizing the radial heat transport, which is due primarily to turbulence.

In order to address the challenges associated with turbulent
transport, it is generally believed that one must take into account
the close coupling between the slow ($\sim 1 \ s$) evolution of
large-scale ($\sim 1 \ m$) variations in equilibrium density,
temperature, and flow profiles and the rapid ($\sim 1 \ MHz$)
fluctuations of small-scale ($\sim 10^{-5} \ m$) plasma turbulence.
This interaction of vastly disparate temporal and spatial scales
renders direct numerical and analytical approaches intractable;
instead, more sophisticated multiscale models are required.  One such
model is derived in Chapter~\ref{chap:hierarchy}, with the notable absence of
equations describing the evolution of equilbrium flows.  These flows are believed
to play a critical role in the formation of the edge pedestal and internal transport barriers,
thus limiting the immediate applicability of \verb#Trinity# to core plasmas.

To overcome the difficulty associated 
with the presence of a wide range of scales, different models have typically been applied to 
address turbulence and transport separately~\cite{TRANSP1,TRANSP2,MMM,ONETWO,ASTRA1,ASTRA2,JETTO}. 
Slowly-evolving, large-scale plasma transport 
has been widely modeled as a diffusive process, with theoretically
and numerically derived diffusion coefficients.  The magnetic equilibrium is typically modeled with the 
equations of magnetohydrodynamics (MHD), which treat the plasma as a single magnetized fluid.  These
approaches are generally inadequate for accurately describing the rapidly-evolving, small-scale turbulence
responsible for anomalous transport in fusion devices.  The instabilities driving microturbulence arise, in part,
due to the development of nontrivial structure in the distribution of plasma particle velocities (which can
be present due to the long collisional mean free path in hot fusion plasmas).  Since this
structure is not easily captured by conventional fluid models or tractable analytical approaches, 
a numerical description of kinetic, small-scale plasma turbulence is necessary.  An example illustrating this
point is provided in Chapter~\ref{chap:linzp}.

\begin{table}[h]
\begin{tabular}{|p{1.2in}|p{2.2in}|p{2in}|}
\hline
{\large\bf Physics}  & {\large\bf Space scale} & {\large\bf Time Scale} \\ \hline
\raggedright Electron Energy Transport from 
  & \raggedright Scale perpendicular to ${\bf B}$ is $\qquad
    \sim\rho_e-\rho_i\sim 0.001 \, {\rm cm} - 0.1 \, {\rm cm}$ 
  & 
\\  ETG modes 
 & \raggedright Scale parallel to  ${\bf B}$ is $\qquad qR\sim 15$ m 
  & $\omega^*_e \sim 500 \, {\rm kHz}-5\, {\rm MHz}$  
\\   \hline
\raggedright Ion Energy Transport from 
 & \raggedright Scale perpendicular to ${\bf B}$ is $\qquad \sim\rho_i
   - \sqrt{\rho_i L_T} \sim 0.1 \, {\rm cm} - 8 \, {\rm cm}$ &
\\ ITG modes 
 & \raggedright Scale parallel to  ${\bf B}$  is $qR\sim 15$ m 
 &  $\omega^*_i \sim 10-100 \, {\rm kHz}$ 
\\ \hline
\raggedright Transport Barriers 
  & \raggedright Unknown scaling of perpendicular scales. Measured 
    scales suggest width $\sim 1-10\, {\rm cm}$ 
  & Lifetime $100$ s or more in core? Relaxation oscillations for edge barrier
    with unknown frequency.
\\ \hline 
\raggedright Magnetic islands,  
  & \raggedright Island width $\sim 10\rho_i \sim 1$ cm. 
  & Growth time $\sim 1-100$ s. $\qquad \qquad $ 
\\
\raggedright Tearing modes and NTMs. 
 & Eigenfunction extent $\sim L_p \sim 100$ cm.  Turbulent correlation
   length near island $\sim 1$ cm?
 & Island frequency $\sim
    100 \, {\rm Hz} - 1 \,{\rm kHz}$. 
    Turbulent frequency near island $\sim 100 \, {\rm kHz}$   
\\ \hline
\end{tabular}
\caption{Some important tokamak space and time scales.  Numerical values refer to ITER.}
\label{multimak1}
\end{table}

\begin{table}[h]
\begin{tabular}{|p{1.2in}|p{2.2in}|p{2in}|}
\hline
{\large\bf Physics}  & {\large\bf Space scale} & {\large\bf Time Scale} \\ \hline
\raggedright Sawteeth 
  & \raggedright Reconnection layer width $\sim 0.05 \, {\rm cm}
    \qquad \qquad$      
  & Crash time $50 \, \mu s - 100\, \mu s$ $\qquad \qquad\qquad$ Real frequency
    $\sim 100\, {\rm Hz} - 1\, {\rm khz}.\qquad$
\\
  & Eigenfunction extent $\sim L_p \sim 100$ cm.$\qquad \qquad $ 
  & Ramp time $1-100$s 
\\ \hline 
Discharge 
  & Profile scales $L_p \sim 100$ cm 
  & Energy confinement time $2-4$s 
\\
Evolution 
 &
 & Burn time unknown
\\ \hline
\end{tabular}
\label{multimak2}
\caption{Some important tokamak space and time scales.  Numerical values refer to ITER.}
\end{table}

\section{Kinetic nature of magnetized plasma turbulence}

In order to address the complexities of plasma turbulence with
existing computer technology, the full kinetic description must be
simplified.  This can be accomplished by exploiting the separation of
time and space scales in fusion plasmas.  In this thesis, we employ
the widely-used $\delta f$ gyrokinetic
model~\cite{antonPoF80,friemanPoF82,howesApJ06}, which takes advantage of the
following scale separations: the turbulence and resultant fluxes are
calculated in a stationary equilibrium, exploiting the separation of
the fast turbulence time scale and the slow profile evolution time
scale; the variation of equilibrium gradient scale lengths
perpendicular to the magnetic field line is ignored (local
assumption), exploiting the separation of the short perpendicular
turbulence scale and the long perpendicular profile scale; and the
dynamics of the turbulence itself is calculated assuming the particles
gyrate about the ambient magnetic field lines infinitely fast,
exploiting the difference in time scales between the dynamics of
interest and a host of much faster processes that occur in magnetized
plasmas.  Furthermore, a distinction is made between fluctuations
along the equilibrium magnetic field, which are assumed to have long
(device size) wavelengths, and cross-field fluctuations, which have
short (Larmor radius) wavelengths.  Finally, the experimentally
observed and theoretically well-founded expectation that the turbulent
correlation lengths in the directions perpendicular to the magnetic
field are small compared to the device dimensions (for large enough
devices, high enough magnetic fields, and suitable distances from edge
boundary layers) allows one to simulate small volumes of plasma
surrounding individual magnetic field lines, called flux tubes, and to
extrapolate the results from these small volumes to nearby flux tubes~\cite{beerPoP95}.
[See \secref{sec:multiscale} for a more detailed discussion.]
This is an assumption of statistical homogeneity among patches of
plasma that are many turbulent correlation lengths apart.  It is a
particularly well-motivated and unsurprising approach for axisymmetric
confinement devices such as tokamaks.  It would be unwise to
ignore this opportunity to reduce the simulation effort,
choosing instead to simulate a large number of statistically identical
regions of plasma, absent an expectation of something such as important intermittent
fluctuations.

\begin{figure}
\centering
\includegraphics[height=5.0in]{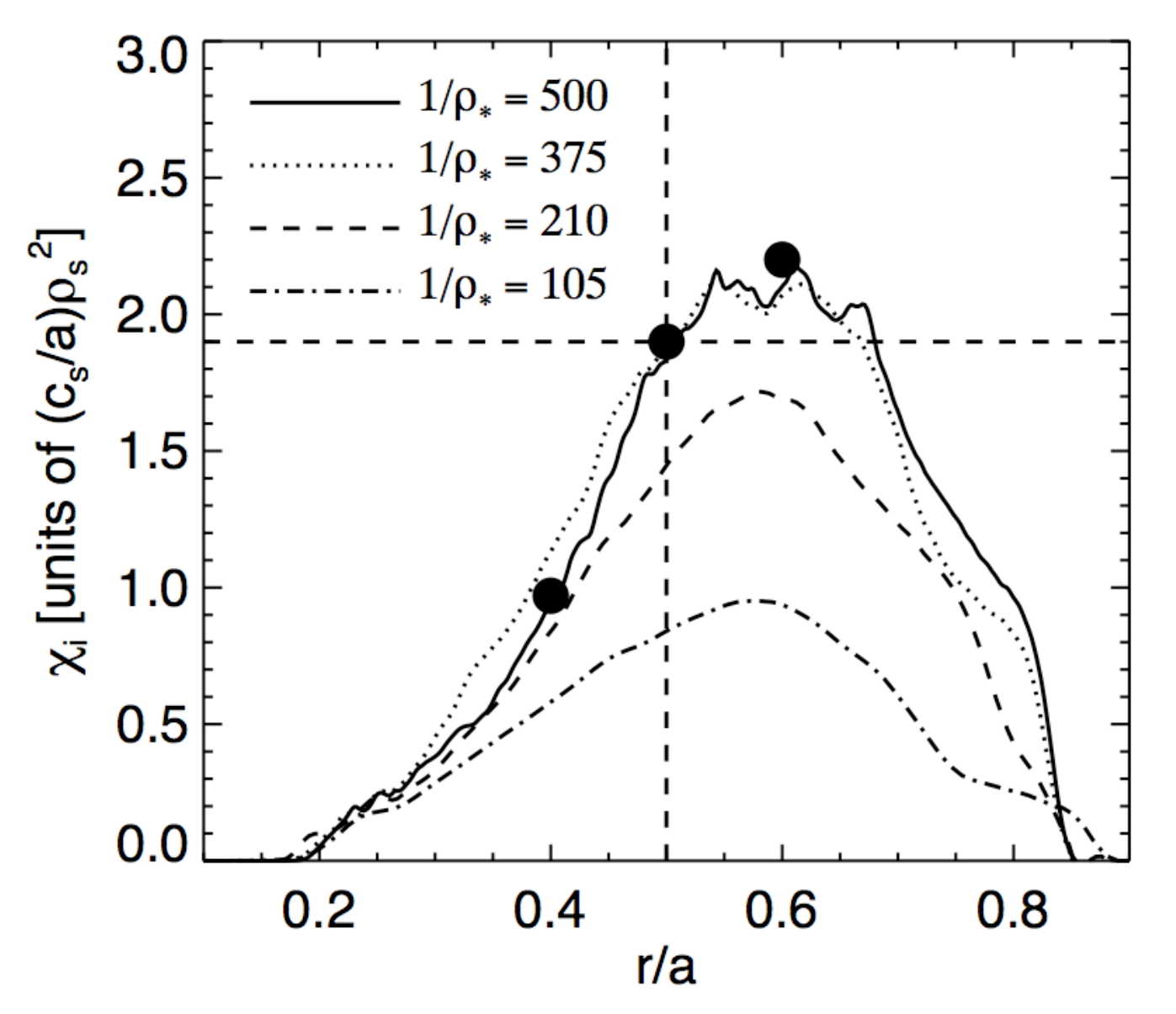}
\caption{Comparison of ion thermal diffusivity $\chi_{i}$
calculated from local (\texttt{GS2}) and global (\texttt{GYRO}) simulations 
as a function of $\rho_{*}\equiv \rho / a$, where $\rho$ is the gyroradius and $a$ is the 
minor radius of the device.  For sufficiently small $\rho_{*}$, the local and global calculations
of thermal diffusivity are in excellent agreement.  Figure taken from Ref.~\cite{candyPoP04}.}
\label{fig:local}
\end{figure}

These assumptions allow for the reduction of the problem from the
long-time evolution of fast, gyroradius-scale turbulence throughout
the full device, to the slow evolution of a few coupled magnetic flux
tubes, each filled with fast, small-scale turbulent fluctuations.  The
fundamental validity of this approach for sufficiently large device
size ($\rho_{*}\sim 0.003$) has been demonstrated~\cite{candyPoP04} by
comparing results from flux tube simulations with results for the same
cases from global simulations (Fig.~\ref{fig:local}), which allow for radial variation of
equilibrium profiles within a turbulence simulation.

\begin{figure}
\centering
\includegraphics[height=1.8in]{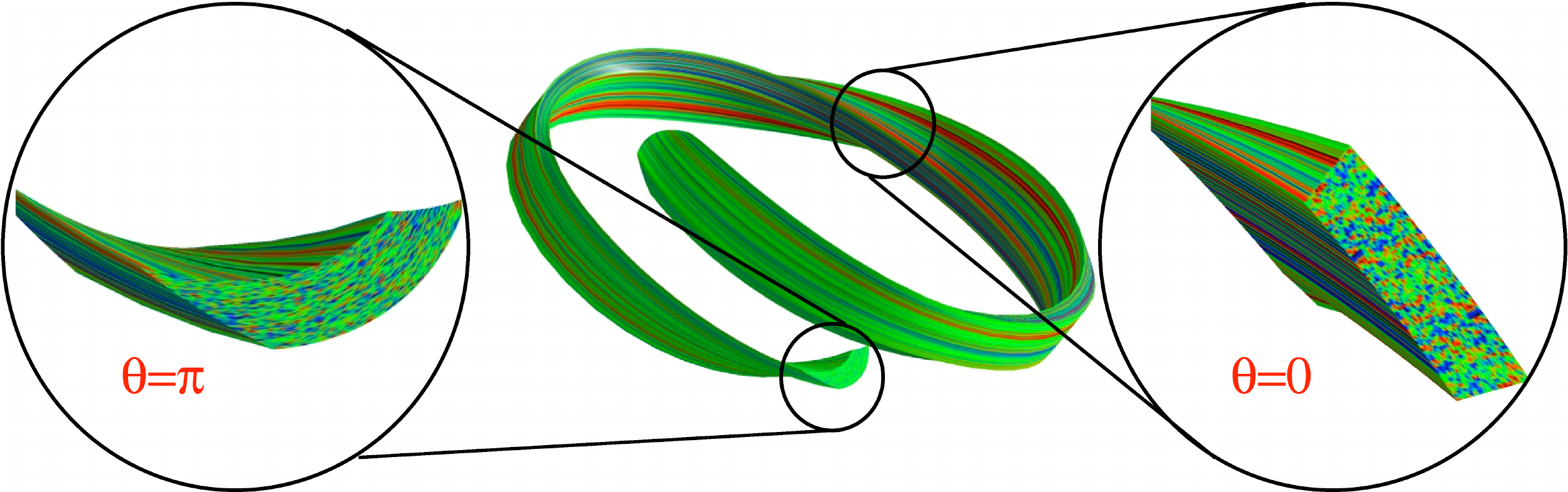}
\caption{Illustration of the flux tube simulation domain used in \texttt{Trinity}.
Colors represent the amplitude of perturbations in the electrostatic potential.  Notice that
the turbulence is long wavelength along the equilibrium magnetic field and short wavelength 
in the plane perpendicular to it.  Graphic courtesy of D. Applegate.}
\label{fig:flxtube}
\end{figure}

\begin{figure}
\centering
\includegraphics[height=2.0in]{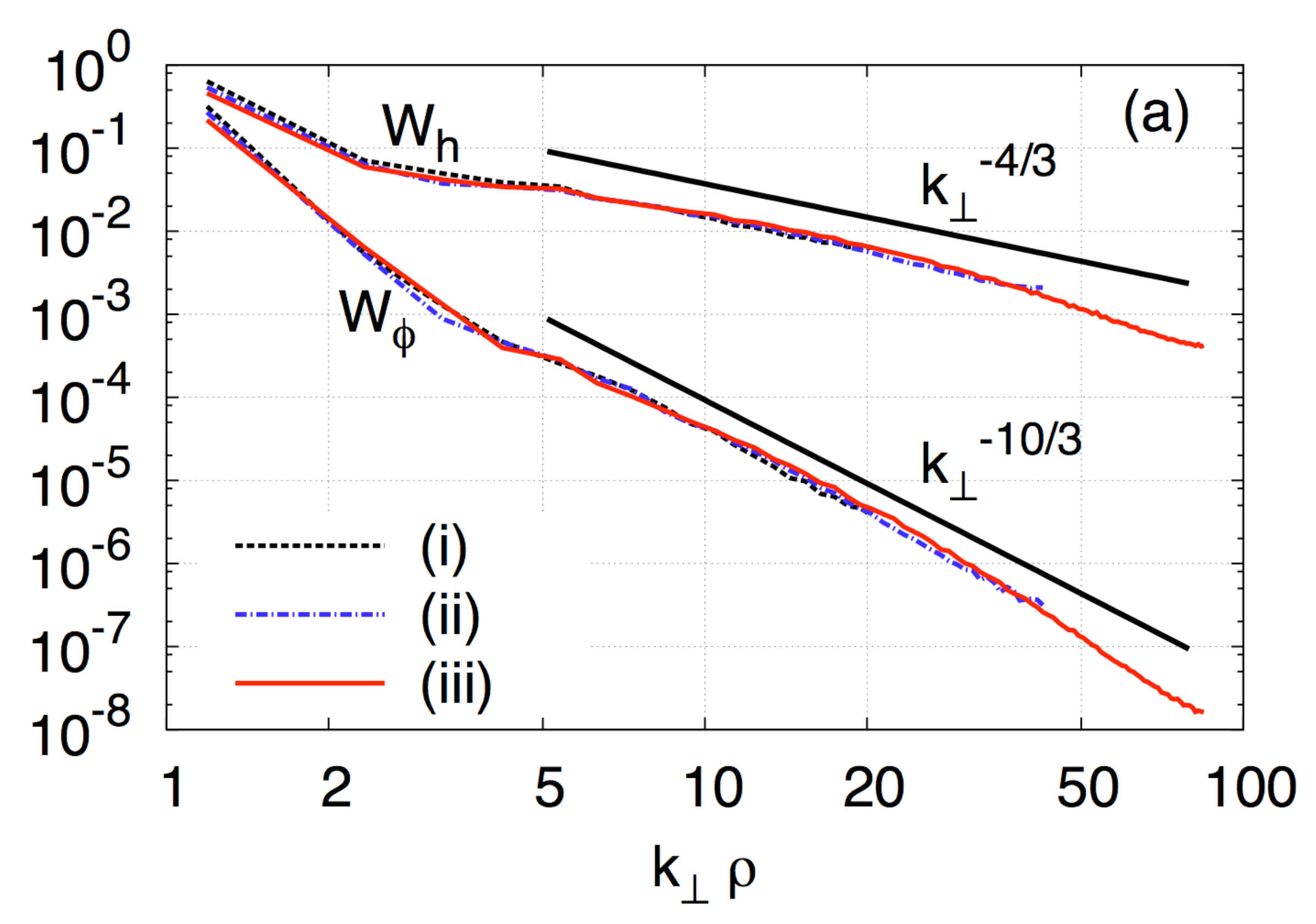}
\includegraphics[height=2.0in]{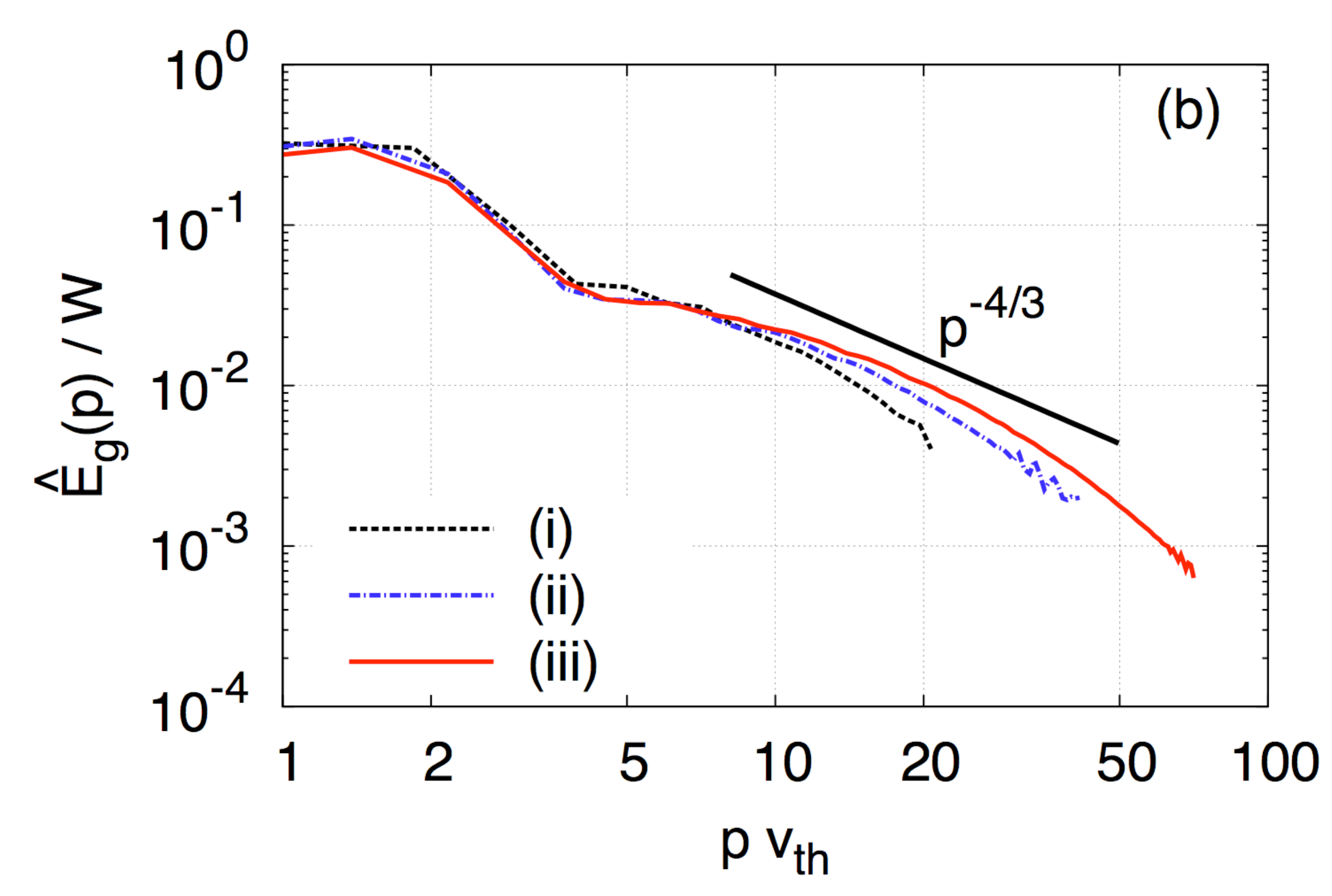}
\caption{(Left): Cascade of entropy from large to small physical space structures 
($W_{\phi}\sim\sum_{\left|\mbf{k}_{\perp}\right|=k_{\perp}} q^{2}n_{0}\left|\Phi_{\mbf{k}}\right|^{2}/2T_{0}$ 
and $W_{h}\sim \sum_{\left|\mbf{k}_{\perp}\right|=k_{\perp}}\int\vvol \ T_{0}\left|h_{\mbf{k}}\right|^{2}/2F_{0}$ 
are the entropy generation arising from the Boltzmann and non-Boltzmann responses of the perturbed 
distribution function, respectively).  Solid black lines are theoretical
predictions~\cite{schekAPJ07}, and colored lines are data taken from 4D ($k_{\parl}=0$),
electrostatic turbulence simulations at different resolutions~\cite{tatsunoPRL08}.  
(Right):  Spectra characterizing the cascade of entropy in velocity space, with 
$\hat{E}_{g}(p)=\sum_{\mbf{k}} p\left|\hat{g}_{\mbf{k}}(p)\right|^{2}$, where
$\hat{g}_{\mbf{k}}(p)=\int\vvol \ J_{0}(pv_{\perp}) g_{\mbf{k}}(\mbf{v})$
is the Hankel transform of the guiding center perturbed distribution function, $g$.  Solid black line
is the theoretical prediction~\cite{plunkJFM08}, and colored lines are data taken from same runs as the
figure on the left.  Figures taken from Ref.~\cite{tatsunoPRL08}.}
\label{fig:spectomo}
\end{figure}

Despite the significant simplifications granted by these gyrokinetic
assumptions, plasma turbulence simulations are still computationally
challenging.  Turbulence in conventional, neutral fluids is already a complex phenomenon;
understanding it has proven to be one of the great scientific
challenges of our time.  Kinetic plasma turbulence, which may be
characterized as particles interacting primarily with electromagnetic
waves and occasionally with one another via collisions, possesses an
additional level of complexity.  For instance, a fundamental concept
in fluid turbulence is the cascade of energy from large-scale to
small-scale spatial structures.  In gyrokinetic turbulence, the
three-dimensional cascade is replaced by a five-dimensional cascade of
entropy from large-scale to small-scale phase space
structures~\cite{dorlandPoP93,schekPPCF08,schekAPJ07,tatsunoPRL08,tatsunoPoP08}.  
This is illustrated in Fig.~\ref{fig:spectomo}.

\begin{figure}
\centering
\includegraphics[height=5.0in]{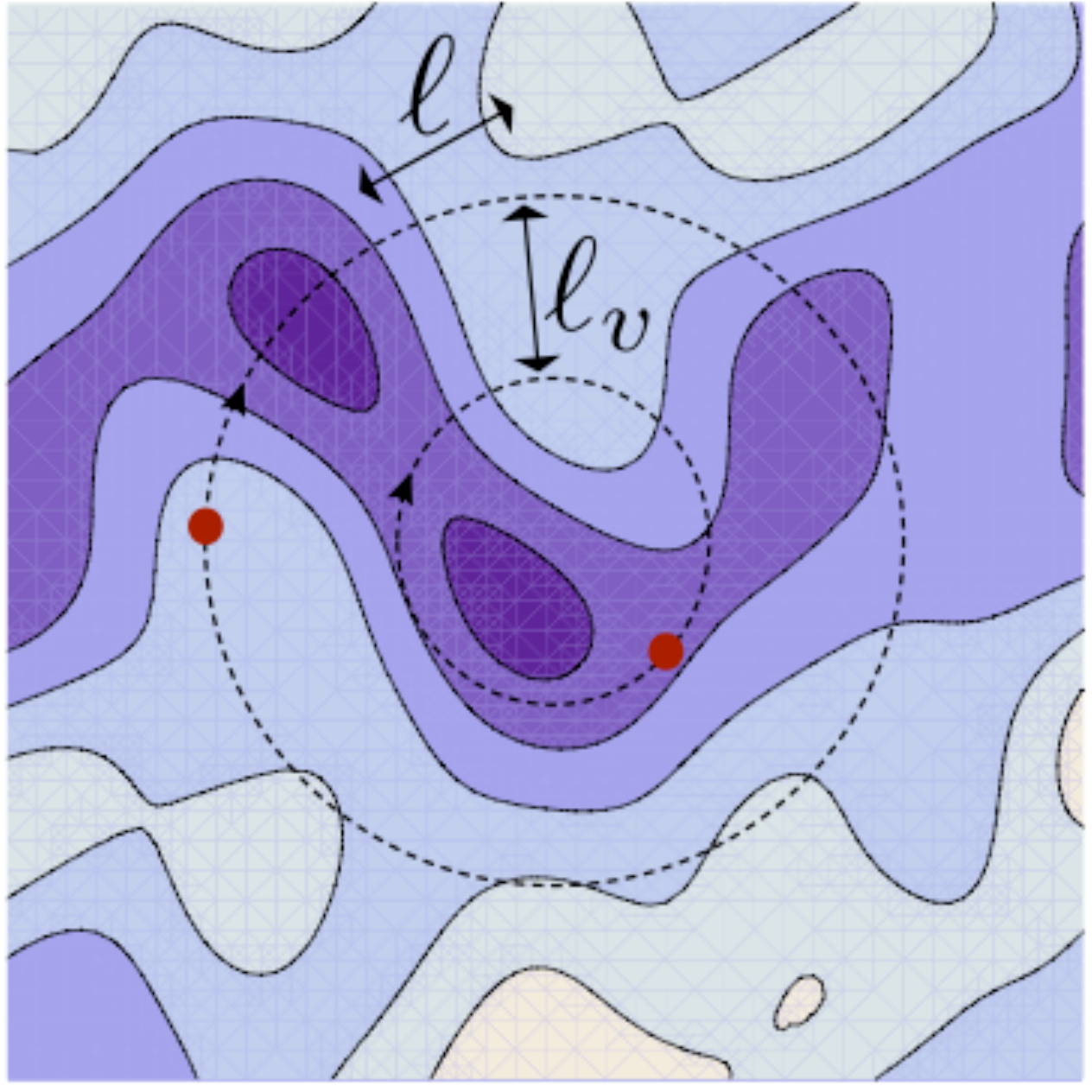}
\caption{Cartoon illustrating nonlinear perpendicular phase mixing, which leads to the
development of small-scale structures in both physical and velocity space.  When the separation
between particle gyroradii with the same guiding center becomes comparable to the characteristic 
wavelength of the turbulence, the motion of the particles become decorrelated.  Since the size of the
gyroradii are proportional to the particles' perpendicular velocities, this indicates a decorrelation in 
velocity space as well, leading to the development of small-scale structure.  Figure taken from Ref.~\cite{plunkJFM08}.}
\label{fig:nlmixing}
\end{figure}

It is well known that in weakly collisional plasmas, Landau and Barnes damping of
electromagnetic fluctuations leads
to the development of small-scale structure in the distribution of particle parallel velocities.
This is a result of mixing in phase space, where particles streaming along magnetic
field lines transfer spatial structure into velocity structure. [The potential development of 
infinitesimally small scales in velocity space is illustrated in \apref{app:landamp},
where we consider the simple case of collisionless Landau damping of the ion acoustic wave.]
In addition to this parallel phase mixing arising from linear convection, there exists a 
perpendicular phase mixing process due to the averaged $E\times B$ particle motion~\cite{dorlandPoP93}.
A cartoon of this process is shown in Fig.~\ref{fig:nlmixing}.  As particles rapidly
gyrate about equilibrium magnetic field lines, they see spatially varying electromagnetic 
fluctuations that are essentially static in time.  The $E\times B$ drift they experience
is thus a result of the gyroaveraged electromagnetic fields they see.  Particles at the same guiding
center position experience different gyroaveraged fields depending on their Larmor radius
(which depends on perpendicular particle velocities) and thus drift with different guiding
center velocities.  This results in a mixing of particles with different perpendicular
velocities in the gyrocenter distribution function and the generation of small scales in the perpendicular velocity 
space.

Because of the tendency of weakly-collisional plasmas to develop fine structures in velocity
space, one must pay careful attention to numerical resolution in velocities.  This is ideally
done by conducting a grid convergence study in velocity space.  However, this is a numerically
expensive process, so it is not always done.  We have developed computationally cheap 
velocity space resolution diagnostics that allow us to monitor resolution at virtually no additional
cost.  When coupled with an adaptive collisionality, we are able to confidently simulate
velocity space dynamics with the approximate minimal number of grid points in velocity space
necessary for resolution.  This is detailed in Chapter 4.

\section{Turbulent heating and the importance of collisions}

The evolution of equilibrium pressure profiles is determined by a balance between transport
processes and local heating.  The net heating consists of contributions from a number of 
sources, including external heating, atomic heating, Ohmic heating, and thermal energy
exchange between species.  Most of these phenomena have been extensively studied,
both analytically and through the use of numerical transport solvers.  However, little
attention has been given to anomalous heating arising from microturbulence.
The gyrokinetic turbulent heating of species $s$ can be defined in various ways.  In 
Refs.~\cite{hintonPoP06} and~\cite{waltzPoP08} it is defined to be
\begin{equation}
\tld{\mathcal{H}}\equiv\int d^{3}\mbf{r} \left(\delta \mbf{J}_{\parl}\cdot\delta\mbf{E}_{\parl}+
\delta\mbf{J}_{D}\cdot\delta\mbf{E}_{\perp}\right),
\label{eqn:gkheat}
\end{equation}
where $\delta\mbf{J}_{\parl}$ is the perturbed parallel current,
$\delta\mbf{J}_{D}$ is the current perturbation due to particle
drifts, and $\delta \mbf{E}_{\parl}$ and $\delta \mbf{E}_{\perp}$ are
the parallel and perpendicular perturbed electric fields,
respectively.  In Chapter 3, we derive an equation for the evolution of the equilibrium
pressure that leads us to a somewhat different definition for the turbulent heating.
We show, however, that both definitions lead to a net (species-summed) turbulent heating
of zero.

While the net turbulent heating is zero, the turbulent heating for each species (or 
equivalently, the turbulent energy exchange between species) is not necessarily zero.  It is 
formally the same order as the heat transport, so there is a possibility of significant turbulent 
energy exchange between species.  In the cases considered in
Ref.~\cite{waltzPoP08}, it was found that the parallel and
perpendicular contributions to the turbulent heating nearly cancel,
giving only a $10\%$ adjustment to the net heating.  However, to our
knowledge, no additional cases have been considered, and the definition used for the turbulent
heating does not contain all of the turbulent heating terms appearing in the equations we
derive in Chapter 3 for the time evolution of the equilibrium pressure.
Turbulent heating therefore deserves further careful study.

In Chapter 3, we also express the turbulent heating as the sum of a
positive-definite quantity describing collisional entropy generation
and a term representing energy exchange between the equilibrium and
the turbulence.  Because the collisional entropy generation term is
positive definite, it is generally easier to obtain a converged
statistical average for it in numerical simulations than for the
$\delta\mbf{J}\cdot\delta\mbf{E}$ terms, which tend to have large
amplitude oscillations associated with particles
``sloshing'' back and forth in plasma waves.  To calculate the
collisional entropy generation, we have developed a model gyrokinetic collision operator
that retains the key properties of physical collisions.

In general, collisional physics are not carefully treated (if treated at all) in gyrokinetic
simulations of turbulence~\cite{abelPoP08,barnesPoP08}.  In principle, one would
like to include the full linearized Landau operator~\cite{landauPZS36}, but numerical implementation
and calculation of the so-called ``field-particle'' part of this operator is quite challenging.
Approximate models for the collision operator have been derived~\cite{hirshmanPoF76,cattoPoF76},
but they do not possess all of the properties one would like a collision operator to possess~\cite{abelPoP08}.
Consequently, we have derived a new model collision operator for use in gyrokinetics, which
is an improvement over previous operators.  Some of the key properties of our operator are:
local conservation of particle number, momentum, and energy; satisfaction of Boltzmann's
$H$-Theorem; efficient smoothing in velocity space; and reduction to the full linearized
Landau operator in the short wavelength limit (where dissipation primarily occurs).  The
derivation of this operator is presented in Chapter 5, and numerical implementation in \verb#Trinity#
and tests are presented in Chapter 6.


\section{Stiff transport}

\begin{figure}
\centering
\includegraphics[height=4.0in]{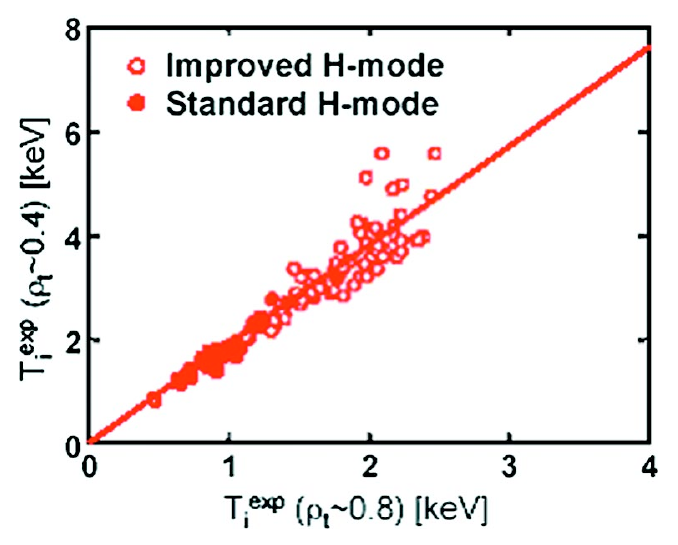}
\caption{Plot of the core temperature as a function of the edge temperature in the
High-confinement mode of operation (H-mode) on ASDEX-U.  Note the linear scaling,
which indicates that the temperature gradient scale length across the device is fixed 
(at the critical gradient) and independent of temperature.  The tendency of profile gradients
to stay near the critical gradient implies a stiff dependence of the heat flux on equilibrium gradients.
Figure taken from Ref.~\cite{gruberPoP05}.}
\label{fig:stiff}
\end{figure}

Kinetic microinstabilities depend on a large number of plasma
parameters.  However, the dominant microinstabilities in most magnetic
confinement fusion devices are driven unstable primarily by
sufficiently strong temperature gradients.  Since these
microinstabilites cause high levels of heat transport, they
effectively limit temperature gradients to be at or below the critical
gradient at which the kinetic modes go unstable (unless the
temperature near the edge of the plasma is low or the external heating
is very
large)~\cite{dorlandIAEA94,kotschPoP95,dorlandIAEA00}.  Below
the critical gradient, there is a low level of transport due to
neoclassical effects (see e.g. Ref.~\cite{hintonRMP76}) that has a
relatively weak dependence on temperature gradient scale length.
Above the critical gradient, the level of transport increases
dramatically because turbulent transport has a stiff dependence on
temperature gradient scale length.  For relatively high temperature
plasmas with reasonable external heating power, this feature (stiff
transport) leads to profiles adjusting so that their gradients are
stuck at the critical gradient.  Consequently, the core temperature
depends sensitively on the temperature at the edge of the device
(Fig.~\ref{fig:stiff}).  Without high edge temperatures, ITER will not
likely achieve its target core temperature, for instance~\cite{dimitsPoP00}. The edge
plasma is not modeled in this thesis because of the complicated
physics involved and because sharp gradients occur near the edge of
the device (in what is known as the edge pedestal), challenging the
applicability of the gyrokinetic ordering we consider.

This stiff dependence of the fluxes on the driving gradients and the sharp transition between 
neoclassical and turbulent transport at the critical gradient have another unfortunate 
consequence: they make turbulent transport simulations very challenging.  Stiff systems
are notoriously difficult to address numerically because the sensitivity of the equations to 
small perturbations can lead to extreme restrictions on the time step size.  In order to avoid (or at least
limit) these restrictions, one should treat the transport equations implicitly.  Developing
such an implicit scheme is a nontrivial
problem since the transport is described by a set of coupled, nonlinear partial differential equations.  However,
implicit techniques for nonlinear equations, such as Newton's method, have 
successfully been applied to plasma transport equations with model fluxes~\cite{jardinJCP08}.
We derive an implicit technique for solving the plasma transport equations with
nonlinear, gyrokinetic fluxes in Chapter 7.

\section{Multiscale simulations of turbulent transport and heating}
\label{sec:multiscale}

\begin{figure}
\centering
\includegraphics[height=4.0in]{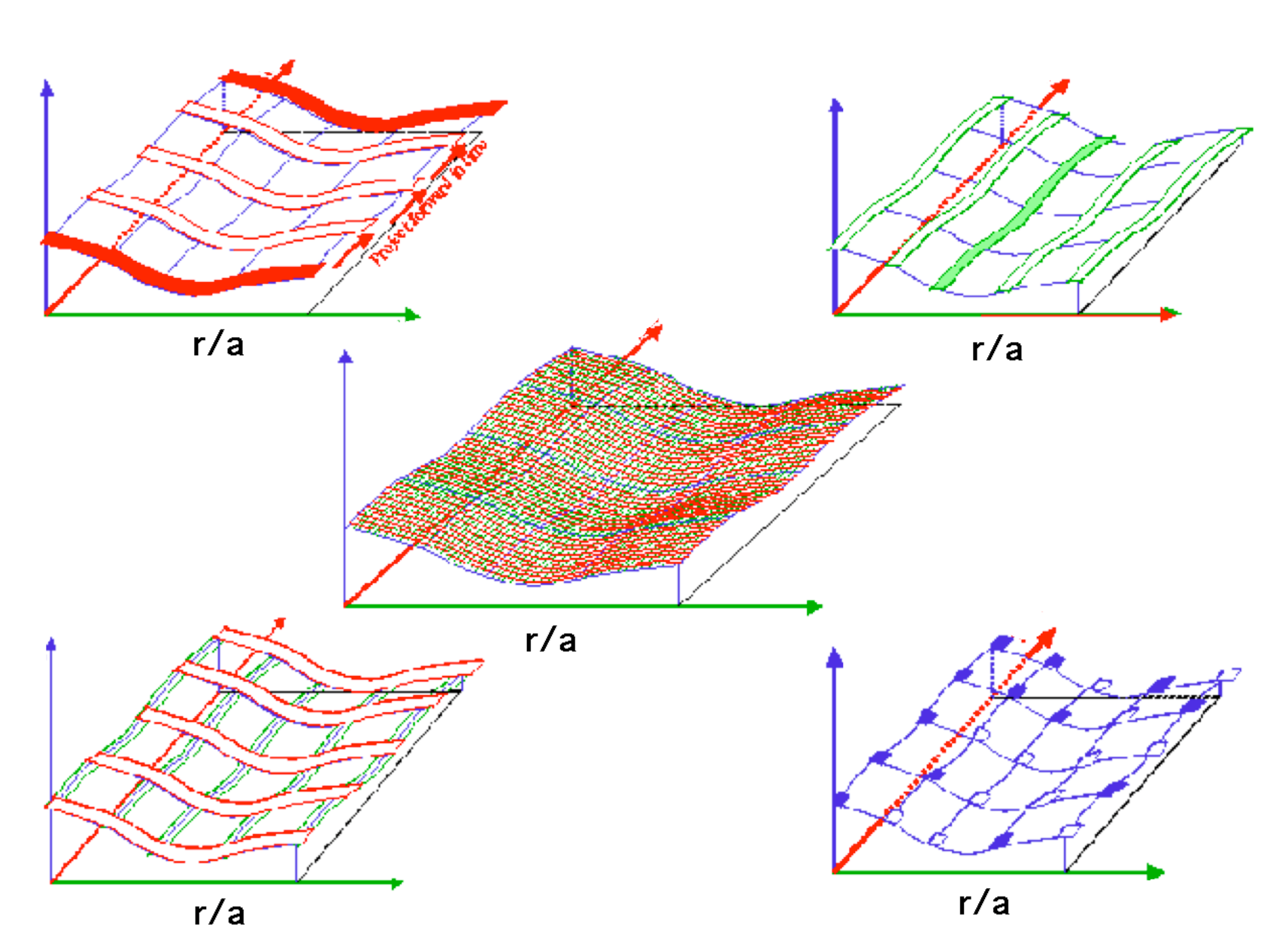}
\caption{(Center): Fine scale grid in space and time. (Top left): Coarse equilibrium grid spacing in time,
with regions of fine grid spacing embedded.  Each horizontal red strip represents simulation of turbulent dynamics
to steady-state, keeping equilibrium quantities constant.  (Top right): Coarse equilibrium grid spacing in radius,
with regions of fine grid spacing embedded.  Each vertical green strip represents simulation of turbulent dynamics
in a narrow flux tube, assuming no radial variation of equilibrium profiles or gradients across the domain.
(Bottom left): Combination of the multiscale space and time grids.  (Bottom right): Small blue squares
are the simulation domain resulting from the multiscale mesh in space and time.}
\label{fig:msgrid}
\end{figure}

\begin{figure}
\centering
\includegraphics[height=2.0in]{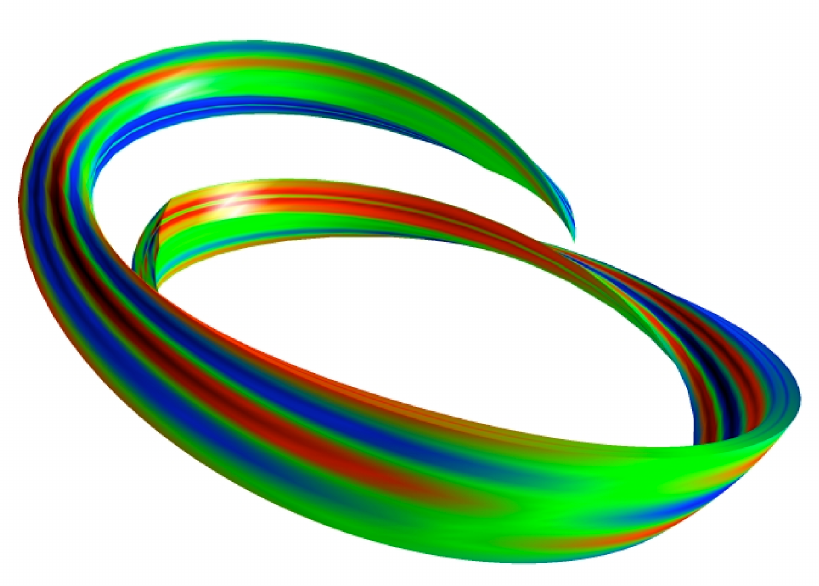}
\includegraphics[height=2.0in]{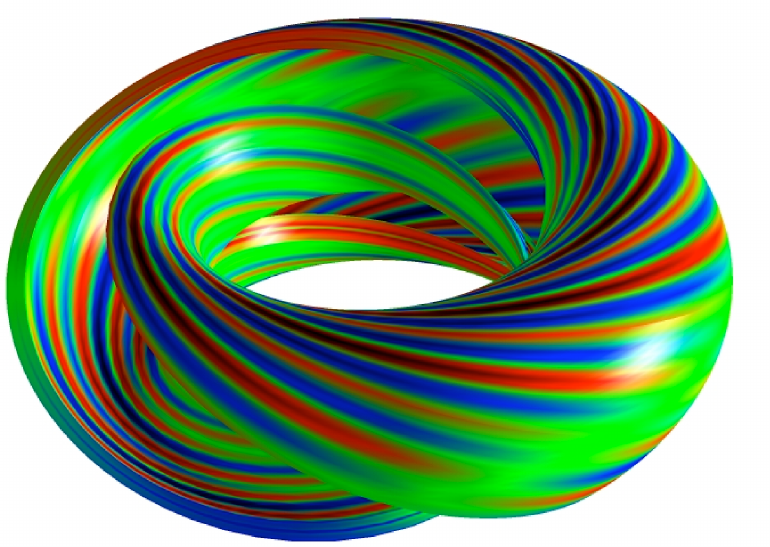}
\includegraphics[height=2.0in]{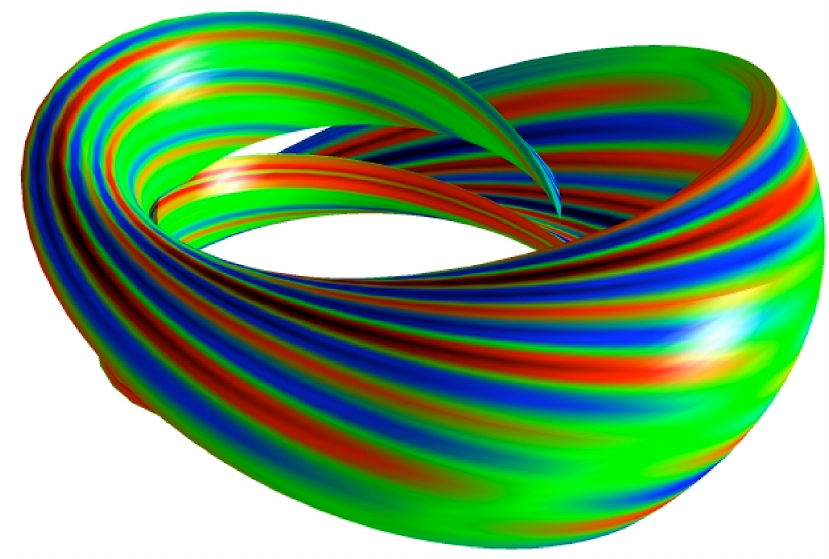}
\includegraphics[height=2.0in]{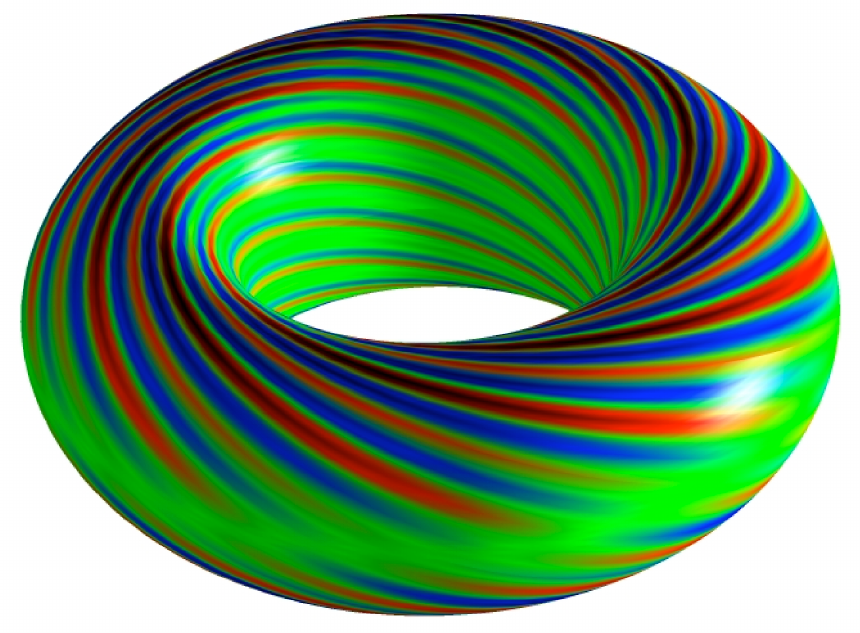}
\caption{Illustration of flux tubes from \texttt{Trinity} simulations.  Using statistical periodicity
of the turbulence, a single flux tube (top left) several decorellation lengths long can be used
to map an entire flux surface (3 flux tubes at top right, 6 at bottom left, and 8 at bottom right).
Colors represent the amplitude of perturbations in the electrostatic potential.  Graphics
courtesy of D. Applegate.}
\label{fig:flxtubes}
\end{figure}

Assuming no intermediate time or space scales are present, a direct numerical simulation 
resolving fine (turbulence) time and space scales throughout the volume of a fusion device 
for an entire discharge is not necessary.  Instead, one can use the separation of scales
embodied in the gyrokinetic turbulence and transport equations derived in Chapter~\ref{chap:hierarchy}
to embed small regions of fine grid spacing in a coarse, equilibrium-scale mesh (Fig.~\ref{fig:msgrid}).
We adopt this approach by calculating turbulent fluxes and heating in a series of flux tubes,
each of which is used to map out an entire magnetic flux surface (Fig.~\ref{fig:flxtubes}).
These flux surfaces are coupled together as radial grid points in the one-dimensional
equations describing the evolution of radial profiles of equilibrium density and pressure.

The computational savings from using our multiscale scheme can be quite large.  The use
of field line-following coordinates decreases the number of grid points necessary along the
equilibrium magnetic field since parallel turbulence wavelengths are much longer than
perpendicular wavelengths.  The use of a flux tube simulation domain 
to map out an entire flux surface decreases the number of grid points 
necessary in the direction perpendicular to the field (but lying near the flux surface).  
Although the radial domain covered by a series of coupled flux tubes is comparable to
the domain of a conventional global approach, the spacing of the radial grid points is
more optimal.  This is because the range of wavenumbers (or equivalently, the grid spacing)
necessary to resolve the turbulent fluctuations varies across the large-scale radial profile due to variations in 
density, temperature, and magnetic geometry.  Each flux tube is naturally able to simulate
a range of wavenumbers independent of the other flux tubes, constituting an adaptive radial
grid.  Finally, evolution of the turbulence and transport on separate time scales using the 
gyrokinetic hierarchy of Chapter 3 allows for simulation of the entire discharge while
sampling only a fraction of the total discharge time.  
[Note that the algorithms derived and implemented here can be used to simulate the 
time-dependent evolution of the equilibrium, even for ``fast'' phenomena, such as 
heat and cold pulses; steady-state transport is not assumed.]  
Taking into account all of these contributions, the rough savings estimate given for ITER in 
Chapter 7 is a factor on the order of $10^{10}$.  These savings can be used to 
include additional physics, such as coupled electron-ion dynamics, electromagnetic 
fluctuations, multiple ion species, {\it etc.,} in each flux tube.  
Furthermore, they place coupled turbulence, transport, and heating calculations 
within reach on current computing resources.

At the time of this writing, it is possible to obtain millions of
CPU-hours on parallel computers with $\mathcal{O}(10^5)$ processors.  For a
global simulation of a steady-state ITER core plasma, \texttt{Trinity}
might require 16 flux tubes, each running turbulence simulations
requiring $\sim 4000$ processors -- enabling multispecies,
electromagnetic turbulence simulations in each flux tube, for example.
The algorithm derived below can spawn 2-4 copies of each flux tube
simultaneously to estimate the fluxes and their main dependencies; the
precise number can be determined at run time to match the available
resources.  Assuming 2 copies of each of the 16 flux tubes, each
running on 4000 processors, such a simulation would utilize 128,000
cores, with nearly perfect linear scaling, and should run to
completion in a few hours.  Thus, the algorithms presented here will
allow routine simulations to study a range of physical conditions and
magnetic configurations on existing computers, not just an annual ``stunt run''
with limited physics content and limited scientific value.

\renewcommand{\thechapter}{2}

\chapter{\textbf{Microstability}}
\label{chap:linzp}
\vspace{+90pt}

\section{Introduction}

Kinetic theory is complicated, but sometimes necessary.  In the hot, magnetized plasmas of 
magnetic confinement fusion experiments, the collisional mean free path can be many 
kilometers -- distances much greater than the device size.  This leads to the development
of nontrivial structure in the distribution of particle velocities, as we discuss in detail in
Chapter~\ref{chap:vspaceres}.  Conventional fluid models do not 
accurately describe drift-type instabilities under these 
circumstances~\cite{hassamPFB90}.  Because drift instabilities typically induce strong 
energy transport when the driving gradient is pushed beyond the threshold of the given 
instability, knowledge of the threshold criterion is key to the interpretation of much 
experimental data.  

In this chapter we illustrate the necessity of a kinetic treatment for the instabilities
leading to small-scale plasma turbulence.  We do so by calculating a kinetic stability
threshold and comparing with stability thresholds from various fluid 
theories~\cite{kesnerPoP00,simakovPoP01,simakovPoP02,kesnerPoP02}.  What we will 
find is that fluid theory 
significantly underestimates the range of instability~\cite{ricciPoP06}.

As our example system, we choose to consider the entropy mode~\cite{kadomtsevSP60} in 
a Z-pinch magnetic field configuration~\cite{freidbergMHD}.
This configuration consists of a current
running through the plasma in the $\unit{z}$ direction, generating
a radially varying equilibrium magnetic field in the $\unit{\phi}$ direction.  
Here, we are using cyclindrical coordinates, i.e. $(R,\phi,z)$.  For strong pressure gradients,
the plasma is unstable to magnetohydrodynamic (MHD) instabilities with fast growth rates~\cite{kesnerPoP00,coppinsPFB89}.
If the pressure gradient is sufficiently weak, the plasma is stable to MHD instabilities, but 
potentially unstable to the entropy mode.  To demonstrate the importance of the kinetic 
approach, we calculate the stability threshold of the low-$\beta$ (electrostatic) entropy 
mode and compare with the results obtained from a number of fluid theories.

\section{Linear stability analysis}

For our linear stability analysis, we will be working within the framework of $\delta f$
gyrokinetics, which is described in detail in Chapter~\ref{chap:hierarchy}.  The distribution 
function $f$ for species $s$ is given by
\begin{equation}
f_{s} = F_{0s} + h_{s} + f_{2s},
\end{equation}
where $h_{s}$ is the non-Boltamann part of the lowest order perturbed distribution function,
$f_{2s}$ contains higher order terms, and $F_{0s}=F_{Ms}\left(1-q_{s}\Phi/T_{0s}\right)$, 
with $F_{M}$ a Maxwellian, $q_{s}$ the particle charge, $\Phi$ the electrostatic potential, and $T_{0s}$
the equilibrium temperature.  With these definitions, the electrostatic version of the 
linear, collisionless gyrokinetic equation is
\begin{equation}
\pd{h_{s}}{t} + \vpa\unit{b}\cdot\nabla h_{s} + \left<\mbf{v}_{E}\right>_{\mbf{R}}\cdot\nabla F_{0s} + \mbf{v}_{B}\cdot\nabla h_{s} = \frac{q_{s}F_{0s}}{T_{0s}}\pd{\left<\Phi\right>_{\mbf{R}}}{t},
\label{eqn:gkeqes}
\end{equation}
where $\unit{b}\equiv \mbf{B}_{0}/B_{0}$ is the unit vector in the direction of the
equilibrium magnetic field, $\mbf{B}_{0}$, 
\begin{equation}
\mbf{v}_{E}\equiv \frac{c}{B_{0}}\unit{b}\times\nabla\Phi
\end{equation}
is the $E\times B$ velocity,
\begin{equation}
\mbf{v}_{B}\equiv \frac{\unit{b}}{\Omega_{0}}\times\left(\vpa^{2}+\frac{\vpe^{2}}{2}\right)\frac{\nabla B_{0}}{B_{0}}-\frac{4\pi}{cB_{0}\Omega_{0}}\vpa^{2}\mbf{J}_{\perp}
\end{equation}
is the sum of the curvature and $\nabla B$ drift velocities, $\Omega_{0}=qB_{0}/mc$ is
the particle gyrofrequency, $\mbf{J}_{\perp}$ is the equilibrium perpendicular current,
and the angled brackets $\left< . \right>_{\mbf{R}}$ denote an average
over gyroangle at fixed guiding center position $\mbf{R}.$

To proceed, we use the form of the equilibrium magnetic field, $\mbf{B}_{0}=B_{0}(r)\unit{\phi}$,
to compute $\mbf{J}_{\perp}$ and to determine an MHD equilibrium condition.  After 
some algebra, we find
\begin{gather}
\frac{4\pi}{c}\mbf{J}_{\perp} = \frac{B_{0}}{R}\left(1+\frac{R}{B_{0}}\pd{B_{0}}{R}\right)\\
\frac{\nabla B_{0}}{B_{0}} = -\frac{1}{R}\left(1-\beta\frac{R}{2L_{p}}\right)\unit{R},
\end{gather}
where 
$\beta=8\pi p_{0} / B_{0}^{2}$ is the plasma beta, $p_{0}$ is the equilibrium pressure, and
$L_{p}^{-1}=-\partial \ln p_{0}/\partial r$ is the inverse pressure gradient scale length.  In the low
$\beta$ limit, we find $\partial B_{0}/\partial R \approx -B_{0}/R$, giving $\mbf{J}_{\perp}\approx0$
and
\begin{equation}
\mbf{v}_{B} \approx \frac{1}{R\Omega_{0}}\left(\vpa^{2}+\frac{\vpe^{2}}{2}\right)\unit{z}.
\end{equation}

We now return to the linear gyrokinetic equation (\ref{eqn:gkeqes}).  For simplicity, we take the
ion and electron temperature gradients to be zero.  Assuming perturbed quantities are of the form 
$h=\tld{h}\exp[ik_{z}z-i\omega t]$, we obtain an algebraic equation for $h$:
\begin{equation}
\left[\omega-\frac{k_{z}}{R\Omega_{0s}}\left(\vpa^{2}+\frac{\vpe^{2}}{2}\right)\right]\tld{h}_{s}=\frac{q\left<\tld{\Phi}\right>_{\mbf{R}}}{T_{0s}}\left(\omega-\omega_{*s}\right)F_{0s},
\end{equation}
where
\begin{equation}
\omega_{*s} = \frac{c k_{z} T_{0s}}{q_{s} B_{0} L_{n}}
\end{equation}
is the diamagnetic drift frequency.
Defining the normalized quantities $\omega_{Ns}\equiv \omega/\omega_{*s}$ and $x\equiv v/v_{th,s}$
and solving for $h_{s}$, we have
\begin{equation}
\tld{h}_{s} = \frac{q\left<\tld{\Phi}\right>_{\mbf{R}}}{T_{0s}}\frac{\omega_{Ns}-Sgn[q_{s}]T_{0s}/T_{0e}}{\omega_{Ns}-Sgn[q_{s}]\la L_{n}/R \ra \left(x_{\parl s}^{2}+x_{\perp s}^{2}/2\right)T_{0s}/T_{0e}}F_{0s}.
\label{eqn:linh}
\end{equation}
We currently have an additional unkown: $\Phi$.  We can obtain an expression for $\Phi$
by using Poisson's equation and asserting quasineutrality (i.e. $\sum_{s}q_{s}n_{s}=0$).
In terms of $h$ and $\Phi$, quasineutrality gives
\begin{equation}
\sum_{s}q_{s}\int \vvol\left(\left< h_{s} \right>_{\mbf{r}} - \frac{q_{s}\Phi}{T_{0s}}F_{Ms}\right) = 0,
\label{eqn:linqn}
\end{equation}
where $\left< . \right>_{\mbf{r}}$ represents a gyroaverage at constant particle position $\mbf{r}$.

Substituting Eq.~(\ref{eqn:linh}) into Eq.~(\ref{eqn:linqn}), we obtain
\begin{equation}
\sum_{s}\frac{q_{s}^{2}\Phi}{T_{0s}}\int\vvol \left(J_{0}(a_{s})^{2}\frac{\omega_{Ns}-Sgn[q_{s}]T_{0s}/T_{0e}}{\omega_{Ns}-Sgn[q_{s}]\la L_{n}/R \ra \left(x_{\parl s}^{2}+x_{\perp s}^{2}/2\right)T_{0s}/T_{0e}}-1\right)F_{Ms}=0,
\label{eqn:linqnh}
\end{equation}
where $J_{0}$ is a Bessel function of the first kind, and $a=k_{z}v_{\perp}/\Omega_{0}$.
The velocity integration in Eq.~(\ref{eqn:linqnh}) is nontrivial.  To simplify the analysis,
we focus on the limit in which $k_{\perp}\rho_{s}\ll 1$ (the drift-kinetic limit).  In this 
case, $J_{0}(a)\approx 1$.  We are then interested in evaluating an integral of the form
\begin{equation}
I \equiv \int \vvol \frac{\exp[-x^{2}]}{\omega_{N}-Sgn[q_{s}]\xi\left(x_{\parl}^{2}+x_{\perp}^{2}/2\right)},
\end{equation}
where $\xi\equiv |L_{n}/R|(T_{0s}/T_{0e})$.  Defining $\tld{\omega}\equiv \omega_{N}/\xi + Sgn[q_{s}]x_{\perp}^{2}/2$
and $I_{N}\equiv -I\xi/2\pi v_{th,s}^{3}$, and using cylindrical velocity space coordinates, our integral 
takes the form
\begin{equation}
I_{N} = \int_{0}^{\infty}dx_{\perp} \ x_{\perp}\exp[-x_{\perp}^{2}]\int_{-\infty}^{\infty} dx_{\parl}\frac{\exp[-x_{\parl}^{2}]}{\tld{\omega}+Sgn[q_{s}]x_{\parl}^{2}}.
\label{eqn:in}
\end{equation}

Focusing on the $x_{\parl}$ integral, we take an aside and consider
\begin{equation}
\tld{I}\equiv\int_{-\infty}^{\infty}dx_{\parl}\frac{\exp[-\alpha x_{\parl}^{2}]}{\tld{\omega}+Sgn[q_{s}]x_{\parl}^{2}},
\label{eqn:itld}
\end{equation}
where $\alpha$ is a velocity-independent parameter.  Differentiating $\tld{I}$ with respect
to $\alpha$ gives
\begin{equation}
\frac{d\tld{I}}{d\alpha}=-\int_{-\infty}^{\infty}dx_{\parl}\ x_{\parl}^{2}\frac{\exp[-\alpha x_{\parl}^{2}]}{\tld{\omega}+Sgn[q_{s}]x_{\parl}^{2}}.
\end{equation}
Now we integrate this expression by parts:
\begin{equation}
\begin{split}
\frac{d\tld{I}}{d\alpha} &= Sgn[q_{s}]\left(-\int_{-\infty}^{\infty}dx_{\parl}\exp[-\alpha x_{\parl}^{2}]+\tld{\omega}\int_{-\infty}^{\infty}dx_{\parl}\frac{\exp[-\alpha x_{\parl}^{2}]}{\tld{\omega}+Sgn[q_{s}]x_{\parl}^{2}}\right)\\
&=Sgn[q_{s}]\left(\tld{\omega}\tld{I}-\sqrt{\frac{\pi}{\alpha}}\right).
\end{split}
\end{equation}
This differential equation has two distinct solutions depending on the value of $Sgn[q_{s}]$.
They are
\begin{gather}
\tld{I}_{-} = \exp[-\alpha\tld{\omega}]\left(C_{1}+\pi\frac{\erfi\left[\sqrt{\alpha\tld{\omega}}\right]}{\sqrt{\tld{\omega}}}\right)
\label{eqn:itldm}
\\
\tld{I}_{+} = \exp[\alpha\tld{\omega}]\left(C_{2}-\pi\frac{\erf\left[\sqrt{\alpha\tld{\omega}}\right]}{\sqrt{\tld{\omega}}}\right),
\label{eqn:itldp}
\end{gather}
where the subscript on $\tld{I}$ denotes the sign of $q_{s}$, $\erf$ is the error function, 
$\erfi$ is the imaginary error function, and $C_{1}$ and $C_{2}$ are unspecified constants.

In order to determine $C_{1}$ and $C_{2}$, we must go back to Eq.~(\ref{eqn:itld}),
set $\alpha=0$, and perform the resulting integral.  We find
\begin{equation}
\begin{split}
\tld{I}(\alpha=0) &= \int_{-\infty}^{\infty}dx_{\parl}\left(\tld{\omega}+Sgn[q_{s}]x_{\parl}^{2}\right)^{-1}\\
&=\pi\sqrt{\frac{Sgn[q_{s}]^{3}}{\tld{\omega}}}.
\end{split}
\end{equation}
Once again, this splits into two solutions depending on the value of $Sgn[q_{s}]$:
\begin{gather}
\tld{I}_{-}(\alpha=0)=Sgn[Im[\omega_{N}]] i \frac{\pi}{\sqrt{\tld{\omega}}}\\
\tld{I}_{+}(\alpha=0)=\frac{\pi}{\sqrt{\tld{\omega}}},
\end{gather}
where $Im[\omega_{N}]$ denotes the imaginary part of $\omega_{N}$.  Applying
these results to Eqs.~(\ref{eqn:itldm}) and~(\ref{eqn:itldp}), we obtain the following
expressions for $C_{1}$ and $C_{2}$:
\begin{gather}
C_{1}=Sgn[Im[\omega_{N}]]i\frac{\pi}{\sqrt{\tld{\omega}}}\\
C_{2}=\frac{\pi}{\sqrt{\tld{\omega}}}.
\end{gather}
The solutions for $\tld{I}_{-}$ and $\tld{I}_{+}$ are then
\begin{gather}
\tld{I}_{-}=\exp[-\alpha\tld{\omega}]\frac{\pi}{\sqrt{\tld{\omega}}}\left(Sgn[Im[\omega_{N}]]i + \erfi[\sqrt{\alpha\tld{\omega}}]\right)\\
\tld{I}_{+}=\exp[\alpha\tld{\omega}]\frac{\pi}{\sqrt{\tld{\omega}}}\erfc[\sqrt{\alpha\tld{\omega}}],
\end{gather}
where $\erfc$ is the complementary error function.  To get the parallel integral in $I_{N}$ [Eq.~(\ref{eqn:in})], we
simply take the limit of $\tld{I}$ as $\alpha\rightarrow 1$.  We then obtain the following:
\begin{gather}
I_{N,-} = \int_{0}^{\infty}dx_{\perp} \ x_{\perp}\exp[-x_{\perp}^{2}]\left[\exp[-\tld{\omega}]\frac{\pi}{\sqrt{\tld{\omega}}}\left(Sgn[Im[\omega_{N}]]i+\erfi[\sqrt{\tld{\omega}}]\right)\right]
\label{eqn:inm}
\\
I_{N,+}=\int_{0}^{\infty}dx_{\perp} \ x_{\perp}\exp[-x_{\perp}^{2}]\exp[\tld{\omega}]\frac{\pi}{\sqrt{\tld{\omega}}}\erfc[\sqrt{\tld{\omega}}].
\label{eqn:inp}
\end{gather}

Each of the terms in Eqs.~(\ref{eqn:inm}) and~(\ref{eqn:inp}) can be evaluated in a straightforward manner (using
a handbook of integrals or a symbolic integration package, for instance).  The resulting equations are
\begin{gather}
I_{N,-}=-\sqrt{\pi^{3}}{2}\exp[-\hat{\omega}]\left(\erfc[\sqrt{-\hat{\omega}}]\right)^{2}\\
I_{N,+}=\sqrt{\pi^{3}}{2}\exp[\hat{\omega}]\left(\erfc[\sqrt{\hat{\omega}}]\right)^{2},
\end{gather}
where $\hat{\omega}\equiv-\omega_{N}/\xi$.
Plugging these expressions into the original integrals of interest from Eq.~(\ref{eqn:linqnh}), we get
\begin{equation}
\int \vvol \frac{\omega_{N}+1}{\omega_{N}+\la L_{n}/R \ra \left(x_{\parl}^{2}+x_{\perp}^{2}/2\right)}F_{Me}=n_{0e}\frac{\pi}{2}\left(\omega_{N}+1\right)\la \frac{R}{L_{n}}\ra\exp[-\hat{\omega}]\left(\erfc[\sqrt{-\hat{\omega}}]\right)^{2}
\label{eqn:ielectronp}
\end{equation}
for electrons and
\begin{equation}
\int \vvol \frac{\omega_{N}-\tau}{\omega_{N}-|L_{n}/R|\tau\left(x_{\parl}^{2}+x_{\perp}^{2}/2\right)}F_{Mi}=-\frac{n_{0i}}{\tau}\frac{\pi}{2}\left(\omega_{N}-\tau\right)\la\frac{R}{L_{n}}\ra\exp[\hat{\omega}]\left(\erfc[\sqrt{\hat{\omega}}]\right)^{2}
\label{eqn:iionp}
\end{equation}
for ions, with $\tau\equiv T_{0i}/T_{0e}$ the ratio of ion to electron temperatures.  Substituting Eqs.~(\ref{eqn:ielectronp})
and~(\ref{eqn:iionp}) into the quasineutrality expression (\ref{eqn:linqnh}) results in the following dispersion relation:
\begin{equation}
\begin{split}
1+\tau-&\left(\tau-\omega_{N}\right)\frac{\pi}{2\tau}\la \frac{R}{L_{n}}\ra\exp[\hat{\omega}]\left(\erfc[\sqrt{\hat{\omega}}]\right)^{2}\\
&-\tau\left(1+\omega_{N}\right)\frac{\pi}{2}\la\frac{R}{L_{n}}\ra\exp[-\hat{\omega}]\left(\erfc[\sqrt{-\hat{\omega}}]\right)^{2}=0.
\end{split}
\end{equation}

The presence of the complementary error functions in the above dispersion relation makes analysis complicated.  We simplify
matters by assuming $\la\hat{\omega}\ra\ll 1$ (i.e. $\la\omega_{N}\ra\ll\la\xi\ra$) and taking $\tau=1$.  To lowest order
in $\omega_{N}$, we find
\begin{equation}
\sqrt{\omega_{N}}=\frac{\pi\la R/L_{n}\ra-2}{\sqrt{8\pi}\la R/L_{n}\ra^{3/2}}\left(1-\sqrt{-1}\right).
\end{equation}
This expression must be treated carefully.  Depending on the choice of branch cut, $\sqrt{-1}=\pm i$.  The choice of branch cut, 
coupled with an assumption about the sign of $Im[\omega_{N}]$, also sets a restriction on the signs of the real and imaginary
parts of $\sqrt{\omega_{N}}$.

First, we take a branch cut along the negative real axis, so that the arguments of complex numbers are defined on the interval $[-\pi,\pi)$.
In this case, $\sqrt{-1}=-i$ and 
\begin{equation}
\sqrt{\omega_{N}}=\frac{\pi\la R/L_{n}\ra -2}{\sqrt{8\pi}\la R/L_{n}\ra^{3/2}}\left(1+i\right).
\label{eqn:omncut1}
\end{equation}
If $Im[\omega_{N}]>0$, then $Re[\sqrt{\omega_{N}}]>0$ and $Im[\sqrt{\omega_{N}}]>0$.  Consequently, Eq.~(\ref{eqn:omncut1})
is only valid when $\la R/L_{n} \ra > 2/\pi$.  If we instead assume that $Im[\omega_{N}]<0$, then we have $Re[\sqrt{\omega_{N}}]>0$
and $Im[\sqrt{\omega_{N}}]<0$.  This is clearly not possible in Eq.~(\ref{eqn:omncut1}) since 
$Re[\sqrt{\omega_{N}}]=Im[\sqrt{\omega_{N}}]$, so no damped waves exist for this choice of branch cut.

Now we take our branch cut along the positive real axis, so that the arguments of complex numbers are defined on the interval $[0,2\pi)$.
For this case, $\sqrt{-1}=i$, and the equation for $\sqrt{\omega_{N}}$ becomes
\begin{equation}
\sqrt{\omega_{N}}=\frac{\pi\la R/L_{n}\ra -2}{\sqrt{8\pi\la R/L_{n}\ra^{3/2}}}\left(1-i\right).
\label{eqn:omncut2}
\end{equation}
If $Im[\omega_{N}]>0$, then $Re[\sqrt{\omega_{N}}]>0$ and $Im[\sqrt{\omega_{N}}]>0$.  Since 
$Re[\sqrt{\omega_{N}}]=-Im[\sqrt{\omega_{N}}]$ in Eq.~(\ref{eqn:omncut2}), no growing modes are allowed
for this choice of branch cut.  For $Im[\omega_{N}]<0$, we get $Re[\sqrt{\omega}_{N}]<0$ and 
$Im[\sqrt{\omega_{N}}]>0$.  This is only satisfied for $\la R/L_{n}\ra < 2/\pi$.

Combining the results of Eqs.~(\ref{eqn:omncut1}) and~(\ref{eqn:omncut2}) and keeping in mind their range
of validity, we obtain our solution for $\omega_{N}$:
\begin{equation}
\omega_{N}=\frac{\left(\pi\la R/L_{n} \ra - 2\right)\la \pi \la R/L_{n} \ra -2\ra}{4\pi\la R/L_{n}\ra^{3}}i,
\end{equation}
which indicates instability for gradients steeper than the critical gradient, given by 
$\la R/L_{n} \ra_{crit}=2/\pi$.  A similar, if somewhat
messier, calculation can be done for arbitrary temperature ratio and with lowest order
finite Larmor radius effects included~\cite{ricciPoP06}.  The result is
\begin{equation}
\omega \frac{\la L_{n} \ra}{v_{th,i}} = \frac{\left[\left(1+\tau\right)\left(\frac{\pi}{2}\la R/L_{n} \ra-1\right)-k_{\perp}^{2}\rho_{i}^{2}\la R/L_{n}\ra\left(\pi/2-1\right)\right]^{2}}{2\pi\left(1+\tau^{3}\right)^{2}\la R/L_{n}\ra^{3}}\left(\tau^{2}-1\pm2\tau^{3/2}i\right)k_{\perp}\rho_{i},
\end{equation}
where the $+$ sign applies for $\la R/L_{n} \ra > \la R/L_{n} \ra_{crit}$ and the $-$ sign applies for $\la R/L_{n}\ra<\la R/L_{n}\ra_{crit}$,
where 
\begin{equation}
\la R/L_{n} \ra_{crit}=\frac{2\left(1+\tau\right)}{\pi\left(1+\tau\right)-k_{\perp}^{2}\rho_{i}^{2}\left(\pi-2\right)}.
\end{equation}

\section{Comparison with fluid theory}

Previous studies~\cite{hassamPoF84,akerstedtJPP90,kesnerPoP00,simakovPoP01,simakovPoP02,kesnerPoP02} of the low-$\beta$ Z-pinch system considered here have identified two
distinct linearly unstable modes.  The first mode is the well-known ideal interchange
mode described by MHD, which is the dominant instability (with growth rate scaling
like $\gamma\sim c_{s}/\sqrt{RL_{n}}$, $c_{s}=\sqrt{(T_{0e}+T_{0i})/m_{i}}$) 
in the strong gradient regime.  However, the ideal interchange mode becomes stable at 
moderate
gradients, leaving only the shorter wavelength entropy mode.  This mode consists of
perturbations to both the plasma density and temperature, but not the plasma pressure.

\begin{figure}
\centering
\includegraphics[height=5.0in]{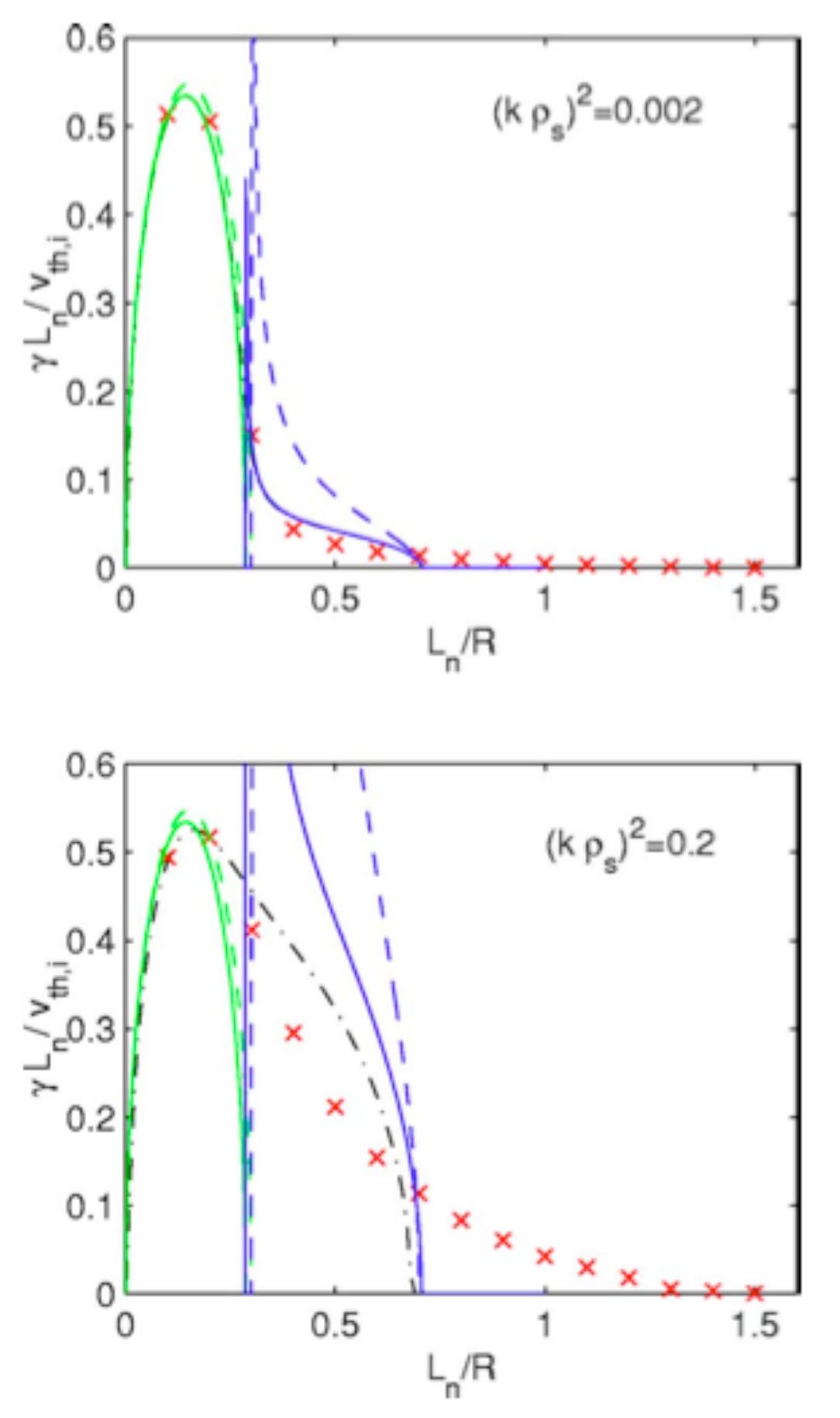}
\caption{Growth rate of the interchange and the entropy mode as a function of 
$L_{n}/R$ for two different values of $k\rho_{s}$ (both with $\tau=1$).  The 
collisionless gyrokinetic
growth rate (red "$\times$" marks) is compared to the growth rates from the gyrofluid 
model (black dotted-dashed line), the ideal collisional (green dashed line) and collisionless
(gree solid line) interchange mode, and the collisional (blue dashed line) and collisionless
(blue solid line) fluid entropy mode.  The kinetic model is necessary to obtain the correct
stability boundary and to obtain the correct growth rate for weak to moderately strong gradients.
Figures taken from Ref.~\cite{ricciPoP06}.}
\label{fig:zpgrowth}
\end{figure}

In Fig.~\ref{fig:zpgrowth}, the dependence of the growth rate on the density gradient scale
length is considered for a number of different models.  Collisional and collisionless MHD
equations accurately predict the growth rate in the regime of strong density
gradient, but they do not describe the dynamics of the entropy mode and thus are not 
applicable for sufficiently weak gradients (where the interchange mode is stabilized).
Fluid models capture the correct qualitative behavior for the entropy mode, but they
do not provide good quantitative agreement with the growth rate and they underestimate
the critical gradient determined from the gyrokinetic model by a factor of approximately two.

\begin{figure}
\centering
\includegraphics[height=5.0in]{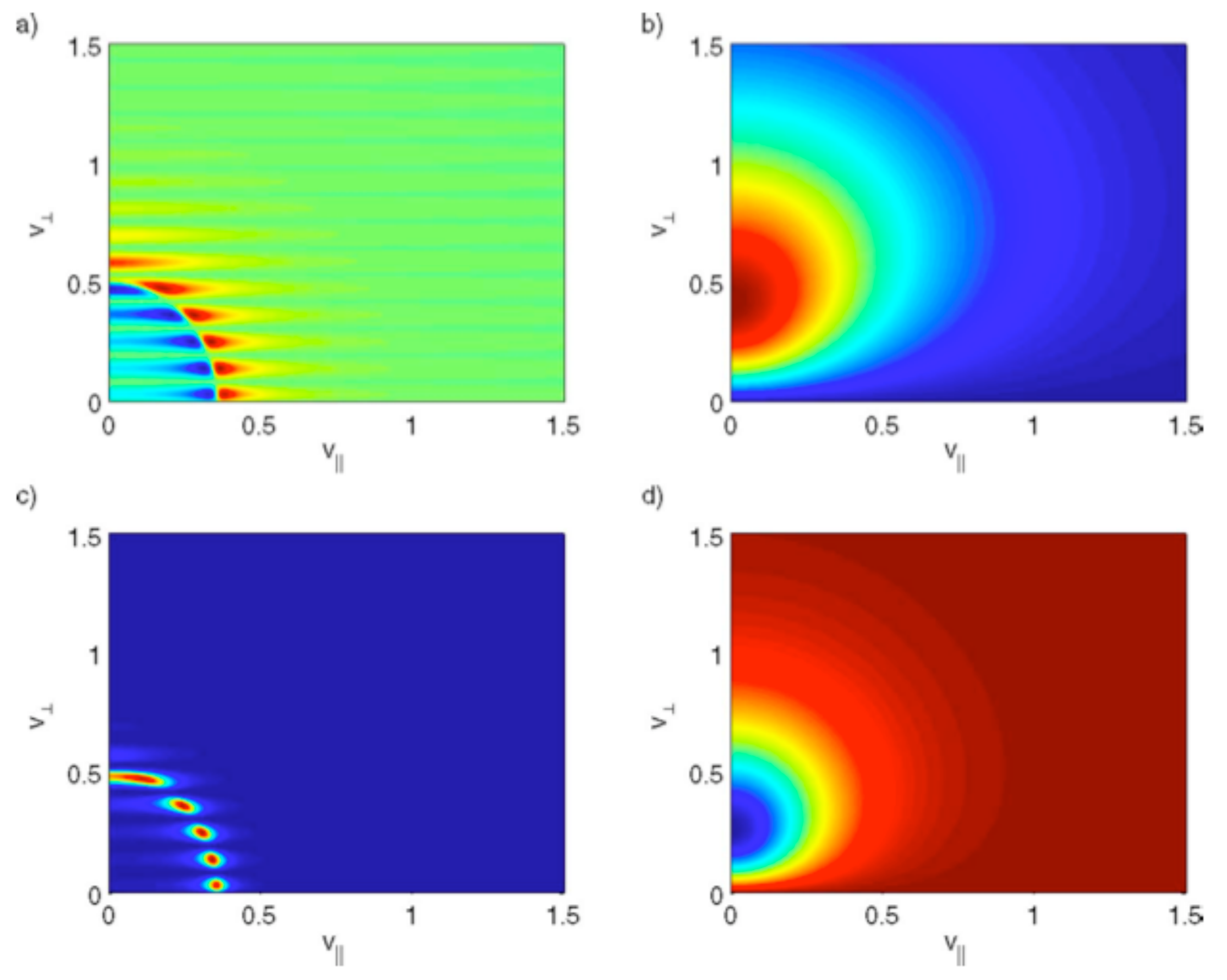}
\caption{Real (a,b) and imaginary (c,d) parts of the ion (a,c) and electron (b,d)
velocity distribution functions for the case of a moderate density gradient ($L_{n}/R = 0.5$)
and large $k\rho_{s}$ ($=38$).  Axes are normalized to $v_{th,s}$.  We see significant
structure in $v_{\perp}$ for ions in addition to an energy resonance arising from the
curvature drift.  Figures taken from Ref.~\cite{ricciPoP06}.}
\label{fig:zpdist}
\end{figure}

We see in Fig.~\ref{fig:zpdist} why a fluid description is not necessarily sufficient for this
system: the distribution of particles in velocity space possesses nontrivial structure.  In
particular, there is an energy resonance in the velocity space arising from the presence
of curvature drifts.  Additionally, fine scale structure is present in the distribution of
perpendicular velocities.  The development of such fine scale structure is discussed in
more detail in Chapter~\ref{chap:vspaceres}.

As a final note, we point out that even for this highly simplified system, the kinetic stability calculation
was quite involved.  To accurately model linear physics for kinetic systems of practical interest to the fusion 
community, we would 
need to retain finite Larmor radius effects, finite temperature gradients, trapped particles, magnetic shear, and a 
host of other important features.  Such calculations are burdensome at best and quite often
analytically intractable.  Furthermore, while linear analysis is undoubtedly useful in providing qualitative insight
into the gross characteristics of plasma behavior, a nonlinear treatment is of course necessary to address
turbulent dynamics.  The inherent difficulties of analytic calculations make computer simulations a critical
component in advancing our understanding of plasma turbulence.

\renewcommand{\thechapter}{3}

\chapter{\textbf{Turbulent transport hierarchy for gyrokinetics}}
\label{chap:hierarchy}
\vspace{+30pt}

\section{Introduction}
\label{sec:hierintro}

As noted in Chapter~\ref{chap:intro}, a full treatment of turbulent transport in weakly
collisional plasmas involves the interaction of slowly evolving, large scale
profiles and rapidly evolving, small scale turbulence.  It is not straightforward to neglect 
either process, because the large scale profiles and small scale turbulence are dynamically 
coupled.  The evolution of the profiles is governed by the turbulence, and the instabilities 
that give rise to the turbulence respond sensitively to the profiles.  The parameter space that 
characterizes these interactions is very large, limiting the ultimate applicability of parametric 
fits~\cite{kotschPoP95, waltzPoP97}.
The analytic and numerical 
difficulties associated with this wide range of time and space scales are compounded by the 
kinetic nature of the instabilities driving the turbulence; in principle, one must consider
a six-dimensional phase space consisting of three dimensions each for physical and velocity
space.  Such a system is both analytically and numerically intractable.  Consequently,
it is necessary to work with a reduced model.

In this chapter, we will derive such a reduced model by taking advantage of the
wide space and time scale separations present in many weakly-collisional plasmas.
In particular, we closely follow the treatment of Refs.~\cite{plunk,wang} to introduce a set 
of ordering assumptions (the $\delta f$ gyrokinetic 
ordering~\cite{antonPoF80,friemanPoF82,howesApJ06})
that leads to a reduction of the phase space and to the development of a hierarchical set of 
equations, in which equations describing the turbulent fluctuations and equilibrium profiles 
are coupled, but evolved separately.  This represents a significant simplification of the 
system, which we show in later chapters
allows for computationally efficient, first-principles simulations of turbulent transport
evolution over long time scales.

\section{Ordering assumptions}
\label{sec:hierorder}

We take the Fokker-Planck equation as the starting point from whence we will derive
our hierarchical set of equations describing turbulent transport:
\begin{equation}
\pd{f_{s}}{t}+\mbf{v}\cdot\pd{f_{s}}{\mbf{r}} + \frac{q_{s}}{m_{s}}\left(\mbf{E}+\frac{\mbf{v}\times\mbf{B}}{c}\right)\cdot\pd{f_{s}}{\mbf{v}}=\sum_{u}C[f_{s},f_{u}],
\label{eqn:fp}
\end{equation}
where $f_{s}$ represents the distribution of particles of species $s$ in position $\mbf{r}$
and velocity $\mbf{v}$, $q_{s}$ is particle charge, $m_{s}$ is particle mass, 
$\mbf{E}$ and $\mbf{B}$ are the electric and magnetic fields, $c$ is the speed of light, and
$C[f_{s},f_{u}]$ is a bilinear operator describing the effect on particles of species $s$ 
of collisions with particles of species $u$.  For convenience of notation, we shall 
henceforth drop the species subscript $s$ where it leads to no ambiguity.

Assuming the potential energy of nearest-neighbor interaction is much less than the kinetic 
energy of the particles, this equation describes the full range of dynamics in a six-dimensional
phase space and time for particles of species $s$ moving in a self-consistent electromagnetic 
field.  It does not take into account the effect of external sources of particles, momentum,
energy, etc.  These will certainly be present in fusion devices, but they are local processes, 
and their time and space scales are associated with thermodynamic processes.  Consequently,
we neglect them in our analysis and insert them ad hoc when we consider the slow evolution
of thermodynamic quantities such as density and temperature.  

Unfortunately, the comprehensive nature of the Fokker-Planck equation makes 
theoretical analysis burdensome and numerical simulations computationally 
infeasible.  To make progress, we simplify our model by adopting a variant~\cite{plunk,wang}
of the $\delta f$ gyrokinetic ordering~\cite{antonPoF80,friemanPoF82,howesApJ06}, which exploits scale separation 
in time and space.  We first separate quantities into equilibrium and fluctuating parts:
\begin{equation}
f=F_{0}+\delta f, \ \ \mbf{B} =\mbf{B}_{0}+\mbf{\delta B}, \ \ \mbf{E} = \mbf{\delta E},
\end{equation}
with
\begin{equation}
\frac{\delta f}{F_{0}} \sim \frac{\delta B}{B_{0}} \sim \epsilon \ll 1.
\end{equation}
Formally, we define the smallness parameter $\epsilon$ as a ratio of small to large spatial
scales within the plasma:
\begin{equation}
\epsilon \equiv \frac{\rho}{L},
\end{equation}
where $L$ is a typical scale length associated with the equilibrium and $\rho$ is the radius
of particle gyration in the equilibrium magnetic field.  We separate spatial scales by assuming cross-field fluctuations vary
on the gyroradius scale, while all other quantities vary on the equilibrium scale:
\begin{gather}
\nabla_{\perp} \delta f \sim \frac{\delta f}{\rho}\\
\nabla F_{0} \sim \frac{F_{0}}{L}, \ \ \nabla_{\parl} \delta f \sim \frac{\delta f}{L}.
\end{gather}
We separate temporal scales by assuming gyromotion is faster than the dynamic frequencies of
interest, which are themselves much faster than the evolution of the equilibrium profiles:
\begin{equation}
\tau^{-1} \sim \epsilon^{2}\omega \sim \epsilon^{3}\Omega_{0i},
\end{equation}
where $\Omega_{0i}=|q|B_{0}/m_{i}c$ is the ion gyrofrequency and
\begin{equation}
\pd{F_{0}}{t} \sim \frac{F_{0}}{\tau}, \ \ \pd{\delta f}{t} \sim \omega \delta f.
\end{equation}
Additionally, we take the the $E\times B$ velocity to be an order smaller than the thermal
velocity:
\begin{equation}
\frac{c \ \delta E}{B} \sim \epsilon v_{th},
\end{equation}
where we have assumed no background electric field (and therefore no equilibrium flow).  
We can express the fields in terms of potentials as follows:
\begin{gather}
\label{eqn:E}
\mbf{E}=-\nabla \Phi - \frac{1}{c}\pd{\mbf{A}}{t}\\
\mbf{B}_{0} = \nabla\times\mbf{A}_{0}\\
\delta\mbf{B} = \nabla\times\delta\mbf{A}
\end{gather}
where $\Phi$ is the electrostatic potential and $\mbf{A}$ is the vector potential.
Note that our ordering requires $\delta\mbf{A}/\mbf{A}_{0}\sim\epsilon^{2}$
(since $\nabla A_{0} \sim A_{0}/L$, $\nabla \delta A \sim \delta A/\rho$, and 
$\delta B/B_{0}\sim\epsilon$),
so that the electric field is primarily electrostatic in nature.

The final ordering assumptions we 
make are that the collision frequency $\nu$ is comparable to the fluctuation frequency
and that the distribution function varies in velocity space on the scale of the thermal
velocity:
\begin{equation}
\nu\sim\omega, \ \ \pd{f_{s}}{\mbf{v}} \sim \frac{f_{s}}{v_{th,s}}.
\label{eqn:vordering}
\end{equation}
Note that this choice of ordering does not prevent us from considering the 
cases $\nu \ll \omega$ and $\nu \gg \omega$ as subsidiary orderings~\cite{schekPPCF08}.
In general, subsidiary orderings can be applied using a large number of plasma parameters
as the expansion parameter.
While most of the potential expansion parameters have order unity variations across many
experiments, some, such as the electron-ion mass ratio, the
plasma beta, and the aspect ratio of the device, are good expansion parameters for a large
class of systems.  However, we are interested in deriving a set of equations that are widely
applicable to turbulent transport studies, so we do not consider such subsidiary expansions,
which would limit the range of validity of our model.

With these ordering assumptions, we can now expand the Fokker-Planck equation (\ref{eqn:fp})
in the smallness parameter $\epsilon$.  What we find is a set of ordered equations that ultimately
provides us with information about the evolution of the equilibrium, instabilities, fluctuations, and transport.  
In what follows, the ordering is taken relative to $v_{th}F_{0}/L$.

\subsection{$F_{0}$ does not depend on gyrophase}

The lowest-order equation in our $\epsilon$ expansion is
\begin{equation}
\frac{q}{mc}\mbf{v}\times\mbf{B_{0}}\cdot\pd{F_{0}}{\mbf{v}}
=-\Omega_{0}\pd{F_{0}}{\vartheta} = 0,
\label{eqn:F0nogyro}
\end{equation}
where $\vartheta$ denotes the gyroangle.  
Thus, the equilibrium distribution function
is independent of gyroangle.  We note that in obtaining the first equality in the above equation,
it is convenient to use cylindrical $(\vpe,\vartheta,\vpa)$ or spherical $(v,\vartheta,\xi)$
coordinates in velocity space ($\xi\equiv \vpa/v$ is the pitch-angle).  Both of these
coordinate systems will be used frequently in this and later chapters.

\subsection{$F_0$ is Maxwellian and $\delta f$ can be decomposed usefully}

At the next order $(\epsilon^0)$, we find
\begin{equation}
\mbf{v}\cdot\nabla F_{0} + \mbf{v}_{\perp}\cdot \nabla f_{1} + \frac{q}{m}
\left(\mbf{E}_{1} + \frac{\mbf{v}\times\mbf{B}_{1}}{c}\right)\cdot\pd{F_{0}}{\mbf{v}}
+ \frac{\mbf{v}\times\mbf{B}_{0}}{c}\cdot\pd{f_{1}}{\mbf{v}}=\sum_{u}C[F_{0},F_{0u}],
\label{eqn:eps0}
\end{equation}
where we are using the notation $\delta f \equiv f_1 + f_2 + ..., \ etc.$, 
with $f_n/F_0\sim \epsilon^n$.  Multiplying this equation by $1+\ln F_{0}$
and manipulating, we have
\begin{equation}
\begin{split}
\nabla \cdot \left(F_{0}\ln F_{0}\right)\mbf{v} &+ \left(1+\ln F_{0}\right)\mbf{v}_{\perp}\cdot \nabla f_{1} +
\pd{}{\mbf{v}} \cdot \left[ \frac{q}{m}\left(\mbf{E}_{1} 
+ \frac{\mbf{v}\times\mbf{B}_{1}}{c}\right)F_{0}\ln F_{0}\right]\\
&=\left(1+\ln F_{0}\right)\left(\Omega_{0}\pd{f_{1}}{\vartheta}+\sum_{u}C[F_{0},F_{0u}]\right).
\label{eqn:eps0start}
\end{split}
\end{equation}
Integrating this expression over all velocities gives
\begin{equation}
\int d^{3}v \left(\nabla \cdot \left(F_{0}\ln F_{0}\right)\mbf{v}+\left(1+\ln F_{0}\right)\mbf{v}_{\perp}\cdot \nabla f_{1} 
= \sum_{u}\ln F_{0}C[F_{0},F_{0u}]\right),
\label{eqn:eps0vint}
\end{equation}
where we have used the divergence theorem to eliminate the third term in Eq.~(\ref{eqn:eps0start}),
and we have used the fact that collisions locally conserve particle number to assert
$\int d^{3}v \ C[F_{0},F_{0u}]=0$.  

\begin{figure}
\centering
\includegraphics[height=3.5in]{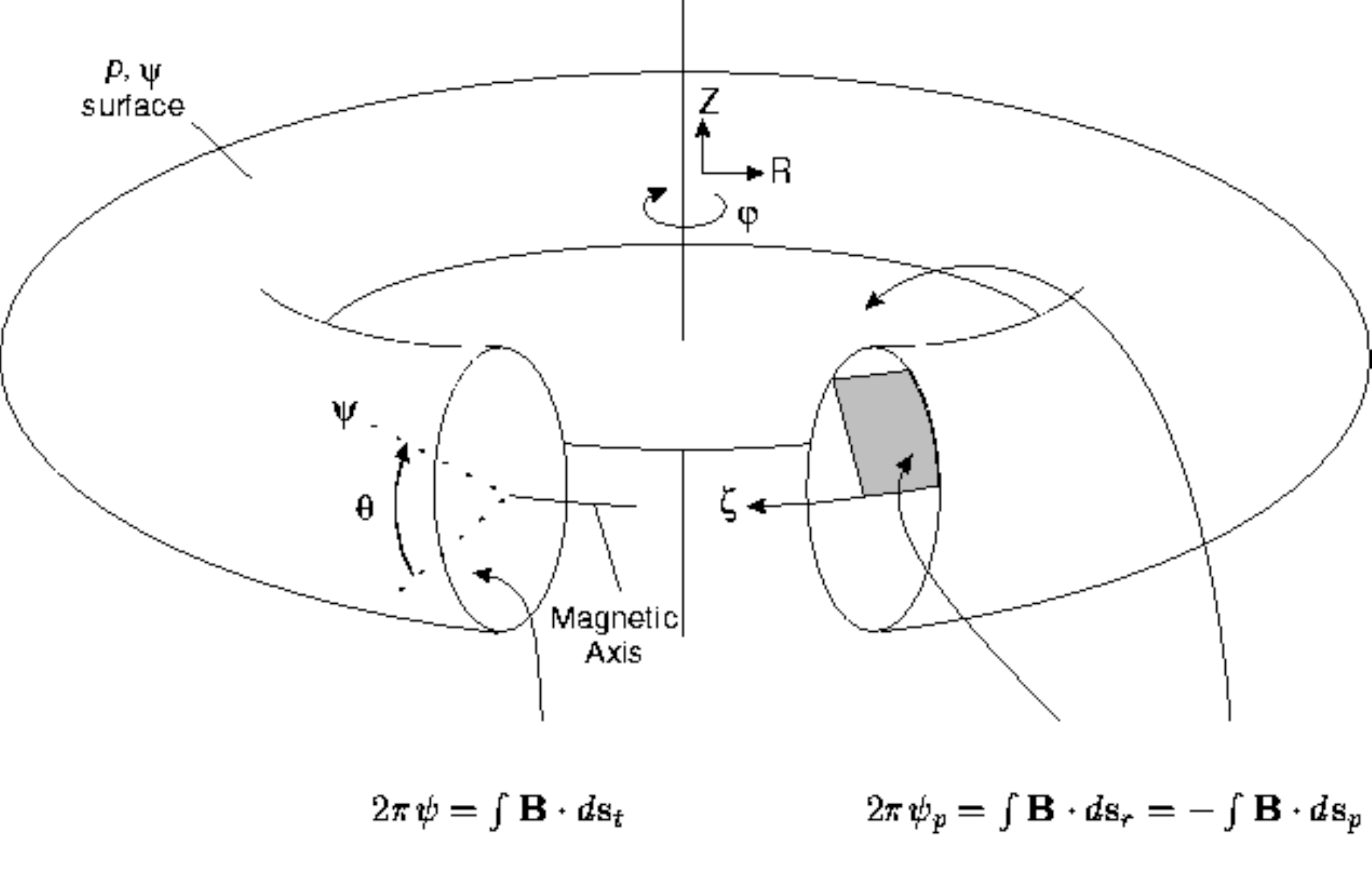}
\caption{Schematic of an axisymmetric magnetic field configuration. The flux surface,
labeled by pressure $p$ or toroidal/poloidal flux $\Psi$ has no variation in the toroidal ($\varphi$) direction.  Figure
taken from Ref.~\cite{pinches}.}
\label{fig:axiB}
\end{figure}

Before proceeding, we need to define an intermediate spatial average, which we will
see is equivalent to a flux-surface average.  We restrict our attention to axisymmetric
equilibrium magnetic field configurations (Fig.~\ref{fig:axiB}), which can be represented in the following general form
(a fuller discussion of the magnetic geometry is given in \apref{app:geo}):
\begin{equation}
\mbf{B}_{0} = I (\psi) \nabla \phi + \nabla \psi \times \nabla \phi,
\label{eqn:axiB}
\end{equation}
where $\psi = (2 \pi)^{-2} \int dV \, \mbf{B}\cdot \nabla \theta$ is the
poloidal flux, $dV$ is the volume element, and $\theta$ and $\phi$ are the physical 
poloidal and toroidal angles respectively.  The quantity $I(\psi) = R B_T$ is 
a measure of the toroidal magnetic field, where $R$ 
is the major radius and $B_T$ is the toroidal magnetic field strength.  

The intermediate spatial average of any quantity $\mathcal{F}(\mbf{r})$ is then defined as follows:
\begin{equation}
\lbb \mathcal{F}(\mbf{r}) \rbb \equiv \frac{1}{V} \int_{0}^{2\pi}d\phi \int_{-\pi}^{\pi}d\theta \int_{\psi_{0}-\Delta\psi/2}^{\psi_{0}+\Delta\psi/2} d\psi \ \mathcal{F}(\mbf{r}),
\label{eqn:spaceavg}
\end{equation}
where
\begin{eqnarray}
\label{eqn:vol}
V&\equiv&\int_{0}^{2\pi}d\phi \int_{-\pi}^{\pi}d\theta \int_{\psi_{0}-\Delta\psi/2}^{\psi_{0}+\Delta\psi/2} d\psi \ J \\
J &\equiv& \left(\nabla \psi \times \nabla \theta \cdot \nabla \phi\right)^{-1},
\end{eqnarray}
and $\psi_{0}$ denotes location in the radial coordinate $\psi$.  
Formally, we define the intermediate spatial length $\Delta\psi$ to be order $\epsilon^{1/2}L$
(i.e. $\rho \ll \Delta\psi \ll L$).
Note that in the limit where the volume $V$ becomes vanishingly small, the spatial average 
of Eq.~(\ref{eqn:spaceavg}) reduces to the usual flux surface average.  We also point
out that the spatial averaging is done about a fixed point in space so that it does not depend
on time.  It is possible to define the spatial average so that it is taken with respect to a fixed
flux surface.  However, since the flux surfaces themselve evolve in time on the equilibrium
time scale, this would require additional terms to account for the time dependence.

We now apply the intermediate spatial average to Eq.~(\ref{eqn:eps0vint}).  The
divergence theorem applied to the first term gives
\begin{equation}
\lbb \nabla \cdot \left(F_{0}\ln F_{0}\right)\mbf{v} \rbb = \int \frac{d\mbf{A}}{V} \cdot \mbf{v}F_{0}\ln F_{0},
\end{equation}
where $d\mbf{A}$ is the area element whose vector direction is normal to the surface bounding the volume 
integral (the flux surface).  Since the magnetic field lies within the flux surface, $d\mbf{A}\cdot \mbf{B}_{0}=0$, so that
\begin{equation}
\int d^{3}\mbf{v}\int \frac{d\mbf{A}}{V} \cdot \mbf{v}F_{0}\ln F_{0} = \int d^{3}\mbf{v}\int \frac{d\mbf{A}}{V} \cdot \mbf{v}_{\perp}F_{0}\ln F_{0} = 0.
\end{equation}
The last equality follows from the fact that $F_{0}$ is independent of gyroangle, and $\mbf{v}_{\perp}$ 
is an odd function of gyroangle.

The second term in Eq.~(\ref{eqn:eps0vint}) can be integrated by parts: the surface term is zero
at this order and the remaining term is dropped in ordering by $\epsilon^{1/2}$.  Consequently, we find
\begin{equation}
\lbb \int d^{3}v \ \ln F_{0} \ \sum_{u}C[F_{0},F_{0u}] \rbb = 0.
\end{equation}
From Boltzmann's $H$-Theorem, the only solution to this equation is $F_{0}=F_{M}$, where
$F_{M}$ is a Maxwellian in velocity.

Plugging $F_{0}=F_{M}$ into Eq.~(\ref{eqn:eps0}), we find
\begin{equation}
\mbf{v}_{\perp}\cdot\nabla f_{1} - \Omega_{0}\pd{f_{1}}{\vartheta} = -\mbf{v}\cdot \nabla F_{M} - \mbf{v}_{\perp} \cdot \nabla \left(\frac{q\Phi}{T_{0}}\right)F_{M},
\label{eqn:eps0r}
\end{equation}
where $T_{0}\equiv mv_{th}^{2}/2$ is the equilibrium temperature, and the $v_{\parl}\nabla_{\parl}\left(\frac{q\Phi}{T_{0}}\right)F_{M}$ term has been neglected at this order.
The velocity space derivatives thus far have implicitly been taken at fixed particle position $\mbf{r}$.  However,
it is now useful to switch to the guiding center variable $\mbf{R}\equiv \mbf{r}-\bm{\rho}$, where $\bm{\rho}=\unit{b}\times\mbf{v}/\Omega_{0}$
is the gyroradius vector:
\begin{equation}
\left(\pd{}{\vartheta}\right)_{\mbf{r}} = \left(\pd{}{\vartheta}\right)_{\mbf{R}}
+ \left(\pd{\mbf{R}}{\vartheta}\right)_{\mbf{r}} \cdot \left(\pd{}{\mbf{R}}\right)_{\vartheta}
= \left(\pd{}{\vartheta}\right)_{\mbf{R}} + \frac{\mbf{v}_{\perp}}{\Omega_{0}}\cdot \nabla,
\label{eqn:varthetarR}
\end{equation}
where subscripts on the derivatives denote quantities that are held constant during the differentiation.
It should be noted that there is no ambiguity in the use of the $\nabla$ notation, since
$\partial/\partial\mbf{r}=\partial/\partial\mbf{R}$.
Using this result in Eq.~(\ref{eqn:eps0r}), we have
\begin{equation}
-\Omega_{0}\left(\pd{f_{1}}{\vartheta}\right)_{\mbf{R}} = -\mbf{v}\cdot\nabla F_{M} - \mbf{v}_{\perp} \cdot \nabla \left(\frac{q\Phi}{T_{0}}\right)F_{M}.
\label{eqn:eps0R}
\end{equation}
The homogeneous solution $h$ satisfies
\begin{equation}
\left(\pd{h}{\vartheta}\right)_{\mbf{R}} = 0,
\end{equation}
telling us that $h$ is independent of gyroangle at fixed guiding center position.

We next proceed to find the particular solution.  Applying Eq.~(\ref{eqn:varthetarR}) 
to the righthand side of Eq.~(\ref{eqn:eps0R}) and
remembering that $F_{0}$ is independent of gyroangle, we find
\begin{equation}
-\Omega_{0}\left(\pd{f_{1}}{\vartheta}\right)_{\mbf{R}} = \Omega_{0}\left(\pd{}{\vartheta}\right)_{\mbf{R}}\left[F_{M}\left(1+\frac{q\Phi}{T_{0}}\right)\right] - v_{\parl}\unit{b}\cdot\nabla F_{M}.
\end{equation}
Upon gyro-averaging, we obtain
\begin{equation}
\unit{b}\cdot \nabla F_{M} = 0,
\label{eqn:F0fluxfn}
\end{equation}
which is simply a statement that the equilibrium distribution function is constant
on a flux surface.  Using this result in Eq.~(\ref{eqn:eps0r}), we have
\begin{equation}
\mbf{v}_{\perp}\cdot\nabla f_{1} - \Omega_{0}\pd{f_{1}}{\vartheta} = -\mbf{v}_{\perp}\cdot \nabla \left[F_{M}\left(1 + \frac{q\Phi}{T_{0}}\right)\right],
\end{equation}
which has the particular solution
\begin{equation}
f_{1p} = -\frac{q\Phi}{T_{0}}F_{M} - \bm{\rho}\cdot \nabla F_{M}.
\end{equation}
If we redefine our equilibrium distribution function by absorbing this term, we see that we
have the beginning of a Maxwell-Boltzmann distribution for guiding centers:
\begin{equation}
F_{0} \equiv F_{M}(\mbf{R}) \exp\left(-\frac{q\Phi}{T_{0}}\right).
\end{equation}
Our solution then has the form
\begin{equation}
f = F_{0} + h + f_{2} + ...
\label{eqn:f}
\end{equation}

\subsection{The gyrokinetic equation}

At order $\epsilon^1$, we derive the gyrokinetic equation.  It is convenient to transform to gyrokinetic variables, which is easier if we start again with the 
Fokker-Planck equation:
\begin{equation}
\pd{f}{t} + \frac{d\mbf{R}}{dt} \cdot \pd{f}{\mbf{R}} + \frac{d\varepsilon}{dt}\pd{f}{\varepsilon}
+ \frac{d\mu}{dt}\pd{f}{\mu} + \frac{d\vartheta}{dt}\pd{f}{\vartheta}= \sum_{u}C[f,f_{u}],
\end{equation}
where $\varepsilon\equiv mv^{2}/2+q\Phi$ is the particle energy and $\mu\equiv mv_{\perp}^{2}/2B$
is the magnetic moment.  The $\mathcal{O}(\epsilon)$ terms yield
\begin{equation}
\pd{h}{t} + \frac{d\mbf{R}}{dt}\cdot\pd{}{\mbf{R}}\left(F_{0}+h\right)
- \frac{F_{0}}{T_{0}}\frac{d\varepsilon}{dt} - \Omega_{0}\pd{f_{2}}{\vartheta}
= C[h + \mbf{\rho}\cdot\nabla F_{M}],
\end{equation}
where we define
\begin{equation}
C[f] \equiv \sum_{u} \left(C[f,F_{M,u}]+C[F_{M},f_{u}]\right).
\end{equation}
We take the gyroaverage of this equation (at constant $\mbf{R}$) to eliminate the $f_{2}$ term:
\begin{equation}
\pd{h}{t} + \left< \frac{d\mbf{R}}{dt} \right>_{\mbf{R}} \cdot \nabla (F_{0}+h) - \frac{F_{0}}{T_{0}}\left<\frac{d\varepsilon}{dt}\right>_{\mbf{R}} = \left< C[h] \right>_{\mbf{R}},
\label{eqn:hgyro}
\end{equation}
where we have used the fact that both $F_{M}$ and $h$ are independent of gyroangle at fixed
guiding center position.  To the order we will need, the guiding 
center velocity and power are 
\begin{equation}
\left< \frac{d\mbf{R}}{dt} \right>_{\mbf{R}} = v_{\parl}\unit{b} + \left< \mbf{v}_{D} \right>_{\mbf{R}}
\label{eqn:dRdt}
\end{equation}
and
\begin{equation}
\left< \frac{d\varepsilon}{dt} \right>_{\mbf{R}} = q \pd{\left< \chi \right>_{\mbf{R}}}{t} - \frac{q}{c}\mbf{v}\cdot\pd{\mbf{A}_{0}}{t},
\label{eqn:dEdt}
\end{equation}
where
\begin{equation}
\left< \mbf{v}_{D} \right>_{\mbf{R}} \equiv \frac{\unit{b}}{\Omega_{0}}\times\left[\frac{q}{m}\nabla\left<\chi\right>_{\mbf{R}} + \vpa^{2}\unit{b}\cdot\nabla\unit{b}+\frac{v_{\perp}^{2}}{2}\frac{\nabla B_{0}}{B_{0}}\right]
\label{eqn:vdrift}
\end{equation}
is the gyroaveraged guiding center drift velocity.
Substituting Eqs.~(\ref{eqn:dRdt}) and~(\ref{eqn:dEdt}) in Eq.~(\ref{eqn:hgyro}),
we obtain
\begin{equation}
\pd{h}{t} + v_{\parl}\unit{b}\cdot\nabla h + \left< \mbf{v}_{D} \right>_{\mbf{R}} \cdot \nabla (F_{0}+h) = \left< C[h] \right>_{\mbf{R}} + \frac{qF_{0}}{T_{0}}\left(\pd{\left< \chi \right>_{\mbf{R}}}{t} - \frac{\mbf{v}}{c}\cdot\pd{\mbf{A}_{0}}{t}\right).
\label{eqn:fullgk}
\end{equation}
To obtain the gyrokinetic equation in a standard form, we first split $h$ into two pieces: one that varies
on the equilibrium spatial scale (called the neoclassical part) and one that varies on the 
gyroradius scale (called the turbulent part):
\begin{equation}
h \equiv h_{t} + h_{nc}.
\end{equation}
We then apply the intermediate spatial average defined previously to obtain an equation
for the evolution of $h_{nc}$:
\begin{equation}
\pd{h_{nc}}{t}+v_{\parl}\unit{b}\cdot\nabla h_{nc} + \mbf{v}_{B}\cdot\nabla F_{0}= \left< C[h_{nc}]\right>_{\mbf{R}}-\frac{qF_{0}}{T_{0}}\frac{\mbf{v}}{c}\cdot\pd{\mbf{A_{0}}}{t},
\label{eqn:neoclassical}
\end{equation}
where
\begin{equation}
\mbf{v}_{B} \equiv \frac{\unit{b}}{\Omega_{0}}\times\left[\vpa^{2}\unit{b}\cdot\nabla\unit{b}+\frac{v_{\perp}^{2}}{2}\frac{\nabla B_{0}}{B_{0}}\right]
\end{equation}
consists of the sum of the curvature and $\nabla B$ drifts.
Subtracting the neoclassical equation (\ref{eqn:neoclassical}) from Eq.~(\ref{eqn:fullgk})
yields the gyrokinetic equation for the evolution of the turbulent distribution function:
\begin{equation}
\pd{h_{t}}{t} + v_{\parl}\unit{b}\cdot\nabla h_{t} + \left< \mbf{v_{\chi}} \right>_{\mbf{R}} \cdot \nabla (F_{0}+h_{t}) +\mbf{v}_{B}\cdot\nabla h_{t}= \left< C[h_{t}] \right>_{\mbf{R}} + \frac{qF_{0}}{T_{0}}\pd{\left< \chi \right>_{\mbf{R}}}{t},
\label{eqn:gkeqn}
\end{equation}
with 
\begin{equation}
\mbf{v_{\chi}} \equiv \frac{c}{B_{0}}\unit{b}\times\nabla\chi
\label{eqn:vexb}
\end{equation}
being the generalized $E\times B$ drift velocity.

\subsection{Transport equations; thermodynamics}

The $\mathcal{O}(\epsilon^{2})$ equation includes terms involving $f_{2}$, for which
we have no expression.  The goal in this subsection is to manipulate the equation to eliminate
second order quantities in favor of products of first order quantities.  We will accomplish
this by taking a moment approach and averaging over intermediate space and time scales.
The result, as we will see, is a closed set of fluid equations for the evolution of the equilibrium
density and temperature profiles.  Because of the time and space averaging, these equations 
are only valid when there are no important time or space scales between the turbulence
and equilibrium scales.  

Before we begin the moment approach, we define the
intermediate time average of a quantity $\mathcal{F}(t)$ in a manner analogous to the intermediate spatial average of 
Eq.~(\ref{eqn:spaceavg}):
\begin{equation}
\overline{\mathcal{F}(\tau_{0})}\equiv\frac{1}{\Delta{\tau}}\int_{\tau_{0}-\frac{\Delta\tau}{2}}^{\tau_{0}+\frac{\Delta\tau}{2}} dt \ \mathcal{F}(t),
\label{eqn:timeaverage}
\end{equation}
where $\omega^{-1} \ll \Delta \tau \ll \tau$, and we formally define $\Delta\tau\sim\epsilon\tau$.

\subsubsection{Slow density profile evolution}

To obtain an equation for the evolution of the equilibrium density profiles, we first take the density moment of the Fokker-Planck equation.  The term involving the Lorentz force vanishes because it is a perfect divergence in velocity space,
and the collisional term vanishes because of local particle conservation.  We are left with the usual continuity equation:
\begin{equation}
\int d^{3}v \left[\pd{f}{t} + \nabla \cdot \left(\mbf{v}f\right)\right] = 0.
\label{eqn:continuity}
\end{equation}
Applying the gyrokinetic ordering from the beginning of the chapter, we see that both $\partial f_{2}/\partial t$
and $\nabla \cdot \left(\mbf{v}f_{2}\right)$ enter at order $\epsilon^{2}$.  Since we do not want to solve for $f_{2}$,
we must eliminate it from the equation.  We will accomplish this by averaging over intermediate scales in space and time.

Performing the intermediate spatial average of Eq.~(\ref{eqn:spaceavg}) and using the divergence theorem,
we can write
\begin{equation}
\lbb \nabla \cdot \left(\mbf{v}f\right) \rbb = \int \frac{d\mbf{A}}{V} \cdot \mbf{v}f = \int \frac{dA}{V}\frac{\nabla \psi}{\la \nabla \psi \ra} \cdot \mbf{v}f.
\label{eqn:vdotpsi1}
\end{equation}
By inspection, one can see that this can in turn be written
\begin{equation}
\int \frac{dA}{V}\frac{\nabla \psi}{\la \nabla \psi \ra} \cdot \mbf{v}f = \frac{1}{V}\pd{}{\psi}\lbb V \nabla \psi \cdot \mbf{v}f \rbb. 
\label{eqn:vdotpsi2}
\end{equation}
We can manipulate this into a more useful form in a couple of steps.  First, we use the expression for the axisymmetric magnetic field given in Eq.~(\ref{eqn:axiB}) to get
\begin{equation}
\left(\mbf{v}\cdot\nabla\psi\right) f = -R^{2}\nabla \phi \cdot \left(\mbf{v}\times\mbf{B}_{0}\right)f.
\label{eqn:vdotpsi2p5}
\end{equation}
Noting that $\mbf{v}\times\mbf{B}_{0}=-\Omega_{0}\pd{\mbf{v}}{\vartheta}$ and integrating the righthand side
of the above expression by parts in gyroangle, we obtain
\begin{equation}
\int d^{3}\mbf{v} \left(\mbf{v}\cdot\nabla\psi\right) f = \int d^{3}\mbf{v} \ \left(R^{2} \nabla \phi \cdot \mbf{v}\right)\left[\left(\mbf{v}\times\mbf{B}_{0}\right)\cdot\pd{f}{\mbf{v}}\right].
\label{eqn:vdotpsi3}
\end{equation}
Using Eqs.~(\ref{eqn:vdotpsi1}),~(\ref{eqn:vdotpsi2}), and~(\ref{eqn:vdotpsi3}) in Eq.~(\ref{eqn:continuity}),
we obtain
\begin{equation}
\lbb \int\vvol \pd{f}{t} \rbb + \frac{1}{V}\pd{}{\psi} \lbb V \int\vvol \left(R^{2}\nabla \phi \cdot \mbf{v}\right)\left[\left(\mbf{v}\times\mbf{B}_{0}\right)\cdot \pd{f}{\mbf{v}}\right]\rbb=0.
\end{equation}
Substituting for $\left(\mbf{v}\times{B}_{0}\right)\cdot\partial f/\partial \mbf{v}$ from the Fokker-Planck
equation (\ref{eqn:fp}), we have
\begin{equation}
\begin{split}
\lbb \int\vvol \pd{f}{t} \rbb &- \frac{1}{V}\pd{}{\psi} \lbb V \int\vvol \frac{mc}{q}\left(R^{2}\nabla \phi \cdot \mbf{v}\right)\left(\pd{f}{t}+\mbf{v}\cdot\nabla{f}\right.\right.\right.\\
&\left.\left.\left.+\frac{q}{m}\left(\mbf{E}+\frac{\mbf{v}\times\delta\mbf{B}}{c}\right)\cdot \pd{f}{\mbf{v}} - \sum_{u}C[f,f_{u}]\right)\rbb=0.
\end{split}
\label{eqn:eps2start}
\end{equation}
We now apply the intermediate time average defined by Eq.~(\ref{eqn:timeaverage}), plug in our expression (\ref{eqn:f}) 
for $f$, and examine each of the resulting terms up to order $\epsilon^{2}$.

The first term in Eq.~(\ref{eqn:eps2start}) becomes
\begin{equation}
\lbb \int \vvol \overline{\pd{f}{t}} \rbb = \lbb \int \vvol \left(\overline{\pd{F_{0}}{t}}+\overline{\pd{h}{t}}+\overline{\pd{f_{2}}{t}}\right) \rbb.
\end{equation}
Since $h$ and $f_{2}$ both vary on the fluctuation time scale, the only term that survives the time average is the one involving $F_{0}$:
\begin{equation}
\lbb \int \vvol \overline{\pd{f}{t}} \rbb = \pd{n_{0}}{t},
\end{equation}
where $n_{0}\equiv\lbb \int\vvol F_{0} \rbb$.  Because of the slow time variation of $F_{0}$, this term
enters at $\mathcal{O}(\epsilon^{2})$.  The other term in Eq.~(\ref{eqn:eps2start}) involving $\partial f/\partial t$
is treated analogously.  However, the prefactor multiplying it is of order $\epsilon$, dropping the overall order to $\epsilon^{3}$.  Consequently, we may neglect it.

We next treat the term containing $\mbf{v}\cdot\nabla f$.  Employing Eq.~(\ref{eqn:f}) for $f$, we have
\begin{equation}
\begin{split}
&\frac{1}{V}\pd{}{\psi} \lbb V \int\vvol \frac{mc}{q}\left(R^{2}\nabla \phi \cdot \mbf{v}\right)\left(\mbf{v}\cdot\nabla{f}\right) \rbb\\
&=\frac{1}{V}\pd{}{\psi} \lbb V \int\vvol \frac{mc}{q}\left(R^{2}\nabla \phi \cdot \mbf{v}\right)\left(\mbf{v}\cdot\nabla\left[F_{0}+h+f_{2}\right]\right) \rbb
\end{split}
\label{eqn:ncpart}
\end{equation}
Since $F_{0}=F_{M}(\psi)$ and $\nabla \phi \cdot \nabla\psi=0$, the term in the integrand containing 
$F_{0}$ is odd in velocity space and therefore integrates to zero.  Integrating the $f_{2}$ term by parts in space, we see that the gradient operator is transferred to the equilibrium
quantity $R$.  Due to the slow cross-field spatial variation of equilibrium quantities, this drops the term to $\mathcal{O}(\epsilon^{3})$, so it does not contribute at this order.  

We can also integrate the $h$ term by parts in space,
and we find that the turbulent piece vanishes due to periodicity.  All that is left is a term involving the 
neoclassical piece of $h$, which we address now.  First we define the neoclassical pressure tensor:
\begin{equation}
\mbf{P} \equiv \int d^{3}v \mbf{v}\mbf{v}h_{nc} = \int d^{3}v \left[\vpa^{2} \unit{b}\unit{b}+\frac{v_{\perp}^{2}}{2}\left(\mbf{I}-\mbf{b}\mbf{b}\right)\right]h_{nc} + \tld{\mbf{P}},
\end{equation}
where $\mbf{I}$ is the identity tensor and $\tld{\mbf{P}}=\int \vvol \left(\mbf{v}_{\parl}\mbf{v}_{\perp}+\mbf{v}_{\perp}\mbf{v}_{\parl}\right)h_{nc}$ 
is an antisymmetric tensor containing the off-diagonal components of the pressure tensor.  
The neoclassical term from Eq.~(\ref{eqn:ncpart}) can then be written
\begin{equation}
\begin{split}
\frac{1}{V}\pd{}{\psi} \Big<\Big< V &\int\vvol \frac{mc}{q}\left(R^{2}\nabla \phi \cdot \mbf{v}\right)\left(\mbf{v}\cdot\nabla h_{nc}\right) \Big>\Big>\\
&=\frac{1}{V}\frac{mc}{q}\pd{}{\psi} \Big<\Big< V \left[\nabla \cdot \left(\mbf{P}\cdot R^{2}\nabla\phi \right)-\mbf{P}:\nabla\left(R^{2}\nabla\phi\right)\right]\Big>\Big>,
\end{split}
\label{eqn:ncpart2}
\end{equation}
where the double dot tensor product is defined
\begin{equation}
\mbf{A}:\mbf{B} \equiv \sum_{i}\sum_{j}A_{ij}B_{ji}.
\end{equation}
The first term on the right in Eq.~(\ref{eqn:ncpart2}) vanishes upon application of the divergence 
theorem due to periodicity in toroidal and poloidal angles.  Because 
$\nabla\left(R\nabla\phi\right)=\left(\mbf{R}\nabla\phi-\nabla\phi \mbf{R}\right)$
is an antisymmetric tensor, the symmetric part of the pressure tensor vanishes in the second term on the
right, leaving
\begin{equation}
\begin{split}
\frac{1}{V}\pd{}{\psi} \Big<\Big< V &\int\vvol \frac{mc}{q}\left(R^{2}\nabla \phi \cdot \mbf{v}\right)\left(\mbf{v}\cdot\nabla h_{nc}\right) \Big>\Big>\\
&=-\frac{1}{V}\frac{mc}{q}\pd{}{\psi} \Big<\Big< V \left[\tld{\mbf{P}}:\nabla\left(R^{2}\nabla\phi\right)\right]\Big>\Big>=0.
\label{eqn:ptanti}
\end{split}
\end{equation}
To obtain the final equality, we made use of the identity $\int d^{3}\mbf{R} \int_{\mbf{R}} \vvol=\int d^{3}\mbf{r}\int_{\mbf{r}}\vvol$,
where the subscripts on the integrals denote the variable to be held fixed.  Since $h_{nc}$
is independent of gyroangle at fixed $\mbf{R}$ and $(\mbf{v}_{\parl}\mbf{v}_{\perp}+\mbf{v}_{\perp}\mbf{v}_{\parl})$
is odd in gyroangle, we see that the $\tld{\mbf{P}}$ inside the spatial average of 
Eq.~(\ref{eqn:ptanti}) vanishes.

Now we examine the term containing the Lorentz force:
\begin{equation}
\begin{split}
\frac{1}{V}\pd{}{\psi}\Big<\Big<V &\int \vvol \left(R^{2}\nabla\phi\cdot\mbf{v}\right)\left(c\delta\mbf{E}+\mbf{v}\times\delta\mbf{B}\right)\cdot\pd{f}{v}\Big>\Big>\\
&=\frac{1}{V}\pd{}{\psi}\Big<\Big<V \int \vvol \left(R^{2}\nabla\phi\cdot\mbf{v}\right)\left(c\delta\mbf{E}+\mbf{v}\times\delta\mbf{B}\right)\cdot\pd{}{v}\left(F_{0}+h+\delta f_{2}\right) \Big>\Big>
\end{split}
\end{equation}
First, we consider the terms containing $F_{0}$.  For the magnetic force term we find
\begin{equation}
\int \vvol \left(R^{2}\nabla\phi\cdot\mbf{v}\right)\left(\mbf{v}\times\delta\mbf{B}\right)\cdot\pd{F_{0}}{\mbf{v}}
\sim \int \vvol \left(R^{2}\nabla\phi\cdot\mbf{v}\right)\left(\mbf{v}\times\delta\mbf{B}\right)\cdot\mbf{v}F_{0} = 0.
\end{equation}
For the electric force term, we have
\begin{equation}
\begin{split}
\int \vvol \left(R^{2}\nabla\phi\cdot\mbf{v}\right)\delta\mbf{E}\cdot\pd{F_{0}}{\mbf{v}}
&= \int \vvol \left(R^{2}\nabla\phi\cdot\mbf{v}\right)\pd{}{\mbf{v}}\cdot\left(F_{0}\delta\mbf{E}\right)\\
&=\int \vvol R^{2}\nabla\phi\cdot\left[\nabla\Phi+\frac{1}{c}\pd{}{t}\left(\mbf{A}_{0}+\delta\mbf{A}\right)\right]F_{0},
\end{split}
\end{equation}
where we have integrated by parts and substituted Eq.~(\ref{eqn:E}) for the electric field between lines one and two.
Upon application of the intermediate time average, the $\delta A$ term drops in ordering.
We next apply the intermediate spatial average:
\begin{equation}
\Big<\Big< R^{2}\nabla\phi \cdot \nabla\Phi F_{0} \Big>\Big> = \Big<\Big< R\pd{\Phi}{\phi} F_{0}\Big>\Big> = 0,
\end{equation}
where the final equality follows from the fact that the equilibrium is axisymmetric and
all other quantities are periodic in torodial angle $\phi$.

Now we consider the term involving $h$.  Integrating by parts in velocity space, we get
\begin{equation}
\begin{split}
\int \vvol \left(R^{2}\nabla\phi\cdot\mbf{v}\right)\left(\delta\mbf{E}+\frac{\mbf{v}\times\delta\mbf{B}}{c}\right)\cdot\pd{h}{\mbf{v}}
&= \int \vvol \left(R^{2}\nabla\phi\cdot\mbf{v}\right) \pd{}{\mbf{v}}\cdot\left[\left(\delta\mbf{E}+\frac{\mbf{v}\times\delta\mbf{B}}{c}\right)h\right]\\
&= -\int \vvol R^{2}\nabla\phi \cdot \left(\delta\mbf{E}+\frac{\mbf{v}\times\delta\mbf{B}}{c}\right)h.
\label{eqn:lorhpart}
\end{split}
\end{equation}
Rewriting $\delta\mbf{E}$ and $\delta\mbf{B}$ in terms of the potentials $\Phi$
and $\mbf{A}$, we have
\begin{equation}
\begin{split}
\delta\mbf{E}+\frac{\mbf{v}\times\delta\mbf{B}}{c}& = -\nabla\Phi-\frac{1}{c}\pd{\mbf{A}}{t} + \frac{\mbf{v}}{c}\times\left(\nabla\times\delta\mbf{A}\right)\\
&=-\nabla\Phi-\frac{1}{c}\pd{\mbf{A}}{t} + \frac{1}{c}\left[\nabla\left(\mbf{v}\cdot\delta\mbf{A}\right)-\left(\mbf{v}\cdot\nabla\right)\delta\mbf{A}\right]\\
&=-\nabla\chi-\frac{1}{c}\pd{\mbf{A}}{t}-\left(\frac{\mbf{v}}{c}\cdot\nabla\right)\delta\mbf{A},
\label{eqn:lorpot}
\end{split}
\end{equation}
where $\chi$ is defined in Eq.~(\ref{eqn:chi}).  The terms involving 
$\vpa\unit{b}\cdot\nabla\delta\mbf{A}$ and
$\partial \mbf{A}/\partial t$ are higher order and can thus be neglected
(the $\partial \mbf{A}/\partial t$ term was retained earlier when multiplied by 
$F_{0}$, but here it is multiplied by $h$).  Further, only the fluctuating part of $h$
contributes, since the intermediate spatial average of Eq.~(\ref{eqn:lorhpart}) drops the order of the $h_{nc}$ part.
The $\mbf{v}_{\perp}\cdot\nabla\delta\mbf{A}$ term vanishes by integrating by
parts in space and using the identity from Eq.~(\ref{eqn:varthetarR}):
\begin{equation}
\begin{split}
\Big<\Big< \int \vvol \ R^{2}\nabla\phi\cdot\left[\left(\mbf{v}_{\perp}\cdot\nabla\right)\delta\mbf{A}\right]h \Big>\Big>
&= -\Big<\Big< \int \vvol \ R^{2}\nabla\phi\cdot\delta\mbf{A}\left(\mbf{v}_{\perp}\cdot\nabla h\right) \Big>\Big>\\
&= -\Big<\Big< \int \vvol \ R^{2}\nabla\phi\cdot\delta\mbf{A}\left(\pd{h}{\vartheta}\right)_{\mbf{r}} \Big>\Big>\\
& = 0,
\end{split}
\end{equation}
where we have used the fact that $h$ is independent of gyroangle at fixed $\mbf{R}$ and
$\delta\mbf{A}$ is independent of gyroangle at fixed $\mbf{r}$.

The only part of the distribution function contributing to the collisional term is
\begin{equation}
\tld{h}_{nc}\equiv h_{nc}-\bm{\rho}\cdot F_{M}
\end{equation}
because $h_{t}$ vanishes upon spatial average and $C[F_{M}]=0$.
Collecting results, we have the equation for the evolution of the equilibrium density:
\begin{equation}
\pd{n_{0}}{t} = \frac{1}{V}\pd{}{\psi}\Big<\Big< V \int \vvol R^{2}\nabla\phi \cdot\left(\pd{\mbf{A}_{0}}{t}F_{0}+h_{t}\nabla\chi - \frac{m\mbf{v}}{q}C[\tld{h}_{nc}]\right)\Big>\Big>.
\label{eqn:nevo}
\end{equation}

\subsubsection{Slow temperature profile evolution}

The equation for the evolution of the equilibrium temperature profiles is derived in a manner 
analagous to the equation for the equilibrium density profiles.  We begin by taking the energy 
moment of the Fokker-Planck equation (\ref{eqn:fp}) and performing the intermediate
spatial average of Eq.~(\ref{eqn:spaceavg}):
\begin{equation}
\begin{split}
\lbb \int \vvol \ \frac{mv^{2}}{2} \right.\right. &\left.\left.\left[\pd{f}{t}+\mbf{v}\cdot\nabla f + \frac{q}{m}\left(\mbf{E}+\frac{\mbf{v}\times\mbf{B}}{c}\right)\cdot\pd{f}{\mbf{v}}\right] \rbb \\
&= \lbb \int \vvol \ \frac{mv^{2}}{2}\sum_{u}C[f,f_{u}] \rbb.
\end{split}
\label{eqn:eps2tstart}
\end{equation}
We then substitute Eq.~(\ref{eqn:f}) for $f$ into Eq.~(\ref{eqn:eps2tstart}), average over 
the intermediate time scale, and consider the order of each of the terms.

As before with density evolution, the terms involving time derivatives of $h$ and $\delta f_{2}$
do not contribute at this order.  The contribution to the first term of Eq.~(\ref{eqn:eps2tstart})
from $F_{0}$ is
\begin{equation}
\lbb \int \vvol \frac{mv^{2}}{2}\pd{f}{t} \rbb = \frac{3}{2}\pd{n_{0}T_{0}}{t},
\end{equation}
where $T_{0}\equiv\lbb \int \vvol \ mv^{2}F_{0} \rbb /2n_{0}$.

The term in Eq.~(\ref{eqn:eps2tstart}) containing $\mbf{v}\cdot\nabla f$ can be 
manipulated as it was in Eqs.~(\ref{eqn:vdotpsi1})-(\ref{eqn:vdotpsi3}) to obtain
\begin{equation}
\lbb \int \vvol \ \frac{mv^{2}}{2}\mbf{v}\cdot\nabla f  \rbb 
= \frac{1}{V}\pd{}{\psi} \lbb V\int \vvol \ \frac{mv^{2}}{2}R^{2}\nabla \phi \cdot \mbf{v} \left[\left(\mbf{v}\times\mbf{B}_{0}\right)\cdot\pd{f}{\mbf{v}}\right] \rbb.
\end{equation}
Substituting for $\left(\mbf{v}\times\mbf{B}_{0}\right)\cdot\partial f/\partial \mbf{v}$
from the Fokker-Planck equation~(\ref{eqn:fp}), we have
\begin{equation}
\begin{split}
\lbb \int\vvol \ \frac{mv^{2}}{2}\mbf{v}\cdot\nabla f  \rbb & = -\frac{1}{V}\pd{}{\psi} \lbb V \int\vvol \ \frac{mv^{2}}{2}\frac{mc}{q}\left(R^{2}\nabla \phi \cdot \mbf{v}\right)\left(\pd{f}{t}+\mbf{v}\cdot\nabla{f}\right.\right.\right.\\
&\left.\left.\left.+\frac{q}{m}\left(\mbf{E}+\frac{\mbf{v}\times\delta\mbf{B}}{c}\right)\cdot \pd{f}{\mbf{v}} - \sum_{u}C[f,f_{u}]\right)\rbb.
\label{eqn:eps2tfok}
\end{split}
\end{equation}
Using the same methods employed to derive the density evolution equation, one can show that the terms
involving $\partial f/\partial t$ and $\mbf{v}\cdot\nabla f$ do not contribute at this order.

We focus first on the Lorentz force term.  As before with density evolution, we use $f=F_{0}+h+\delta f_{2}$
and examine each term.  First, we consider the terms involving $F_{0}$.  The magnetic force
term is zero since $\partial F_{0}/\partial \mbf{v}\sim \mbf{v}F_{0}$ and 
$\left(\mbf{v}\times\delta\mbf{B}\right)\cdot\mbf{v}=0$.  The electric force
term gives
\begin{equation}
\begin{split}
\Big<\Big< \int \vvol &\left(R^{2}\nabla\phi\cdot\mbf{v}\right)\frac{mv^{2}}{2}c\delta\mbf{E}\cdot\pd{F_{0}}{\mbf{v}} \Big>\Big>\\
&=-\Big<\Big< \int \vvol cR^{2}\nabla\phi\cdot\left[\frac{mv^{2}}{2}\delta\mbf{E}+\left(\mbf{v}\cdot\delta\mbf{E}\right)\mbf{v}\right]F_{0}\Big>\Big>,
\end{split}
\end{equation}
where we have integrated by parts in velocity space.  Writing the electric field in terms
of potentials and noting that the intermediate spatial average of fluctuating quantities
is zero, we find
\begin{equation}
\begin{split}
\Big<\Big< \int \vvol &\left(R^{2}\nabla\phi\cdot\mbf{v}\right)\frac{mv^{2}}{2}c\delta\mbf{E}\cdot\pd{F_{0}}{\mbf{v}} \Big>\Big>\\
&= \Big<\Big< \int \vvol R^{2}\nabla\phi\cdot\frac{mv^{2}}{2}\pd{\mbf{A}_{0}}{t}F_{0}\Big>\Big>
\end{split}
\end{equation}

Next we consider the term involving $h$:
\begin{equation}
\begin{split}
\Big<\Big< \int \vvol &\left(R^{2}\nabla\phi\cdot\mbf{v}\right)\frac{mv^{2}}{2}\left(c\delta\mbf{E}+\mbf{v}\times\delta\mbf{B}\right)\cdot\pd{h}{\mbf{v}}\Big>\Big>\\
&= -\Big<\Big< \int \vvol R^{2}\nabla\phi \cdot \left[\frac{mv^{2}}{2}\left(c\delta\mbf{E}+\mbf{v}\times\delta\mbf{B}\right)+mc\mbf{v}\left(\mbf{v}\cdot\delta\mbf{E}\right)\right]h_{t} \Big>\Big>,
\end{split}
\end{equation}
where we have integrated by parts in velocity space and neglected the $h_{nc}$ terms because
the spatial average makes them higher order.  Also, only the electrostatic part of the
electric field contributes at this order.  Rewriting the above expression in terms of potentials,
we have
\begin{equation}
\Big<\Big< \int \vvol R^{2}\nabla\phi\cdot\left[\left(c\nabla\chi+\mbf{v}_{\perp}\cdot\nabla_{\perp}\delta\mbf{A}\right)\frac{mv^{2}}{2}+mc\mbf{v}\left(\mbf{v}\cdot\nabla\Phi\right)\right]h_{t}\Big>\Big>,
\label{eqn:eps2tcancel}
\end{equation}
where the term involving $\vpa\unit{b}\cdot\nabla\delta\mbf{A}$ has been dropped since it is higher order.
The $\mbf{v}_{\perp}\cdot\nabla\delta\mbf{A}$ term can be neglected following the same argument 
given after Eq.~(\ref{eqn:lorpot}).  The part of the collision operator that contributes at $\mathcal{O}(\epsilon^{2})$ 
in Eq.~(\ref{eqn:eps2tfok}) is the same as for particle transport, $C[\tld{h}_{nc}]$.

Next, we consider the Lorentz force term in Eq.~(\ref{eqn:eps2tstart}).  The magnetic
force term can be written as a perfect divergence in velocity space and subsequently vanishes
upon integration.  However, due to the presence of the $v^{2}$ factor, the electric force
term is nonzero.  Integrating this term by parts in velocity gives
\begin{equation}
\int\vvol \frac{mv^{2}}{2}\frac{q}{m}\mbf{E}\cdot\pd{f}{\mbf{v}}=-\int\vvol \ q\mbf{E}\cdot\mbf{v}f.
\label{eqn:edotv}
\end{equation}
Considering the electrostatic part of $\mbf{E}$, we have
\begin{equation}
\begin{split}
q\int\vvol\left(\mbf{v}\cdot\nabla\Phi\right)f &= q\int\vvol\left[\mbf{v}\cdot\nabla\left(\Phi f\right)+\pd{\left(\Phi f\right)}{t}-f\pd{\Phi}{t}-\Phi\left(\pd{f}{t}+\mbf{v}\cdot\nabla f\right)\right]\\
&=q\int\vvol\left[\mbf{v}\cdot\nabla\left(\Phi f\right)+\pd{\left(\Phi f\right)}{t}-f\pd{\Phi}{t}\right],
\end{split}
\end{equation}
where we used the continuity equation (\ref{eqn:continuity}) to obtain the final equality.
The intermediate time average eliminates the second term and the $F_{0}$
piece of the final term on the last line.  Performing the intermediate spatial average
eliminates the neoclassical terms and the $F_{0}$ part of the first term, leaving
\begin{equation}
q\left<\left<\int\vvol\left(\mbf{v}\cdot\nabla\Phi\right)f\right>\right> =q\left<\left<\int\vvol\left[\mbf{v}\cdot\nabla\left(\Phi h_{t}\right)-h_{t}\pd{\Phi}{t}\right]\right>\right>.
\label{eqn:eps2tes}
\end{equation}
We can place the first term in a more convenient form by first using Eqs~(\ref{eqn:vdotpsi1})-(\ref{eqn:vdotpsi3}):
\begin{equation}
\left<\left<\int\vvol \ q\mbf{v}\cdot\nabla\left(\Phi h_{t}\right)\right>\right>=\frac{1}{V}\pd{}{\psi}\left<\left< V\int\vvol \left(R^{2}\nabla\phi\cdot\mbf{v}\right)q\Phi\left[\left(\mbf{v}\times\mbf{B}_{0}\right)\cdot\pd{h_{t}}{\mbf{v}}\right]\right>\right>
\end{equation}
Noting that $(\mbf{v}\times\mbf{B}_{0})\cdot\partial h_{t}/\partial \mbf{v}=-B_{0}(\partial h_{t}/\partial \vartheta)_{\mbf{r}}$
and using Eq.~(\ref{eqn:varthetarR}), we get
\begin{equation}
\left<\left<\int\vvol \ q\mbf{v}\cdot\nabla\left(\Phi h_{t}\right)\right>\right> = -\frac{1}{V}\pd{}{\psi}\left<\left< V\int\vvol \frac{qB_{0}\Phi}{\Omega_{0}}\left(R^{2}\nabla\phi\cdot\mbf{v}\right)\mbf{v}\cdot\nabla h_{t}\right>\right>,
\end{equation}
where we have used $(\partial h_{t}/\partial \vartheta)_{\mbf{R}}=0$.  Integrating
by parts in space then gives
\begin{equation}
\left<\left<\int\vvol \ q \mbf{v}\cdot\nabla\left(\Phi h_{t}\right)\right>\right> = -\frac{1}{V}\pd{}{\psi}\left<\left< V\int\vvol \left(R^{2}\nabla\phi\cdot\mbf{v}\right)mc \ h_{t}\mbf{v}\cdot\nabla \Phi\right>\right>,
\label{eqn:extrajdote}
\end{equation}
which cancels with the last term in Eq.~(\ref{eqn:eps2tcancel}).

We now consider the inductive part of the electric field, $\mbf{E}_{I}$.  Since $F_{M}$ is isotropic in 
velocity space, we have
\begin{equation}
\int \vvol \ q\left(\mbf{v}\cdot\mbf{E}_{I}\right)F_{M}\left(1-\frac{q\Phi}{T_{0}}\right) = 0.
\end{equation}
The term involving $f_{2}$ does not enter at this order, so we are left with
\begin{equation}
\int \vvol \ q\left(\mbf{v}\cdot\mbf{E}_{I}\right)\left(\bm{\rho}\cdot\nabla F_{M}-h\right) = \int \vvol \ \frac{q}{c}\left(\mbf{v}\cdot\pd{\mbf{A}}{t}\right)\left(h-\bm{\rho}\cdot\nabla F_{M}\right).
\end{equation}
Applying the intermediate spatial average, we obtain
\begin{equation}
\begin{split}
\Big<\Big< &\int \vvol \ q\left(\mbf{v}\cdot\mbf{E}_{I}\right)\left(\bm{\rho}\cdot\nabla F_{M}-h\right)\Big>\Big>\\
&= \left<\left<\int \vvol \ \frac{q}{c}\left[\left(\mbf{v}\cdot\pd{\delta\mbf{A}}{t}\right)h_{t}+\left(\mbf{v}\cdot\pd{\mbf{A_{0}}}{t}\right)\tld{h}_{nc}\right]\right>\right>.
\end{split}
\end{equation}
Combining this expression with the one remaining term from the electrostatic part 
(i.e. the last term in Eq.~(\ref{eqn:eps2tes})) gives
\begin{equation}
\left<\left<\int \vvol \ \frac{q}{c}\left[-\pd{\chi}{t}h_{t}+\left(\mbf{v}\cdot\pd{\mbf{A_{0}}}{t}\right)\tld{h}_{nc}\right]\right>\right>.
\label{eqn:hdchidt}
\end{equation}

The final term to evaluate in Eq.~(\ref{eqn:eps2tstart}) is the inter-species collisional energy 
exchange.  It has the following standard form (see e.g. Ref.~\cite{helander}):
\begin{equation}
\left<\left<\vvol \frac{mv^{2}}{2}\sum_{u}C[f,f_{u}]\right>\right> = \sum_{u}n_{0}\nu_{\epsilon}^{su}\left(T_{0u}-T_{0}\right),
\end{equation}
where 
\begin{equation}
\begin{split}
\nu_{\epsilon}^{su} &\equiv  6.88 \frac{\left(m_{s}m_{u}\right)^{1/2}q_{s}^{2}q_{u}^{2}n_{u}\ln \Lambda_{su}}{\left(m_{s}T_{u}+m_{u}T_{s}\right)^{3/2}}\\
&=1.55 \nu_{su}\left(\frac{m_{u}}{m_{s}}\right)^{1/2}\left(\frac{T_{u}}{T_{s}}+\frac{m_{u}}{m_{s}}\right)^{-3/2},
\label{eqn:nueq}
\end{split}
\end{equation}
with
\begin{equation}
\nu_{su} \equiv \frac{4\pi n_{u}q_{u}^{2}q_{s}^{2}\ln \Lambda_{su}}{m_{s}^{1/2}\left(2T_{s}\right)^{3/2}}
\end{equation}
the collision frequency, $q$ the particle charge, and $\ln \Lambda_{su}$ the Coulomb logarithm.
Collecting results, we have an equation for the evolution of the equilibrium pressure, $p_{0}=n_{0}T_{0}$:
\begin{equation}
\begin{split}
\frac{3}{2}\pd{p_{0}}{t} &= \frac{1}{V}\pd{}{\psi}\left<\left< V\int\vvol \frac{mv^{2}}{2}R^{2}\nabla\phi\cdot\left[\pd{\mbf{A}_{0}}{t}F_{M}
+h_{t}\nabla\chi-\frac{m\mbf{v}}{q}C[\tld{h}_{nc}]\right]\right>\right>\\
&+\left<\left< \int\vvol \ q\left[h_{t}\pd{\chi}{t}-\mbf{v}\cdot\pd{\mbf{A}_{0}}{t}\tld{h}_{nc}\right]\right>\right>+\sum_{u}n_{0}\nu_{\epsilon}^{su}\left(T_{0u}-T_{0}\right)
\label{eqn:pevo}
\end{split}
\end{equation}

\subsubsection{Species-summed pressure equation and turbulent heating}

The term $\lbb\int\vvol \ q_{s} h_{t,s}(\partial \chi / \partial t)\rbb$ in
Eq.~(\ref{eqn:pevo}) describes turbulent heating of the equilibrium.
In order to illuminate the nature of this turbulent heating, we
sum Eq.~(\ref{eqn:pevo}) over species and consider the evolution of the total pressure, 
$p_{T}=\sum_{s}p_{0s}$:
\begin{equation}
\begin{split}
\frac{3}{2}\pd{p_{T}}{t} &= \frac{1}{V}\pd{}{\psi}\lbb V \left(-\mbf{Q}_{T}\cdot\nabla\psi+p_{T}R^{2}\nabla\phi\cdot\pd{\mbf{A}_{0}}{t}\right)\rbb\\
&-\sum_{s}\frac{1}{V}\pd{}{\psi}\lbb V \int\vvol\frac{m_{s}v^{2}}{2} \frac{R^{2}m_{s}}{q_{s}}\left(\nabla\phi\cdot\mbf{v}\right)C[\tld{h}_{nc,s}]\rbb\\
&+\sum_{s}\lbb\int\vvol q_{s}\left[h_{t}\pd{\chi}{t}-\mbf{v}\cdot\pd{\mbf{A}_{0}}{t}\tld{h}_{nc}\right]\rbb,
\end{split}
\label{eqn:ptotevo}
\end{equation}
where 
\beq
\mbf{Q}_{T} \equiv \sum_{s}\int \vvol \ \frac{m_{s}v^{2}}{2}\mbf{v}_{\chi}h_{t,s}
\eeq
is the species-summed turbulent heat flux.  We now proceed to show that species-summed
turbulent heating term, $\sum_{s}\left<\left<\int\vvol \ q_{s} h_{t,s}(\partial \chi / \partial t)\right>\right>$
is zero for steady-state turbulence.

First, we consider the electrostatic component of $\chi$:
\begin{equation}
\lbb \pd{\Phi}{t}\sum_{s}q_{s}\int \vvol \ h_{t,s}\rbb = \lbb\frac{q_{s}^{2}}{2T_{0s}}\pd{\Phi^{2}}{t}\rbb,
\end{equation}
where we have used quasineutrality.  Upon time averaging, this term vanishes.  Next
we consider the inductive component of $\chi$:
\begin{equation}
\begin{split}
\lbb \frac{1}{c}\pd{\delta\mbf{A}}{t}\cdot\sum_{s} q_{s}\int \vvol \ \mbf{v}h_{t,s} \rbb &= \lbb \frac{1}{c}\pd{\delta\mbf{A}}{t} \cdot \delta\mbf{J}\rbb\\
&=-\lbb \delta\mbf{E}_{I} \cdot \delta\mbf{J}\rbb,
\end{split}
\label{eqn:einddotj}
\end{equation}
where $\delta \mbf{J}$ is the perturbed current, and $\delta\mbf{E}_{I}$ is the inductive
part of the fluctuating electric field.  Applying Ampere's Law and using the Coulomb
gauge, Eq.~(\ref{eqn:einddotj}) becomes
\beq
\begin{split}
\lbb \frac{1}{c}\pd{\delta\mbf{A}}{t}\cdot\sum_{s} q_{s}\int \vvol \ \mbf{v}h_{t,s} \rbb &= -\lbb\frac{1}{c}\pd{\delta\mbf{A}}{t} \cdot \nabla^{2}\delta\mbf{A}\rbb\\
&= \lbb \frac{1}{2} \pd{}{t}\left(\nabla\delta\mbf{A}:(\nabla\delta\mbf{A})^{T}\right)\rbb,
\end{split}
\eeq
where the superscript $T$ denotes the matrix transpose.  This term also vanishes upon time averaging.
Consequently, we have
\begin{equation}
\sum_{s}\lbb\int\vvol q_{s}h_{t,s}\pd{\chi}{t}\rbb =0,
\label{eqn:noturbheat}
\end{equation}
where the usual intermediate time average is implied.  Therefore there is no species-summed
turbulent heating.  Using Eq.~(\ref{eqn:noturbheat}) in the pressure evolution equation (\ref{eqn:ptotevo}), 
we obtain
\beq
\begin{split}
\frac{3}{2}\pd{p_{T}}{t} &= \frac{1}{V}\pd{}{\psi}\lbb V \left(-\mbf{Q}_{T}\cdot\nabla\psi+p_{T}R^{2}\nabla\phi\cdot\pd{\mbf{A}_{0}}{t}\right)\rbb\\
&-\sum_{s}\frac{1}{V}\pd{}{\psi}\lbb V \int\vvol\frac{m_{s}v^{2}}{2} \frac{R^{2}m_{s}}{q_{s}}\left(\nabla\phi\cdot\mbf{v}\right)C[\tld{h}_{nc,s}]\rbb.
\end{split}
\eeq

While there is no net turbulent heating, this does not rule out the possiblity of
significant turbulent energy exchange between species (the turbulent heating species by 
species is the same order as the turbulent heat transport term, for instance).  Few studies
have been conducted investigating the effects of turbulent energy exchange~\cite{hintonPoP06,waltzPoP08},
and in these studies, the turbulent heating was defined as $\lbb\delta\mbf{J}\cdot\delta\mbf{E}\rbb$.
Examining Eqs.~(\ref{eqn:edotv})-(\ref{eqn:hdchidt}), we see that this is not quite equivalent
to the turbulent heating defined here (the difference between them is the additional term appearing in
Eq.~(\ref{eqn:extrajdote})).  Consequently, the impact of the turbulent heating term in 
Eq.~(\ref{eqn:pevo}) on the evolution of equilibrium pressure profiles deserves further
study.

As an aside, we note that the net turbulent heating defined in Refs.~\cite{hintonPoP06} 
and~\cite{waltzPoP08} is also zero.  From Poynting's Theorem (using low-frequency 
Maxwell's equations for the fluctuating fields), we have
\begin{equation}
\pd{}{t}\int d^{3}\mbf{r} \ \frac{\delta B^{2}}{8\pi} + \frac{c}{4\pi}\oint d\mbf{S} \cdot \left(\delta\mbf{E}\times\delta\mbf{B}\right) = -\int d^{3}\mbf{r} \ \delta\mbf{J}\cdot\delta \mbf{E}.
\end{equation}
Using statistical periodicity of the fluctuations to eliminate the surface integral and 
applying the intermediate time average, we find
\beq
\int d^{3}\mbf{r} \ \delta\mbf{J}\cdot\delta\mbf{E} = 0.
\eeq
Therefore, net turbulent heating of the equilibrium defined in this alternate way (as 
$\int d^{3}\mbf{r} \ \delta\mbf{J}\cdot\delta\mbf{E}$) is also zero.

In numerical simulations, it is often the case that the turbulent heating term as written in 
Eq.~(\ref{eqn:pevo}) has large amplitude oscillations in time, making it
difficult to quickly obtain a steady-state time average.
Before we finish our derivation, we would like to rewrite the turbulent heating term
in a more convenient form for simulation.  Making use of the fact that $h_{t}$ is independent of gyroangle
at fixed $\mbf{R}$, we can change variables from $\mbf{r}$ to $\mbf{R}$ in our phase
space integration to get
\begin{equation}
\int \frac{d^{3}\mbf{r}}{V}\int\vvol \ q h_{t}\pd{\chi}{t} = \int \frac{d^{3}\mbf{R}}{V}\int\vvol \ q h_{t} \pd{\left<\chi\right>_{\mbf{R}}}{t}.
\end{equation}
Multiplying the gyrokinetic equation (\ref{eqn:gkeqn}) by $h_{t} T_{0}/F_{0}$,
averaging over phase space (with $\mbf{R}$ as our spatial variable), and averaging
over the intermediate time scale, we find that most terms do not contribute at this order.
We are left with
\begin{equation}
\left<\left< \int\vvol \ q\pd{\chi}{t}h_{t}\right>\right> = \left<\left<\int\vvol \ \frac{h_{t}T_{0}}{F_{0}}\left(\mbf{v_{\chi}}\cdot\nabla F_{0}-\left< C[h_{t}] \right>_{\mbf{R}}\right)\right>\right>.
\label{eqn:turbheat}
\end{equation}
The first term on the righthand side in Eq.~\ref{eqn:turbheat} is the energy exchange 
between the equilibrium and the turbulence, 
which is generally cooling the equilibrium.  The second term is the collisional heating,
or entropy generation, which is a positive-definite quantity.  This sign-definiteness facilitates
quick calculation of converged steady-state values for the turbulent heating. 

Using these results in Eq.~(\ref{eqn:pevo}), we obtain the final 
form of our equation for the equilibrium pressure evolution for each species:
\begin{equation}
\begin{split}
\frac{3}{2}\pd{p_{0}}{t} &= \frac{1}{V}\pd{}{\psi}\left<\left< V\int\vvol \frac{mv^{2}}{2}R^{2}\nabla\phi\cdot\left[\pd{\mbf{A}_{0}}{t}F_{M}
+h_{t}\nabla\chi-\frac{m\mbf{v}}{q}C[\tld{h}_{nc}]\right]\right>\right>\\
&+\left<\left< \int\vvol \ \frac{h_{t}T_{0}}{F_{0}}\left(\mbf{v_{\chi}}\cdot\nabla F_{0}-\left< C[h_{t}] \right>_{\mbf{R}}\right)-q\mbf{v}\cdot\pd{\mbf{A}_{0}}{t}\tld{h}_{nc}\right>\right>\\
&+\sum_{u}n_{0}\nu_{\epsilon}^{su}\left(T_{0u}-T_{0}\right)
\label{eqn:pevo2}
\end{split}
\eeq

\section{Summary}
\label{sec:hiersum}

In this chapter, we began with the Fokker-Planck equation and introduced a set of ordering 
assumptions that allowed us to derive a closed set of equations describing the self-consistent
evolution of turbulence and thermodynamics processes (transport and heating).  The
ordering assumptions are a variant of the standard $\delta f$ gyrokinetic ordering.  In 
particular, we assumed that all quantities could be split into (well-separated) slowly and 
rapidly varying parts in space and time, choosing a definite space-time ordering in terms
of the expansion parameter $\rho / L$.  The amplitude of
of the rapidly varying parts were assumed much smaller than that of the slowly varying ones.
Furthermore, we chose a definite ordering for the scale of velocity space structures ($\delta v \sim v_{th}$)
and for the collision frequency ($\nu \sim \omega$).  Our ordering and procedure for
deriving the hierarchical equations follows closely the treatment of Refs.~\cite{plunk}
and~\cite{wang}.  A similar hierarchical set of equations, with the inclusion of 
low Mach-number, long-wavelength flows, is derived in Ref.~\cite{parraPPCF08}.

For convenience in later chapters, we collect the key results from our calculations here.  
The single-particle distribution function is written in the form
\begin{equation}
f = F_{0} + h + f_{2},
\end{equation}
where $h$ is the non-Boltzmann part of the lowest-order perturbed distribution function,
$f_{2}$ represents higher-order corrections, and
\begin{equation}
F_{0} = F_{M}(\mbf{R})\exp\left[-\frac{q\Phi}{T_{0}}\right],
\end{equation}
with $F_{M}$ a Maxwellian in velocity space.  
The slowly-varying, equilibrium part of the distribution function, $F_{M}$, is
independent of gyroangle and does not vary spatially within a magnetic flux surface:
\begin{gather}
\pd{F_{M}}{\vartheta} = 0,\\
\unit{b}\cdot\nabla F_{M} = 0.
\end{gather}
The non-Boltzmann part of the lowest-order perturbed distribution function (representing
rapid fluctuations in space and time), $h$, is also independent of gyroangle at fixed
guiding center position, $\mbf{R}$:
\begin{equation}
\left(\pd{h}{\vartheta}\right)_{\mbf{R}} = 0.
\end{equation}
The evolution of the part of $h$ associated with turbulent dynamics is given by the
well-known gyrokinetic equation:
\begin{equation}
\pd{h}{t} + v_{\parl}\unit{b}\cdot\nabla h + \left< \mbf{v_{\chi}} \right>_{\mbf{R}} \cdot \nabla (F_{0}+h) +\mbf{v}_{B}\cdot\nabla h= \left< C[h_{t}] \right>_{\mbf{R}} + \frac{qF_{0}}{T_{0}}\pd{\left< \chi \right>_{\mbf{R}}}{t}.
\end{equation}
To close the system, we need information about the electromagnetic fields and the
evolution of equilibrium (thermodynamic) quantities.  In the gyrokinetic ordering used
here, the low-frequency Maxwell's equations become:
\begin{gather}
\sum_{s}q_{s}\int d^{3}v \left(\left< h_{s} \right>_{\mbf{r}} - \frac{q_{s}\Phi}{T_{0s}}F_{Ms}\right) = 0\\
\nabla^{2}\delta A_{\parl} = -\frac{4\pi}{c}J_{\parl} = -\frac{4\pi}{c}\sum_{s}q_{s}\int\vvol \ \vpa \left< h_{s} \right>_{\mbf{r}}\\
\nabla_{\perp}^{2}\delta B_{\parl} = -\frac{4\pi}{c}\unit{b}\cdot\left[\nabla_{\perp}\times\sum_{s}q_{s}\int \vvol \left< \mbf{\vpe}h_{s}\right>_{\mbf{r}}\right].
\end{gather}
The first expression above is Poisson's equation, with the assumption of quasineutrality, and
the remaining expessions are the parallel and perpendicular components of Ampere's Law.
Finally, the equations describing the evolution of the equilibrium thermodynamic
quantities (assuming no equilibrium flows) are
\begin{equation}
\pd{n_{0}}{t} = \frac{1}{V}\pd{}{\psi}\Big<\Big< V \int \vvol R^{2}\nabla\phi \cdot\left(\pd{\mbf{A}_{0}}{t}F_{0}+h_{t}\nabla\chi - \frac{m\mbf{v}}{q}C[\tld{h}_{nc}]\right)\Big>\Big>,
\end{equation}
\begin{equation}
\begin{split}
\frac{3}{2}\pd{p_{0}}{t} &= \frac{1}{V}\pd{}{\psi}\left<\left< V\int\vvol \frac{mv^{2}}{2}R^{2}\nabla\phi\cdot\left[\pd{\mbf{A}_{0}}{t}F_{M}
+h_{t}\nabla\chi-\frac{m\mbf{v}}{q}C[\tld{h}_{nc}]\right]\right>\right>\\
&+\left<\left< \int\vvol \ \frac{h_{t}T_{0}}{F_{0}}\left(\mbf{v_{\chi}}\cdot\nabla F_{0}-\left< C[h_{t}] \right>_{\mbf{R}}\right)-q\mbf{v}\cdot\pd{\mbf{A}_{0}}{t}\tld{h}_{nc}\right>\right>\\
&+\sum_{u}n_{0}\nu_{\epsilon}^{su}\left(T_{0u}-T_{0}\right).
\end{split}
\end{equation}


\renewcommand{\thechapter}{4}

\chapter{{\bf Resolving velocity space dynamics in gyrokinetics}}
\label{chap:vspaceres}
\vspace{+30pt}

\section{Introduction}
\label{sec:vresintro}

Many plasmas of interest to the astrophysical and fusion communities are weakly collisional.
For such plasmas, velocity space dynamics are often important, and a kinetic description
is necessary.  Since the kinetic description requires a six-dimensional phase space, simulating
weakly collisional plasma processes can be computationally challenging.
Employing the gyrokinetic ordering~\cite{antonPoF80,friemanPoF82,howesApJ06} reduces the 
dimensionality by eliminating gyrophase dependence, but we are still left with a relatively 
high-dimensional system. Consequently, one would like to know how many grid points
are necessary along each dimension, particularly in velocity space, in order to resolve a 
given simulation.

In the absence of collisions or some other form of dissipation, the distribution of 
particles in velocity space can develop arbitrarily small-scale structrues.  Clearly, this
presents a problem for gyrokinetic simulations, as an arbitrarily large number of grid
points would be necessary to resolve such a system.  Of course, all physical systems
possess a finite collisionality, which sets a lower bound on the size of velocity space
structures and, therefore, an upper bound on the number of grid points required for
resolution.  We would like to know how sensitive the plasma dynamics are to the 
magnitude and form of the velocity space dissipation.  In particular, we would like
answers to the following set of questions:  Given a fixed number of grid points,
how much dissipation is necessary to ensure a resolved simulation?  Alternatively, 
given a fixed amount
of dissipation, how many grid points are necessary to ensure a resolved simulation?
Futhermore, what measurable effect, if any, does the addition of dissipation have on
collisionless plasma dynamics?

These questions have been addressed for very few plasma processes~\cite{watanabePoP04,candyPoP06}, 
in large part due to the computational expense involved with such a study.  In this
chapter, we propose computationally efficient diagnostics for monitoring velocity
space resolution and apply these diagnostics to a range of weakly-collisional plasma
processes using the continuum gyrokinetic code \verb#GS2#~\cite{kotschCPC95}.  
With the aid of these diagnostics, we have implemented an adaptive collision
frequency that allows us to resolve velocity space dynamics with the approximate minimal 
necessary physical dissipation~\cite{barnesSFTC07}.  We find that the velocity space dynamics
for growing modes are well resolved with few velocity space grid points, even in the 
collisionless limit.  Including a small amount of collisions ($\nu \ll \omega$) is necessary
and often sufficient to adequately resolve nonlinear dynamics and the long-time behavior of 
linearly damped modes.

The chapter is organized as follows.  In Sec.~\ref{sec:gkv} we discuss velocity space dynamics 
in gyrokinetics and provide examples illustrating the development of small-scale structure
in collisionless plasmas.  Sec.~\ref{sec:gs2v} contains a brief overview of the \verb#Trinity# velocity 
space grid and its dissipation mechanisms.  We describe diagnostics for monitoring velocity 
space resolution in Sec.~\ref{sec:vresdiag} and apply them to a number of plasma 
processes.  In Sec.~\ref{sec:adaptC}, we introduce an adaptive collision frequency and 
present numerical results.  We discuss our findings in Sec.~\ref{sec:vspacesum}.

\section{Gyrokinetic velocity space dynamics}
\label{sec:gkv}

\verb#Trinity# solves the coupled system consisting of the low-frequency Maxwell's equations and
the nonlinear, electromagnetic gyrokinetic equation with a model Fokker-Planck 
collision operator:
\begin{equation}
\pd{ h }{t}
+ \overbrace{\left(v_{\parl}\unit{b}+ \mbf{v_{\chi}} +\mbf{v_{B}}\right) \cdot \nabla h}^{\mathcal{K}} 
= \left< C[h] \right>_{\mbf{R}} \\
+ \underbrace{\frac{q F_{0}}{T}\pd{\left< \chi \right>_{\mbf{R}}}{t} 
- \mbf{v_{\chi}} \cdot \nabla F_{0}}_{\mathcal{S}},
\label{eqn:gke}
\end{equation}
where
\begin{equation}
h = f_{1} + \frac{q \Phi}{T}F_{M}
\end{equation}
is the non-Boltzmann part of the perturbed distribution function,
\begin{equation}
\mbf{v_{B}} = \frac{\unit{b}}{\Omega} \times \left[ v_{\parl}^{2} \ \left(\unit{b}\cdot\nabla\right) \unit{b}
+\frac{v_{\perp}^{2}}{2}\frac{\nabla B_{0}}{B_{0}}\right]
\end{equation}
is the sum of the curvature and $\nabla B$ drift velocities,
\begin{equation}
\mbf{v_{\chi}} = \frac{c}{B_{0}}\unit{b} \times \nabla \chi
\end{equation}
is the generalized $E\times B$ velocity,
\begin{equation}
\chi = \Phi - \frac{\mbf{v}}{c} \cdot \mbf{A}
\label{eqn:chi}
\end{equation}
is the generalized electromagnetic potential, $\left< \cdot \right>_{\mbf{R}}$ denotes a gyro-average 
at fixed guiding center position $\mbf{R}$, and
\begin{equation}
F_{0} = F_{M}\left(1 - \frac{q\Phi}{T}\right)
\end{equation}
is the lowest order expansion of a Maxwell-Boltzmann distribution.  The
exact form of the collision operator, $C[h]$, used in \verb#Trinity# is discussed briefly in 
Sec.~\ref{sec:gs2v} and described in detail in Chapters 5 and 6.

We can group the various terms in the gyrokinetic equation (\ref{eqn:gke}) into
three distinct categories:  source terms, labeled by $\mathcal{S}$, which typically
drive large-scale structures in velocity space; convection terms, labeled by
$\mathcal{K}$, which lead to phase-mixing and the development of small-scale 
structures in velocity space; and dissipation, given by the collision operator, which smooths
the distribution function towards a shifted Maxwellian velocity distribution.  In general, the 
structure that develops from the balancing of these terms can be quite complicated.  However,
we can gain insight into how small-scale velocity structures develop by considering 
simplified collisionless systems.  

In the absence of collisions, arbitrarily small scales can develop in velocity space.  This
is a result of phase-mixing, arising due to convection in real 
space~\cite{krommesPoP99,schekAPJ07}.
As a simple example of this phenomenon, we include 
in Appendix~\ref{app:landamp} a calculation of the perturbed 
distribution function for the collisionless ion acoustic wave in a slab.  The result,
quoted here, illustrates the tendency of collisionless plasma processes to drive
small-scale velocity space structures:
\begin{equation}
\bar{f}_{1}(z,v_{\parl},t) = e^{ik_{\parl}\left(z-v_{\parl}t\right)}G(v_{\parl}) + H(z,v_{\parl},t),
\label{eqn:iawf}
\end{equation}
where the overbar on $f_{1}$ indicates an average over perpendicular velocities.
The quantities $G$ and $H$ are explicitly derived in Appendix~\ref{app:landamp}.  Here, it is
sufficient to note that both $G$ and $H$ are smooth functions of the parallel velocity.  
The presence of the oscillatory factor
 $e^{-ik_{\parl}v_{\parl}t}$ in the first term (often called the ballistic term)
leads to the development of a characteristic wavelength in velocity space that decreases 
inversely with time.
The amplitude of this ballistic term remains comparable to the second term in Eq.~(\ref{eqn:iawf}) 
for all time, leading to the development of large amplitude oscillations
of the distribution function at arbitrarily small-scales in velocity space.  A snapshot
of this behavior at $t=10\left(k_{\parl}v_{t,i}\right)^{-1}$ is shown in Fig.~\ref{fig:iaw}.

\begin{figure}
\centering
\includegraphics[height=5.0in]{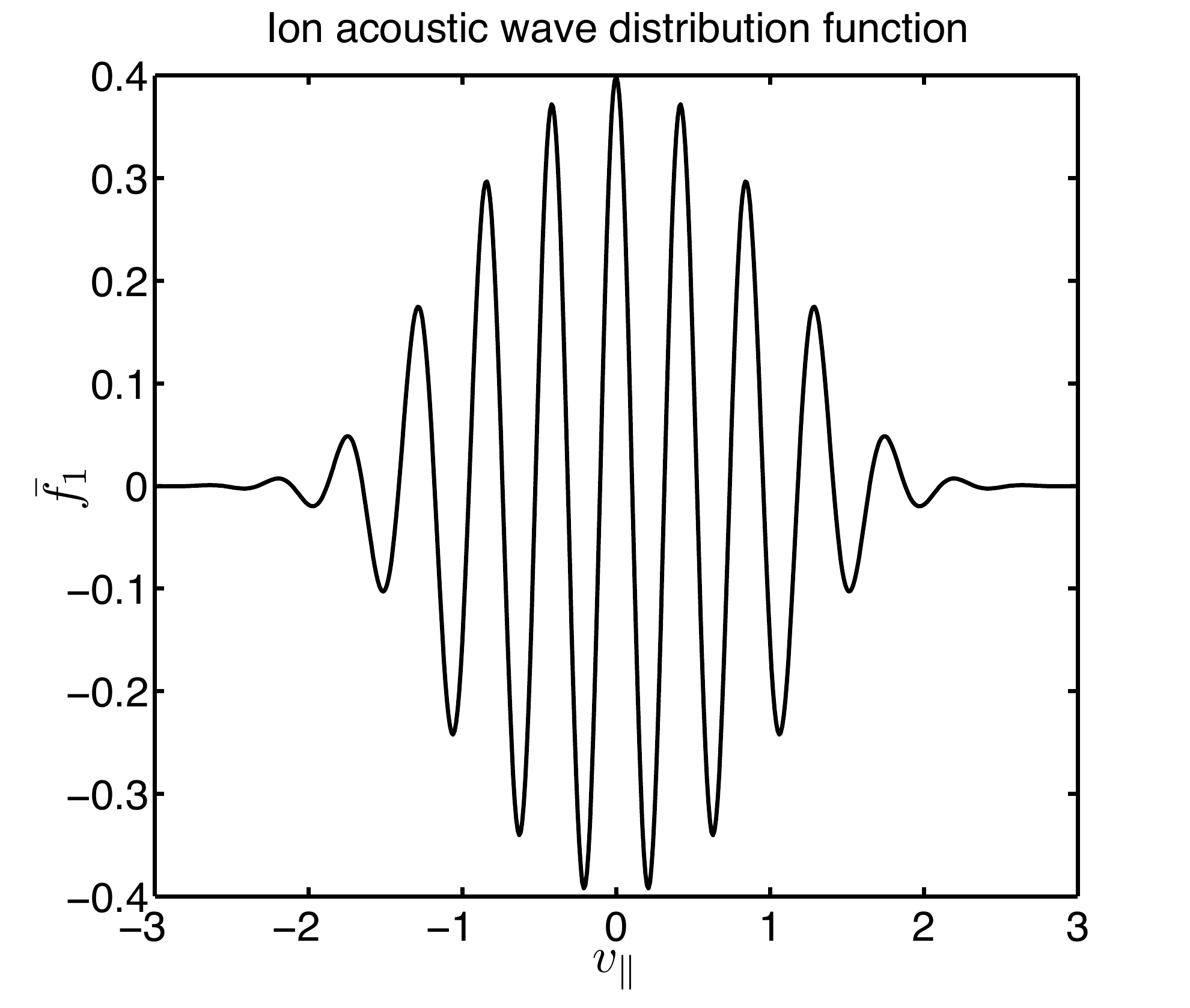}
\caption{Plot of $\bar{f}(v_{\parl})$ (normalized by $F_{0}$) at $t=10 \left(k_{\parl}v_{t,i}\right)^{-1}$.
The parallel velocity on the horizontal axis is normalized by $v_{th}$ and
$\bar{f}(v_{\parl})$ was initially a Maxwellian.}
\label{fig:iaw}
\end{figure}

The same calculation carried out for the collisionless ITG mode in a slab yields a 
distribution function with a similar ballistic term component.  However, since this 
mode is linearly
unstable, there is also a term describing large-scale structure in velocity space whose
amplitude grows in time to dominate the distribution function.  As a result, no 
significant small-scale structure develops.  This is a typical feature of linearly growing
modes in the collisionless limit.

Of course, all physical systems have a finite collisionality.  The dissipation arising
from this collisionality is critically important:  It is a necessary requirement for the 
existence of a statistically steady state~\cite{krommesPoP94,krommesPoP99}, and it sets a lower bound 
on the scale-size of structures in velocity space~\cite{schekAPJ07}.  A simple estimate 
for the scale-size
of velocity space structures can be obtained by assuming a steady state and balancing
the collisional term with the other terms in the gyrokinetic equation.  Noting that
$C\sim\nu v_{th}^{2} \partial_{v}^{2}$ (see e.g. Ref~\cite{hirshmanPoF76} 
or Ref~\cite{cattoPoF76}), we find
\begin{equation}
\frac{\delta v}{v_{th}} \sim \sqrt{\frac{\nu}{\omega}},
\label{eqn:delvorder}
\end{equation}
where $\nu$ is the collision frequency, $\omega$ is the dynamic frequency of interest,
$v_{th}\equiv\sqrt{2T/m}$ is the thermal velocity, and $\delta v$ is the scale-size of fluctuations
in velocity space.  This estimate predicts that velocity space structures much smaller than 
the thermal velocity develop in the weakly collisional limit, $\nu \ll \omega$, as we 
would expect from our consideration of simplified collisionless systems.


\section{\texttt{Trinity} velocity space}
\label{sec:gs2v}

In order to understand the velocity space resolution diagnostics described 
in later sections, it is necessary for the reader to have a basic knowledge of the 
way in which velocity space dynamics are treated in \verb#Trinity#.  To that purpose, 
we now give a brief explanation of the velocity space coordinates and dissipation
mechanisms employed in \verb#Trinity#.

\subsection{Velocity space coordinates}

Only two velocity space coordinates
are necessary in gyrokinetics because gyroaveraging has eliminated any gyrophase
dependence.  Fundamentally, \verb#Trinity# uses kinetic energy, $\varepsilon$, and a quantity related to magnetic 
moment, $\lambda=\mu/\varepsilon$, as its velocity space coordinates.  This choice
eliminates all velocity space derivatives from the collisionless gyrokinetic equation and
simplifies the discretization of derivatives in the model collision operator.  Consequently,
the spacing of the velocity space grid points is chosen to provide
accurate velocity space integrals while satisfying the necessary boundary condition
at particle bounce points.  

\subsubsection{Energy grid}

The volume element in velocity space can be written
\begin{equation}
\int d^{3}v = \frac{B_{0}}{2}\sum_{\sigma}\int_{0}^{2\pi} d\vartheta \int_{0}^{1/B_{0}}\frac{d\lambda}{\sqrt{1-\lambda B_{0}}} \int_{0}^{\infty}dv \ v^{2}
\end{equation}
where $\vartheta$ is the gyroangle and $\sigma$ denotes the sign of $v_{\parl}$.
Until recently, the energy grid in \verb#GS2# followed the treatment of Ref.~\cite{candyJCP03},
which places energy integrals in a convenient form by a change of variables to
\begin{equation}
X(x) = -\frac{2}{\sqrt{\pi}}x e^{-x^{2}} + \erf\left[x\right],
\end{equation}
where $x\equiv v / v_{th}$.
This transforms the range of integration from $x\in [0,\infty)$ to $X\in[0,1)$:
\begin{equation}
\int d^{3}v = \frac{\sqrt{\pi}}{8}B_{0} v_{th}^{3}\sum_{\sigma}\int_{0}^{2\pi} d\vartheta \int_{0}^{1/B_{0}}\frac{d\lambda}{\sqrt{1-\lambda B_{0}}} \int_{0}^{1}dX \ e^{x^{2}}.
\end{equation}
The integration domain is split into the subintervals $[0,X_{0})$ and $[X_{0},1)$,
with the perturbed distribution function assumed to be
approximately Maxwellian on $[X_{0},1)$.  
Gauss-Legendre quadrature rules~\cite{hilde} are then used to determine the location of the 
grid points in the interval $[0,X_{0})$.

This energy grid provides spectrally accurate energy integrals (i.e. error
$\sim (1/N)^{N}$, where $N$ is the number of energy grid points), provided the integrand
is analytic over the integration domain (see e.g. Ref.~\cite{boyd}).  Unfortunately, this is seldom the 
case.  To understand why, we consider the functional form of x(X).  Taylor
expanding $X$ about $x=0$, we find $X\sim x^{3}$, or equivalently, $x\sim X^{1/3}$.
This indicates a branch cut in $x$ originating from $X=0$, so that most functions of $x$ are 
non-analytic at $X=0$.  In Fig~\ref{fig:xX}, we examine $x(X)$.  We see that not only is $x$ 
non-analytic at $X=0$, but also at $X=1$, where $x\rightarrow \infty$.  The fact that 
$x$ possesses singularities 
at the endpoints of the domain in $X$ means that the integration scheme is not
spectrally accurate for most integrands of interest (especially since the Bessel 
functions $J_{0}(k_{\perp}v_{\perp}/\Omega)$ and $J_{1}(k_{\perp}v_{\perp}/\Omega)$, 
which are non-analytic at $X=0$ and $X=1$, appear in all integrals of the
distribution function at fixed particle position $\mbf{r}$).  This is demonstrated in 
Fig.~\ref{fig:j0newvold}, where we examine the accuracy of the numerical integral 
of $h(\mbf{R})=F_{M}$ (at fixed $\mbf{r}$) as we vary the number
of velocity space grid points.

\begin{figure}
\centering
\includegraphics[height=4.0in]{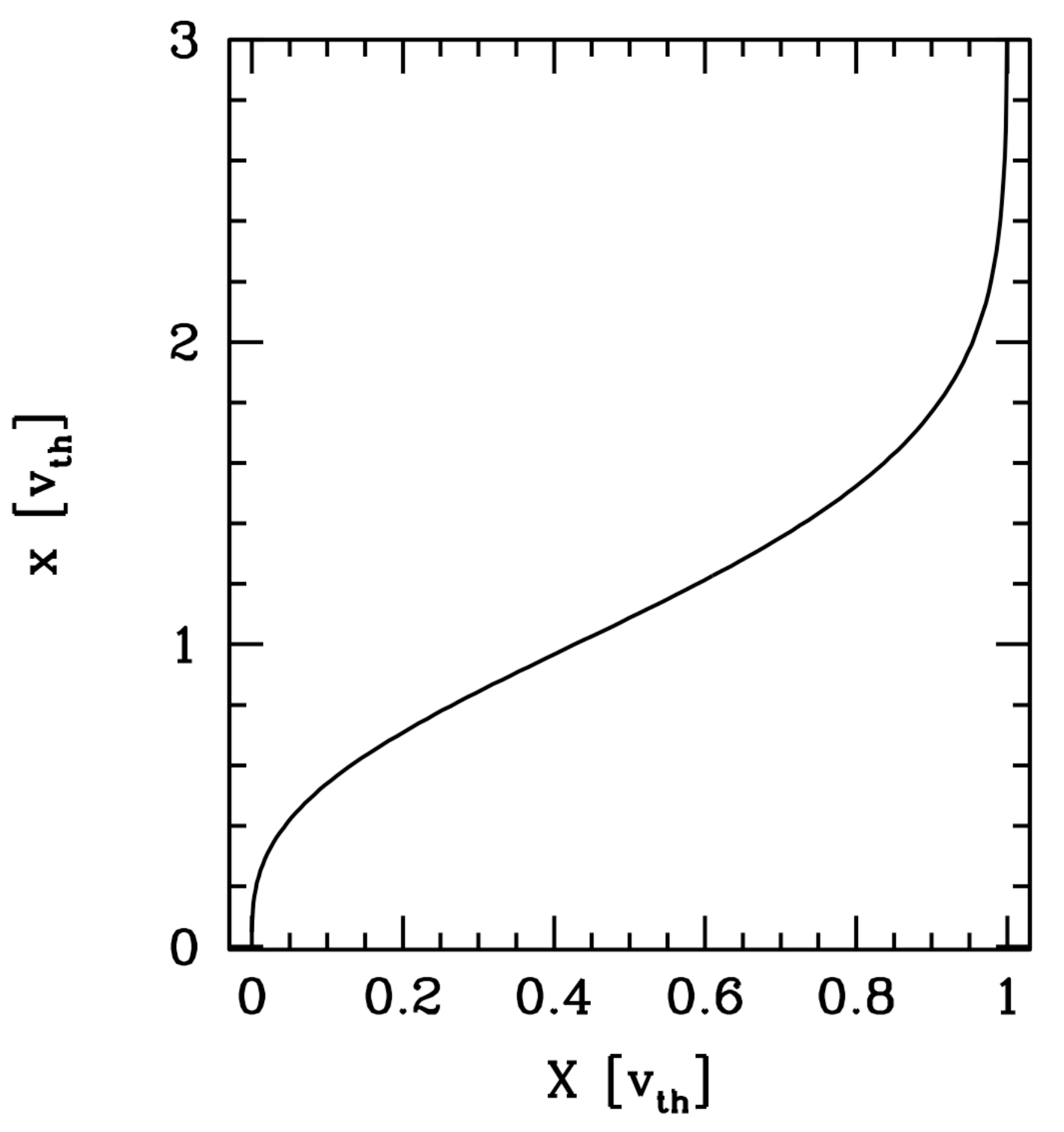}
\caption{Plot of normalized velocity $x$ over the entire $X$ domain.  The function
$x(X)$ has singularities at the boundaries of the domain due to a branch cut originating at 
$X=0$ and to $x$ going to $\infty$ at $X=1$.}
\label{fig:xX}
\end{figure}

\begin{figure}
\centering
\includegraphics[height=4.0in]{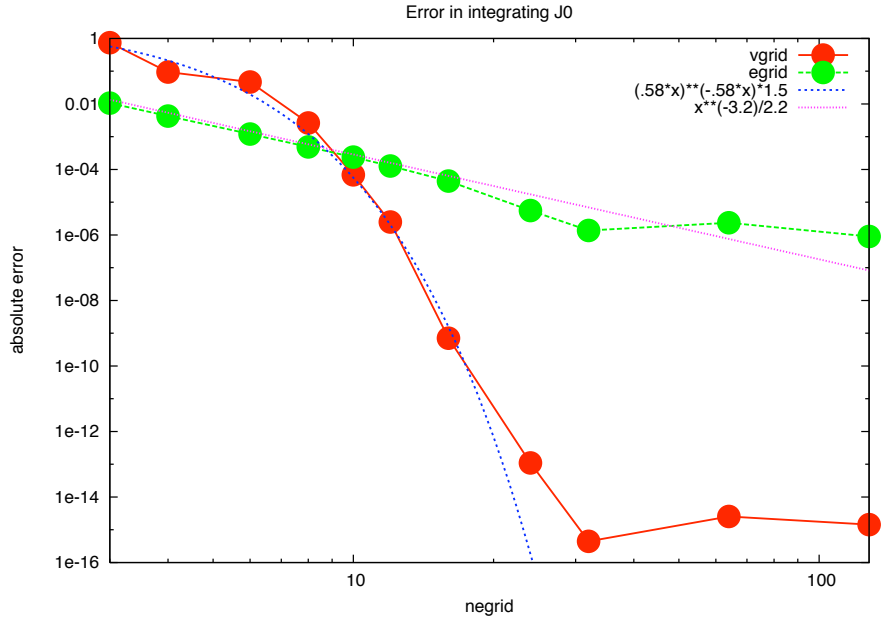}
\caption{Plot showing absolute error in numerical integral of $J_{0}(\frac{k_{\perp}v_{\perp}}{\Omega_{0}})$.
The integration scheme of Ref.~\cite{candyJCP03} has error proportional to $negrid^{-3.25}$, while our scheme
has error proportional to $0.6*negrid^{-0.6*negrid}$.  Note that the minimum error in the non-spectral scheme 
is on the order of $10^{-6}$, while in our scheme it is on the order of $10^{-16}$, which is a limitation imposed
by double precision evaluation of the Bessel function.}
\label{fig:j0newvold}
\end{figure}

In order to achieve spectral accuracy, we have implemented a new energy grid.  We begin by
splitting the velocity integration into two separate integrals:
\begin{equation}
\int_{0}^{\infty} dx \ x^{2} G(x) = \int_{0}^{x_{0}} dx \ x^{2} G(x) + \int_{x_{0}}^{\infty} dx \ x^{2} G(x),
\end{equation}
where $x_{0}$ is a free parameter and $G(x)$ is the function we wish to integrate.  On the 
first interval, ($0, x_{0}$), we use
Gauss-Legendre quadrature rules in $x$ to obtain grid locations.  Note that use of $x$
as our integration variable ensures that the integrand $x^{2} G(x)$
will be analytic as long as $G$ is analytic in $x$ over the interval.

For the interval $(x_{0},\infty)$ we make the change of variable 
$y\equiv x^{2}-x_{0}^{2}$ to transform the integral to
\begin{equation}
\int_{x_{0}}^{\infty} dx \ x^{2} G(x)
= \frac{1}{2}\int_{0}^{\infty} dy \ e^{-y}\left[ e^{y}\sqrt{y+x_{0}^{2}} G(x)\right].
\end{equation}
We then use Gauss-Laguerre quadrature rules in $y$ to obtain grid locations.
Note that the volume element is analytic within the domain of integration, as is 
$x(y)=\sqrt{x_{0}^{2}+y}$, so that the integrand will be analytic
as long as $G$ is an analytic function of $x$.

Our use of spectral integration techniques (i.e. Gaussian quadrature), coupled with
the analyticity of our integrand for well-behaved functions $G(x)$, ensures
the spectral accuracy of our integration scheme.  While an exponential order of convergence
is assured, the rate of convergence depends on the exact nature of the integrand and our
choice of the parameter $x_{0}$.  In general we choose $x_{0}\gtrsim 2.5$ so that the branch
cut at $y=-x_{0}^{2}$ is sufficiently far from the domain of integration in $y$ to minimally
impact the rate of convergence.  We demonstrate the spectral
accuracy of the scheme and determine the rate of convergence for a number
of test functions in Figs.~\ref{fig:j0newvold} and~\ref{fig:specconv}.  It is worthwhile
to note that for few grid points ($\lesssim 8$ in Fig.~\ref{fig:j0newvold}) 
the grid given in Ref.~\cite{candyJCP03} may be more accurate.  This is because 
the energy variable $X$
eliminates velocity-dependence of the volume element (when solving for the normalized 
distribution function $\tilde{h}\equiv h/F_{0}$), while the new v-space integrals 
described here have the velocity-dependent volume element $x^{2}e^{-x^{2}}$ that must be 
integrated regardless of the form of $\tilde{h}$.

\begin{figure}
\centering
\includegraphics[height=5.0in]{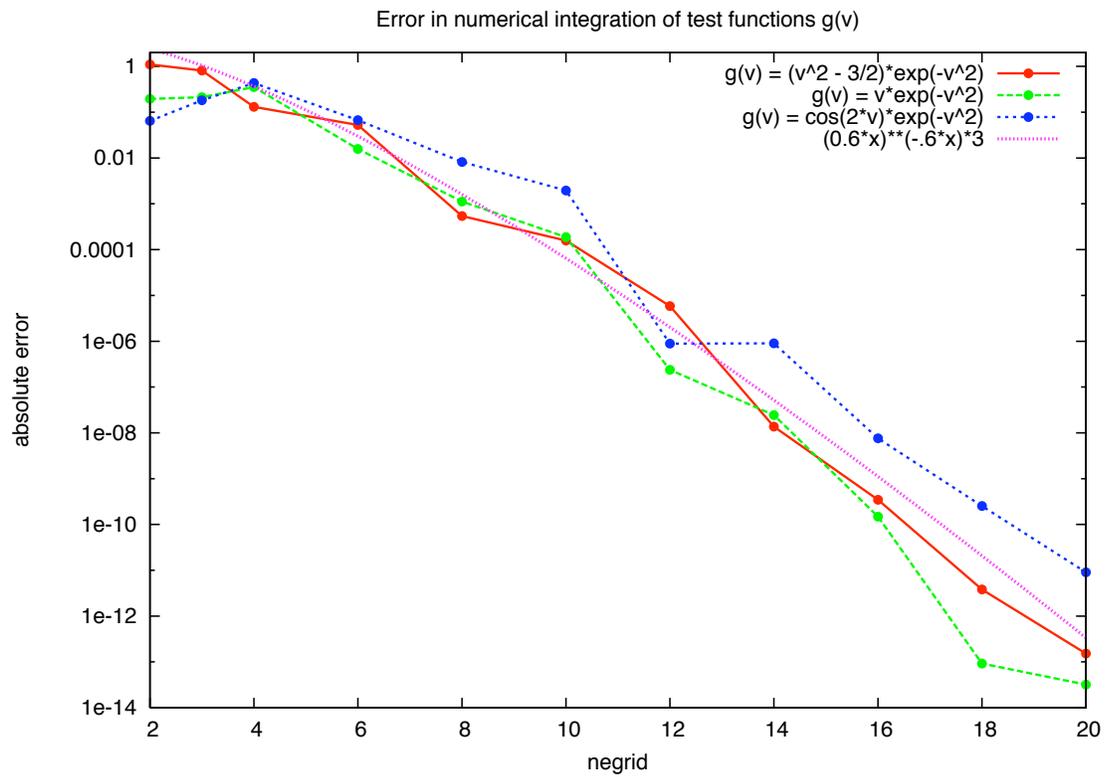}
\caption{Plot showing absolute error in numerical integral of a number of test functions, $g(v)$.
Our integration scheme has a rate of convergence proportional to approximately $0.6*negrid^{-0.6*negrid}$.  
Note that the minimum error approaches $10^{-16}$, which is a limitation imposed
by double precision arithmetic.}
\label{fig:specconv}
\end{figure}

\subsubsection{Lambda grid}

For systems with curved magnetic field lines, special care is also required when dealing with 
$\lambda$~\cite{kotschCPC95}.  There are two
reasons for this:  the grid points provided by Gaussian quadrature rules are concentrated
near the endpoints of the domain, whereas one would like them to be concentrated at
the trapped-passing boundary; and one must ensure that the proper boundary condition
(i.e. $f(v_{\parl}=0^{+})=f(v_{\parl}=0^{-})$) is satisfied at each of the bounce points.  Consequently, the $\lambda$-grid is divided
into two regions corresponding to trapped and untrapped particles, respectively.

For values of $\lambda$ such that $0 \leq \lambda < 1/B_{max}$, the
corresponding particles are untrapped by the magnetic potential well.  In this region
of velocity space, the integration variable $\tilde{\xi}\equiv\sqrt{1-\lambda B_{max}}$ is 
chosen.  It is similar to pitch-angle, but it has no spatial dependence.  Similarly to the
energy, Gauss-Legendre quadrature rules are used to obtain the location of grid points
in  $\tilde{\xi}$.  This naturally provides a concentration of gridpoints near the 
trapped-passing boundary.

For values of $\lambda$ such that $1/B_{max} < \lambda < 
1/B_{min}$, the corresponding particles are trapped by the magnetic potential well.
In the trapped region, grid points are chosen to fall on bounce points in order to
allow for the enforcement of boundary conditions.  Mathematically, this means that
for each value of $\theta$, there must be a corresponding 
$\lambda$ such that
\begin{equation}
\xi(\theta)=\frac{v_{\parl}(\theta)}{v}=\sqrt{1-\lambda B_{0}(\theta)} = 0,
\end{equation}
where $\theta$ gives the position along the unperturbed
magnetic field line and $\xi$ is the pitch-angle.  This choice of $\lambda$ values also leads 
to a concentration 
of grid points near the trapped-passing boundary.  A typical \verb#Trinity# grid layout for a system 
with trapped particles is shown in Fig.~\ref{fig:GS2v}.  It should be noted that
the $\lambda$ integrals, like the energy integrals, are spectrally accurate, 
provided the distribution function is analytic in $\xi$.

\begin{figure}
\centering
\includegraphics[height=5.0in]{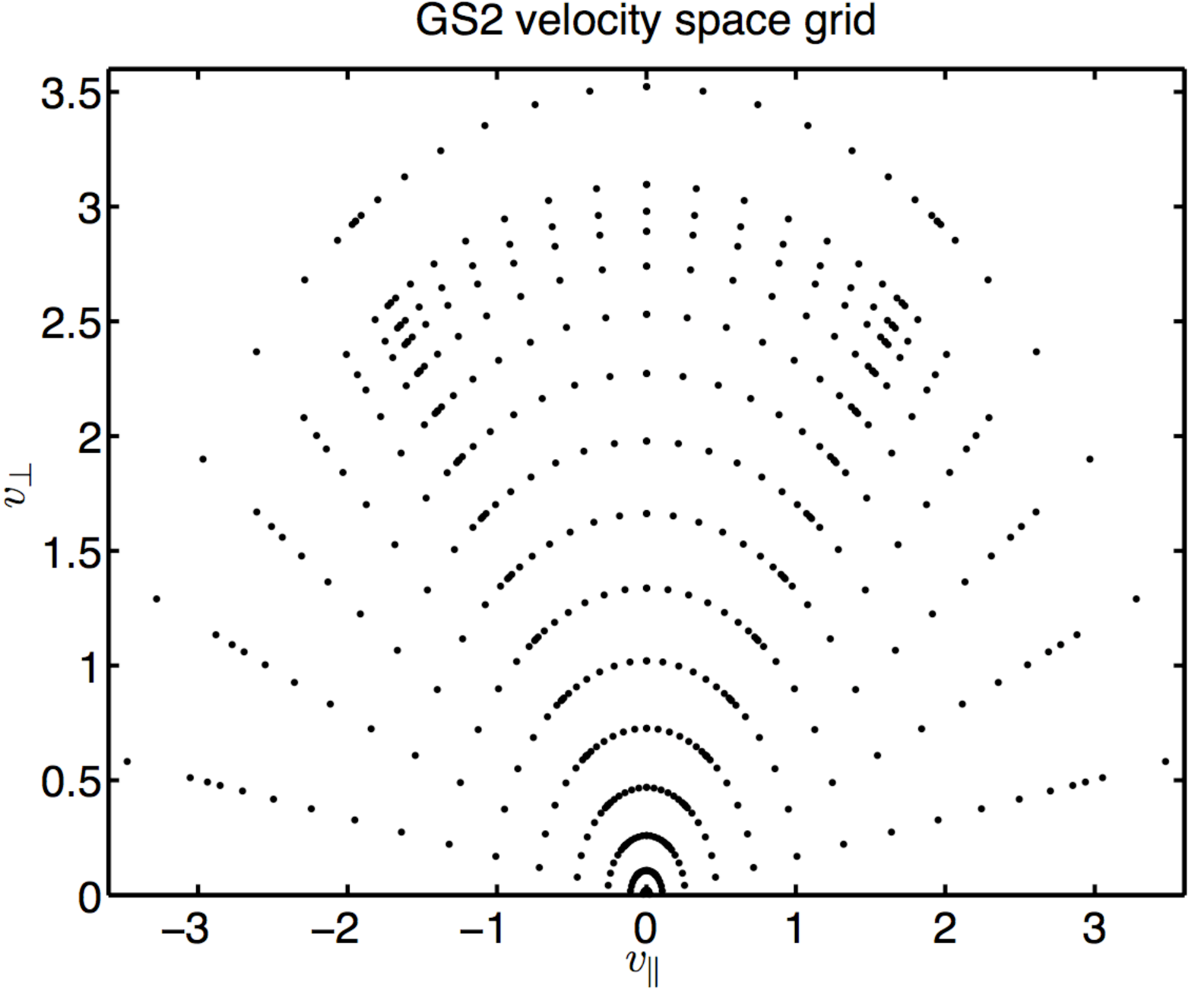}
\caption{Typical velocity space grid used in \texttt{Trinity}.  Grid points are concentrated near
the trapped-passing boundary (whose location varies with $\theta$) and at lower
energy values where the Maxwellian weighting dominates.}
\label{fig:GS2v}
\end{figure}

\subsection{Velocity space dissipation}

Some form of dissipation is often necessary to prevent the formation
of arbitrarily small-scale structures in velocity space.  This can be achieved either through
artificial numerical dissipation or through implementation of a model collision operator.
Both options are available in \verb#Trinity#.

\subsubsection{Model collision operator}

\verb#Trinity# uses a model Fokker-Planck collision operator that includes the effects of pitch-angle
scattering and energy diffusion while satsifying Boltzmann's H-Theorem and conserving 
particle number, momentum, and energy~\cite{abelPoP08, barnesPoP08}:
\begin{equation}
C[h] = \mathcal{L}[h] + \mathcal{D}[h] + \mathcal{M}[h],
\label{eqn:collop}
\end{equation} 
where
\begin{equation}
\mathcal{L}[h] = \frac{\nu_{D}}{2}\left(\pd{}{\xi}\left(1-\xi^{2}\right)\pd{h}{\xi} + \frac{1}{1-\xi^{2}}\pd{^{2}h}{\vartheta^{2}}\right)
\end{equation}
is the Lorentz collision operator,
\begin{equation}
\mathcal{D}[h] = \frac{1}{4x^{2}} \pd{}{x}\left(\nu_{s}x^{2}F_{0} \pd{}{x}\frac{h}{F_{0}}\right)
\end{equation}
is the energy diffusion operator, and $\mathcal{M}[h]$
contains momentum- and energy-conserving corrections.  We defer a detailed discussion
of the collision operator to Chapter 5.  Here we simply present the gyroaveraged
collision operator in spectral form:
\begin{equation}
\begin{split}
\left< C[h] \right>_{k} &= \frac{\nu_{D}}{2}\pd{}{\xi}\left(1-\xi^{2}\right)\pd{h_{k}}{\xi} + \frac{v_{th}^{2}}{4v^{2}}\pd{}{v}\left(\nu_{s}v^{2}F_{0}\pd{}{v}\frac{h_{k}}{F_{0}}\right) \\
&- \frac{k_{\perp}^{2}\rho^{2}}{8\Omega_{0}^{2}}\left(\frac{2v^{2}}{v_{th}^{2}}\nu_{D}\left(1+\xi^{2}\right)+\nu_{s}\left(1-\xi^{2}\right)\right)h_{k}+\nu_{E} v^{2} J_{0}(a) F_{0} \frac{\int d^{3}v \ \nu_{E} v^{2} J_{0}(a) h_{k}}{\int d^{3}v \ \nu_{E} v^{4} F_{0}}\\
&+\nu_{D}F_{0}\Big(J_{0}(a)v_{\parl}
\frac{\int d^{3}v \ \nu_{D}v_{\parl}J_{0}(a)h_{\mbf{k}}}{\int d^{3}v \ \nu_{D}v_{\parl}^{2}F_{0}} 
+ J_{1}(a)v_{\perp}\frac{\int d^{3}v \ \nu_{D}v_{\perp}J_{1}(a)h_{\mbf{k}}}{\int d^{3}v \ \nu_{D}v_{\parl}^{2}F_{0}}\Big)\\
&-\Delta \nu F_{0}\Big(J_{0}(a)v_{\parl}
\frac{\int d^{3}v \ \Delta \nu v_{\parl}J_{0}(a)h_{\mbf{k}}}{\int d^{3}v \ \Delta \nu v_{\parl}^{2}F_{0}} 
+ J_{1}(a)v_{\perp}\frac{\int d^{3}v \ \Delta \nu v_{\perp}J_{1}(a)h_{\mbf{k}}}{\int d^{3}v \ \Delta \nu v_{\parl}^{2}F_{0}}\Big)
\end{split}
\label{gyroC}
\end{equation}
where $k$ is the wavenumber and $a\equiv kv_{\perp}/\Omega_{0}$.  The velocity-dependent
collision frequencies $\nu_{s}$, $\nu_{D}$, $\nu_{E}$, and $\Delta\nu$ are given by
\begin{gather}
\nu_{s} = \frac{2\nu_{s u}}{x^{3}}\left(\erf[x] - \frac{2xe^{-x^{2}}}{\sqrt{\pi}}\right)\\
\nu_{D} = \frac{1}{x^{2}}\left(\nu_{s u}\frac{\erf[x]}{x} - \frac{\nu_{s}}{4}\right)\\
\nu_{E} = \frac{2\nu_{s u}}{x^{3}}\left(\erf[x] - \frac{4xe^{-x^{2}}}{\sqrt{\pi}}\right)\\
\Delta\nu = \nu_{D}-\nu_{s},
\end{gather}
with $\nu_{su}$ the frequency of collisions of particles of species $s$
with particles of species $u$.  Details on numerical implementation of the collision 
operator (\ref{gyroC}) are given in Chapter 6.

\subsubsection{Numerical dissipation}

Numerical dissipation enters in \verb#Trinity# through two mechanisms.  The first is the optional 
decentering of spatial and temporal finite differences, as described in Ref.~\cite{kotschCPC95}.  
The lowest order contribution to dissipation due to decentering in time and space is
\begin{equation}
\pd{^{2}h}{t \partial \theta}\left[\Delta \theta \left(\delta - \frac{1}{2}\right)+ \left(v_{\parl}\right)_{j+1/2}\Delta t\left(\beta - \frac{1}{2}\right)\right]
-\pd{^{2}\left<\chi\right>_{\mbf{R}}}{t \partial \theta}\left[\Delta \theta \left(\delta - \frac{1}{2}\right)\frac{qF_{0}}{T}\right],
\end{equation}
where $\Delta \theta$ is the grid spacing along the field line, $\Delta t$ is the time step
size, and $\delta$ and $\beta$ are parameters that allow for the variation of the
spatial and temporal discretization schemes between fully explicit ($\delta$ or $\beta=0$)
and fully implicit ($\delta$ or $\beta=1$)\footnote{\texttt{Trinity} actually uses $\tilde{\beta}=\beta-1/2$, 
but we choose to use $\beta$ here for simplicity.}

In order to see how this term leads to
dissipation, we consider the simplified system governed by the equation
\begin{equation}
\pd{h}{t} + v\pd{h}{\theta}= 0.
\end{equation}
Finite differencing this equation using the scheme given in Ref.~\cite{kotschCPC95}, we
find that numerically we are solving the equation
\begin{equation}
\pd{h}{t} + v\pd{h}{\theta} \approx -\pd{^{2}h}{t \partial \theta}\left[\Delta \theta \left(\delta - \frac{1}{2}\right)+ v\Delta t\left(\beta - \frac{1}{2}\right)\right].
\end{equation}
Assuming $h=\tilde{h}(t) e^{ik\theta}$, we obtain the solution
\begin{equation}
\tilde{h}(t) \sim \exp\left[\frac{kvt}{i-k\left(\Delta \theta \left(\delta-1/2\right) 
+ v\Delta t\left(\beta-1/2\right)\right)}\right],
\label{eqn:numdiss}
\end{equation}
which is damped unless $\beta=\delta=1/2$, as show in Fig.~\ref{fig:numdiss}.
While decentering of finite differences can sometimes improve numerical stability, care must
be taken to ensure such artificial dissipation does not lead to unphysical behavior.  This
is typically done by monitoring the ratio of artificial to physical dissipation, which, ideally, 
should be small.

\begin{figure}
\centering
\includegraphics[height=4.5in]{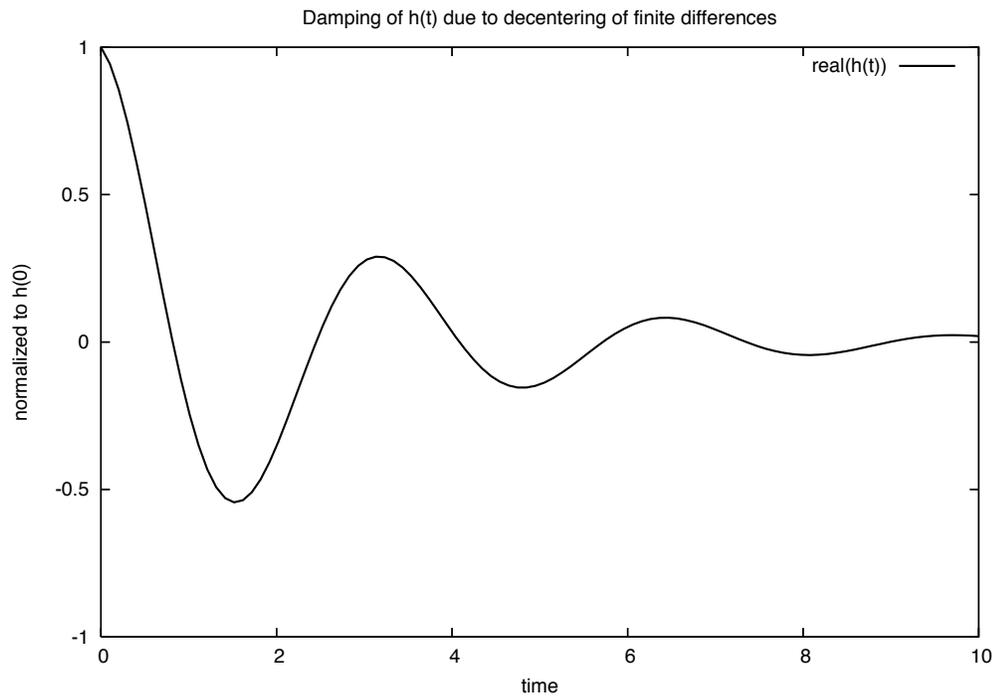}
\caption{Damping of the real part of the distribution function $h$ [Eq.~(\ref{eqn:numdiss})]
as a result of decentered finite differences in space and time.  Here, we are considering 
$v=1$, $k=2$, $\Delta x=\Delta t=0.1$, and $\beta=\delta=1.0$ (fully implicit).}
\label{fig:numdiss}
\end{figure}

The second source of numerical
dissipation arises in systems with sheared magentic fields due to the necessity of a
'twist-and-shift' parallel boundary 
condition~\cite{beerPoP95}.  This non-periodic boundary condition couples modes at 
opposite ends of the simulation domain 
along the field line.  Since only a finite number of modes can be kept in a simulation,
some modes will eventually couple to modes that are not present, and this information is
lost.  The information that is lost is replaced by a smoothed distribution function, leading
to a loss of entropy.  If a sufficiently large number of modes are kept in the simulation, the
energy contained in the highest modes, and therefore the lost entropy, should be negligible.
Currently, this entropy loss is not regularly diagnosed in \verb#Trinity#.  In
principle, it could (and should) be diagnosed in order to verify that the entropy lost is
small compared to the entropy generated by collisions.

\section{Velocity space resolution diagnostics}
\label{sec:vresdiag}

There are numerous ways in which one could try to determine whether or not a particular
simulation is well-resolved in velocity space.  
Ideally, one would perform a grid convergence study for each simulation; if 
quantities of interest are unchanged by doubling the number of grid points, one can feel
relatively confident in the simulation results.  However, this process is computationally
expensive, as it involves running a simulation multiple times with an excessive number
of grid points.  Consequently, it is not desirable to perform a grid convergence study for every
simulation.  In practice, one tests convergence for a problem thought to be resolution intensive 
and posits that other simulations, which likely require fewer grid points, are therefore 
resolved.  Unfortunately, one seldom knows in advance how fine the structure in
velocity space will become, so one can't be fully confident that every simulation is resolved.

An alternative approach that has recently gained popularity in the computational plasma 
physics community
involves monitoring entropy balance in the system~\cite{watanabePoP04,candyPoP06}. 
Multiplying the gyrokinetic equation (\ref{eqn:gke}) by $hT_{0}/F_{0}$ and
integrating over all phase space gives the desired relation for entropy balance:
\begin{equation}
\frac{1}{2} \pd{S}{t} =  X + \Gamma + H,
\label{eqn:sbalance}
\end{equation}
where
\begin{equation}
S=\int d^{3}\mbf{r}\int d^{3}\mbf{v} \ \frac{T_{0}}{F_{0}}h^{2}
\end{equation}
is a lowest order expression for the perturbed entropy,
\begin{equation}
X = \int d^{3}\mbf{r}\int d^{3}\mbf{v} \ q\pd{\left< \chi \right>}{t}h
\end{equation}
describes turbulent heating,
\begin{equation}
\Gamma=-\int d^{3}\mbf{r}\int d^{3}\mbf{v} \ hT_{0}
\mbf{v_{\chi}}\cdot \nabla \ln F_{0}
\end{equation}
is the entropy flux due to background inhomogeneity, and
\begin{equation}
H=\int d^{3}\mbf{r}\int d^{3}\mbf{v} \ \frac{hT_{0}}{F_{0}}\left< C[h] \right>
\end{equation}
describes entropy change due to collisional heating.

Since the gyrokinetic equation itself is automatically satisfied by a gyrokinetic solver,
the only possible sources of inbalance in Eq.~(\ref{eqn:sbalance}) come from numerical 
dissipation and errors in the numerical approximations to phase space integrals.  If
the change in entropy due to numerical dissipation is also diagnosed and included
in the entropy balance, as is often the case, then we are left with errors due only
to phase space integration.  Since the errors in these particular integrals are not directly
related to errors in the calculation of the distribution function at the newest timestep,
they do not necessarily correlate with the simulation resolution.  In
particular, one could easily define a poorly-resolved system for which this diagnostic
predicts perfect entropy balance.  One such example is the linear, collisionless ion acoustic 
wave in a slab (treated in detail in Appendix~\ref{app:landamp}).  For this case, we numerically find 
entropy balance despite the fact that the numerical damping rate goes bad due to poor 
resolution in velocity space.

Of course, one could simply produce plots or movies of the distribution function in 
velocity space over the course of the simulation to see if structure develops at the gridscale. 
This is undoubtedly useful and possibly sufficient in some cases.  However, what exactly 
one sees depends on how the data is visualized; for data on irregularly spaced grids, the
interpolation scheme used to generate the images often introduces erroneous or misleading
structure.  Furthermore, for simulations involving non-trivial spatial structure,  one would 
have to examine movies of the distribution function at each point in physical space.  
This is a memory- and time-intensive approach that is rarely feasible.

We would like to have computationally cheap diagnostics that provide real-time 
information on velocity space resolution that is easy to analyze and interpret.  
In the following subsections, 
we present two such diagnostics developed for implementation in \verb#Trinity# that could easily
be adapted for use in other continuum kinetic simulations.

\subsection{Integral error estimates}

Upon consideration of the collisionless gyrokinetic-Maxwell's system of equations, one 
finds that the only
nontrivial operation in velocity space is integration, which enters in the calculation of the
electromagnetic fields.  Consequently, resolution in velocity space is limited only by the
accuracy with which the velocity space integrals are calculated.  By calculating the error
in our numerical integration, we are thus able to monitor velocity space resolution.

In particular, when we discretize the gyrokinetic equation, we obtain an equation of the form
\begin{equation}
g_{j+1} = G\left[g_{j},\Phi_{j},\Phi_{j+1},\chi_{j},\chi_{j+1}\right],
\label{eqn:gnew}
\end{equation}
where $g\equiv\left< f_{1} \right>$ is the perturbed, guiding center distribution function, $\Phi$ is the 
electrostatic potential, $\chi$ is the generalized
electromagnetic potential defined in Eq.~(\ref{eqn:chi}), $G$ is 
a function that depends on the details of the numerical scheme, and the subscript denotes
the timestep.  We assume that the 
time-converged solution for $g$ is independent of the initial condition.  Since
using the calculated $g_{j}$ and $\Phi_{j}$ is equivalent to specifying a new initial 
condition, we find that the time-converged solution is independent of errors in $g$ and 
$\Phi$ at earlier timesteps.  This is convenient because it means we can monitor
resolution merely by calculating the error made in the latest timesteps of a time-converged
simulation.

Ideally, we would accomplish this 
by calculating estimates for the error in $\Phi_{j+1}$ and $\chi_{j+1}$ and plugging
these into Eq.~(\ref{eqn:gnew}) to obtain an error estimate for $g_{j+1}$.  This
might be feasible for linear systems, but the presence of nonlinear terms makes this approach
computationally prohibitive.  Consequently, we must define an alternative quantity whose 
error estimate is cheaper to compute, but that can still be used as a means
of monitoring velocity space resolution.  There are numerous possible candidates; we 
choose to compute two quantities, $v_{\Phi}$ and $v_{A}$, related to $\nabla_{\perp}\Phi$
and $\nabla_{\perp}A_{\parl}$:
\begin{equation}
 \left(\begin{array}{c} v_{\Phi}\\
v_{A}\end{array}\right) = \textnormal{max}\{k_{x},k_{y}\}  \left(\begin{array}{c} \Phi(\theta, k_{x},k_{y})\\
A_{\parl}(\theta,k_{x},k_{y})\end{array}\right),
\label{eqn:errquant}
\end{equation}
where $k_{x}$ and $k_{y}$ are the wavenumbers corresponding to the $x-y$ coordinates 
$x\equiv \left(\psi - \psi_{0}\right)q_{0}/B_{0}r_{0}$ and
$y\equiv-\left(\alpha - \alpha_{0}\right)r_{0}/q_{0}$~\cite{beerPoP95}.
Here, $\psi$ is the 
poloidal flux, $\alpha$ is the field line label, $B_{0}$ is the background magnetic field at 
the magnetic axis,
$r_{0}$ is the distance from the magnetic axis to the center of the simulation domain, and
$q_{0}$ is the safety factor on the field line of interest, labeled by 
$(\psi_{0},\alpha_{0})$.
The quantities in Eq.~(\ref{eqn:errquant}) were chosen because, with the exceptions of the 
parallel convection term and one source term,
$\Phi$ and $A_{\parl}$ always enter the gyrokinetic equation for $g$ multiplied by
either $k_{x}$ or $k_{y}$.
Therefore, it is reasonable that this $k$-weighted quantity is most likely to be responsible
for errors in $g_{j+1}$.  Although not considered here, the expression (\ref{eqn:errquant})
could potentially be improved by including $k_{\parl}$ in the max operator.  This would
take into account the effect of the parallel convection term.  However, there is recent
theoretical~\cite{schekAPJ07} and numerical~\cite{tatsunoPRL08,tatsunoPoP08} evidence which
suggests that velocity space structure is primarily generated by nonlinear perpendicular phase 
mixing (instead of linear, parallel phase mixing).

Having chosen appropriate
indicators of velocity space resolution, we must devise a method for estimating the
error in these quantities.  This error depends on the particular numerical 
integration scheme used.  For
the energy and untrapped $\lambda$ integrals, which use Gaussian
quadrature, the error, $\epsilon_{G}$, is given by
\begin{equation}
\epsilon_{G} = \gamma_{m}f^{(2m)}(\zeta),
\label{eqn:gauss}
\end{equation}
where $f$ is the integrand, $m$ is the number of grid points, and $\zeta$ is some 
unkown point in the interval of integration.  The quantity $\gamma_{m}$ is
\begin{equation}
\gamma_{m} = \frac{2^{2m+1}\left(m!\right)^{4}}{\left(2m+1\right)\left[\left(2m\right)!\right]^{3}}
\label{eqn:gaussleg}
\end{equation}
for the untrapped $\lambda$ and finite domain energy integrals that use Gauss-Legendre
quadrature and
\begin{equation}
\gamma_{m} = \frac{\left(m!\right)^{2}}{\left(2m\right)!}
\label{eqn:gausslag}
\end{equation}
for the semi-infinite domain energy integral that uses Gauss-Laguerre quadrature.
The error, $\epsilon_{L}$, for the trapped $\lambda$ integrals, 
which use a newly upgraded integration scheme based on Lagrange interpolating 
polynomials (see, e.g. Ref.~\cite{hilde}), is given by
\begin{equation}
\epsilon_{L} = \frac{1}{m!}\int f^{(m)}(\zeta)  \pi(x)dx,
\label{eqn:lagrange}
\end{equation}
where
\begin{equation}
\pi(x)=\prod_{i=1}^{m}\left(x-x_{i}\right),
\end{equation}
with $x_{i}$ the $i^{th}$ grid point.  It should be noted that  $\zeta$ in Eqn 
(\ref{eqn:lagrange}) is an unknown function of $x$ whose domain is some subset
of the interval of integration.

From Eqns (\ref{eqn:gauss}) and (\ref{eqn:lagrange}), we see that Gaussian
quadrature gives exact results for polynomials of degree less than $2m$, while the 
Lagrangian method gives exact results only for polynomials of degree less than $m$. 
We say that the two schemes have degrees of precision $2m-1$ and $m-1$, respectively.
This difference arises because the
grid points in the Lagrangian method are fixed by boundary conditions, whereas the 
grid points in Gaussian quadrature are free parameters optimally chosen to 
improve the scheme's degree of precision.

Unfortunately, the formal error expressions ($\ref{eqn:gauss}$) and (\ref{eqn:lagrange})
are not very useful in practice:  they require information about high-order derivatives of
the distribution function, which is unavailable.  
As an alternative estimate for the error, we choose to compare multiple integral 
approximations computed with different degrees of precision, a common technique
in numerical analysis.

\subsubsection{General description of the scheme}

Given the value of a function $f(x)$ at $N$ fixed points on the interval $[a,b]$, 
we would like to find two different approximations to the integral $\int_{a}^{b}f(x)dx$.
In our earlier discussion, we stated that an approximation with degree of precision $N-1$
can be found using a technique based on Lagrange interpolation; we call this approximation
$A_{h}$.  If we instead choose
to use only $M$ of the given functional values ($M<N$), we can use the same technique to 
find another integral approximation, $A_{l}$, with degree of precision $M-1$.  An 
estimate for the absolute error $\epsilon_{a}$ in the less accurate of these two 
approximations is obtained by taking the difference between the two:
\begin{equation}
\epsilon_{a} = \left|A_{h} - A_{l}\right|.
\end{equation}
Making the reasonable assumption that the approximation with higher degree of precision
is more accurate, $\epsilon_{a}$ represents the error in $A_{l}$.  However, it can
also be used as a more conservative error estimate for $A_{h}$.

If the $N$ points are chosen according to Gaussian quadrature rules, then one can find
an integral approximation with degree of precision $2N-1$.  As before, a second 
approximation can be obtained by using only $M$ of the $N$ grid points.  However,
due to the uniqueness of the grid points used for Gaussian quadrature, the $M$-point
grid no longer satisfies Gaussian quadrature rules.  As a result, this second 
approximation once again has degree of precision $M-1$.  Since the degrees of
precision of the two approximations differ by greater than a factor of two, the resulting
error estimate is likely to be very conservative when applied to $A_{h}$.  The factor
of approximately two difference in degree of precision makes this error estimate similar
to that obtained by comparing results from runs with $N$ and $N/2$ grid points, respectively
(for which the degrees of precision would be $2N-1$ and $N-1$).


The conservative nature of the error estimate for $A_{h}$ depends upon our assumption 
that a higher degree of precision results in a more accurate integral approximation.  
For Gaussian quadrature, it can be shown that the error in the integral approximation can be
made arbitrarily small by choosing the degree of precision large enough~\cite{hilde}.  
The same result
does not necessarily hold for the Lagrangian method with arbitrary grid spacing because
the weights in this case are not all guaranteed to be positive.  However, the 
error $\epsilon_{M}$ in an $M$-point integral approximation satisfies
\begin{eqnarray}
\epsilon_{M} &\leq 2& \epsilon \sum_{i=1}^{M}\left|w_{i}^{(M)}\right| \\
&\leq& 2 \epsilon M \max_{i=1,M} \left|w_{i}\right| \\
&=& 2 \epsilon M \kappa(M),
\end{eqnarray}
where $\epsilon$ can be chosen arbitrarily small for large enough $M$, and $w_{i}^{(M)}$
is the weight corresponding to the $i^{th}$ grid point out of $M$.  From this result,
we see that as long as $\kappa$ is bounded when $M\rightarrow\infty$, then 
$\epsilon_{M}\rightarrow0$ as $M\rightarrow \infty$.  This cannot be verified in
advance, but one can gain confidence by checking a posteriori.  In practice, we calculate
$\kappa$ for the chosen $M$ and subdivide the integration domain into subintervals
with fewer points if $\kappa$ is larger than some reasonable value.
  
 \subsubsection{Implementation in \texttt{Trinity}} 
  
  In \verb#Trinity#, we must compute two-dimensional integrals over energy and $\lambda$.  As
  stated in \secref{sec:gs2v}, each of these integrals is effectively separated into two by splitting
 the $\lambda$ integration into trapped and untrapped regions.  Since the number of grid
 points in energy and both $\lambda$ regions can be varied independently of each other,
 we wish to monitor resolution in each of these three variables individually.  This entails
 computing three separate integral
error estimates: one for energy integrals, one for
 untrapped $\lambda$ integrals, and one for trapped $\lambda$ integrals.
 
These integral error estimates are calculated using the technique described in the previous
subsection.  For energy and untrapped $\lambda$ integrals, Gaussian quadrature is used
to obtain the two-dimensional integral approximation $A_{h}$.  This approximation
has degree of precision $2N_{\varepsilon}-1$ for the energy integration and $2N_{u}-1$
for the untrapped $\lambda$ integration, where $N_{\varepsilon}$ and $N_{u}$ are the number 
of energy and untrapped $\lambda$ grid points, respectively.  To obtain the second
approximation, $A_{l}$, we fix the grid and weights for one variable and drop one grid point for the other variable,
recomputing the weights.  As an example, we choose to drop an untrapped $\lambda$ grid 
point.  The degree of precision for $A_{l}$ is then $2N_{\varepsilon}-1$ for the energy integration
and $N_{u}-2$ for the untrapped $\lambda$ integration.
Since there is nothing special about the particular grid point we drop, we repeat the process a
total of $N_{u}$ times, each time dropping a different point and computing a different set of 
weights.  The final error estimate is an average of these error estimates.

For the trapped $\lambda$ integrals, Lagrangian quadrature is used to obtain $A_{h}$,
which has degree of precision $N_{t}-1$.  We obtain the approximation $A_{l}$
by dropping two points symmetrically about $v_{\parallel}=0$, as shown in 
Fig.~\ref{fig:droptrap}.  We drop an additional point here because it provides a
slightly more conservative error estimate and because maintaining the symmetry of the 
grid points provides better stability for the weights associated with the Lagrange 
interpolation scheme.  As before, we repeat this process for each possible grid point 
pair and take the average of the individual error estimates to get the final error estimate.

\begin{figure}
\centering
\includegraphics[height=5.0in]{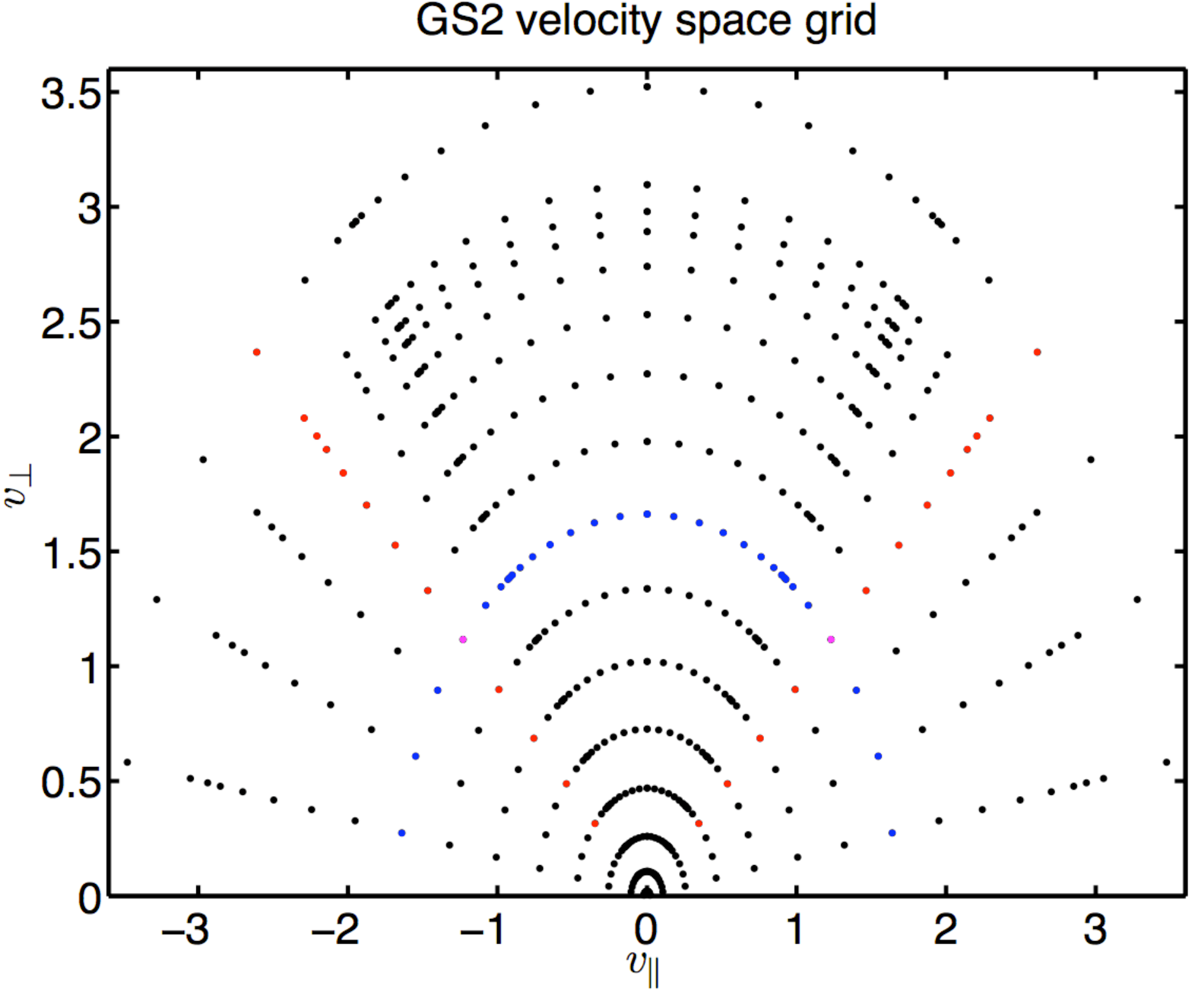}
\caption{Red grid points are sample trapped $\lambda$ grid points that are dropped
when calculating integral approximation with lower degree of precision.}
\label{fig:droptrap}
\end{figure}
  
All modified grids and weights necessary for the integral error estimates are computed once
at initialization and need not be computed again.  The additional integrations necessary to 
obtain our error estimates are computationally
cheap when compared to the expense of solving for the distribution function and fields
at each time step.  Furthermore, we do not need an error estimate at each time step, 
so the diagnostic can be used sparingly.  Consequently, our error estimate comes
at essentially no extra cost.

\subsection{Spectral method}

An alternative method for testing v-space resolution is to expand the velocity space 
distribution function in an appropriate basis set and monitor the amplitude of the basis
function coefficients.  Whenever the highest mode number coefficients that can be
accurately calculated in the simulation acquire appreciable
amplitudes, we can no longer feel confident that the simulation is resolved.  
Since we choose our grid points according to Gauss-Legendre
quadrature, it is convenient (and most accurate) to choose the Legendre polynomials as 
our basis functions.  The coefficient of the $m^{th}$ Legendre polynomial in the
expansion of $h$ is given by
\begin{eqnarray}
c_{m} &=& \frac{2m+1}{2}\int_{-1}^{1} h(s) P_{m}(s) ds \\
&\approx& \frac{2m+1}{2} \sum_{i=1}^{n_{\epsilon}-1} w_{i} h(s_{i}) P_{m}(s_{i}),
\label{eqn:cm}
\end{eqnarray}
where $P_{m}$ is the $m^{th}$ Legendre polynomial, and $\{w_{i}\}$ are the weights
associated with Gauss-Legendre quadrature.
The integral approximation in Eq.~(\ref{eqn:cm}) has degree of precision $2N-1$.  
Assuming $h$ has a degree of at least $m$ (otherwise $c_{m}=0$), our approximation
for $c_{m}$ is only exact for $m<N$.

There are various ways in which one could use these $\{c_{m}\}$ to estimate the error
in velocity space resolution.  We assume locality of interaction between the various 
modes so that we only have to monitor the amplitudes of the few highest modes.  At each 
($\theta$, $k_{x}$, $k_{y}$)-point, we find the maximum amplitude of the three highest
mode number spectral coefficients, $c_{h,max}$,
and the maximum amplitude of all the spectral coefficients, $c_{max}$.  We then use
the following normalized sum as a relative estimate for the error:
\begin{equation}
\epsilon_{c} =  \sum_{\theta,k_{x},k_{y}} c_{h,max}(\theta,k_{x},k_{y}) \bigg/
\sum_{\theta,k_{x},k_{y}} \ c_{max}(\theta,k_{x},k_{y}).
\label{eqn:specErr}
\end{equation}
When the normalized amplitude $\epsilon_{c}$ grows too large, we can no
longer be confident that the simulation is resolved.  Of course, how large $\epsilon_{c}$
can get before resolution suffers varies from problem to problem.  As before with
the integral method, we determine a scaled estimate of the error based on empirical
evidence from a wide range of simulation data.

\subsection{Application of error diagnostics}

We have applied both the integral and spectral error diagnostics to a diverse set of 
simulations, including:
linearly growing modes such as the electron drift wave and the ITG mode; 
linearly damped modes such as the ion acoustic wave and kinetic Alfven wave;
neoclassical transport;
and nonlinear dynamics of slab ETG and toroidal ITG modes.  From these simulations, 
we have determined empirical scaling factors for our conservative error estimates.  Here,
we present typical results from a cross-section of the above simulations.

\begin{figure}
\centering
\includegraphics[height=3.0in]{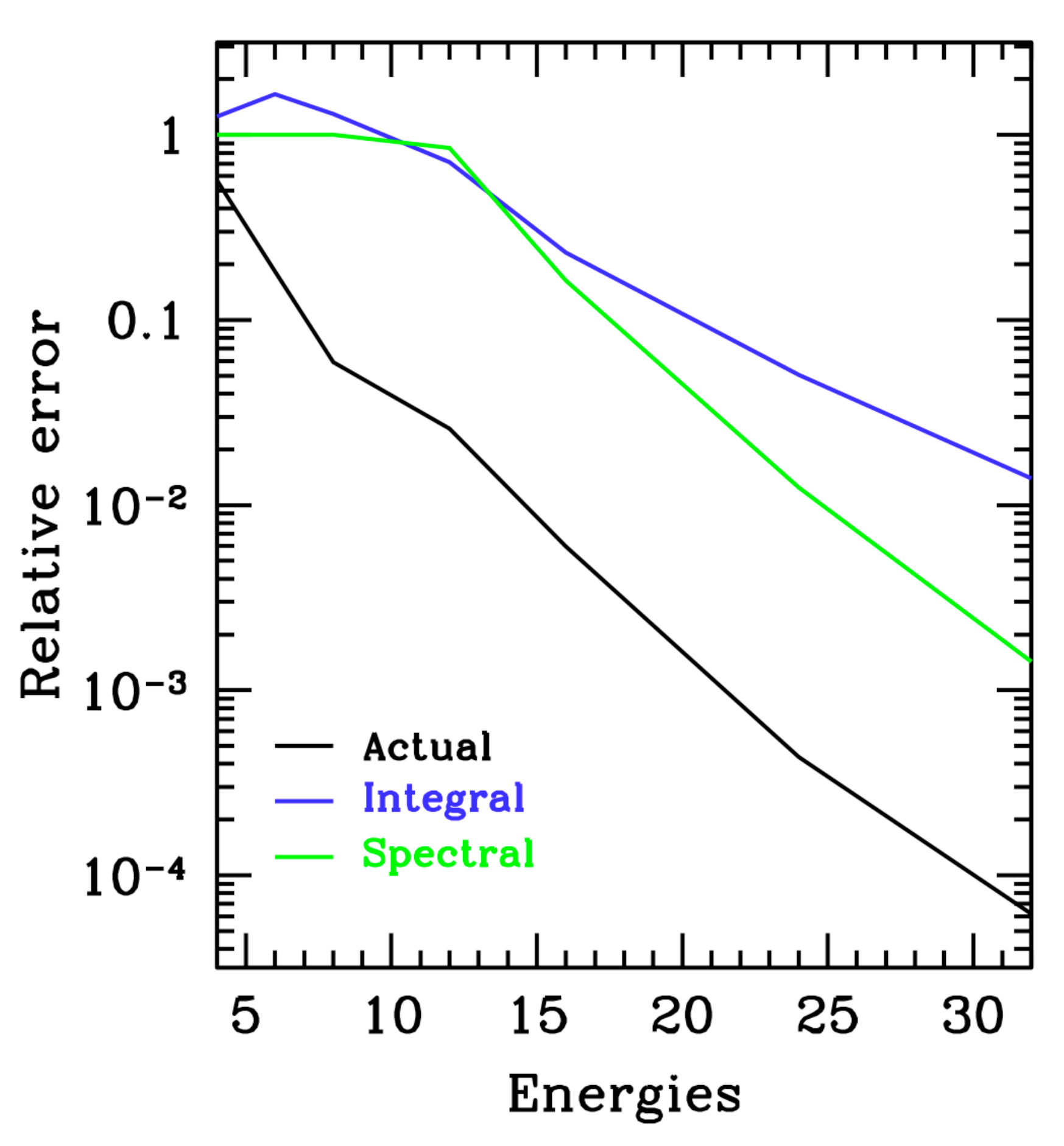}
\includegraphics[height=3.0in]{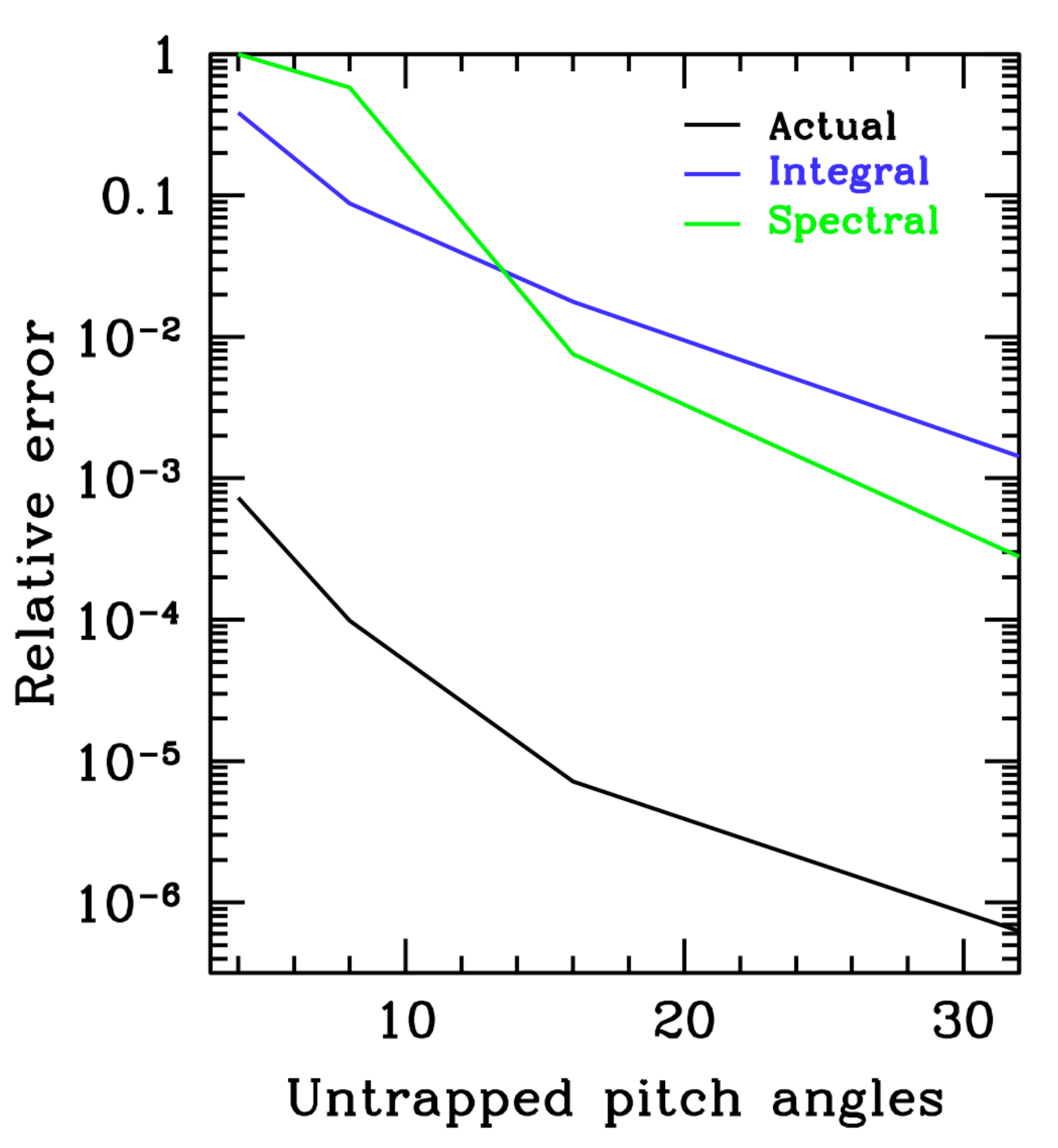}
\includegraphics[height=3.0in]{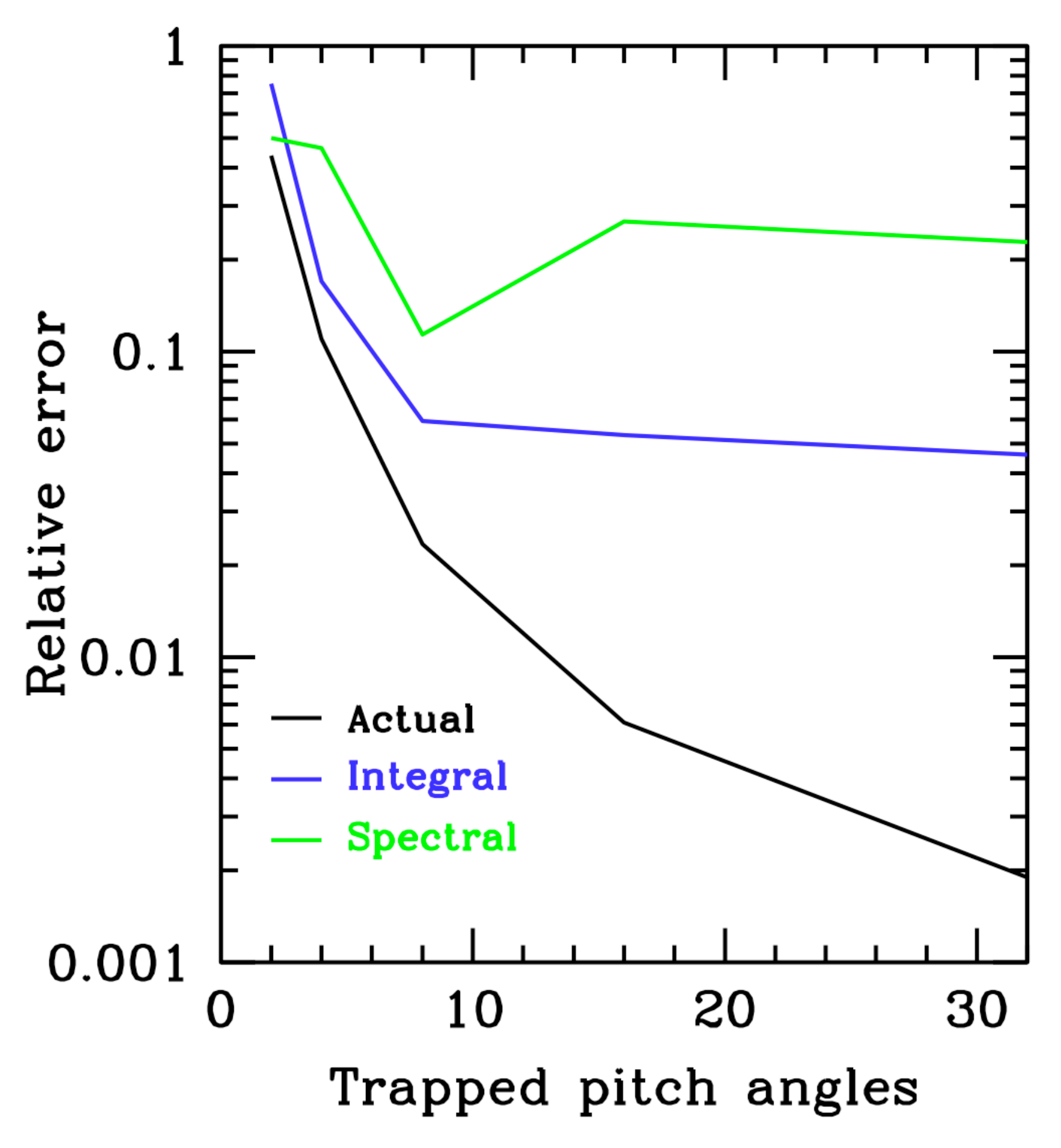}
\caption{Comparison of actual and (unscaled) estimated error in wave frequency due to insufficient
resolution in energy (top left), untrapped $\lambda$ (top right), and trapped $\lambda$ (bottom).
The actual wave frequency, $\omega$, is determined
from a higher resolution run with 64 grid points in energy and both trapped and untrapped
$\lambda$.  The actual relative error, $\epsilon$, is then defined to be $\epsilon=
\sqrt{\frac{\left|\omega-\omega_{n}\right|^{2}}{\left|\omega\right|^{2}}}$,
where $\omega_{n}$ is the approximation to $\omega$ obtained from a run with $n$
grid points.}
\label{fig:lincycerr}
\end{figure}

Fig.~\ref{fig:lincycerr} compares the unscaled error estimates 
in energy and $\lambda$ with the actual errors in
growth rate as we vary the number of grid points in a linear simulation of the collisionless
toroidal ITG mode (using Cyclone base case parameters~\cite{dimitsPoP00}).  
The simulation remains well-resolved down to very few 
grid points, and, qualitatively, the error estimates agree well with the 
actual error.  The error due to resolution in untrapped $\lambda$ is still small for as little as
four grid points due to our choice of velocity variables, as illustrated by the snapshot
of the distribution function shown in Fig.~\ref{fig:lincycdist}.



\begin{figure}
\centering
\includegraphics[height=4.0in]{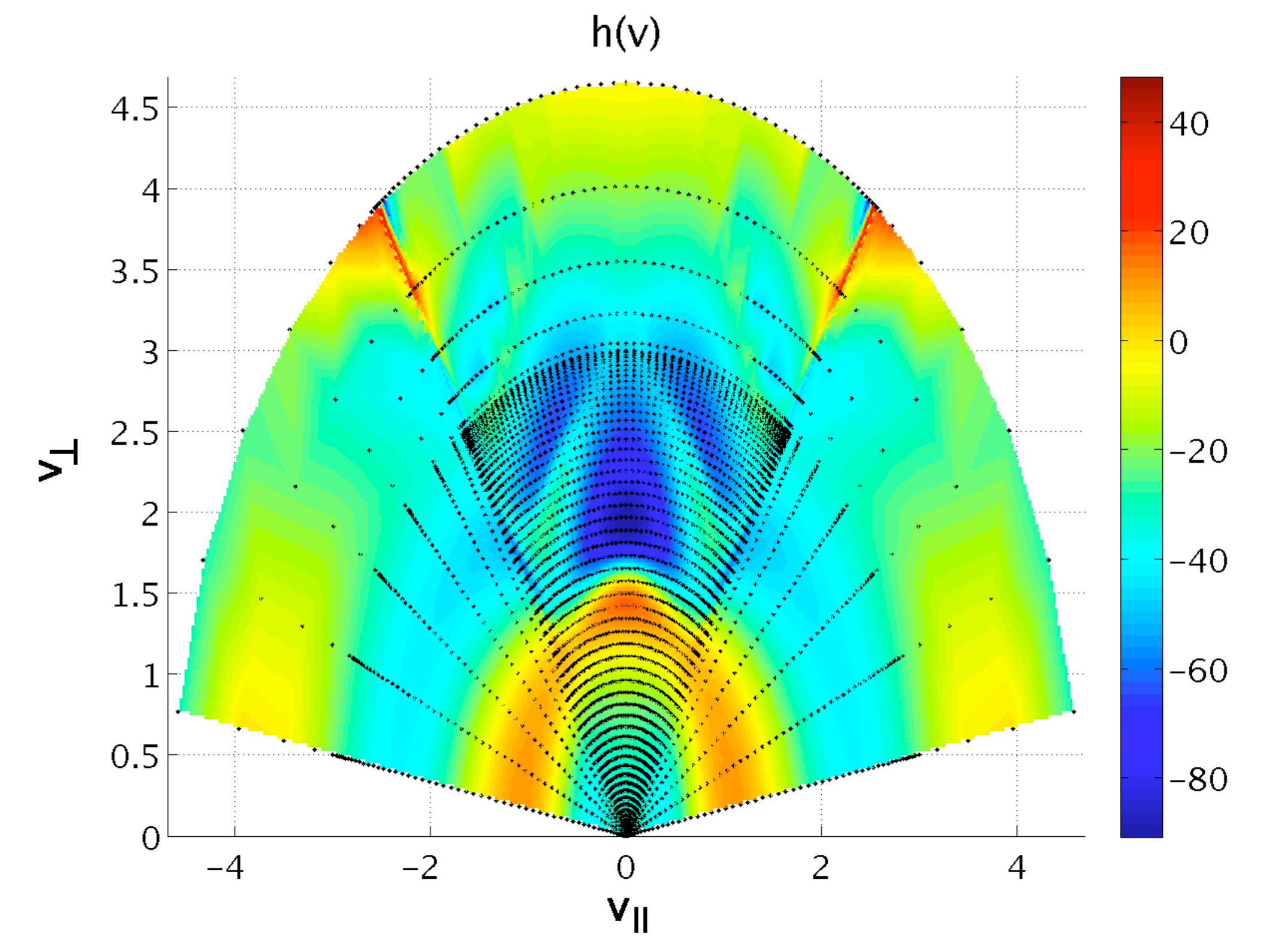}
\caption{Non-Boltzmann part of the perturbed distribution function, normalized by
$F_{0}(a/\rho_{i})$.  The use of a polar grid in velocity space minimizes the number of
grid points necessary for resolution.}
\label{fig:lincycdist}
\end{figure}

Figs.~\ref{fig:kawdamp} and~\ref{fig:kawerr} show the damping of $A_{\parl}$ and
the corresponding scaled error estimates for the simulation of a collisionless 
kinetic Alfven wave.
The collisionless damping rate in Fig.~\ref{fig:kawdamp} agrees with theory until 
sub-gridscale structure develops in velocity space, at which point damping ceases.  The
onset of sub-gridscale structure corresponds to the peak in scaled error in 
Fig.~\ref{fig:kawerr}.  The addition of a small 
collisionality prevents sub-gridscale structure, as shown in Fig.~\ref{fig:kawdamp}, 
where the damping rate of $A_{\parl}$
agrees well with theory indefinitely.  This is accurately predicted by the error estimates of 
Fig.~\ref{fig:kawerrcoll}, which never reach appreciable magnitude.

\begin{figure}
\centering
\includegraphics[height=3.0in]{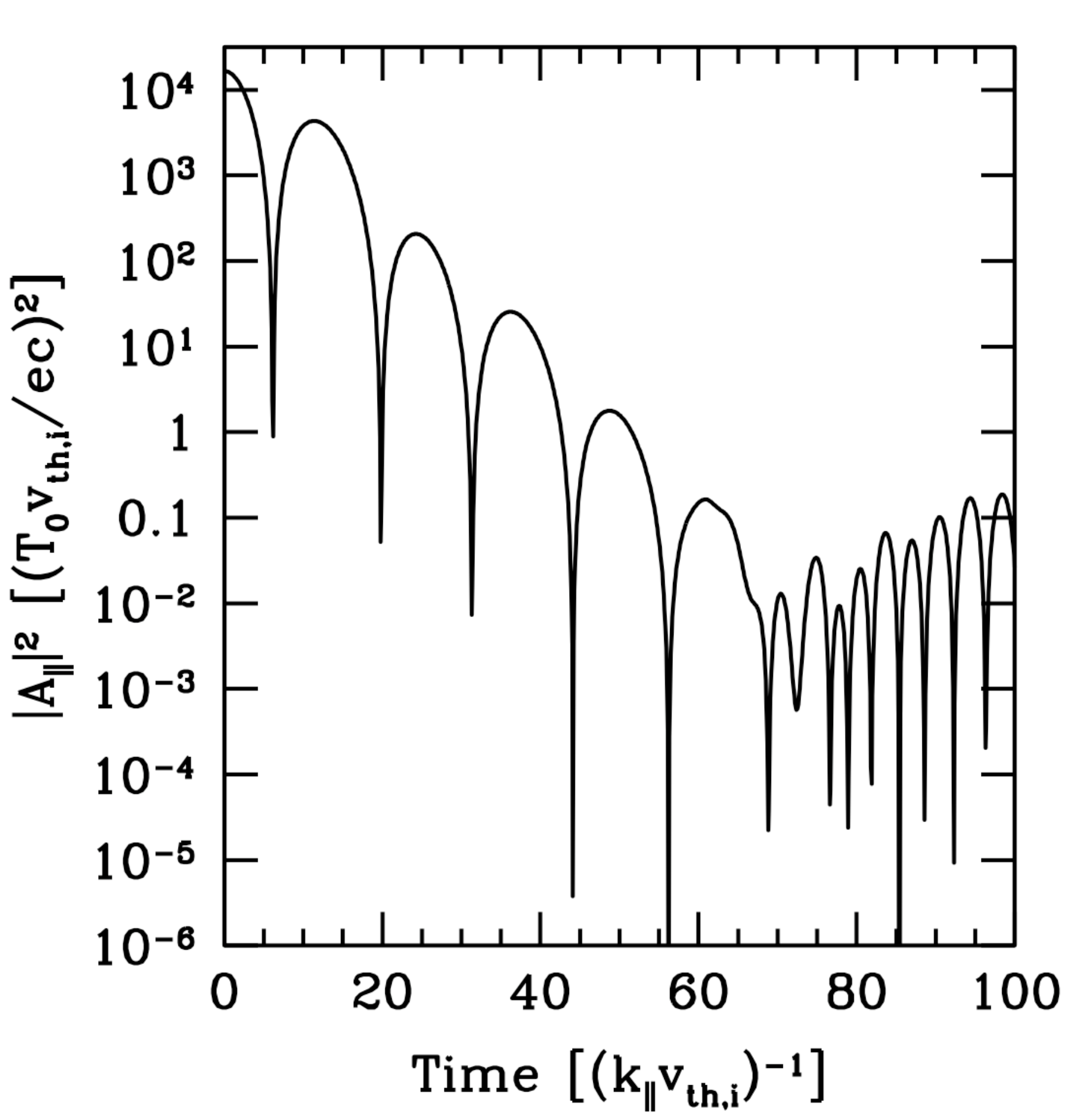}
\includegraphics[height=3.0in]{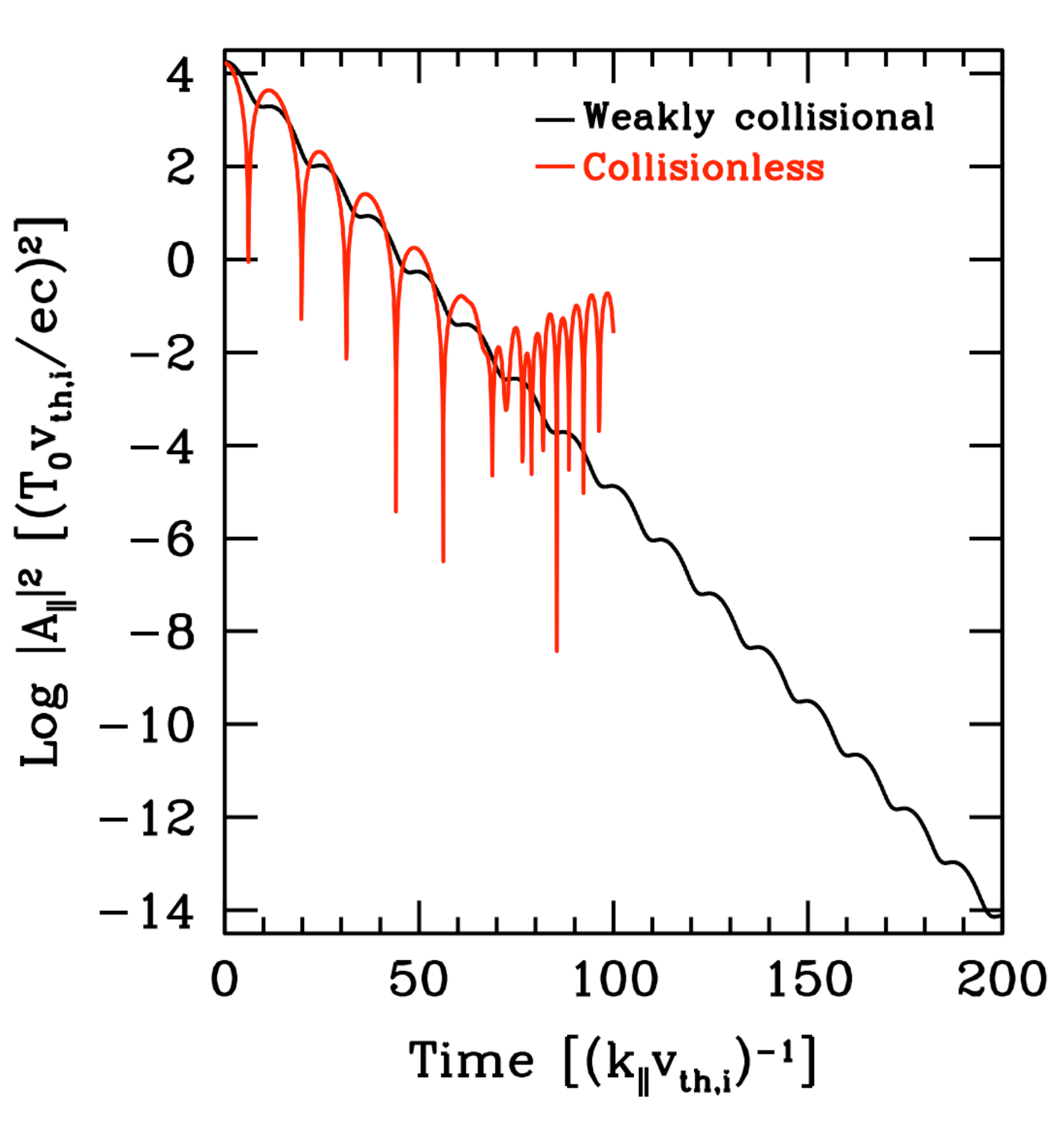}
\caption{Barnes damping of the kinetic Alfven wave.  In the absence of collisions (left),
sub-grid scale structures develop in velocity space, and the damping rate goes bad.
A small collisionality ($\nu \ll \gamma$) prevents the development of
sub-grid scale structures in velocity space, and the damping rate remains correct
indefinitely (right).}
\label{fig:kawdamp}
\end{figure}

\begin{figure}
\centering
\includegraphics[height=3.2in]{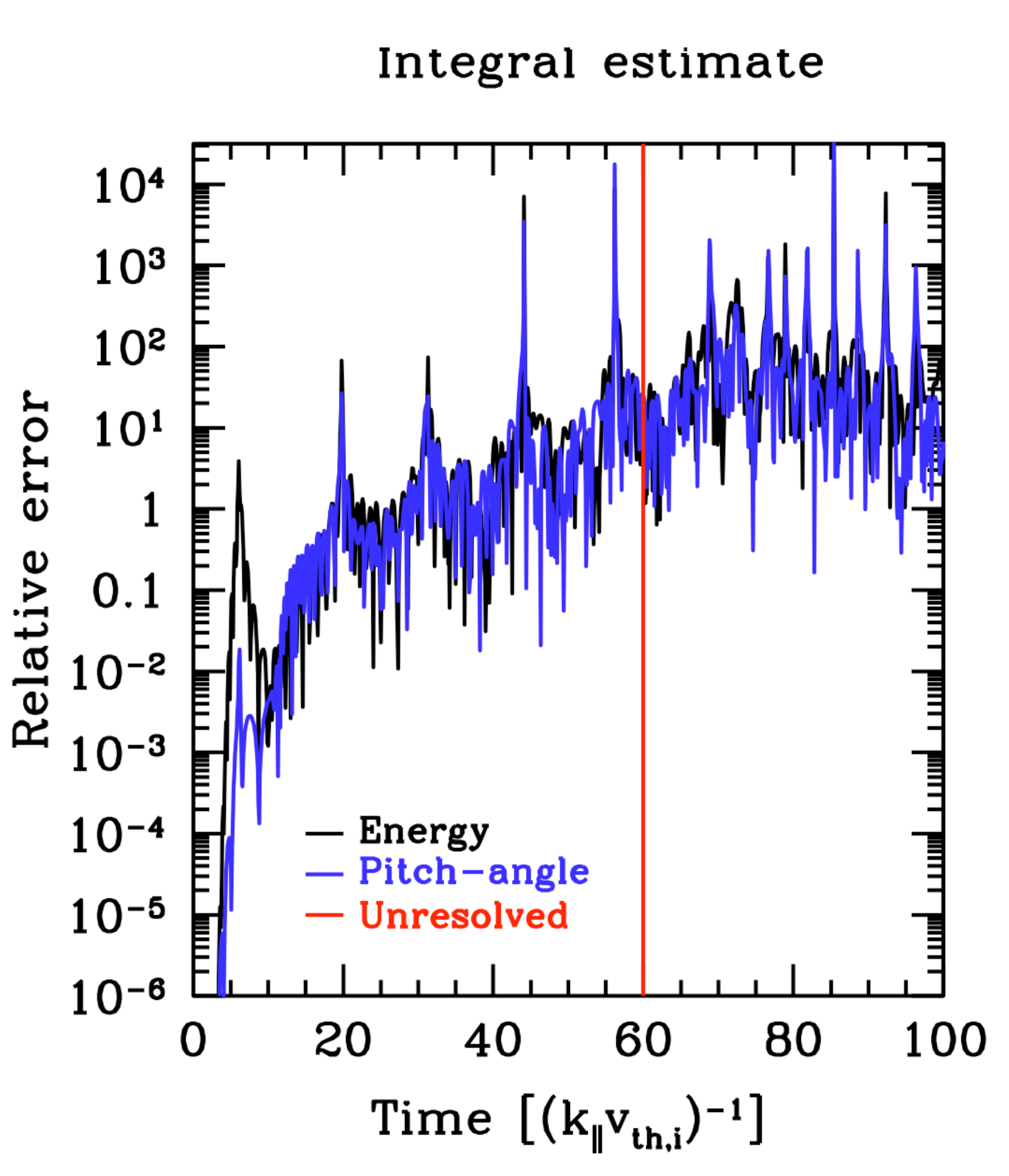}
\includegraphics[height=3.2in]{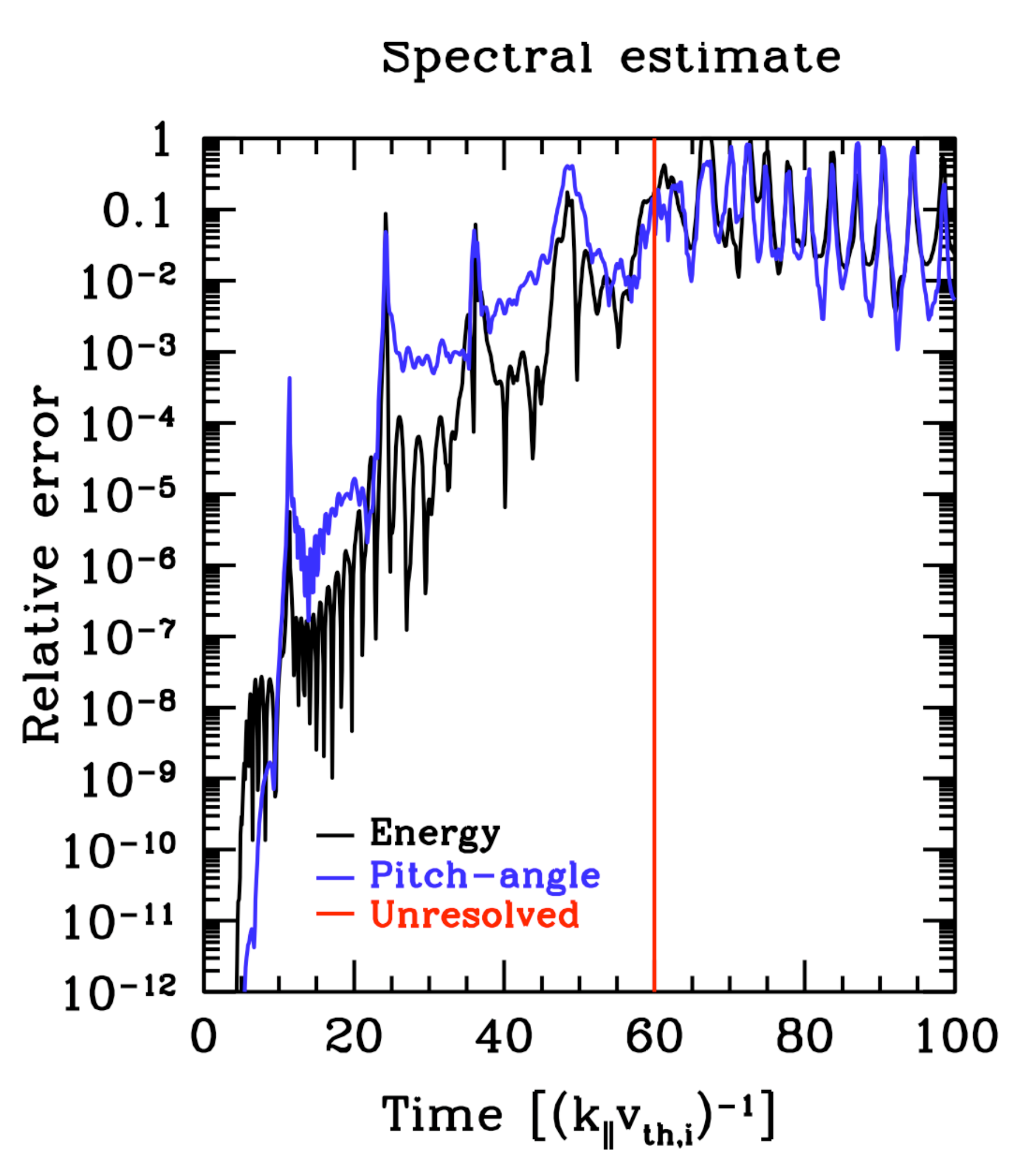}
\caption{Integral (left) and spectral (right) error estimates for the collisionless kinetic Alfven wave.}
\label{fig:kawerr}
\end{figure}



\begin{figure}
\centering
\includegraphics[height=3.2in]{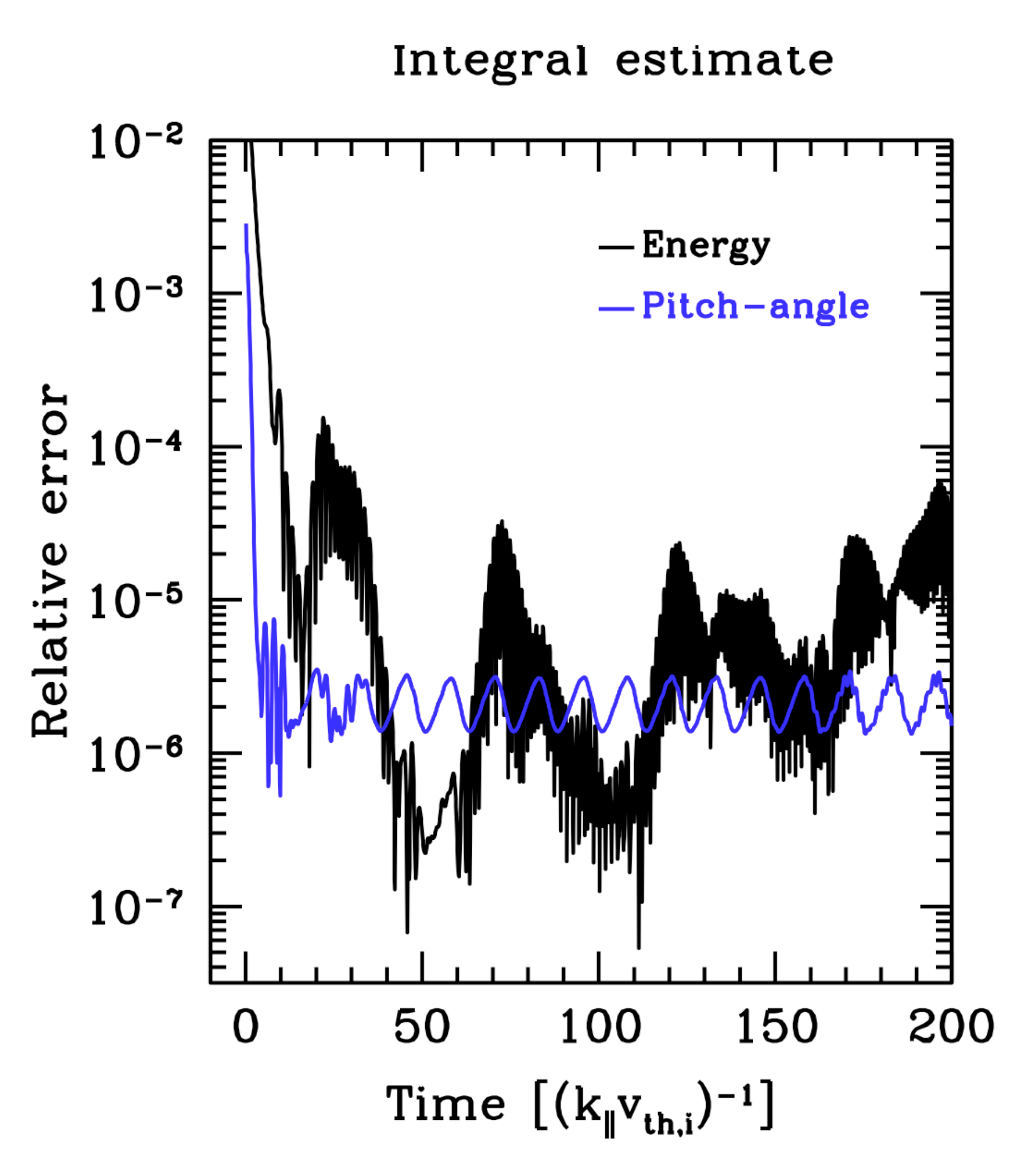}
\includegraphics[height=3.2in]{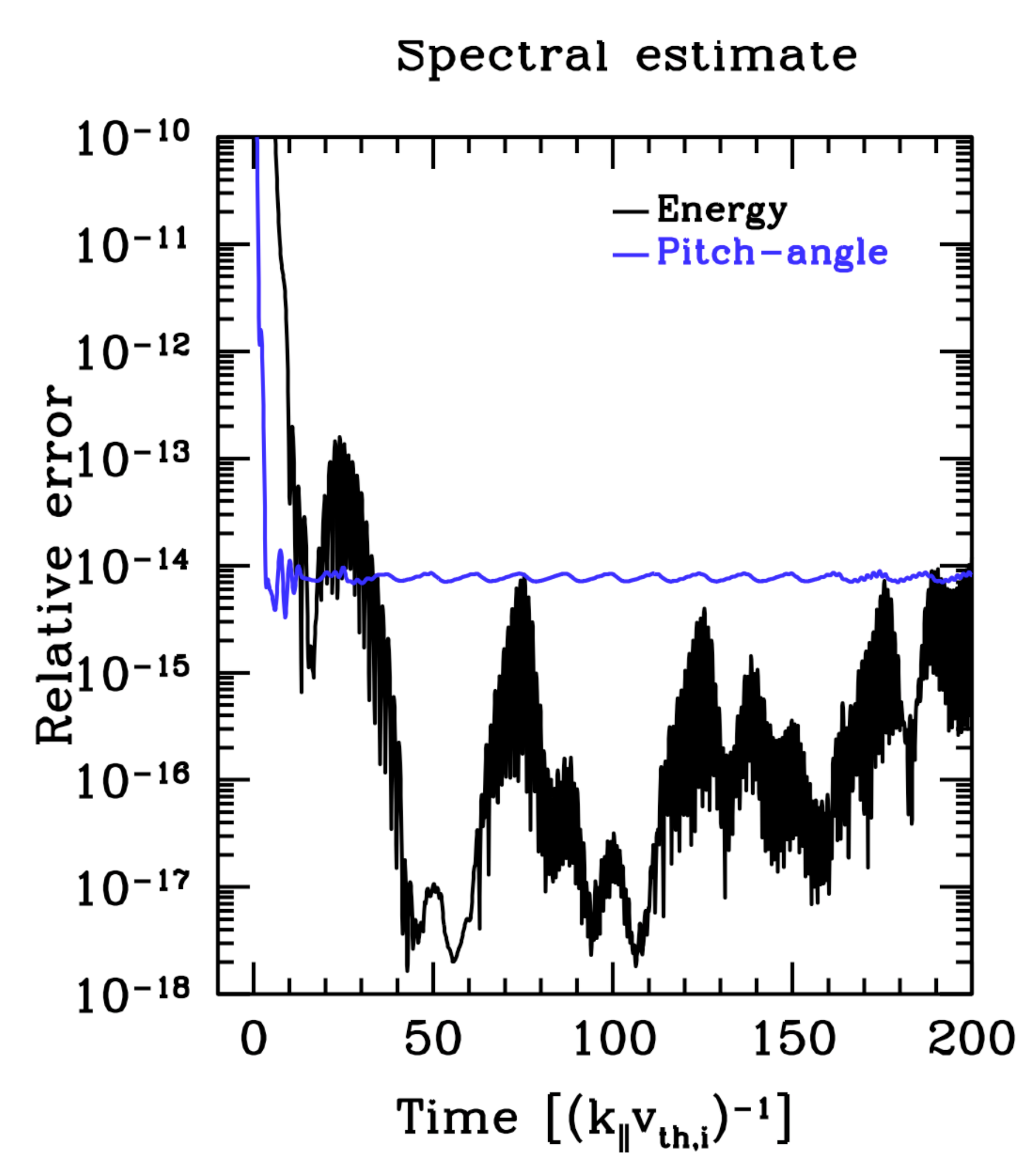}
\caption{Integral and spectral error estimates correctly indicate that the weakly collisional kinetic
Alfven wave simulation is well resolved.}
\label{fig:kawerrcoll}
\end{figure}



\section{Adaptive collision frequency}
\label{sec:adaptC}

As stated earlier, we would like to know what combination of dissipation and grid spacing
is necessary for a resolved simulation.  One way to approach this problem is
to fix the dissipation and vary the number of grid points to find how many are required
to get an accurate result.  This is the general idea behind the error estimation diagnostics
described in the previous section.  However, if we wanted to use this approach to ensure
that the simulation remains resolved, we would have to implement an adaptive grid,
which is difficult to do for massive, multi-processor calculations.  

Instead, we choose an alternative approach: we fix the number of grid points and vary
the dissipation until we have a well-resolved result.  In particular, we have implemented
an adaptive collision frequency in \verb#Trinity# that allows for the independent variation of 
the collisionality associated with pitch-angle scattering and energy diffusion.  Given
an acceptable error tolerance for velocity space calculations, a scaled version of the 
integral error estimate described in the previous section is used to determine whether
or not the simulation is well-resolved.  The collision frequency is then adjusted using
a feedback process until the scaled estimate of the error converges to within some
pre-specified window of the desired error tolerance.  In this way, the approximate 
minimum possible
dissipation is used to achieve an acceptable degree of resolution in velocity space.

Of course, the amount of dissipation necessary to resolve a simulation at a fixed number
of grid points may be quite large if a coarse grid is used.  Consequently, the collisionless
dynamics may be modified.  As a result, it is necessary to compare the converged 
collision frequency with dynamic frequencies of interest in the problem.

As an example we consider a nonlinear simulation of electron temperature gradient (ETG)
turbulence in slab geometry (i.e. straight background magnetic field).  In the nonlinear
phase, small scales are expected to develop in velocity space, potentially challenging numerical 
resolution.  In Fig.~\ref{fig:adaptetg}, we see that this is indeed the case.  Our velocity
space resolution diagnostics indicate that the errors in velocity space begin to increase
sharply during the transition from linear instability to turbulence.  However, our use of
an adaptive collision frequency prevents the estimated error from exceeding the user-defined
relative error tolerance (in this case, $0.01$).  We see that the error remains on the threshold
of the error tolerance, while the collision frequency for energy diffusion increases to a 
steady-state value of $\nu\approx0.27 \ k_{\parl}v_{th,e}$, which is well below the
dynamic frequency in the system.  Consequently, the collisionless dynamics are unaltered.

\begin{figure}
\centering
\includegraphics[height=3.0in]{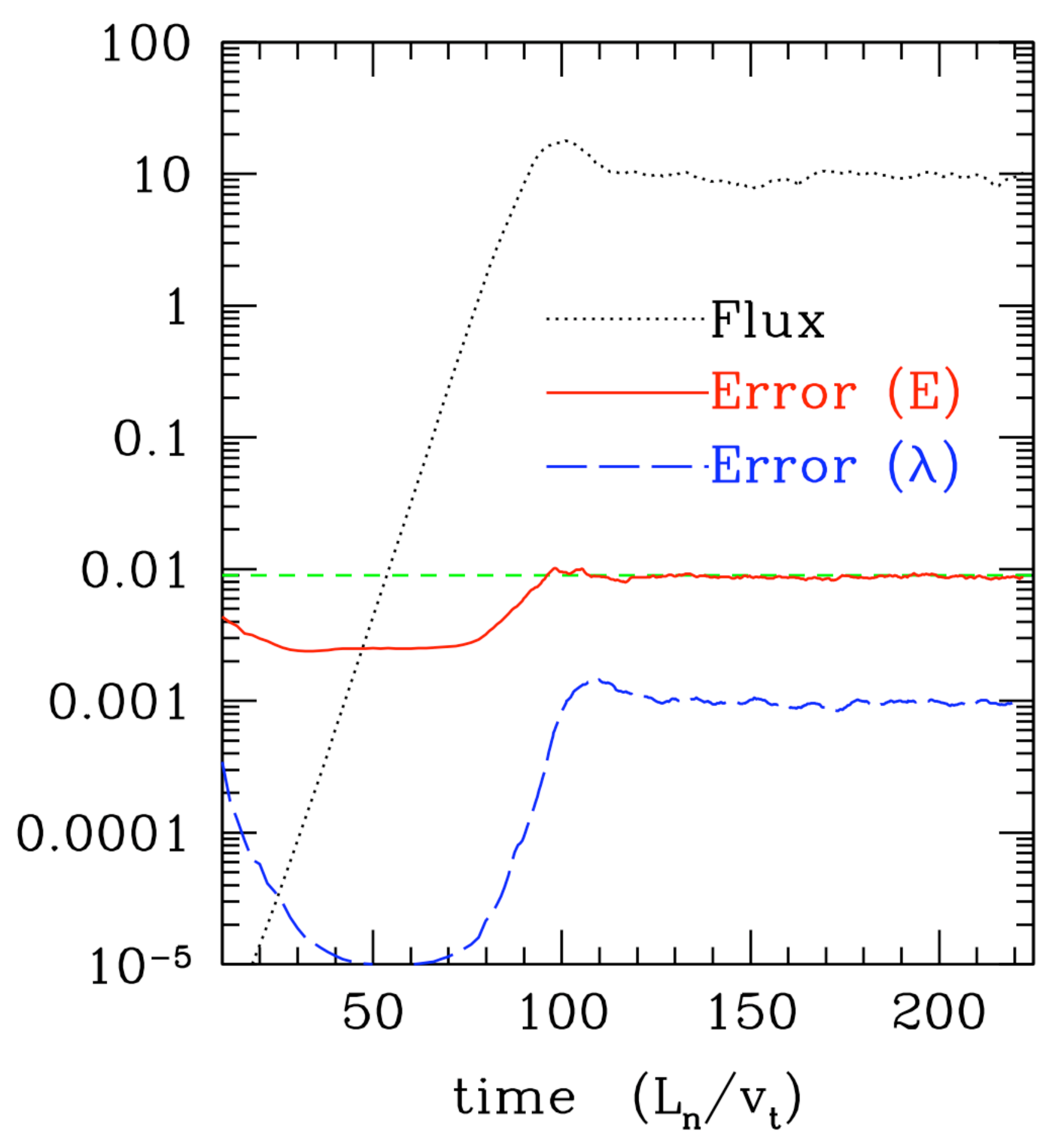}
\includegraphics[height=3.0in]{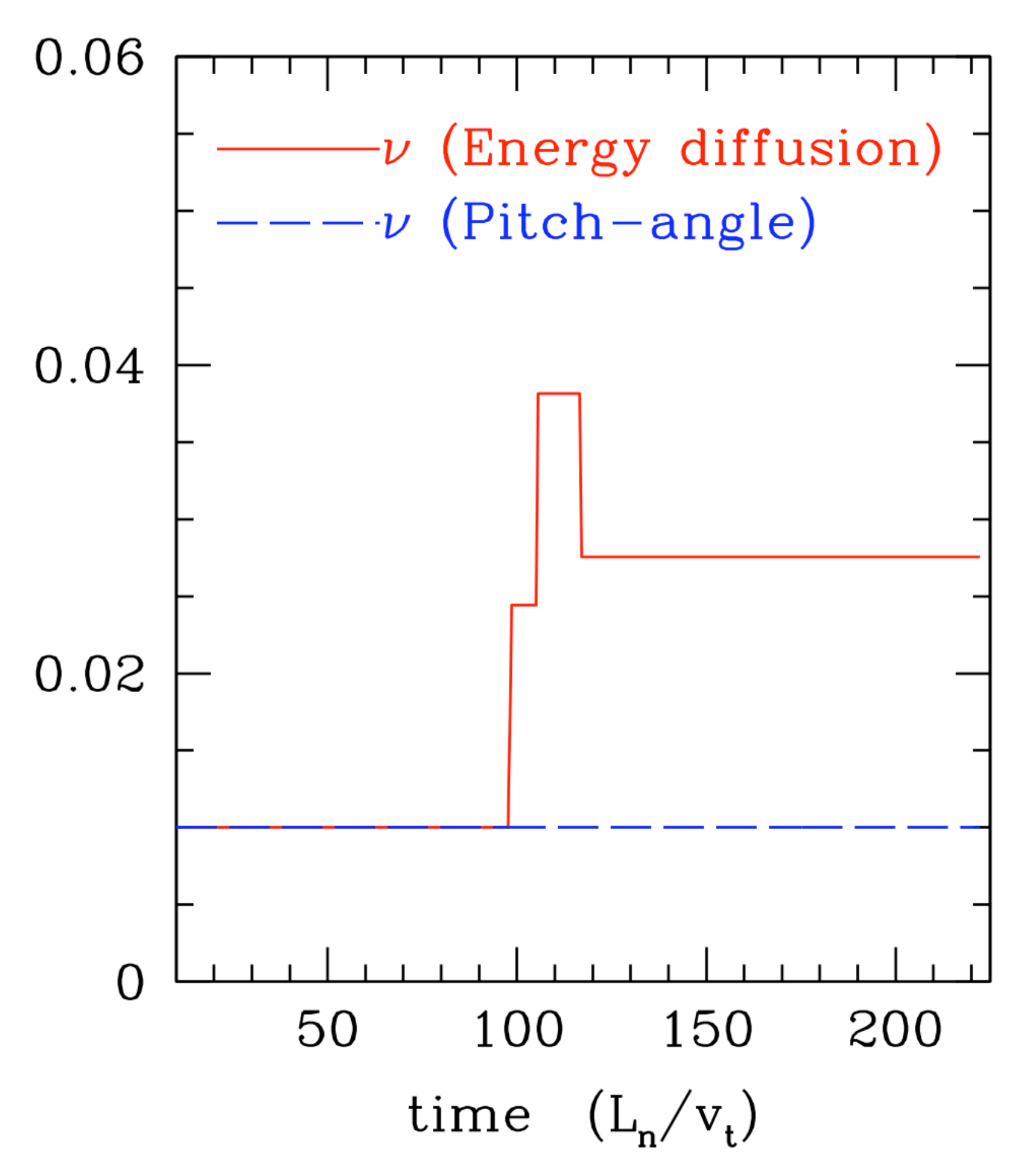}
\caption{(Left): Normalized electron heat flux vs. time for a nonlinear simulation
of ETG turbulence.  Scaled estimates of the error in energy and $\lambda$ resolution
increase during nonlinear saturation, but are kept within the specified error tolerance of $0.01$
with the use of an adaptive collision frequency.  (Right): Collision frequency (normalized 
by $k_{\parl}v_{th,e}$) vs. time.}
\label{fig:adaptetg}
\end{figure}



\section{Summary}
\label{sec:vspacesum}

In this chapter, we discussed the development of small-scale structure in velocity space,
presented a set of velocity space resolution diagnostics for use in gyrokinetic simulations,
and introduced an adaptive collisionality that allows us to resolve simulations with
an approximate minimal necessary dissipation for a fixed number of grid points in velocity space.  
In \secref{sec:gkv}
we demonstrated the tendency of collisionless plasmas to develop increasingly fine
scales in the distribution of particle velocities and discussed the phase mixing processes
that lead to such behavior.

In \secref{sec:gs2v} we described the treatment of velocity space in the gyrokinetic
code \verb#Trinity#.  We gave details on the choice of velocity space variables
(energy and pitch-angle) and discretization scheme, which is chosen to minimize
the error of the numerical integrals necessary to obtain the electromagnetic fields.  
This included presentation of a newly implemented energy grid, which provides spectrally 
accurate integrals over particle energies.  Additionally, we gave a brief discussion of
both the physical and numerical dissipation mechanisms available for use in \verb#Trinity#.

We discussed common approaches to monitoring velocity space resolution in \secref{sec:vresdiag}
and the difficulties associated with each.  We then proposed two new measures of velocity
space resolution and detailed implementation in \verb#Trinity#.  One of the proposed
resolution diagnostics involves obtaining estimates for the error in field integrals by
comparing numerical integrals obtained using integration schemes with differing degrees
of precision.  The other resolution diagnostic involves decomposing the perturbed distribution
function into spectral components in velocity space and monitoring the amplitude of the
spectral coefficients.  Both diagnostics should be quite conservative.

We then applied our resolution diagnostics to a number of example problems, including
Landau damping of the ion acoustic wave, Barnes damping of the kinetic Alfven wave,
and linear instability of the toroidal ITG mode.  We found that both diagnostics do well
in qualitatively estimating errors due to limited velocity space resolution.  Due
to their conservative nature, an empirical scaling factor was necessary to obtain correct
quantitative predictions.  

In \secref{sec:adaptC} we coupled the error estimates from our resolution diagnostics
with a model physical collision operator to develop an adaptive collision frequency.  This
adaptive collision frequency allowed us to resolve velocity space while using an approximate
minimal necessary amount of dissipation.  When using the adaptive collision frequency, one
must monitor the ratio of the collision frequency to the dynamic frequency to ensure
that one is still within the weakly collisional regime.

In conclusion, we found that dissipation was not necessary to resolve linear 
instabilities, but it was necessary to resolve nonlinear dynamics and linearly damped waves.  
For the nonlinear cases considered here (slab ETG and toroidal ITG), the required 
collisionality for resolution obtained with the adaptive collision frequency was
found to be no larger than the physical collisionality used in modern fusion experiments.


\renewcommand{\thechapter}{5}

\chapter{{\textbf{Linearized model Fokker-Planck collision operator for gyrokinetics: theory}}}
\label{ch:colltheory}
\vspace{+90pt}

\section{Introduction}
\label{sec_intro}

It has long been known that in many turbulent systems the difference between vanishingly 
small dissipation and no dissipation is striking, and that this can be 
linked theoretically to the non-interchangeability of limits $t \rightarrow \infty$ and 
$\nu \rightarrow 0$, where $\nu$ is e.g. viscosity, resistivity or collision frequency. 
Turbulence transfers energy from scales at which it is injected into the system to scales where 
it is dissipated, leading to heating. When the dissipation 
coefficients are small, the system has to generate very fine-scale fluctuations in order to 
transfer the energy to scales at which dissipation becomes efficient.  However, with
finite $\nu$, there will always exist a scale at which the injected energy is dissipated.

In plasma turbulence, all dissipation (meaning any effect that leads to irreversible heating) is 
ultimately collisional, so the transfer of energy generally occurs in phase space --- i.e., both 
in the position and velocity space (see extended discussion of energy cascade in plasma 
turbulence in Ref.\ \cite{schekPPCF08} and references therein). There are a number 
of specific mechanisms, both linear and nonlinear, that give rise to phase-space 
mixing~\cite{krommesPoP94,krommesPoP99,schekPPCF08,hammettPFB91,dorlandPFB93,watanabePoP04,schekAPJ07}.
It is the resulting large gradients in the velocity space that eventually bring collisions into 
play however small the collision frequency (such small-scale velocity-space structure has, 
e.g., been found and explicitly measured in gyrokinetic simulations~\cite{watanabePoP04,howesPoP08,barnesPoP08,tatsunoPRL08}). 
Thus, in any plasma turbulence simulation, some effective collisionality should be present
to smooth the small-scale structure in the velocity. 

While one may take the view that the numerical grid can play the role of effective 
collisions~\cite{candyPoP06}, we consider it to be a safer course of action to model 
collisional physics in a controlled fashion. In order to explain why, we would like to 
emphasize that, besides velocity-space smoothing, there is another key reason why collisions 
must be included. Collisions, through the dissipation of small-scale fluctuations in phase 
space, provide the link between irreversible plasma heating (macroscopic transport) and 
turbulence, so they are necessary in order for the system to converge to a statistically steady 
state. We shall now explain this statement.

Consider the $\delta f$ kinetics detailed in Chapter 2.  This model assumes that it is 
physically reasonable to split the 
distribution function into a slowly (both spatially and temporally) varying equilibrium part 
and a rapidly varying fluctuating part: $f= F_0+\delta f$. We saw in Chapter 2 that $F_0$ 
is a Maxwellian distribution, $F_0=(n_0/\pi^{3/2}\vth^3)\exp(-v^2/\vth^2)$, where 
$n_0$ is density, $\vth=(2T_0/m)^{1/2}$ is the thermal speed, $T_0$ is temperature and $m$ 
is particle mass. This will be the case if collisions are not extremely weak (for the weakly 
collisional formulation of $\delta f$ gyrokinetics, see Ref.~\cite{howesApJ06}). One 
can show that the fundamental energy balance governing the evolution of the turbulent 
fluctuations is~\cite{krommesPoP94,schekPPCF08,watanabePoP04,schekAPJ07,howesApJ06,sugamaPoP96,hallatschekPRL04,scottPoP07}
\begin{equation}
\label{balance}
\begin{split}
&\frac{d}{dt} \left( -\sum_s T_{0s}\delta S_s + U \right) =\\
&\qquad P + \sum_s\iint\frac{T_{0s}\delta f_s}{F_{0s}}\,C[\delta f_s]
d \bm{v} d \bm{r},
\end{split} 
\end{equation}
where $s$ is the species index, 
$\delta S = -\iint d\bm{r}d\bm{v}\,\delta f^2/2F_0$ is the entropy of the fluctuations, 
$U=\int d\bm{r}\,(E^2+B^2)/8\pi$ is the energy of the (fluctuating) electromagnetic field, 
$P$ is the input power (energy source of the turbulence), 
and $C[\delta f]$ is the linearized collision operator. 
In many types of plasma turbulence studied in fusion contexts, the input power 
$P$ is proportional to the heat flux and it is the parameter dependence of 
the mean value of this quantity in the statistically stationary state 
that is sought as the principal outcome of the simulations. 
We can see immediately from the above equation that collisions are required 
to achieve such a steady state (as has been shown in numerical 
simulations~\cite{candyPoP06,krommesPoP99,watanabePoP04,watanabePoP02}) 
and that in this steady state, $P$ must be balanced on the average by the collisional dissipation 
term.

The key property of the collision operator required for this transfer of 
energy from turbulence to the equilibrium distribution to work correctly and, therefore, for 
the heat fluxes to converge to correct steady-state values, is that the collision term in 
\eqref{balance} must be negative-definite: 
\begin{equation}
\label{Htheorem}
\iint \frac{\delta f}{F_{0}}\, C[\delta f] d\bm{r}d\bm{v} \le 0. 
\end{equation}
This ensures that heating is irreversible and that collisions cannot decrease entropy, the latter 
being the statement of Boltzmann's $H$-theorem~\cite{boltzmann}. Any spurious sink of 
entropy will adversely affect the balance between turbulent fluxes and dissipation, so it is 
clear that any model for collisional dissipation must respect the $H$-theorem. 

In view of the above discussion, we can formulate a reasonably restrictive set of criteria for 
any model collision operator: providing dissipation at small scales; obeying the 
$H$-theorem [\eqref{Htheorem}]; locally conserving particle number, 
momentum, and energy; and vanishing on a (local, perturbed) Maxwellian distribution.  
While these properties are analytically convenient, for numerical simulations the operator 
should also be efficiently implementable and carry these properties (at least approximately) 
over to the numerical scheme.

The effect of small angle Coulomb collisions on an arbirtrary distribution function was 
originally calculated by Landau~\cite{landauPZS36}.  In the $\delta f$ kinetic model 
we naturally consider the linearized Landau operator~\cite{helander}.  However, it is 
sufficiently complex that it would exceed the limits on numerical resources that can be 
realistically expended on modeling the collisional physics. 
Consequently, several simplified model collision operators have been developed, both for 
analytical and computational convenience, that try to capture the qualitative essence, if not 
the quantitative detail, of the physics 
involved~\cite{rutherfordPRL70,hirshmanPoF76,cattoPoF76}. This course of action is, 
indeed, eminently sensible: from \eqref{balance}, it seems plausible that, at least as far as 
calculating integral characteristics such as the turbulent fluxes is concerned, neither the exact 
functional form of the collision operator (provided it satisfies the criteria discussed above) 
nor the exact value of the collision frequency (provided it is sufficiently small) should be 
important. All we need is a physically reasonable dissipation mechanism.

For these purposes, it has often been deemed sufficient to use the pitch-angle-scattering 
(Lorentz) operator, sometimes adjusted for momentum 
conservation~\cite{rutherfordPRL70,helander}.  However, in kinetic turbulence, 
there is no reason that small-scale velocity-space structure should be restricted to pitch 
angles. In fact, standard phase-mixing mechanisms applied to gyrokinetics produce structure 
in $v_\parallel$~\cite{hammettPFB91,watanabePoP04}, and there is also nonlinear 
gyrokinetic phase mixing that gives rise to structure in $v_\perp$, which may be an even 
faster and more efficient process~\cite{dorlandPFB93,schekAPJ07,schekPPCF08}. 
Thus, {\it a priori} 
one expects to see small scales both in pitch angle and in the energy variable ($\xi$ and $v$). 
Indeed, it has been confirmed in simulations~\cite{barnesPoP08} that with only Lorentz 
scattering, structure rapidly forms at the grid scale in energy. Thus, a numerically suitable 
model collision operator must include energy diffusion. 

In this chapter,\footnote{This chapter is taken
from a co-authored paper currently in press~\cite{abelPoP08}} we propose such an operator (other operators including energy diffusion have 
been previously suggested~\cite{hirshmanPoF76,cattoPoF76}; we include a detailed 
comparison of our operator with these in \apref{app:ccomp}). Our model operator for 
like-particle collisions, including both pitch-angle scattering and energy diffusion and 
satisfying all of the physical constraints discussed above, is given in \secref{sec_newop} 
(the proof of the $H$-theorem for it is presented in \apref{app:hthm}). In \secref{sec_gk}, 
it is converted (gyroaveraged) into the form suitable for use in gyrokinetic simulations --- a 
procedure that produces some nontrivial modifications. In \secref{sec_ei}, we explain how 
interspecies (and, in particular, electron-ion) collisions can be modeled in gyrokinetic 
simulations to ensure that such effects as resistivity are correctly 
captured. \Secref{sec_summary} contains a short summary and a discussion of the 
consequences of the work presented here. 

The anlytical developments presented in this chapter form the basis for the numerical 
implementation of collisions in the publicly available gyrokinetic code \verb#Trinity#. This 
numerical implementation, as well as a suite of numerical tests are presented in Chapter 6. 

\section{A New Model Collision Operator}
\label{sec_newop}

In this section, we present a new model collision operator for like-particle collisions that 
satisfies the criteria stated above. The interpecies collisions will be considered 
in \secref{sec_ei}.

Let us start by introducing some standard notation. In discussing collision operators on phase 
space, we shall denote $\bm{r}$ the position variable in the physical space and use 
the $(v,\xi,\vartheta)$ coordinates in velocity space, where $v = |\bm{v}|$ is the energy 
variable, $\xi = v_\parallel/v$ is the pitch-angle variable, and $\vartheta$ the gyroangle 
about the equilibrium magnetic field. One can easily adapt the operators presented here to 
unmagnetized plasmas, but as we are interested in gyrokinetic plasmas, we shall concentrate 
on the strongly magnetized case. Taking the notation of Ref.~\cite{helander} as 
the standard, we introduce the normalized velocity variable $x=v/\vth$ and a set of 
velocity-dependent collision frequencies for like-particle collisions: 
\begin{eqnarray}
\nu_{D}(v) &=& \nu\, \frac{\erf( x ) - G( x ) }{ x^3}, \\
\nu_s(v) &=& \nu\,\frac{4G( x ) }{  x},\\
\nu_{\parallel}(v) &=& \nu\, \frac{2G( x ) }{ x^3},\\
\nu_{E}(v) &=& 2\nu_s(v) - 2 \nu_D(v) - \nu_\parallel(v),
\end{eqnarray}
where $\erf(x) = (2/\sqrt{\pi})\int^x_0 e^{-y^2} dy$ is the error function, 
$G(x) = [\erf(x)-x\erf'(x)]/2x^2$ is the Chandrasekhar function, and
$\nu = \sqrt{2}\pi n_0 q^4\ln\Lambda\, T_0^{-3/2}m^{-1/2}$ is the dimensional 
like-particle collision frequency (here $\ln\Lambda$ is the Coulomb logarithm and $q$ is 
the particle charge). 

If one wishes to construct a model linearized collision operator, the following general form 
constitutes a natural starting point
\begin{equation}
C[\delta f] = \pd{}{\bm{v}} \cdot \left[\hat D(\bm{v})\cdot \pd{}{\bm{v}} 
\frac{\delta f}{F_0}\right] + P[\delta f](\bm{v}) F_{0},
\end{equation}
where the first term is the ``test-particle'' collision operator and the second term the 
``field-particle'' operator. Most model operators can be obtained by picking a suitably 
simple form for the velocity-space diffusion tensor $\hat D$ and the functional $P$, 
subject to the constraints that one chooses to impose on the model operator. 

In constructing our model 
operator, we retain the exact form of $\hat D$ for the linearized Landau collision 
operator~\cite{helander}:
\begin{equation}
C[\delta f] = \nu_D L[\delta f] + \frac{1}{v^2} \pd{}{v}\left(\frac{1}{2} v^4 \nu_\parallel F_{0} \pd{}{v}\frac{\delta f}{F_0} \right) + P[\delta f](\bm{v}) F_{0},
\end{equation}
where we have explicitly separated the energy-diffusion part (the second term) and 
the angular part (the first term), which includes pitch-angle scattering and 
is described by the Lorentz operator: 
\begin{equation}
\label{def_L}
L[\delta f] = \frac{1}{2}\left[ \pd{}{\xi}(1-\xi^2)\pd{\delta f}{\xi} + 
\frac{1}{ 1-\xi^2}\pd{^2\delta f}{\vartheta^2}\right].
\end{equation}

Our modeling choice is to pick $P$ to be of the form
\begin{equation}
\label{pres}
P[\delta f](\bm{v}) = \nu_s\, \frac{2\bm{v} \cdot \bm{U}[\delta f]}{\vth^2}
+ \nu_E\, \frac{v^2}{\vth^2}\, Q[\delta f].
\end{equation}
One can view this prescription as first expanding $P$ in spherical harmonics (one can easily 
show that they are eigenfunctions of the full field-particle operator), retaining only the first 
two terms, and then arbitrarily factorizing the explicit $v$ and $\delta f$ dependence of 
each harmonic. The functionals $\bm{U}[\delta f]$ and $Q[\delta f]$ are mandated to 
have no explicit velocity dependence. In this ansatz the $v$ dependence is chosen so that the 
final operator is self adjoint and also to ensure automatic particle conservation by the 
field-particle operator: $\int P[\delta f](\bm{v}) F_0\,d\bm{v}=0$. Indeed the first term 
in \eqref{pres} gives a vanishing contribution to this integral because it is proportional 
to $\bm{v}$, and so does the second term 
because $v^{4}\nu_{E}F_0 = -(\partial/\partial v)(v^5\nu_{\parallel} F_{0})$. 
The functionals $\bm{U}[\delta f]$ and $Q[\delta f]$ are now uniquely chosen so as to 
ensure that the model operator conserves momentum and energy: a straightforward 
calculation gives
\begin{align}
\label{def_U}
\bm{U}[\delta f] &= \frac{3}{2}\frac{\int \nu_s \bm{v} \delta f\, d\bm{v}}
{\int \left(v/\vth\right)^2 \nu_s F_{0}\, d\bm{v}}, \\
\label{def_Q}
Q[\delta f] &= \frac{\int v^2 \nu_E \delta f\, d\bm{v}}
{\int v^2 \left(v/\vth\right)^2 \nu_E F_{0}\, d\bm{v}}.\quad
\end{align} 
These are in fact just the standard correction expressions used for the model 
pitch-angle-scattering operator~\cite{rutherfordPRL70,helander} and for more 
complex operators including energy diffusion~\cite{hirshmanPoF76}.

To summarize, we now have the following model operator for like particle collisions:
\begin{equation}
\begin{split}
\label{model}
C[\delta f] &= \frac{\nu_D}{2}\left[ \pd{}{\xi}(1-\xi^2)\pd{\delta f}{\xi} + 
\frac{1}{ 1-\xi^2}\pd{^2\delta f}{\vartheta^2}\right]
+ \frac{1}{v^2}\pd{}{v} \left( \frac{1}{2} v^4\nu_\parallel F_{0} \pd{}{v}\frac{\delta f}{F_0}\right) \\
&+ \nu_s\,\frac{2\bm{v} \cdot \bm{U}[\delta f] }{\vth^2} F_{0} 
+ \nu_E\,\frac{v^2}{\vth^2}\, Q[\delta f] F_{0},
\end{split}
\end{equation}
where the functionals $\bm{U}[\delta f]$ and $Q[\delta f]$ are given by 
\eqsref{def_U}{def_Q}. The modeling choice of the field-particle operator that we have 
made [\eqref{pres}] means that, in order to compute our collision operator, we have only 
to calculate definite integrals over the entirety of the velocity space --- a significant 
simplification in terms of computational complexity and ease of use in numerical 
simulations (see Chapter 6). 

As we have shown above, our operator conserves particles, momentum and energy by 
construction.  It is also not hard to see that it vanishes precisely when 
$\delta f / F_0 = (1, \bm{v}, v^2)$ and linear combinations thereof, i.e. if $\delta f$ is a 
perturbed Maxwellian.  From this and the fact that the operator is self adjoint, it can be 
shown that the operator {\it only} conserves particles, momentum and energy and 
that no spurious conservation laws have been introduced by our model.  Because the operator 
contains the exact test-particle part, it provides velocity-space diffusion both in energy and in 
angle and thus will efficiently dissipate small-scale structure.  Finally, it satisfies the 
$H$-theorem, as proved in \apref{app:hthm}. 

Our operator thus fulfills the criteria set forth in \secref{sec_intro} to be satisfied by a 
physically reasonable model operator. We now proceed to convert this operator into a form 
suitable for use in gyrokinetics.

\section{Collisions in Gyrokinetics}
\label{sec_gk}

The gyrokinetic theory is traditionally derived for a collisionless 
plasma~\cite{friemanPoF82,brizard2007}. However, as we have argued in 
\secref{sec_intro}, even when the collision frequency is small, collisions should be included 
in order to regularize the phase space and to ensure convergence of fluxes to statistically 
stationary values. Mathematically, collisions can be included in gyrokinetics if the collision 
frequency is formally ordered to be comparable to the fluctuation 
frequency~\cite{howesApJ06},  $\nu\sim\omega\sim k_\parallel\vth$ --- the weakly 
collisional limit (collisionality larger than this leads simply to fluid equations).  In practice, 
collision frequency tends to be smaller than the fluctuation frequency, but this need not upset 
the formal ordering as long as it is not too small: the cases $\nu\gg\omega$ and 
$\nu\ll\omega$ can be treated as subsidiary limits~\cite{schekAPJ07}.

Under the formal ordering $\nu\sim\omega$ (and, in fact, also under an even less restrictive 
ordering allowing for even smaller collisions\footnote{S.~C.~Cowley, unpublished}), we
have shown in Chapter 2 that the equilibrium distribution function (lowest order in the gyrokinetic 
expansion) is a Maxwellian, and the full distribution function can be 
represented as 
\begin{equation}
f = \left(1 - \frac{q \Phi}{T_{0}}\right) F_{0} + h(t,\bm{R},\mu,\varepsilon), 
\end{equation}
where $F_{0}$ is a Maxwellian, $\Phi$ the electrostatic potential (a fluctuating quantity) 
and $h$ the (perturbed) distribution function of the particle guiding centers.  
Here $\varepsilon=mv^2/2$ is the particle energy, $\mu=mv_\perp^2/2B_0$ the first 
adiabatic invariant, $B_0$ the strength of the equilibrium magnetic field,  
$\bm{R} = \bm{r} - \bm{\rho} = \bm{r} - \unit{b}\times\bm{v} / \Omega$ the guiding 
center position, $\Omega$ the cyclotron frequency, and $\unit{b}=\bm{B}_0/B_0$. The 
gyrokinetic equation, written in general geometry and including the collision operator is then
\begin{eqnarray}
\nonumber
&&\pd{h}{t} + (v_\parallel\unit{b} + \bm{v}_D)\cdot\pd{h}{\bm{R}} 
+ \frac{c}{B_0}\left\{\gyroR{\chi},h\right\}\\ 
&&\quad= -q\,\pd{F_0}{\varepsilon}\pd{\gyroR{\chi}}{t} 
+ \frac{c}{B_0}\left\{F_0,\gyroR{\chi}\right\} + \Cgk[h],\quad
\end{eqnarray}
where $\chi = \Phi - \bm{v}\cdot\bm{A}/c$ the gyrokinetic potential, 
$\gyroR{\chi} = (1/2\pi) \int \chi(\bm{R}+\bm{\rho})\,{d}\vartheta$ is 
an average over gyroangles holding $\bm{R}$ fixed (the ``gyroaverage''), 
$\bm{v}_{D}$ is the guiding center drift velocity defined in Eq.~(\ref{eqn:vdrift})
of Chapter~\ref{chap:hierarchy}. 

The gyrokinetic collision operator $\Cgk[h]$ is the gyroaverage of the linearized collision operator. 
The latter acts on the perturbed distribution $h$ holding the particle position $\bm{r}$
(not the guiding center $\bm{R}$) fixed. The latter nuance must be kept in mind when
working out the explicit form of $\Cgk[h(\bm{R})]$ from the unaveraged linearized operator 
$C[h(\bm{r}-\bm{\rho})]$. 

Let us restrict our consideration to local simulations, which are carried out in a flux tube of 
long parallel extent, but short perpendicular extent. In such simulations, one assumes that the 
equilibrium profiles are constant across the tube, but have non-zero gradients across the tube 
so as to keep all the appropriate drifts and instabilities. This permits one to use periodic 
boundary conditions and perform the simulations spectrally perpendicular to field 
lines~\cite{beerPoP95}.  Thus
\begin{equation}
h = \sum\limits_{\bm{k}} e^{i\bm{k}\cdot\bm{R}} h_{\bm{k}}(l,v,\mu),
\end{equation}
where $l$ is a coordinate along the field line and the Fourier transform is understood to be 
only with respect to the perpendicular components of 
$\bm{R}$, i.e., $\bm{k}\equiv\bm{k}_\perp$. Treating the perpendicular coordinates 
spectrally confines all dependence on the gyroangle $\vartheta$ to the exponent, thus we 
can compute the gyroangle dependence explicitly and carry out 
the gyroaveraging of the collision operator in a particularly 
transparent analytical way~\cite{cattoPoF76,schekAPJ07}:
\begin{equation}
\begin{split}
\Cgk[h] &= \gyroR{C\left[\sum_{\bm{k}} e^{i\bm{k}\cdot\bm{R}}\hk\right]}
= \sum_{\bm{k}} \gyroR{e^{i\bm{k}\cdot\bm{r}}C[e^{-i\bm{k}\cdot\bm{\rho}}\hk]}\\
&= \sum_{\bm{k}} e^{i\bm{k}\cdot\bm{R}}\gyroR{e^{i\bm{k}\cdot\bm{\rho}}
C[e^{-i\bm{k}\cdot\bm{\rho}}\hk]},
\end{split}
\end{equation}
where $\bm{\rho}=\unit{b}\times\bm{v}_\perp/\Omega$. 
Thus, in Fourier space
\begin{equation} 
\label{Cgk_gen}
\Cgk[\hk]= \left<e^{i\bm{k}\cdot\bm{\rho}}C[e^{-i\bm{k}\cdot\bm{\rho}}\hk]\right>,
\end{equation} 
where $\left<\dots\right>$ refers to the explicit averaging over the $\vartheta$ dependence. 
Some general properties of this operator are discussed in Appendix B of Ref.~\cite{schekAPJ07}. 

We now apply the general gyroaveraging formula \eqref{Cgk_gen} 
to our model operator given by \eqref{model}. 
The gyrokinetic transformation of variables 
$(\bm{r},v,\xi,\vartheta)\to(\bm{R},\mu,\varepsilon,\vartheta)$ 
mixes position and velocity space. 
However, in the collision operator, to the lowest order 
in the gyrokinetic expansion, we can neglect spatial dependence of $\mu$ that comes via 
the equilibrium magnetic field $B_0(\bm{r})$ and thus use the $(v,\xi)$ velocity variables.  
After some straightforward algebra, which involves converting velocity derivatives at constant $\bm{r}$ 
to those at constant $\bm{R}$ and evaluating the arising gyroaverages as detailed in 
\apref{app:gyroC}, we arrive at the following model gyrokinetic collision operator
\begin{equation}
\begin{split}
\Cgk[\hk] &= \frac{\nu_D}{2}\,\pd{}{\xi}(1-\xi^2)\pd{\hk}{\xi}
+ \frac{1}{v^2}\pd{}{v} 
\left( \frac{1}{2} v^4\nu_\parallel F_{0} \pd{}{v}\frac{\hk}{F_0}\right)\\
&- \frac{1}{4}\left[\nu_D(1+\xi^2) + \nu_\parallel(1-\xi^2)\right]\frac{v^2}{\vth^2}\,
k_\perp^2\rho^2 \hk\\
& + 2\nu_s\,\frac{v_\perp J_1(a)U_\perp[\hk] + v_\parallel J_0(a)U_\parallel[\hk]}{\vth^2}\,F_0 
+ \nu_E\,\frac{v^2}{\vth^2}J_0(a)Q[\hk]F_0,
\label{gyroav}
\end{split}
\end{equation}
where $\rho=\vth/\Omega$ is the thermal Larmor radius (not to be confused with the 
velocity-dependent $\bm{\rho}$), $a = k_\perp v_\perp / \Omega$, 
$J_0$ and $J_1$ are Bessel functions and 
\begin{eqnarray}
\label{Uperp}
U_\perp[\hk] &=& \frac{3}{2}\int \nu_s v_\perp J_1(a) \hk\, d\bm{v} \left/ 
\int { \left(v/\vth\right)^2 \nu_s F_{0}\, d\bm{v} }, \right. \\
\label{Upar}
U_\parallel[\hk] &=& \frac{3}{2}\int \nu_s v_\parallel J_0(a) \hk\, d\bm{v} \left/ 
\int { \left(v/\vth\right)^2 \nu_s F_{0}\, d\bm{v} }, \right. \\
\label{Qhk}
Q[\hk] &=& \int v^2 \nu_E J_0(a) \hk\, d\bm{v} \left/ 
\int { v^2 \left(v/\vth\right)^2 \nu_E F_{0}\, d\bm{v} }. \right.
\end{eqnarray}
Note that since the position and velocity space are mixed by the gyrokinetic 
transformation of variables, $\bm{R}=\bm{r}-\bm{\rho}$, the collision operator now 
contains not just pitch-angle and $v$ derivatives but also a spatial perpendicular 
``gyrodiffusion'' term. 

It is important to make sure that the operator we have derived behaves in a physically 
sensible ways in the long- and short-wavelength limits. When $k_\perp\rho\ll1$, 
all finite-Larmor-radius effects disappear, and we 
end up with pitch-angle scattering and energy diffusion corrected for energy and 
parallel momentum conservation --- the drift-kinetic limit. 
In the opposite limit, $k_\perp\rho\gg1$, we can estimate the behavior of our operator 
by adopting the scaling of the velocity derivatives based on the nonlinear perpendicular 
phase mixing mechanism for gyrokinetic turbulence proposed in Ref.~\cite{schekPPCF08}: 
this produces velocity-space structure with characteristic gradients 
$\vth\partial/\partial v_\perp \sim k_\perp\rho$ 
(see also Refs.~\cite{dorlandPFB93,schekAPJ07}). 
With this estimate, we see that all the field-particle terms in the operator are subdominant 
by a factor of $(k_\perp \rho)^{-3}$. 
Thus the operator reduces to the gyrokinetic form of the test-particle Landau operator 
in this limit. All diffusive terms are also equally large in this scaling, 
supporting our supposition that energy diffusion needs to be included.
These considerations give us some confidence that we correctly model the diffusive aspects 
of the collisional physics in a short-wavelength turbulent regime. Indeed, if one applies the 
same estimates to the full linearized Landau operator, the Rosenbluth potentials of the 
perturbation are small when $k_\perp \rho\gg1$ because they are integrals of a rapidly 
oscillating function, so the dominant effect does, indeed, come entirely from the test-particle 
part of the operator. 

The gyrokinetic collision operator given by \eqref{gyroav} respects the $H$-theorem:
\footnote{This can either be shown directly from \eqref{gyroav} (analogously to 
the proof in \apref{app:hthm})
or inferred from \eqref{Htheorem} by transforming to gyrokinetic variables}
\begin{equation}
\iint \frac{h}{F_0}\,\Cgk[h]\,d\bm{R}d\bm{v}\le 0,
\end{equation}
which is the what has to be satisfied in order for heating and transport in gyrokinetics to be 
correctly calculated~\cite{howesApJ06,schekAPJ07}.  The operator also manifestly diffuses small-scale 
structure both in velocity and in (perpendicular) position space. 
One cannot, however, perform the conservation-law tests upon this operator because one 
cannot separate the position- and velocity-space dynamics, and hence collisional and 
collisionless dynamics, in the gyrokinetic phase space.
Thus, we take the view that the conservation laws are guaranteed for the gyrokinetic collision 
operator in the sense that they were guaranteed for the original model operator from which it 
was derived. For practical numerical applications, this leaves the question of how this 
operator is best discretized and implemented. This is addressed in Chapter 6, 
where we also demonstrate the correct performance of our model operator on a number of 
test problems and show that all its new components (energy diffusion, gyrodiffusion, 
conservation terms) are necessary to avoid unphysical results. 
 
\section{Electron-Ion Collisions}
\label{sec_ei}

Let us now turn to the collisions between different species and focus on a plasma containing 
only electrons and one species of ions with a mass ratio $m_e/m_i\ll1$. The smallness of the 
mass ratio allows for a significant simplification of the interspecies collision terms. 
Since ion-electron collisions are subdominant to ion-ion collisions~\cite{helander},
$\nu_{ie}/\nu_{ii}\sim(m_e/m_i)^{1/2}$, it is safe for most physical purposes 
to neglect the ion-electron collisions and effects associated with them (such as the 
small slow collisional change in the mean ion momentum). Thus, the ion collisions 
can be modeled using the like-particle operator proposed above [\eqref{gyroav}]. 

The situation is different for the electron-ion collisions, which are the same order in 
mass ratio as the electron-electron collisions~\cite{helander}, $\nu_{ei}\sim\nu_{ee}$. 
Thus, the full electron collision operator has two parts: 
\begin{equation}
C[\delta f_e]=C_{ee}[\delta f_e] + C_{ei}[\delta f_e].
\end{equation} 
The electron-electron operator $C_{ee}[\delta f_e]$ can be modeled by the like-particle 
operator proposed above [\eqref{model}], the electrom-ion collision operator can be 
expanded in the mass ratio and to lowest order reads~\cite{helander} 
\begin{eqnarray}
C_{ei}[\delta f_e] &=& \nu_D^{ei} \left(L[\delta f_e] + 
\frac{2 \bm{v}\cdot\bm{u}_i }{{\vth^2}_{e}} F_{0e} \right),\\ 
\nu_D^{ei}(v) &=& \nu_{ei}\left(\frac{{\vth}_e}{v}\right)^3
\end{eqnarray}
where 
$\nu_{ei} = \sqrt{2}\pi n_{0i} Z^2 e^4 \ln \Lambda\,T_{0e}^{-3/2} m_e^{-1/2}$ 
is the dimensional electron-ion collision frequency, $Z=q_i/e$, $e$ is the 
fundamental charge, 
$L$ is the Lorentz operator given by \eqref{def_L}, and 
\begin{equation}
\bm{u}_i = \frac{1}{n_{0i}} \int \bm{v} \delta f_i\,d\bm{v}
\end{equation}
is the ion flow velocity. Thus, the electron-ion collisions are correctly modeled to lowest 
order in the mass ratio by electron pitch-angle scattering off static ions plus electron drag 
against the bulk ion flow. Note that the ion drag term is necessary to correctly capture 
electron-ion friction and hence resistivity; failure to include it leads to incorrect results, with 
mean electron momentum relaxed towards zero rather than towards equality with the mean 
ion momentum. 

Performing the conversion of $C_{ei}$ to the gyroaveraged form in a way analogous 
to what was done in \secref{sec_gk} and \apref{app:gyroC}, we get 
\begin{equation}
\begin{split}
\Cgk^{ei}[h_{e\bm{k}}] &= \nu_D^{ei} \left[ 
\frac{1}{2}\,\pd{}{\xi}(1-\xi^2)\pd{h_{e\bm{k}}}{\xi}
- \frac{1}{4}(1 + \xi^2) \frac{v^2}{{\vth^2}_e}\, k_\perp^2 \rho_e^2 h_{e\bm{k}}
+ \frac{2 v_\parallel J_0(a_e) u_{\parallel i\bm{k}} }{{\vth^2}_{e}} F_{0e} 
\right.\\
&\left.
- \frac{Zm_e}{m_i}\frac{v_\perp^2}{{\vth^2}_e} \frac{J_1(a_e)}{a_e} F_{0e} 
k_\perp^2 \rho_i^2\,\frac{1}{n_{0i}}\int \frac{2{v'_\perp}^2}{{\vth^2}_i} 
\frac{J_1(a_i')}{a_i'} h_{i\bm{k}}(\bm{v}') d\bm{v}'\right],
\label{Cgk_ei}
\end{split}
\end{equation}
where 
\begin{equation}
u_{\parallel i\bm{k}} = \frac{1}{n_{0i}}\int v_\parallel J_0(a_i) h_{i\bm{k}}\,d\bm{v}
\label{upari}
\end{equation} 
and $a_s=k_\perp v_\perp/\Omega_s$ for species $s$ and the rest of the notation 
as the same as in previous sections, with species indices this time. 

Let us estimate the size of the four terms in \eqref{Cgk_ei} at the ion (long) 
and electron (short) scales. The first term (pitch-angle scattering) is always 
important. At the ion scales, $k_\perp\rho_i\sim1$, the third term (parallel ion drag) is 
equally important, while the second term (electron gyrodiffusion) and 
the fourth term are subdominant by a factor of $m_e/m_i$. 
At the electron scales, $k_\perp\rho_e\sim1$, the pitch-angle scattering 
and the electron gyrodiffusion (the first two terms) are both important. 
Since at these scales $k_\perp\rho_i\sim(m_i/m_e)^{1/2}\gg1$, 
the third and fourth terms are subdominant by a factor (resulting from the 
Bessel functions under the velocity integrals) of 
$1/\sqrt{k_\perp\rho_i}\sim(m_e/m_i)^{1/4}$. 
In fact, they are smaller than this estimate because at these short wavelengths, 
the ion distribution function has small-scale structure in velocity space 
with characteristics scales $\delta v_\perp/{\vth}_i\sim1/k_\perp\rho_i$, 
with leads to the reduction of the velocity integrals by another factor of  
$1/\sqrt{k_\perp\rho_i}$. Thus, at the electron scales, the third and fourth terms 
in \eqref{Cgk_ei} are subdominant by a factor of $(m_e/m_i)^{1/2}$. 

These considerations mean that the fourth term in \eqref{Cgk_ei} is always negligible 
and can safely be dropped. The full model gyrokinetic electron collision operator is 
therefore
\begin{equation}
\begin{split}
\label{model_e}
\Cgk[h_{e\bm{k}}] &= \nu_D^{ei} \left[ 
\frac{1}{2}\,\pd{}{\xi}(1-\xi^2)\pd{h_{e\bm{k}}}{\xi}
- \frac{1}{4}(1 + \xi^2) \frac{v^2}{{\vth^2}_e}\, k_\perp^2 \rho_e^2 h_{e\bm{k}}
+ \frac{2 v_\parallel J_0(a_e) u_{\parallel i\bm{k}} }{{\vth^2}_{e}} F_{0e} \right]\\
&+\Cgk^{ee}[h_{e\bm{k}}],
\end{split}
\end{equation}
where the electron-electron model operator $\Cgk^{ee}[h_{e\bm{k}}]$ is given by \eqref{gyroav} 
and $u_{\parallel i\bm{k}}$ by \eqref{upari}. 

Finally, we note that since $e n_{0e}(u_{\parallel i}-u_{\parallel e}) = j_\parallel$ 
is the parallel current, 
the parallel Amp\`ere's law can be used to express $u_{\parallel i}$ in \eqref{model_e} 
in a form that does not contain an explicit dependence on the ion distribution function: 
\begin{equation}
u_{\parallel i\bm{k}} = \frac{1}{n_0e}\int v_\parallel J_0(a_e)h_{e\bm{k}}\, d\bm{v} 
+ \frac{c}{4\pi e n_{0e}}\,k_\perp^2 A_{\parallel\bm{k}}.
\end{equation}
This turns out to be useful in the numerical implementation of the electron operator, 
detailed in Chapter 6.

\section{Summary}
\label{sec_summary}

In \secref{sec_intro} we have argued the necessity of dissipation in turbulence 
simulations, justified the direct modeling of collisions in order to provide such dissipation 
and postulated a set of constraints for a physically reasonable model collision operator. 
Previously used model operators were deemed unsatisfactory, in part because the majority
of them do not contain a mechanism for energy diffusion.
Two of the well-known existing model operators that contain energy diffusion are detailed in 
Refs.~\cite{cattoPoF76} and~\cite{hirshmanPoF76}. However, the former 
does not satisfy the $H$-theorem [\eqref{Htheorem}], and the latter incorrectly captures the 
smallest scales. These problems are demonstrated and discussed in detail in \apref{app:ccomp}. 

In \secref{sec_newop} we presented a new operator [\eqref{model}] that successfully 
introduces energy diffusion while maintaining the $H$-Theorem and conservation laws, thus 
satisfying the conditions set forth in the introduction. This operator is then transformed into 
gyrokinetic form in \secref{sec_gk}, correctly accounting for the gyrodiffusive terms and 
FLR effects [\eqref{gyroav}]. In order to provide a complete recipe for modeling the 
collisional effects in simulations, the same gyroaveraging procedure is applied in 
\secref{sec_ei} to electron--ion collisions, somewhat simplified by the mass-ratio 
expansion [\eqref{model_e}]. This leaves us with a complete picture of collisions in 
gyrokinetic simulations, capturing gyrodiffusion, resitivity and small-scale energy diffusion.

When we discussed the gyroavergaing procedure in \secref{sec_gk} we presented the 
specific case of the application to Eulerian flux-tube $\delta f$ gyrokinetic 
simulations~\cite{jenkoPoP00,candyJCP03}.  However, the form presented in 
\eqref{model} is suitable for inclusion in most $\delta f$ kinetic systems and even 
amenable to use in Lagrangian codes by applying the methods of Refs.~\cite{xu1991nsi} 
or \cite{dimits1994cop} to the gyroaveraged operator given by \eqref{gyroav}. Indeed, 
by suitable discretization of the gyroaveraging procedure~\cite{candyJCP03} it would 
also be usable in a global Eulerian code.

We conclude by noting that the final arbiter of the practicality and effectiveness of this 
collision model is the numerical implementation and testing performed in 
Chapter 6, where our operator is integrated into the \verb#Trinity# code. 
The battery of tests shows that our operator not only reproduces the correct physics in the 
weakly collisional regime but even allows a gyrokinetic code to capture correctly the 
collisional (reduced-MHD) limit.


\renewcommand{\thechapter}{6}
\label{chap_collision}
\chapter{{\textbf{Linearized model Fokker-Planck collision operator for gyrokinetics: numerics}}}
\vspace{+90pt}

\section{Introduction}
\label{sec:cnumintro}

Collisions play an important role in gyrokinetics.  An accurate collision operator is important
for calculation of neoclassical transport~\cite{boltonPoF83,xuPRE08} and the growth rate of 
instabilities such as trapped electron modes~\cite{ernstPoP04,ernstIAEA06}, dissipative
drift waves~\cite{federiciPoF87,rewoldtPoF90,kotschCPC95}, and microtearing modes~\cite{applegatePPCF07} in 
moderate collisionality regimes.
Collisions can also affect the damping of zonal flows~\cite{xiaoPoP07} and other modes
that provide a sink for turbulent energy.
In their absence, arbitrarily fine scales can develop in phase 
space~\cite{krommesPoP94,krommesPoP99,schekAPJ07,schekPPCF08,barnes2PoP08},
which can in some cases pose challenges for discrete numerical algorithms, especially 
in the long-time limit~\cite{grantPoF67,nevinsPoP05}; even a modest amount of collisions
can make accurate numerical calculation much easier.  

Furthermore, inclusion of a small 
collisionality keeps the distribution function smooth enough in velocity space that the 
standard gyrokinetic ordering~\cite{friemanPoF82} for velocity space gradients is satisfied.
For example, the parallel nonlinearity~\cite{villardFE04,candy2PoP06}, given by
\begin{equation}
-\pd{}{v_{\parl}}\left[h\left(\frac{q}{m}\unit{b} + v_{\parl}\frac{\unit{b}\times \nabla B}{B^{2}}\right)\cdot \nabla \left< \Phi \right>\right]
\end{equation}
enters at the same order as the other terms in the gyrokinetic equation if 
the typical scale of parallel velocity fluctuations, $\delta v_{\parl}$, is one order 
smaller in the gyrokinetic expansion parameter $\rho/L$ ($\rho\equiv\textnormal{gyroradius}$ 
and $L\equiv\textnormal{background scale length}$) than the thermal speed, $v_{th}$.
Here, $h$ is the non-Boltzmann part of the perturbed distribution function (defined
more rigorously in the next section), $\Phi$ is
the electrostatic potential, $B$ is the magnetic field strength, $\unit{b}\equiv \mbf{B}_{0}/B_{0}$,
$q$ is particle charge, $m$ is particle mass, and $\left< \ . \ \right>$ denotes the gyroaverage at fixed
guiding center position $\mbf{R}$.

While such a situation is possible in the collisionless limit, a small collisionality prohibits
the formation of structures with $\delta v_{\parl}\sim (\rho/L)v_{th}$.  The level of
collisionality necessary to negate the importance of the parallel nonlinearity can be calculated
by assuming a balance between collisions and fluctation dynamics:
\begin{equation}
\pd{h}{t} \sim C[h] \Rightarrow \omega h \sim  \nu v_{th}^{2} \pd{^{2}h}{v^{2}},
\end{equation}
where $C[h]$ describes the effect of collisions on $h$, $\omega$ is the fluctuation frequency,
and $\nu$ is the collision frequency.  From the above expression, we see that scales 
in velocity space become small enough for the parallel nonlinearity to be important only
when the collision frequency satisfies $\nu \sim (\rho/L)^{2} \omega$.  Such
low collisionalities are not present in most fusion plasmas of interest.
Furthermore, if such an ordering had to be adopted, the lowest order distribution function 
could become strongly non-Maxwellian.  This is clearly a problem for $\delta f$ 
codes that assume an equilibrium Maxwellian.


In light of the above considerations, it is important to include an accurate treatment of dissipation in gyrokinetic
simulations.  In order to faithfully represent gyrokinetic plasma dynamics at reasonable
numerical expense, we take the view that the form of the 
dissipation should be such that it: ensures satisfaction of the standard gyrokinetic ordering;
locally conserves particle number, momentum, and energy; satisfies Boltzmann's $H$-Theorem;
and efficiently smooths phase space structure.  The first of these requirements has already
been discussed in the context of the parallel nonlinearity.  Conservation properties have been 
found to be important, for instance, in calculations of the neoclassical ion thermal conductivity~\cite{hintonRMP76},
as well as in a wide range of problems in fluid dynamics.  The existence of an $H$-Theorem
is critical for entropy balance~\cite{krommesPoP94,sugamaPoP96,howesApJ06} and for the 
dynamics of the turbulent phase space cascade~\cite{schekAPJ07,schekPPCF08}.
Efficient smoothing of phase space structures is necessary to resolve numerical simulations
at reasonable computational expense.  

A commonly employed dissipation mechanism in gyrokinetic simulations is
artificial (hyper) dissipation, often in physical (position) space~\cite{kotschCPC95,jenkoPoP00,candyPoP06,barnes2PoP08,chenPoP07}.
Ideally, the form of the artificial dissipation should be chosen to satisfy the requirements
listed above and should be tested for convergence to the collisionless result.  Of course,
artificial dissipation alone is unable to capture the correct dynamics for moderate to strongly
collisional systems where turbulent fluxes and other observable quantities depend
sensitively on collisionality; for such systems, a physical dissipation model is desired.

A number of such model physical collision operators are employed in gyrokinetic codes~\cite{kotschCPC95,jenkoPoP00,candyPoP06,navkalAPS06}.
These range in complexity from the Krook operator~\cite{rewoldtPoF86} to the 
Rutherford-Kovrizhnikh operator~\cite{rutherfordPRL70} to the Catto-Tsang operator~\cite{cattoPoF76},
all of which have previously been implemented in \verb#GS2# (see, e.g. Ref.~\cite{navkalAPS06}).
However, none of these satisfy all of the properties we require of a good collision operator (See
Appendix~\ref{app:ccomp} for a fuller discussion of this point).
Here, we discuss numerical implementation in \verb#Trinity# of an improved model 
operator which: includes the effects of both pitch-angle scattering and energy diffusion 
(i.e. efficiently smooths in phase space and ensures gyrokinetic ordering); 
conserves particle number, momentum, and energy;
satisfies Boltzmann's $H$-Theorem; and reduces to the linearized Landau test-particle operator 
in the large $k_{\perp}\rho$ limit.  A full description of this operator and a discussion
of its desirable properties is given in Chapter 5.
We will focus on 
how such an operator can be implemented efficiently in gyrokinetic codes while maintaining
the properties listed above and on how our gyrokinetic dissipation scheme (or any other)
might be tested against a number of plasma physics problems.

This chapter\footnote{This chapter taken from Ref.~\cite{barnesPoP08}.} is organized 
as follows: in \secref{sec:cgkprop}, we present the gyroaveraged 
collision operator derived in Chapter 5 and examine properties
that should be taken into account when using it in numerical simulations;
in \secref{sec:cnumimp}, we describe our numerical implementation of the collision operator; 
in \secref{sec:cnumres} we present numerical results for a number of tests demonstrating the 
ability of our collision operator implementation to reproduce correct collisional and 
collisionless physics; and in \secref{sec:cnumsum}, we summarize our findings.  


\section{Properties of the gyroaveraged collision operator}
\label{sec:cgkprop}

In order to include collisions in gyrokinetics, we follow the treatment of Ref.~\cite{howesApJ06} 
and assume the collision frequency, $\nu$, to be the same order 
in the gyrokinetic ordering as the characteristic fluctuation
frequency, $\omega$.\footnote{Note that this ordering does not prevent one from considering
the cases of $\nu \ll \omega$ and $\nu \gg \omega$ as subsidiary orderings~\cite{schekAPJ07}}.
As was shown in Chapter 2, this leads to the requirement that the distribution of particles in
velocity space is Maxwellian to lowest order and allows us to represent the total distribution
function through first order in $\rho/L$ (where $\rho$ is ion gyroradius and $L$ is the scale
length of equilibrium quantities) as
\begin{equation}
f(\mbf{r},\mu,\varepsilon,t) = F_{0}(\varepsilon)\left(1-\frac{q\Phi(\mbf{r},t)}{T_{0}}\right) + h(\mbf{R},\mu,\varepsilon,t),
\end{equation}
where $\mbf{r}$ is particle position, $\mbf{R}=\mbf{r}-\unit{b}\times\mbf{v}/\Omega_{0}$ is guiding center position, 
$\mu\equiv mv_{\perp}^{2}/2B_{0}$ is magnetic moment, 
$\varepsilon\equiv mv^{2}/2$ is particle energy, $F_{0}$ is a Maxwellian,
$\Phi$ is the electrostatic potential, $B_{0}$ is the magnitude of the background 
magnetic field, $T_{0}$ is the background temperature, $q$ is particle charge, and
$\Omega_{0}=qB_{0}/mc$.  
The gyrokinetic equation governing the evolution of $h$ is given by
\begin{equation}
\begin{split}
\pd{h}{t}& + \left(v_{\parl}\unit{b} + \mbf{v}_{D}\right)\cdot \pd{h}{\mbf{R}} 
+ \frac{c}{B_{0}}\{\left<\chi\right>_{\mbf{R}},h\}\\
&= -q\pd{F_{0}}{\varepsilon}\pd{\left<\chi\right>_{\mbf{R}}}{t}
+\frac{c}{B_{0}}\{F_{0},\left<\chi\right>_{\mbf{R}} \} + \left< C[h] \right>_{\mbf{R}},
\end{split}
\end{equation}
where $\unit{b}\equiv \mbf{B}_{0}/B_{0}$, $\mbf{v}_{D}$ is the drift velocity of
guiding centers, $\chi\equiv \Phi - \mbf{v}\cdot\mbf{A}/c$, $\mbf{A}$ is the 
vector potential, $\{a,b\}$ is the Poisson bracket of $a$ and $b$, $\left< a \right>_{\mbf{R}}$ is 
the gyroaverage of $a$ at constant $\mbf{R}$, and $\left< C[h] \right>_{\mbf{R}}$ is the 
gyroaveraged collision operator.  

For $\left< C[h] \right>_{\mbf{R}}$, we restrict our attention to the model collision operator 
presented in Chapter 5.
We work within the framework of the continuum gyrokinetic code \verb#GS2#~\cite{kotschCPC95}, 
which assumes periodicity in the spatial 
directions perpendicular to $\mbf{B}_{0}$ in order to reduce the simulation volume to a
thin flux tube encompassing a single magnetic field line.  Consequently, we require a spectral 
representation of $\left< C[h] \right>_{\mbf{R}}$:
\begin{equation}
\left< C[h] \right>_{\mbf{R}} \equiv \sum_{\mbf{k}} e^{i\mbf{k}\cdot\mbf{R}}C_{GK}[h_{\mbf{k}}],
\end{equation}
where $\mbf{k}$ is the perpendicular wavevector.
For convenience, we reproduce the expression for the same-species part of $C_{GK}[h_{\mbf{k}}]$ 
from Chapter 5 in operator form:
\begin{equation}
C_{GK}[h_{\mbf{k}}] \equiv L[h_{\mbf{k}}] + D[h_{\mbf{k}}]
+ U_{L}[h_{\mbf{k}}] + U_{D}[h_{\mbf{k}}] + E[h_{\mbf{k}}],
\label{eqn:cop}
\end{equation}
where
\begin{equation}
L[h_{\mbf{k}}] \equiv \frac{\nu_{D}}{2}\pd{}{\xi}\left(1-\xi^{2}\right)\pd{h_{\mbf{k}}}{\xi} - \frac{k^{2}v^{2}}{4\Omega_{0}^{2}}\nu_{D}\left(1+\xi^{2}\right)h_{\mbf{k}}
\end{equation}
and
\begin{equation}
D[h_{\mbf{k}}] \equiv \frac{1}{2v^{2}}\pd{}{v}\left(\nu_{\parl}v^{4}F_{0}\pd{}{v}\frac{h_{\mbf{k}}}{F_{0}}\right) - \frac{k^{2}v^{2}}{4\Omega_{0}^{2}}\nu_{\parl}\left(1-\xi^{2}\right)h_{\mbf{k}}
\end{equation}
are the gyroaveraged Lorentz and energy diffusion operators (which together form the 
test-particle piece of the linearized Landau operator, as shown in Refs.~\cite{cattoPoF76}
and~\cite{abelPoP08}),
\begin{equation}
U_{L}[h_{\mbf{k}}] \equiv \nu_{D}F_{0}\Big(J_{0}(a)v_{\parl}
\frac{\int d^{3}v \ \nu_{D}v_{\parl}J_{0}(a)h_{\mbf{k}}}{\int d^{3}v \ \nu_{D}v_{\parl}^{2}F_{0}} 
+ J_{1}(a)v_{\perp}\frac{\int d^{3}v \ \nu_{D}v_{\perp}J_{1}(a)h_{\mbf{k}}}{\int d^{3}v \ \nu_{D}v_{\parl}^{2}F_{0}}\Big)
\label{eqn:scriptuL}
\end{equation}
and
\begin{equation}
U_{D}[h_{\mbf{k}}] \equiv -\Delta \nu F_{0}\Big(J_{0}(a)v_{\parl}
\frac{\int d^{3}v \ \Delta \nu v_{\parl}J_{0}(a)h_{\mbf{k}}}{\int d^{3}v \ \Delta \nu v_{\parl}^{2}F_{0}} 
+ J_{1}(a)v_{\perp}\frac{\int d^{3}v \ \Delta \nu v_{\perp}J_{1}(a)h_{\mbf{k}}}{\int d^{3}v \ \Delta \nu v_{\parl}^{2}F_{0}}\Big)
\label{eqn:scriptuD}
\end{equation}
are the gyroaveraged momentum-conserving corrections to the Lorentz and energy diffusion
operators, and
\begin{equation}
E[h_{\mbf{k}}] \equiv \nu_{E} v^{2} J_{0}(a) F_{0} \frac{\int d^{3}v \ \nu_{E} v^{2} J_{0}(a) h_{\mbf{k}}}{\int d^{3}v \ \nu_{E} v^{4} F_{0}}
\label{eqn:scripte}
\end{equation}
is the gyroaveraged energy-conserving correction (the conserving terms are an approximation 
to the field-particle piece of the linearized Landau operator).  The electron
collision operator has the following additional term to account for electron-ion collisions:
\begin{equation}
C_{GK}^{ei}[h_{e,\mbf{k}}] = \nu_{D}^{ei} \Big(\frac{1}{2}\pd{}{\xi}\left(1-\xi^{2}\right)\pd{h_{e,\mbf{k}}}{\xi}-\frac{k^{2}v^{2}}{4\Omega_{0_{e}}^{2}}\left(1+\xi^{2}\right)h_{e,\mbf{k}}
+\frac{2v_{\parl}u_{\parl}[h_{i,\mbf{k}}]}{v_{th_{e}}^{2}}J_{0}(a_{e})F_{0e}\Big),
\label{eqn:Cei}
\end{equation}
where $\xi\equiv v_{\parl} / v$ is the pitch angle, $a\equiv kv_{\perp}/\Omega_{0}$, $J_{0}$ and $J_{1}$
are Bessel functions of the first kind, $v_{th}\equiv\sqrt{2T_{0}/m}$ 
is the thermal velocity, and $u_{\parl}[h_{i,\mbf{k}}]$ is the perturbed parallel ion flow 
velocity.  Expressions for the velocity-dependent collision frequencies $\nu_{D}$, $\Delta \nu$, 
$\nu_{\parl}$, and $\nu_{E}$ are given in Chapter 5 (which follows the notation of
Ref.~\cite{hirshmanPoF76}).

Having specified the form of our collision operator, we now discuss some of its
fundamental properties that guide our choice of numerical implementation.

\subsection{Collision operator amplitude}

Even when the collisionality approaches zero, $C_{GK}[h_{\mbf{k}}]$ can have 
appreciable amplitude.  There are two reasons for this: first, the velocity dependence
of $\nu_{D}$, $\nu_{E}$, and $\Delta \nu$ is such that each go to infinity 
as $v\rightarrow0$ (so low-velocity particles are always collisional); and second,
we expect the distribution function to develop increasingly smaller scales in $v$ and $\xi$ as 
collisionality decreases, so that 
the amplitude of the terms proportional to $\partial^{2}h/\partial\xi^{2}$ and $\partial^{2}h/\partial v^{2}$ may
remain approximately constant\footnote{This is analagous to the result in fluid turbulence 
where the dissipation rate remains finite as viscosity becomes vanishingly small.} 
(i.e. $C_{GK}[h_{\mbf{k}}] \not\rightarrow 0$ as $\nu \rightarrow 0$)~\cite{krommesPoP94, krommesPoP99, schekAPJ07,barnes2PoP08}.
The fact that $C_{GK}[h_{\mbf{k}}]$ can be quite large even at very low collisionalities
means that it should be treated implicitly if one wants to avoid a stability limit on the
size of the time step, $\Delta t$.  In \secref{sec:cnumimp}, we describe our fully implicit 
implementation of the collision operator.

\subsection{Local moment conservation}

Since collisions locally conserve particle density, momentum,
and energy, one would like these properties to be guaranteed by the discrete version
of the collision operator.  Mathematically, this means that the density, momentum, and energy moments
of the original (un-gyroaveraged) collision operator must vanish (for same-species
collisions).  However, the non-local nature of the gyroaveraging operation introduces finite 
Larmor radius (FLR) effects that lead to nonzero values for the analogous moments of 
$\left< C[h] \right>_{\mbf{R}}$.  Since this is the quantity we employ
in gyrokinetics, we need to find the pertinent relations its moments must satisfy in 
order to guarantee local conservation properties.

This is accomplished by Taylor expanding the Bessel functions $J_{0}$ and $J_{1}$.
In particular, one can show that~\cite{abelPoP08}
\begin{equation}
\int d^{3}v 
\begin{pmatrix}
1 \\
\mbf{v}\\
v^{2}
\end{pmatrix} 
\left<\left< C[h] \right>_{\mbf{R}}\right>_{\mbf{r}}
=\sum_{\mbf{k}}e^{i\mbf{k}\cdot\mbf{r}} \int d^{3}v
\begin{pmatrix}
1 \\
v_{\parl}\unit{b}\\
v^{2}
\end{pmatrix}
C_{GK}^{0}[h_{\mbf{k}}] - \nabla \cdot \Gamma_{C},
\end{equation}
where $\left< . \right>_{\mbf{r}}$ denotes a gyroaverage at fixed $\mbf{r}$,
$C_{GK}^{0}[h_{\mbf{k}}]$ is the operator of Eq.~(\ref{eqn:cop}) with $k\rho=0$
(neglecting FLR terms but retaining nonzero subscripts $\mbf{k}$ for $h$), and $\Gamma_{C}$ 
is the collisional flux of number, momentum, and energy
arising from FLR terms.  Consequently, the density, momentum,
and energy moments of the gyrokinetic equation can be written in the conservative form
\begin{equation}
\pd{\mathcal{M}}{t} + \nabla \cdot \Gamma_{\mathcal{M}} = \int d^{3}v  
\begin{pmatrix}
1 \\
v_{\parl}\unit{b}\\
v^{2}
\end{pmatrix} 
C_{GK}^{0}[h_{\mbf{k}}],
\end{equation}
where $\mathcal{M}\equiv \left(\delta n \ n\delta \mbf{u}_{\parl} \ \delta p\right)^{T}$
represents the perturbed number, momentum, and energy densities, the
superscript $T$ denotes the transpose, and $\Gamma_{\mathcal{M}}$ contains
both the collisional flux, $\Gamma_{C}$, and the flux arising from all other
terms in the gyrokinetic equation (for a more detailed discussion, see Ref.~\cite{abelPoP08}).
Thus, local conservation properties are assured in gyrokinetics as long as the density,
momentum, and energy moments of $C_{GK}^{0}[h_{\mbf{k}}]$ vanish:
\begin{equation}
\int d^{3}v
\begin{pmatrix}
1 \\
v_{\parl}\\
v^{2}
\end{pmatrix}
C_{GK}^{0}[h_{\mbf{k}}]=0.
\label{eqn:momcons}
\end{equation}  
We describe how this is accomplished numerically in \secref{sec:cnumimp}.


\subsection{$H$-Theorem}

In contrast with local conservation properties, the statement of the $H$-Theorem
is unmodified by gyroaveraging the collision operator.  Defining the entropy as 
$S=-f \ln f$, Boltzmann's $H$-Theorem tells us
\begin{equation}
\pd{S}{t} = -\int \frac{d^{3}\mbf{r}}{V}\int d^{3}v \ \ln[f] C[f] \ge 0,
\end{equation}
where $V\equiv \int d^{3}\mbf{r}$ and the double integration spans phase space (the velocity integration is taken at constant 
particle position $\mbf{r}$).  
Expanding the distribution function as before, we find to lowest order in the gyrokinetic ordering
\begin{equation}
\int \frac{d^{3}\mbf{r}}{V}\int d^{3}v \ \frac{h}{F_{0}} C[h] \le 0.
\end{equation}
Changing variables from particle position $\mbf{r}$ to guiding center position $\mbf{R}$,
we obtain
\begin{equation}
\int d^{3}v \int \frac{d^{3}\mbf{R}}{V} \ \frac{h}{F_{0}}  \left< C[h] \right>_{\mbf{R}} \le 0,
\label{eqn:gkdSdt}
\end{equation}
where now the velocity integration is taken at constant $\mbf{R}$.  In this case, the non-locality
of the gyroaveraging operation leads to no modification of the $H$-Theorem because of the definition
of entropy as a phase-space averaged quantity (as opposed to local conservation properties,
which involve only velocity-space averages).  Therefore, one can easily diagnose entropy
generation and test numerical satisfaction of the $H$-Theorem in gyrokinetic simulations, as
we show in \secref{sec:cnumres}.


\section{Numerical implementation}
\label{sec:cnumimp}

It is convenient for numerical purposes to separately treat collisional and collisionless
physics.  Thus, we begin by writing the gyrokinetic equation in the form
\begin{equation}
\pd{h_{\mbf{k}}}{t} = C_{GK}[h_{\mbf{k}}] + \mathcal{A}[h_{\mbf{k}}],
\end{equation}
where $\mathcal{A}[h_{\mbf{k}}]$
represents the rate of change of $h_{\mbf{k}}$ due to the collisionless physics.  
In order to separate these terms, we utilize Godunov dimensional 
splitting~\cite{godunovMS59}, which is accurate to first order in the timestep 
$\Delta t$:
\begin{gather}
\frac{h^{*}_{\mbf{k}}-h^{n}_{\mbf{k}}}{\Delta t} 
= \mathcal{A}[h^{n}_{\mbf{k}},h^{*}_{\mbf{k}}] \label{eqn:hstar}\\
\frac{h^{n+1}_{\mbf{k}}-h^{*}_{\mbf{k}}}{\Delta t}
=C_{GK}[h^{n+1}_{\mbf{k}}], 
\label{eqn:dhdtC}
\end{gather}
where $n$ and $n+1$ are indices representing the current and future time steps,
and $h^{*}_{\mbf{k}}$ is defined by Eq.~(\ref{eqn:hstar}) -- it is the 
result of advancing the collisionless part of the gyrokinetic equation.  With $h^{*}_{\mbf{k}}$
thus given, we restrict our attention to solving Eq.~(\ref{eqn:dhdtC}).
For notational convenience, we suppress all further $\mbf{k}$ subscripts, as we will be
working exclusively in $k$-space.

As argued in \secref{sec:cgkprop}, we must treat the collision operator implicitly to avoid a stability 
limit on the size of $\Delta t$.
We use a first order accurate backward-difference scheme in time 
instead of a second order scheme (such as Crank-Nicholson~\cite{richtmyer})
because it is well known that the Crank-Nicholson scheme
introduces spurious behavior in solutions to diffusion equations
when taking large timesteps (and because Godunov splitting is only first order
accurate for multiple splittings, which will be introduced shortly).

With this choice, $h^{n+1}$ is given by
\begin{equation}
h^{n+1} = \left(1-\Delta t  C_{GK}\right)^{-1}h^{*}.
\end{equation}
In general, $C_{GK}$ is a dense matrix, with both energy and pitch-angle
indices.  Inversion of such a matrix, which is necessary to solve for $h^{n+1}$ in our
implicit scheme, is computationally expensive.  We avoid this by taking two additional
simplifying steps.  First we employ another application of the Godunov splitting technique,
which, combined with the choice of a $(\xi,v)$ grid in \verb#Trinity#~\cite{barnes2PoP08}, allows us to consider energy 
and pitch-angle dependence separately:~\footnote{We note that an implicit treatment of the Catto-Tsang operator (including energy diffusion)
has independently been implemented in \texttt{GS2} using the same splitting technique~\cite{navkalAPS06}.}
\begin{gather}
h^{**} = \left[1-\Delta t \left( 
L + U_L\right)\right]^{-1} h^{*}
\label{eqn:Lsplit}
\\
h^{n+1}=\left[1 - \Delta t \left(D
+ U_D+ E\right)\right]^{-1}h^{**},
\label{eqn:DMEsplit}
\end{gather} 
The $h^{*}$ and $h^{**}$ are vectors whose 
components are the values of $h$ at each of the $(\xi,v)$ grid points.  In Eq.~(\ref{eqn:Lsplit})
we order the components so that
\begin{equation}
h \equiv \left(h_{11},h_{21},...,
h_{N1},h_{12},...,h_{NM}\right)^{T},
\end{equation}
where the first index represents pitch-angle, the second represents energy, and 
$N$ and $M$ are the number of pitch-angle and energy grid points, respectively.
This allows for a compact representation in pitch-angle.  When solving Eq.~(\ref{eqn:DMEsplit}),
we reorder the components of $h$ so that
\begin{equation}
h \equiv \left(h_{11},h_{12},...,
h_{1N},h_{21},...,h_{NM}\right)^{T},
\end{equation}
allowing for a compact representation in energy.

\subsection{Conserving terms}

The matrices $1-\Delta t L$ and 
$1-\Delta t D$ are chosen
to be tridiagonal by employing three-point stencils for finite differencing in $\xi$ and $v$.  
This permits computationally
inexpensive matrix inversion.  However, the full matrices to be inverted include the
momentum- and energy-conserving operators, $U$ and $E$, which are dense
matrices.  We avoid direct inversion of these matrices by employing the Sherman-Morrison
formula~\cite{shermanAMS49,shermanAMS50}, which gives $\mbf{x}$ in the matrix 
equation $M\mbf{x}=\mbf{b}$,
as long as $M$ can be written in the following form:
\begin{equation}
M = A + \mathbf{u}\otimes\mathbf{v},
\end{equation} 
where $\otimes$ is the tensor product.  The solution is then given by
\begin{equation}
\mbf{x} = \mbf{y} - \left[\frac{\mbf{v}\cdot\mbf{y}}{1+\mbf{v}\cdot\mbf{z}}\right]\mbf{z},
\end{equation}
where $\mbf{y}=A^{-1}\mbf{b}$, $\mbf{z}=A^{-1}\mbf{u}$,
and the dot products represent integrals over velocity space.  If $A^{-1}$ is known or
easily obtainable (as in our case), this formulation provides significant computational
savings over the straighforward method of directly inverting the dense matrix $M$.

Details of the application of the Sherman-Morrison formula to Eqs.~(\ref{eqn:Lsplit}) and
(\ref{eqn:DMEsplit}) are given in \apref{app:sm}.  Here, we state the main points.  
The matrix operators $L$ and $D$ are to be identified with $A$, and the integral conserving
terms $U$ and $E$ can be written in the form of the tensor product, $\mbf{u}\otimes\mbf{v}$. 
Identifying
$h^{**}$ and $h^{n+1}$ with $\mbf{x}$, we find that multiple applications of the 
Sherman-Morrison formula give
\begin{equation}
h = \mbf{y}_{2} - \left[\frac{\mbf{v}_{2}\cdot \mbf{y}_{2}}{1+\mbf{v}_{2}\cdot\mbf{z}_{2}}\right]\mbf{z}_{2},
\label{eqn:hx}
\end{equation}
where 
\begin{gather}
\mbf{y}_{2} = \mbf{y}_{0} - \left[\frac{\mbf{v}_{0}\cdot\mbf{y}_{0}}{1+\mbf{v}_{0}\cdot\mbf{s}_{0}}\right]\mbf{s}_{0} - \left[\frac{\mbf{v}_{1}\cdot\mbf{y}_{0}}{1+\mbf{v}_{1}\cdot\mbf{w}_{0}}\right]\mbf{w}_{0}, \label{eqn:sm1}\\
\mbf{z}_{2} = \mbf{z}_{0} - \left[\frac{\mbf{v}_{0}\cdot\mbf{z}_{0}}{1+\mbf{v}_{0}\cdot\mbf{s}_{0}}\right]\mbf{s}_{0}- \left[\frac{\mbf{v}_{1}\cdot\mbf{z}_{0}}{1+\mbf{v}_{1}\cdot\mbf{w}_{0}}\right]\mbf{w}_{0}. \label{eqn:sm2}
\end{gather}
The quantities $\mbf{v}_{0}$, $\mbf{v}_{1}$, $\mbf{v}_{2}$, $\mbf{z}_{0}$, $\mbf{s}_{0}$, $\mbf{w}_{0}$, and $\mbf{y}_{0}$
are specified in Table~\ref{tab:sm} in \apref{app:sm}.  With the exception
of $\mbf{y}_{0}$, each of these quantities
is time-independent, so they need be computed only once at the
beginning of each simulation.  Consequently, inclusion of the conserving terms in our
implicit scheme comes at little additional expense.

We note that when Eq.~(\ref{eqn:hx}) is applied to computing the inverse matrix in 
Eq.~(\ref{eqn:Lsplit}), the corresponding $\mbf{v}_{2}$ 
is nonzero only for the electron collision operator.
This term arises by using the parallel component of Ampere's law to rewrite the electron-ion
collison operator of Eq.~(\ref{eqn:Cei}) as
\begin{equation}
\begin{split}
C_{GK}^{ei}[h_{e}] &= \nu_{D}^{ei}\Bigg(\frac{1}{2}\pd{}{\xi}\left(1-\xi^{2}\right)\pd{h_{e}}{\xi} - \frac{k^{2}v^{2}}{4\Omega_{0,e}^{2}}\left(1+\xi^{2}\right)h_{e}\\
&+\frac{2v_{\parl}}{v_{th_{e}}^{2}}J_{0}(a_{e})F_{0e}\left[u_{\parl}[h_{e}]+\frac{ck^{2}}{4\pi e n_{0,e}}A_{\parl}\right]\Bigg),
\end{split}
\end{equation}
where $e$ is the magnitude of the electron charge, $u_{\parl}[h_{e}]$ is the parallel
component of the electron 
fluid velocity, and $n_{0,e}$ is the equilibrium electron density.
For electron collisions, the $A_{\parl}$ term is absorbed into $h^{*}$ so that we use
the modified quantity
\begin{equation}
\tilde{h}_{e}^{*} = h_{e}^{*} + \nu_{D}^{ei}\Delta t\frac{ck^{2}v_{\parl}}{2\pi e v_{th_{e}}^{2}n_{0,e}}A_{\parl}J_{0}(a_{e})F_{0,e}
\end{equation}
when applying the Sherman-Morrison formula, where $A_{\parl}$ from the $n+1$ time
level is used (for details on the implicit calculation of $A_{\parl}$, see Ref.~\cite{kotschCPC95}).

\subsection{Discretization in energy and pitch angle}

We still must specify our choice of discretization for $C_{GK}$.
Ideally, we would like the discrete scheme to guarantee the 
conservation properties and $H$-Theorem associated with $C$.  
As discussed in \secref{sec:cgkprop}, the former is equivalent to requiring that the $k\rho=0$ component
of $C_{GK}$, $C_{GK}^{0}$, satisfy Eq.~(\ref{eqn:momcons}).  We now
proceed to show that this requirement is satisfied by carefully discretizing the conserving
terms and by employing a novel finite difference
scheme that incorporates the weights associated with our numerical integration scheme.

We begin by writing $C_{GK}^{0}[h]$ for same-species collisions:
\begin{equation}
\begin{split}
C_{GK}^{0}[h]& = \frac{\nu_{D}}{2}\pd{}{\xi}\left(1-\xi^{2}\right)\pd{h}{\xi}
+\frac{1}{2v^{2}}\pd{}{v}\left(\nu_{\parl}v^{4}F_{0}\pd{}{v}\frac{h}{F_{0}}\right)
+ \nu_{D}v_{\parl}F_{0}\frac{\int d^{3}v \ \nu_{D} v_{\parl}h}{\int d^{3}v \ \nu_{D}v_{\parl}^{2}F_{0}}\\
&- \Delta\nu v_{\parl}F_{0}\frac{\int d^{3}v \ \Delta\nu v_{\parl}h}{\int d^{3}v \ \Delta\nu v_{\parl}^{2}F_{0}}
+ \nu_{E}v^{2}F_{0}\frac{\int d^{3}v \ \nu_{E}v^{2}h}{\int d^{3}v \ \nu_{E}v^{4}F_{0}}.
\label{eqn:cgk0}
\end{split}
\end{equation}
With $C_{GK}^{0}$ thus specified, we now consider numerical evaluation of the relevant moments of Eq.~(\ref{eqn:cgk0}).  
To satisfy number conservation ($\int d^{3}v \ C_{GK}^{0}[h]=0$), velocity space
integrals of each of the terms in Eq.~(\ref{eqn:cgk0}) should vanish individually.  For
integrals of the first two terms to vanish, we require a finite difference scheme that satisfies 
a discrete analog of
the Fundamental Theorem of Calculus (i.e. conservative differencing); for the last three
terms, we must have a discrete integration scheme satisfying 
$\int d^{3}v \ \nu_{D}v_{\parl}F_{0}=\int d^{3}v \ \Delta \nu v_{\parl}F_{0} 
= \int d^{3}v \ \nu_{E}v^{2}F_{0} = 0$. 
The requirement that $\int d^{3}v \ \nu_{D}v_{\parl}F_{0}=\int d^{3}v \ \Delta\nu v_{\parl}F_{0}=0$ 
is satisfied by any integration scheme with velocity space grid points and associated 
integration weights symmetric about $v_{\parl}=0$, which 
is true for the $(\xi,v)$ grid used in \verb#Trinity#.  By substituting for $\nu_{E}$ everywhere using
the identity
\begin{equation}
\nu_{E}v^{2}F_{0}=-\frac{1}{v^{2}}\pd{}{v}\left(\nu_{\parl}v^{5}F_{0}\right),
\label{eqn:nue1}
\end{equation}
the other integral constraint
($\int d^{3}v \ \nu_{E}v^{2}F_{0}=0$) reduces to the requirement that finite difference
schemes must satisfy the Fundamental Theorem of Calculus.

Parallel momentum conservation ($\int d^{3}v \ v_{\parl}C_{GK}^{0}[h]=0$) 
introduces the additional requirements that:
\begin{equation}
\int d^{3}v \ v_{\parl}\left(\frac{\nu_{D}}{2}\pd{}{\xi}\left(1-\xi^{2}\right)\pd{h}{\xi}
+ v_{\parl}\nu_{D}F_{0}\frac{\int d^{3}v \ \nu_{D}v_{\parl}h}{\int d^{3}v \ \nu_{D}v_{\parl}^{2}F_{0}}\right)=0
\label{eqn:vpaL}
\end{equation}
and
\begin{equation}
\int d^{3}v \ \frac{v_{\parl}}{2v^{2}}\pd{}{v}\left(\nu_{\parl}v^{4}F_{0}\pd{}{v}\frac{h}{F_{0}}\right)
= \int d^{3}v \ \Delta\nu v_{\parl}^{2}F_{0}\frac{\int d^{3}v \ \Delta\nu v_{\parl}h}{\int d^{3}v \ \Delta\nu v_{\parl}^{2}F_{0}}.
\label{eqn:vpaD}
\end{equation}
If the finite difference scheme used for all differentiation possesses a discrete version of 
integration by parts (upon double 
application), then Eqs.~(\ref{eqn:vpaL}) and $(\ref{eqn:vpaD})$ are numerically 
satisfied as long as:
$v_{\parl}\nu_{D}h$ in the second term of Eq.~(\ref{eqn:vpaL}) is expressed in the form
\begin{equation}
v_{\parl}\nu_{D}h = -\frac{1}{2}\left(\pd{}{\xi}\left(1-\xi^{2}\right)\pd{v_{\parl}}{\xi}\right)\nu_{D}h,
\label{eqn:vpa}
\end{equation}
$\Delta \nu$ on the righthand side of Eq.~(\ref{eqn:vpaD}) is expressed using the identity
\begin{equation}
2\Delta \nu v^{3}F_{0} = \pd{}{v}\left(\nu_{\parl}v^{4}F_{0}\pd{v}{v}\right),
\label{eqn:delnu}
\end{equation}
and all integrals are computed using the same numerical integration scheme
(if analytic results for the integral denominators in terms three and four of Eq.~(\ref{eqn:cgk0}) 
are used, then the necessary exact cancellation in Eqs.~(\ref{eqn:vpaL}) and (\ref{eqn:vpaD})
will not occur).

The only additional constraint imposed by the energy conservation requirement
($\int d^{3}v \ v^{2}C_{GK}^{0}[h]=0$) is that the form of Eq.~(\ref{eqn:nue1})
be slightly modified so that
\begin{equation}
\nu_{E}v^{2}F_{0} = -\frac{1}{v^{2}}\pd{}{v}\left(\nu_{\parl}v^{4}F_{0}\pd{v^{2}}{v}\right),
\label{eqn:nue}
\end{equation}
which still satisfies the number conservation contraint.
Using the forms given by Eqs.~(\ref{eqn:vpa})-(\ref{eqn:nue}), conservation properties
are guaranteed as long as one
employs a finite difference scheme for pitch-angle scattering and energy diffusion that
satisfies discrete versions of the Fundamental Theorem of Calculus and integration by parts.

For the case of equally spaced grid points in $v$ and $\xi$, there is a straightforward 
difference scheme, accurate to secord order in the grid spacing, that satisfies both 
requirements~\cite{degondNM94}:
\begin{equation}
\pd{}{x}G\pd{h}{x} \approx \frac{G_{j+1/2}\left(h_{j+1}-h_{j}\right)-G_{j-1/2}\left(h_{j}-h_{j-1}\right)}{\Delta x^{2}},
\end{equation}
where $x$ is a dummy variable representing either $v$ or $\xi$, $\Delta x$ is the grid
spacing, $h_{j}$ is the value of $h$ evaluated at the grid point $x_{j}$, 
$x_{j\pm 1/2}\equiv (x_{j}+x_{j\pm 1})/2$, and $G$ is either $1-\xi^{2}$ (for 
pitch-angle scattering) or $\nu_{\parl}v^{4}F_{0}$ (for energy diffusion).
However, in order to achieve higer order accuracy in the calculation
of the velocity space integrals necessary to obtain electromagnetic fields,
\verb#GS2#~\cite{barnes2PoP08} and a number of other
gyrokinetic codes~\cite{candyJCP03} use grids with unequal spacing in $v$ and $\xi$ and integration weights
that are not equal to the grid spacings.  

Given the constraints
of a three-point stencil on an unequally spaced grid, we are forced to choose between a 
higher order scheme (a second order accurate scheme can be obtained with compact 
differencing~\cite{durran}, as described in \apref{app:cmpctdiff}) that does not satisfy our two requirements and a lower order scheme that does.
Since our analytic expression for $C_{GK}$ was designed in large part to satisfy conservation 
properties (and because the conserving terms are only a zeroth order accurate approximation
to the field-particle piece of the linearized Landau operator~\cite{hirshmanPoF76}), we 
choose the lower order scheme, given here, as the default:
\begin{equation}
\pd{}{x}G\pd{h}{x} \approx \frac{1}{w_{j}}\left(G_{j+1/2}\frac{h_{j+1}-h_{j}}{x_{j+1}-x_{j}}-G_{j-1/2}\frac{h_{j}-h_{j-1}}{x_{j}-x_{j-1}}\right),
\label{eqn:fd}
\end{equation}
where $w_{j}$ is the integration weight associated with $x_{j}$.  

\begin{figure}
\includegraphics[height=3.0in]{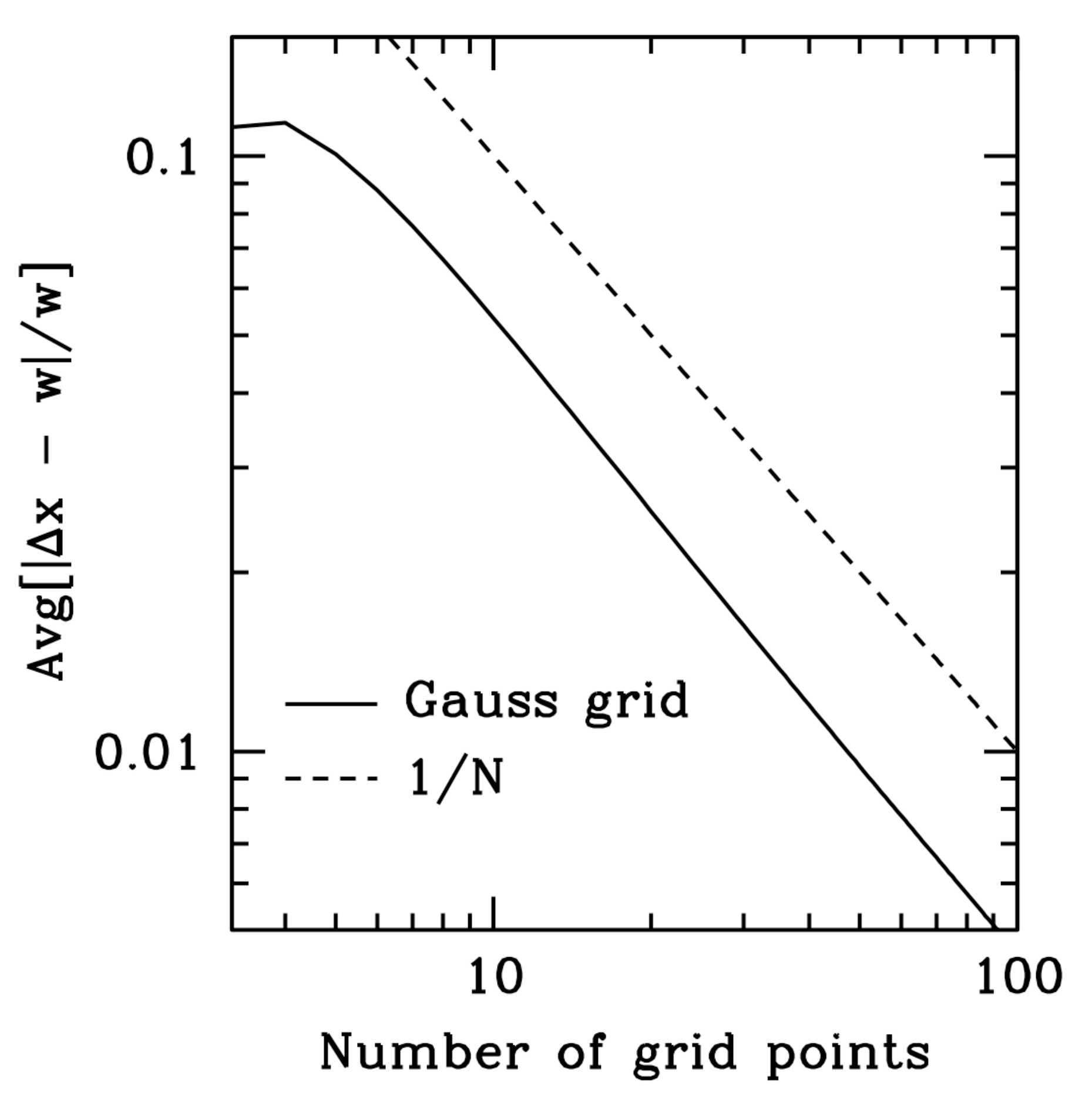}
\includegraphics[height=3.0in]{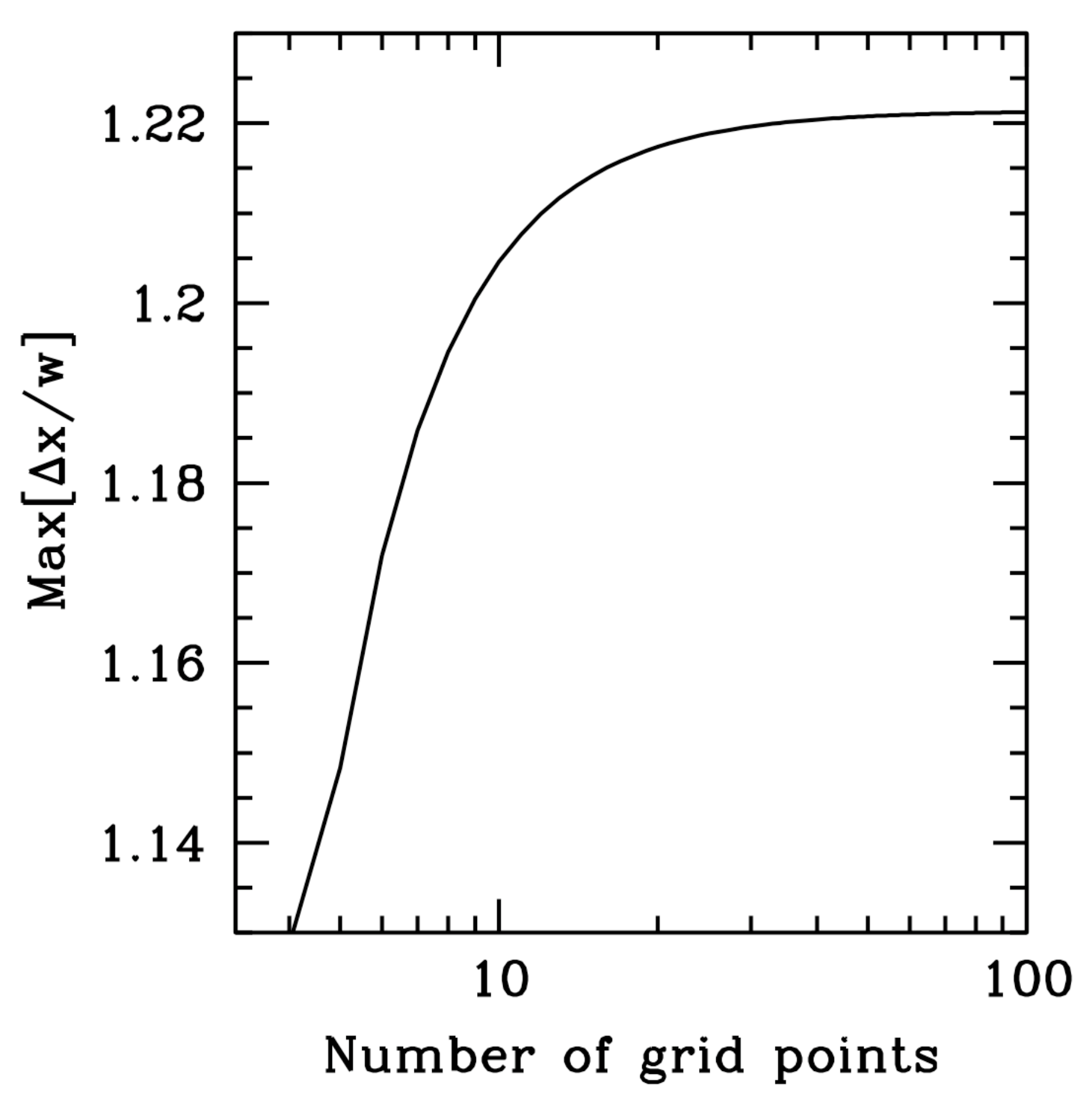}
\caption{(Left): Solid line indicates the scaling of the leading order error, averaged over 
all grid points, of the conservative finite difference scheme for a Gauss-Legendre grid (the 
grid used in \texttt{Trinity}).  The slope of the dotted line corresponds to a first order scheme.
(Right): factor by which the conservative finite difference scheme
of Eq.~(\ref{eqn:fd}) amplifies the true collision operator amplitude at the boundaries of 
the Gauss-Legendre grid.}
\label{fig:errfd}
\end{figure}

\begin{figure*} 
\begin{center}
\includegraphics[height=2.6in]{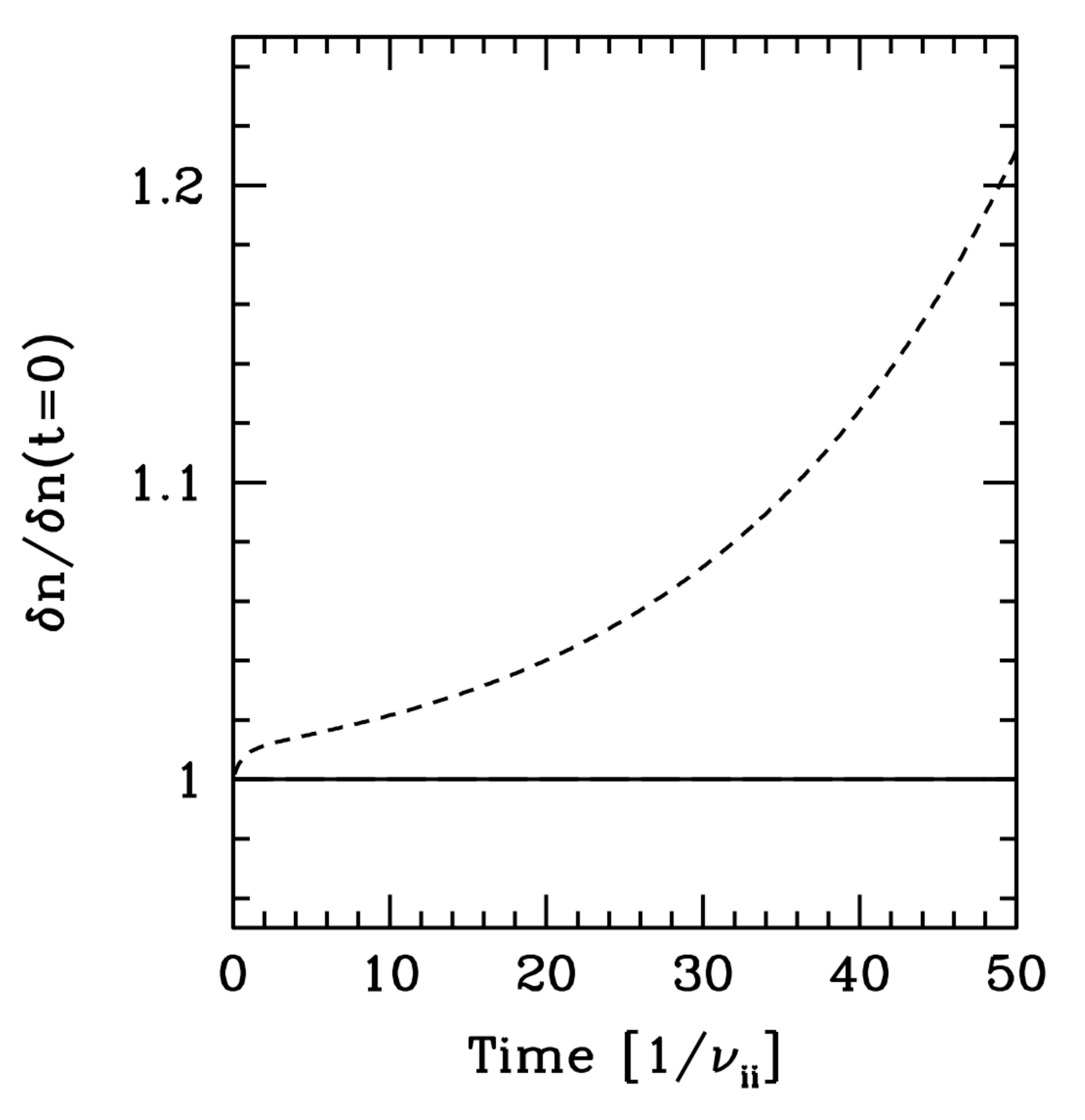}
\includegraphics[height=2.6in]{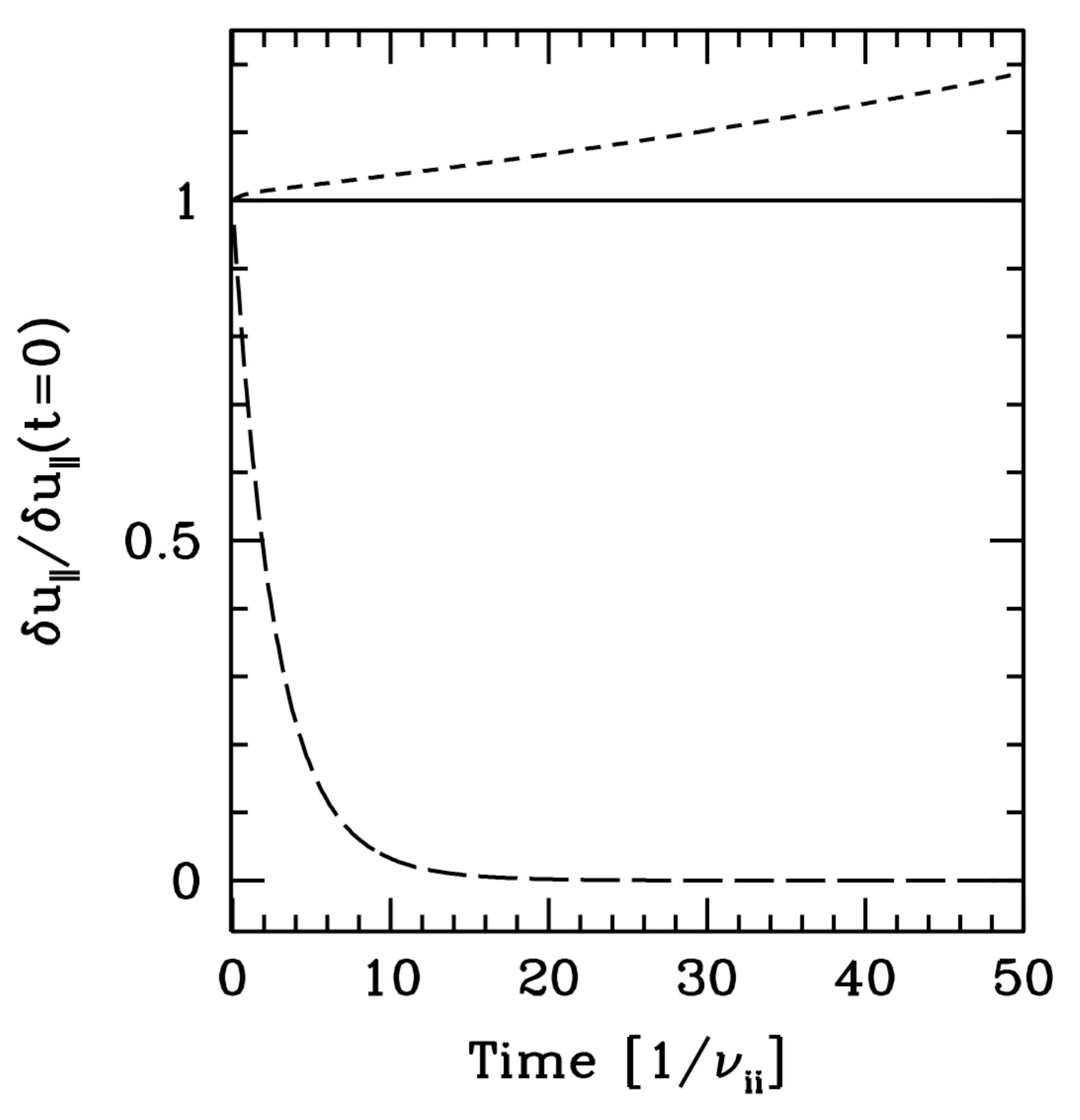}
\includegraphics[height=2.6in]{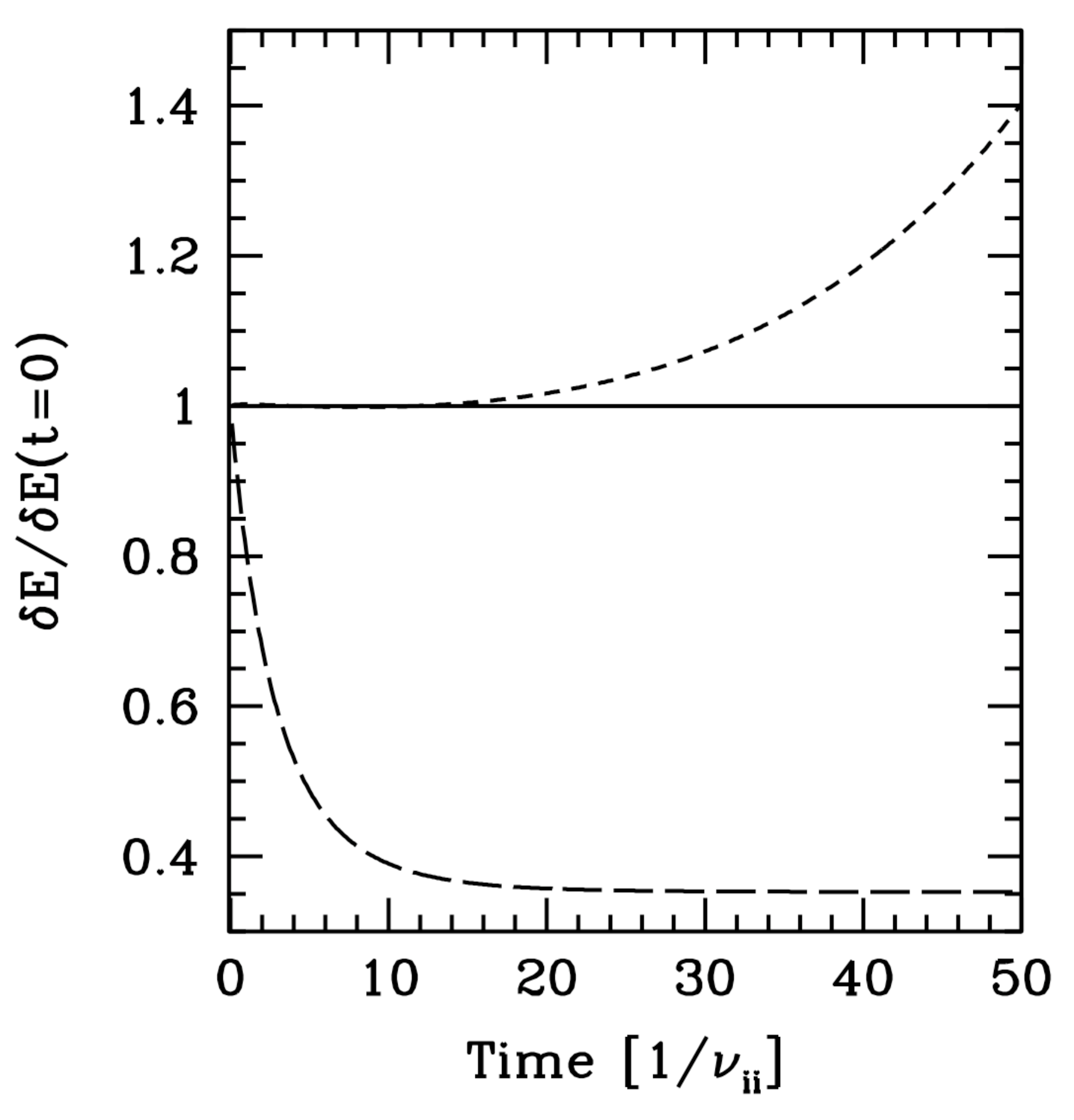}
\end{center}
\caption{Plots showing evolution of the perturbed local density, parallel momentum, and energy 
over fifty collision times.  Without the conserving terms (\ref{eqn:scriptuL})-(\ref{eqn:scripte}), 
both parallel momentum and energy decay significantly over a few collision times (long dashed lines).  
Inclusion of conserving terms with the conservative scheme detailed in \secref{sec:cnumimp} leads to exact 
moment conservation (solid lines).  Use of a non-conservative scheme leads to inexact 
conservation that depends on grid spacing (short dashed lines).}
\label{fig:conserve}
\end{figure*}

Defining
$\Psi\equiv Gh'$, with the prime denoting differentiation with respect to $x$, Taylor
series can be used to show
\begin{equation}
\frac{\Psi_{j+1/2}-\Psi_{j-1/2}}{w_{j}} = \Psi'_{j}\frac{\Delta x_{j}}{w_{j}} + \mathcal{O}\left(\frac{\left(\Delta x\right)_{j}^{2}}{w_{j}}\right ),
\label{eqn:fderr}
\end{equation}
where $\Delta x_{j}=x_{j+1/2}-x_{j-1/2}$.  With the exception of pitch angles
corresponding to trapped particles~\cite{barnes2PoP08,kotschCPC95}, the grid points
$\{x_{j}\}$ and associated integration weights $\{w_{j}\}$ in \verb#Trinity#
are chosen according to Gauss-Legendre quadrature rules~\cite{hilde}.  For this case, we show 
numerically in Fig.~\ref{fig:errfd} that 
\begin{equation}
\frac{1}{N}\sum_{j=1}^{N}\frac{\Delta x_{j}}{w_{j}}
=1+ \mathcal{O}\left(\frac{1}{N}\sum_{j=1}^{N}\Delta x_{j}\right)
=1 + \mathcal{O}\left(\frac{1}{N}\right),
\end{equation}
and
\begin{equation}
\max_{j= 2,...,N-1}\left| 1 - \frac{\Delta x_{j}}{w_{j}} \right| = \mathcal{O}\left(\frac{1}{N}\right),
\end{equation}
where $N$ is the number of grid points in $x$.

The boundary points ($j=1,N$) are excluded from the $\max$ operator above.  This
is because $\Delta x / w$ (the factor multiplying $\Psi_{j}'$ in Eq.~(\ref{eqn:fderr}))
converges to approximately 1.2 for the boundary points 
as the grid spacing is decreased (Fig.~\ref{fig:errfd}).  For the energy diffusion operator, we can make use of
the property that $G(x)=G(x)'=0$ at $x=0$ and $x=\infty$ to show
\begin{equation}
\pm \frac{\Psi_{j\pm1/2}}{w_{j}} = \Psi_{j}' + \mathcal{O}\left(\frac{\left(\Delta x\right)_{j}^{2}}{w_{j}}\right),
\end{equation}
with the plus sign corresponding to $j=1$ and the minus sign to $j=N$.  This is not true
for the Lorentz operator, so we are forced to accept an approximately twenty percent
magnification of the Lorentz operator amplitude at $\xi=\pm1$ and at the trapped-passing
boundaries.  We find that this relatively 
small error at the boundaries has a negligible effect on measureable (velocity space averaged)
quantities in our simulations.

\section{Numerical tests}
\label{sec:cnumres}

We now proceed to demonstrate the validity of our collision operator implementation.  In
particular, we demonstrate conservation properties, satisfaction of Boltzmann's
$H$-Theorem, efficient smoothing in velocity space, and recovery of theoretically expected
results in both collisional (fluid) and collisionless limits.  While we do not claim that the
suite of tests we have performed is exhaustive, it constitutes a convenient set of numerical
benchmarks that can be used for validating collision operators in gyrokinetics.

\subsection{Homogeneous plasma slab}

We first consider the long wavelength limit of a homogeneous plasma slab with Boltzmann
electrons and no variation along the background magnetic field ($k_{\parl}=0$).
The gyrokinetic equation for this system simplifies to
\begin{equation}
\pd{\left(\delta f\right)}{t} \approx C_{GK}^{0}[h],
\end{equation}
which means local density, momentum, and energy should be conserved.  In Fig.~\ref{fig:conserve}
we show numerical results for the time evolution of the local density, momentum, and energy for this system.

Without inclusion of the conserving terms (\ref{eqn:scriptuL})-(\ref{eqn:scripte}), we 
see that density is conserved, as guaranteed by the conservative differencing scheme,
while the momentum and energy decay away over several collision times.
Inclusion of the conserving terms provides us with exact (up to numerical precision)
conservation of number, momentum, and energy.  To illustrate the utility of our
conservative implementation, we also present results from a numerical scheme that does not
make use of Eqs.~(\ref{eqn:vpa})-(\ref{eqn:nue}) and that employs a finite difference
scheme that does not possess discrete versions of the Fundamental Theorem of Calculus
and integration by parts.  Specifically, we consider a first order accurate finite difference
scheme similar to that given by Eq.~(\ref{eqn:fd}), with the only difference being
that the weights in the denominator are replaced with the local grid spacings.  In this case,
we see that density, momentum, and energy are not exactly conserved (how well they
are conserved depends on velocity space resolution, which is 16 pitch angles and 16 energies
for the run considered here).

The rate at which our collision operator generates entropy in the homogenous plasma slab 
is shown in Fig.~\ref{fig:htheorem}.  As required by the $H$-Theorem, the rate of entropy 
production is always nonnegative and approaches
zero in the long-time limit as the distribution function approaches a shifted Maxwellian.
We find this to hold independent of both the grid spacing in velocity space and the
initial condition for the distribution function (in Fig.~\ref{fig:htheorem}, the values
of $h(\xi,v)$ were drawn randomly from the uniform distribution on the interval $[-1/2,1/2]$).

\begin{figure} 
\begin{center}
\includegraphics[height=3.3in]{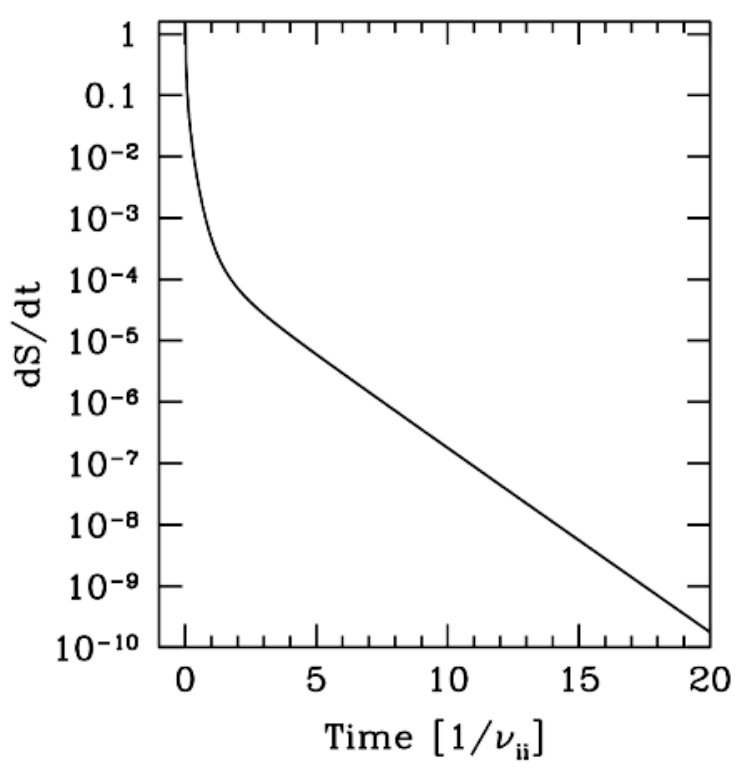}
\end{center}
\caption{Plot of the evolution of entropy generation for the homogeneous plasma slab
over twenty collision times.  Our initial distribution in velocity space is random noise, and 
we use a grid with 16 pitch angles and 8 energies.  The 
entropy generation rate is always nonnegative and approaches zero in the long-time limit.}
\label{fig:htheorem}
\end{figure}

\subsection{Resistive damping}

We now modify the system above by adding a finite $A_{\parl}$.  From fluid theory, 
we know that collisional friction between electrons and ions provides
resistivity which leads to the decay of current profiles.  Because the resistive time is long
compared to the collision time, one can neglect $\partial (\delta f) / \partial t$.  However, since
$A_{\parl}\sim k^{-2}$, and we are considering $k\ll 1$, $\partial A_{\parl} / \partial t$ must 
be retained.  The resulting electron equation is of the form of the classical Spitzer problem (see, e.g., Ref.~\cite{helander}):
\begin{equation}
C_{GK}^{0}[h_{e}] = -\frac{eF_{0,e}}{T_{0,e}}\frac{v_{\parl}}{c}\pd{A_{\parl}}{t}.
\end{equation}
The parallel current evolution for this system is given by
\begin{equation}
J_{\parl}(t) = J_{\parl}(t=0)e^{-\eta t},
\label{eqn:djdt}
\end{equation}
where $\eta=1/\sigma_{\parl}$ is the resistivity, $\sigma_{\parl}=1.98 \tau_{e}n_{e}e^{2}/m_{e}c^{2}$ is
the Spitzer conductivity, and $\tau_{e}=3\sqrt{\pi} / 4 \nu_{ei}$ is the electron collision time.

We demonstrate that
the numerical implementation of our operator correctly captures this resistive damping
in Figs.~\ref{fig:apardamp} and~\ref{fig:upardamp}.  We also see in these figures that in the 
absence of the ion drag term from Eq.~(\ref{eqn:Cei}), the electron flow is incorrectly damped 
to zero (instead of to the ion flow), leading to a steady-state current.  

\begin{figure} 
\begin{center}
\includegraphics[height=3.3in]{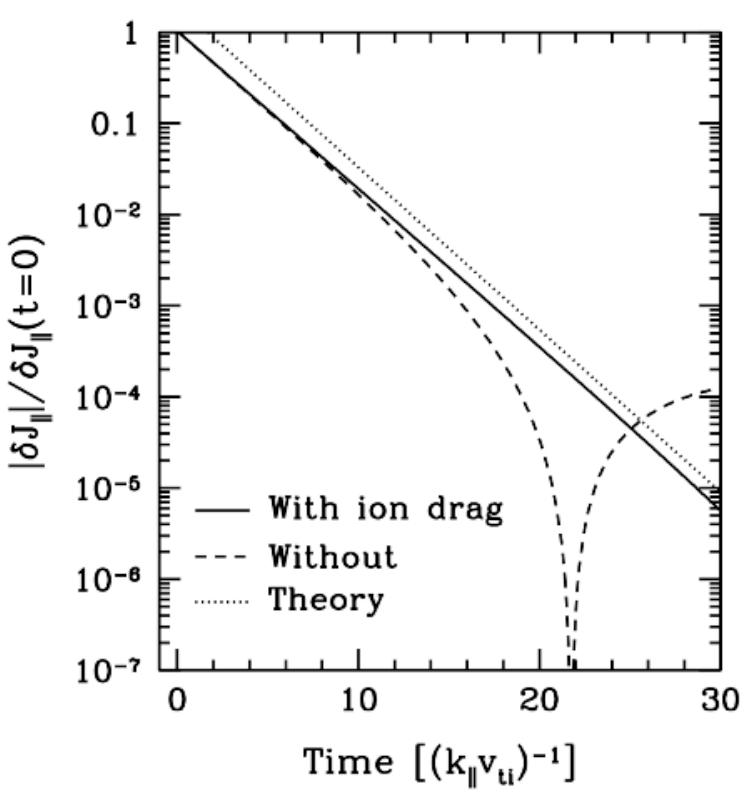}
\end{center}
\caption{Evolution of $|J_{\parl}|$ for the electromagnetic
plasma slab with $\beta=10^{-4}$, $k_{y}\rho_{i}=0.1$, and $\nu_{ei}=10 k_{\parl}v_{th,i}$.  Inclusion
of the ion drag term in the electron-ion collision operator leads to the theoretically predicted 
damping rate for the parallel current given in Eq.~(\ref{eqn:djdt}) .  Without the ion 
drag term, the parallel current decays past zero (at $t\approx22$) and converges to 
a negative value as the electron flow damps to zero.}
\label{fig:apardamp}
\end{figure}

\begin{figure} 
\begin{center}
\includegraphics[height=3.3in]{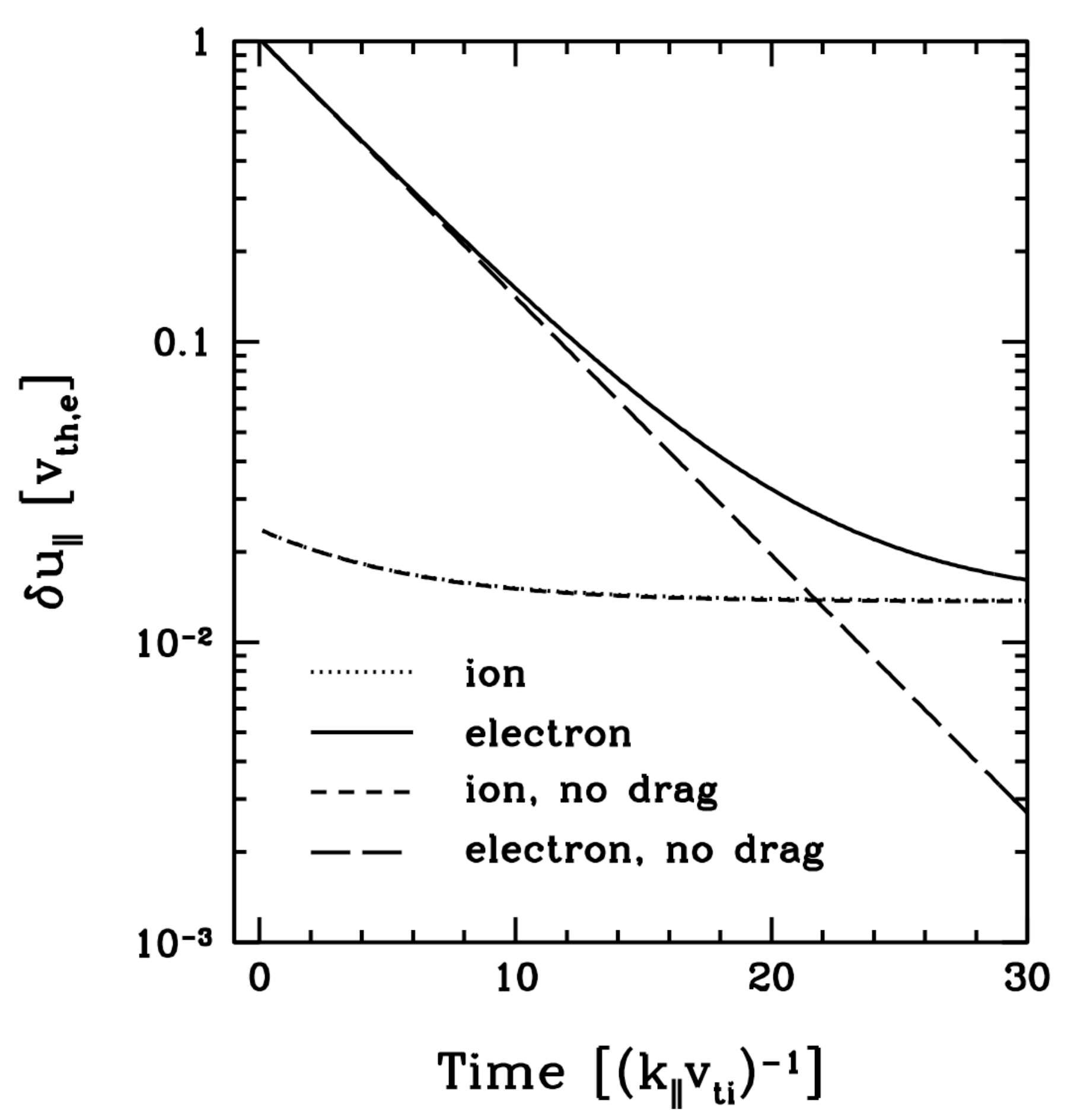}
\end{center}
\caption{Evolution of perturbed parallel flow for the electromagnetic
plasma slab with $\beta=10^{-4}$, $k_{y}\rho_{i}=0.1$, and $\nu_{ei}=10 k_{\parl}v_{ti}$.  
Without inclusion of the ion drag term in Eq.~(\ref{eqn:Cei}), the electron flow is 
erroneously damped to zero (instead of to the ion flow).}
\label{fig:upardamp}
\end{figure}

\subsection{Slow mode damping}

We next consider the damping of the slow mode in a homogenous plasma slab
as a function of collisionality.  In the low $k_{\perp}\rho_{i}$, high $\beta_{i}$
limit, one can obtain analytic expressions for the damping rate in both the collisional 
($k_{\parl}\lambda_{mfp}\ll 1$) and collisionless ($k_{\parl}\lambda_{mfp}\gg 1$) 
regimes, where $\lambda_{mfp}$ is the ion mean free path (see e.g. 
Ref.~\cite{schekAPJ07}).  The expressions are
\begin{equation}
\omega = \pm k_{\parl}v_{A}\sqrt{1-\left(\frac{\nu_{\parl,i}k_{\parl}}{2v_{A}}\right)^{2}}-i\frac{\nu_{\parl,i}k_{\parl}^{2}}{2}
\label{eqn:collisional}
\end{equation}
for $k_{\parl}\lambda_{mfp}\ll 1$, and
\begin{equation}
\omega = -i\frac{\left|k_{\parl}\right|v_{A}}{\sqrt{\pi\beta_{i}}}
\label{eqn:collisionless}
\end{equation}
for $k_{\parl}\lambda_{mfp}\gg 1$.  Here, $v_{A}=v_{th,i}/\sqrt{\beta_{i}}$ is
the Alfven speed, and $\nu_{\parl,i}$ is the parallel ion viscosity, which is inversely 
proportial to the ion-ion collision frequency, 
$\nu_{ii}$: $\nu_{\parl,i}\propto v_{th,i}^{2}/\nu_{ii}$.  As one would expect, the 
damping in the strongly collisional regime [Eq.~(\ref{eqn:collisional})] is due primarily to viscosity, while the collisionless
regime [Eq.~(\ref{eqn:collisionless})] is dominated by Barnes damping~\cite{barnesPoF66}.


In Fig.~\ref{fig:nuscan}, we plot the collisional dependence of the damping rate of the 
slow mode obtained numerically
using the new collision operator implementation in \verb#Trinity#.  In order to isolate the slow mode
in these simulations, we took $\Phi=A_{\parl}=\delta n_{e}=0$ and measured
the damping rate of $\delta B_{\parl}$.  This is possible because $\delta B_{\parl}$
effectively decouples from $\Phi$ and $A_{\parl}$ for our system, and $\delta n_{e}$
can be neglected because $\beta_{i}\gg1$~\cite{schekAPJ07}.
We find quantitative agreement with
the analytic expressions (\ref{eqn:collisional}) and (\ref{eqn:collisionless}) in
the appropriate regimes.  In particular, we recover the correct
viscous behavior in the $k_{\parl}\lambda_{mfp}\ll 1$ limit (damping rate
proportional to $\nu_{\parl,i}$), the correct collisional damping in the $k_{\parl}\lambda_{mfp}\sim 1$
limit (damping rate inversely proportial to $\nu_{\parl,i}$), and the correct collisionless 
(i.e. Barnes) damping in the $k_{\parl}\lambda_{mfp}\gg 1$ limit.

\begin{figure} 
\begin{center}
\includegraphics[height=3.3in]{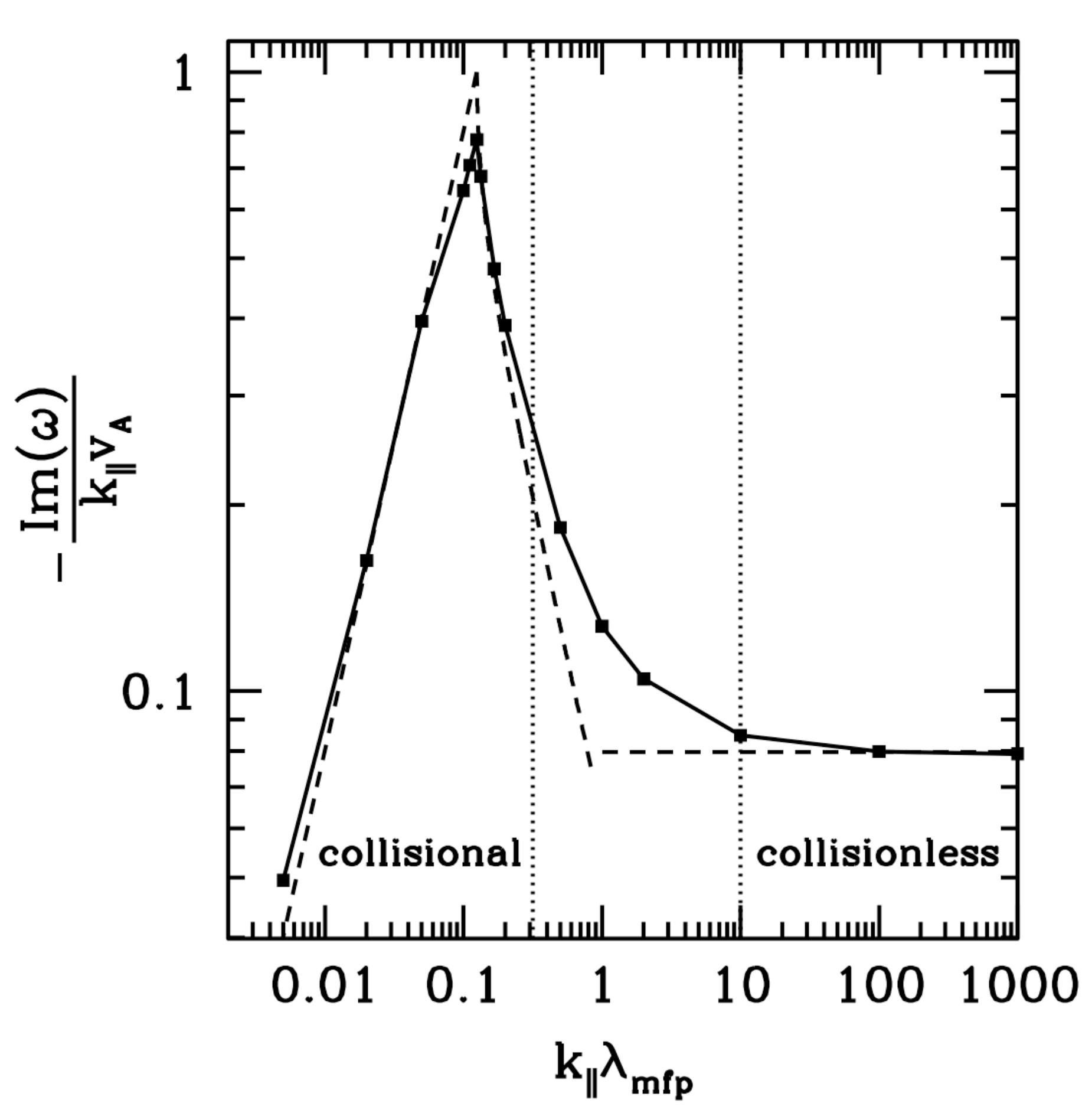}
\end{center}
\caption{Damping rate of the slow mode for a range of collisionalities spanning
the collisionless to strongly collisional regimes.  Dashed lines correspond to the 
theoretical prediction for the damping rate in the collisional ($k_{\parl}\lambda_{mfp}\ll 1$)
and collisionless ($k_{\parl}\lambda_{mfp}\gg 1$) limits.  The solid line is the result 
obtained numerically with \texttt{Trinity}.  Vertical dotted lines denote approximate regions 
(collisional and collisionless) for which the analytic theory is valid.}
\label{fig:nuscan}
\end{figure}


\subsection{Electrostatic turbulence}

Finally, we illustrate the utility of our collision operator in a nonlinear simulation of
electrostatic turbulence in a Z-pinch field configuration~\cite{freidbergMHD}.  We consider the Z-pinch
because it contains much of the physics of toroidal configurations (i.e. curvature)
without some of the complexity (no particle trapping).  At relatively weak pressure gradients,
the dominant gyrokinetic linear instability in the Z-pinch is the entropy 
mode~\cite{kadomtsevSP60,kesnerPoP00,simakovPoP01,simakovPoP02,kesnerPoP02,ricciPoP06},
which is nonlinearly unstable to secondary instabilities such as Kelvin-Helmholtz~\cite{ricciPRL06}.

In previous numerical investigations of linear~\cite{ricciPoP06} and nonlinear~\cite{ricciPRL06}
plasma dynamics in a Z-pinch, collisions were found to play an important role in the damping
of zonal flows and in providing an effective energy cutoff at short wavelengths.  
However, as pointed out in Ref.~\cite{ricciPRL06}, the Lorentz collision operator 
used in those investigations provided insufficient damping of short wavelength structures
to obtain steady-state fluxes.  Consequently, a model hyper-viscosity had to be employed.

We have reproduced a simulation from Ref.~\cite{ricciPRL06} using our new
collison operator, and we find that hyper-viscosity is no longer necessary to obtain
steady-state fluxes (Fig.~\ref{fig:nlzp}).  This can be understood by examining the linear growth
rate spectrum of Fig.~\ref{fig:lzp}.  We see that in this system energy diffusion is much
more efficient at suppressing short wavelength structures than pitch-angle scattering.
Consequently, no artificial dissipation of short wavelength structures is necessary.

\begin{figure} 
\begin{center}
\includegraphics[height=3.3in]{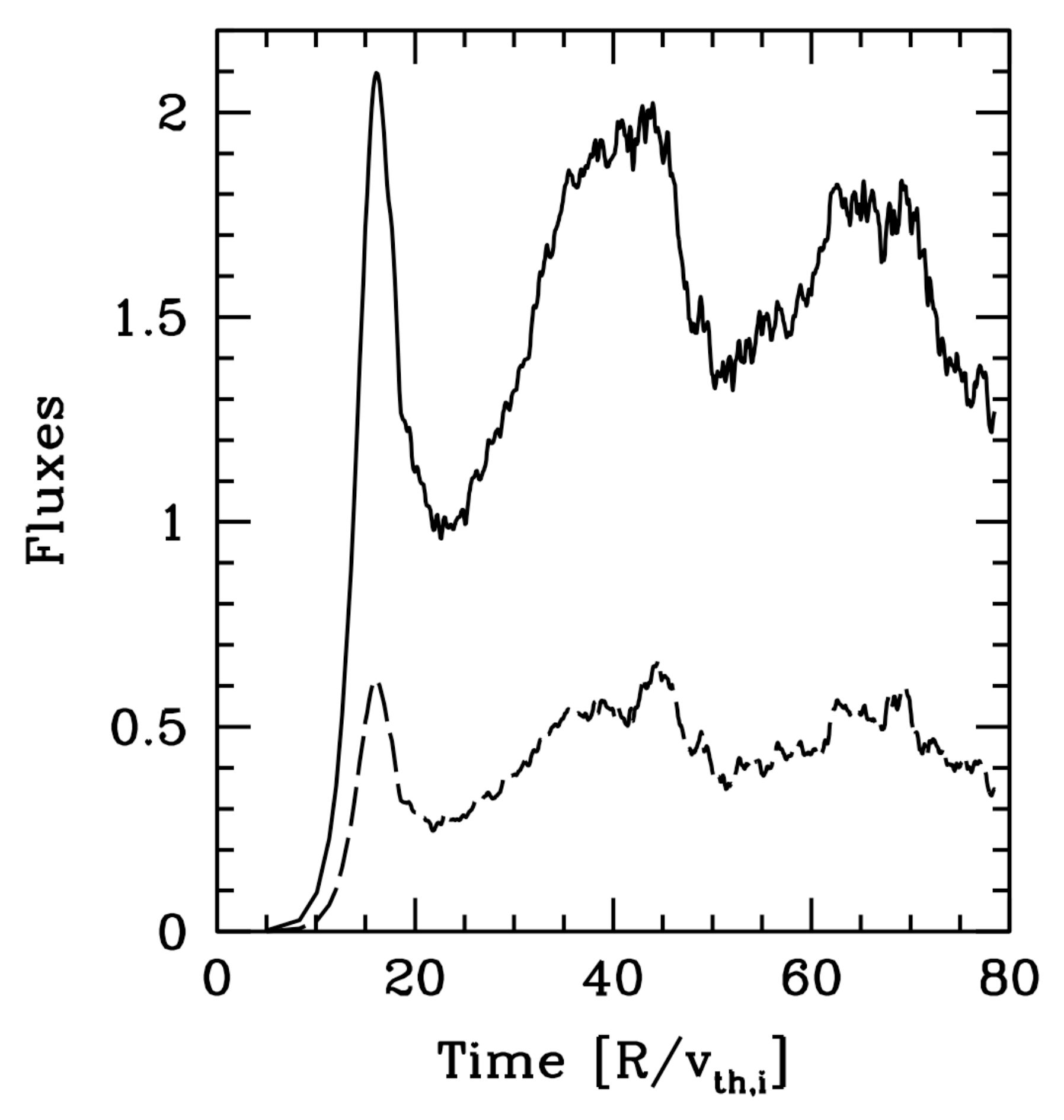}
\end{center}
\caption{Evolution of ion particle and heat fluxes for an electrostatic, 2-species Z-pinch 
simulation.  We are considering $R/L_{n}=2.0$ and $\nu_{ii}=0.01 v_{th,i}/R$.  The particle flux
is indicated by the solid line and is given in units of $(\rho / R) n_{0,i}v_{th,i}$.  The heat 
flux is indicated by the dashed line and is given in units of $(\rho/R) n_{0,i}v_{th,i}^{2}$.
We see that a steady-state is achieved for both fluxes without artificial dissipation.}
\label{fig:nlzp}
\end{figure}

\begin{figure} 
\begin{center}
\includegraphics[height=3.3in]{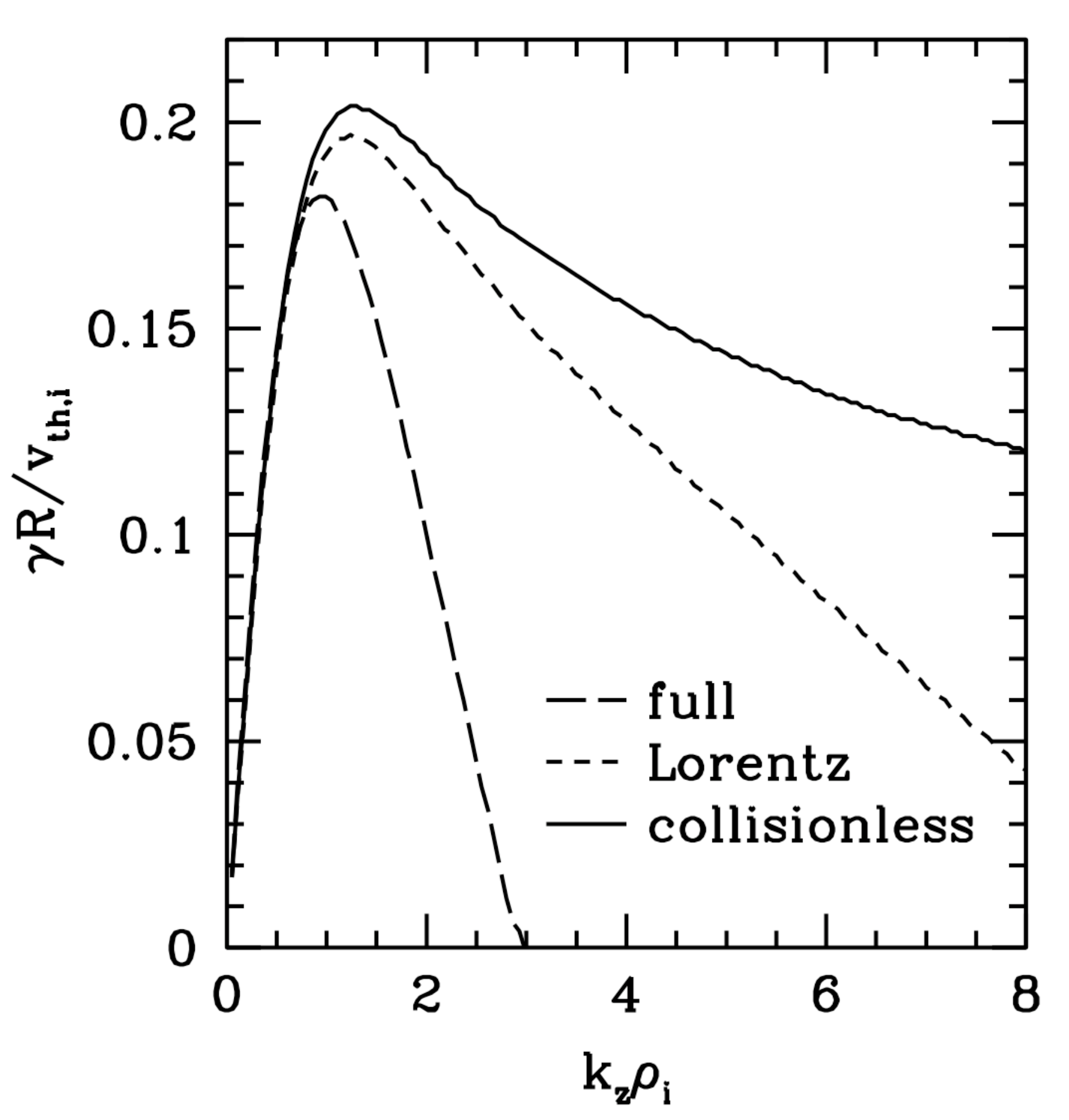}
\end{center}
\caption{Linear growth rate spectrum of the entropy mode in a Z-pinch for $R/L_{n}=2.0$, 
where $R$ is major radius and $L_{n}$ is density gradient scale length.  The solid line
is the collisionless result, and the two dashed lines represent the result of including collisions.
The short dashed line corresponds to using only the Lorentz operator, while the long dashed
line corresponds to using our full model collision operator.  Both collisional cases were
carried out with $\nu_{ii} = 0.01 v_{th,i}/R$.}
\label{fig:lzp}
\end{figure}

\section{Summary}
\label{sec:cnumsum}

In \secref{sec:cnumintro} we proposed a set of key properties that an ideal dissipation scheme
for gyrokinetics should satisfy.  Namely, the scheme should: limit the scale size of
structures in phase space in order to guarantee the validity of the gyrokinetic ordering 
and to provide numerical resolution at reasonable expense;
conserve particle number, momentum, and energy; and satisfy Boltzmann's $H$-Theorem.
While commonly employed simplified collision operators or
hyperviscosity operators may be adequate for some calculations~\cite{dimitsPoP00},
it is important to be able to use the more complete collision operator described in this
paper, which preserves all of these desirable dissipation properties.

In \secref{sec:cgkprop} we presented the model collision operator derived in Chapter 5 and discussed
some of its features that strongly influence our choice of numerical implementation.
In particular, we noted that local conservation properties are guaranteed as long as the
($1$, $v_{\parl}$, $v^{2}$) moments of the $k\rho=0$
component of the gyroaveraged collision operator vanish.  Further, we argued that
the collision operator should be treated implicitly because in some regions of phase space,
its amplitude can be large even at very small collisionalities.

Our numerical implementation of the collision operator was described in \secref{sec:cnumimp}.
We separate collisional and collisionless physics through the use of Godunov 
dimensional splitting and advance the collision operator in time using a backwards
Euler scheme.  The test particle part of the collision operator is differenced using
a scheme that possesses discrete versions of the Fundamental Theorem of Calculus
and integration by parts (upon double application).  These properties are necessary in order 
to exactly satisfy the desired conservation properties in the long wavelength limit.  The
field particle response is treated implicitly with little additional computational expense
by employing repeated application of the Sherman-Morrison formula, as detailed
in \apref{app:sm}

In \secref{sec:cnumres} we presented numerical tests to demonstrate that our implemented
collision operator possesses the properties required for a good gyrokinetic 
dissipation scheme.  In addition to these basic properties, we showed that the
implemented collision operator allows us to correctly capture physics phenomena ranging
from the collisionless to the strongly collisional regimes.  In particular, we provided
examples for which we are able to obtain quantitatively correct results for 
collisionless (Landau or Barnes), resistive, and viscous damping.  

In conclusion, we note that resolution of the collisionless (and collisional) physics in our
simulations was 
obtained solely with physical collisions; no recourse to any form of artificial numerical 
dissipation was necessary.

\renewcommand{\thechapter}{7}

\chapter{\textbf{Numerical framework for coupled turbulent transport calculations}}
\label{chap:msnum}
\vspace{+90pt}

\section{Overview}

As discussed in Chapter 1, any realistic model of turbulent transport and heating in hot,
magnetized plasmas must account for the interaction of small-scale, rapid fluctuations
and large-scale, slowly evolving equilibrium profiles.  The wide range of scales that must
be resolved makes direct numerical simulation prohibitively expensive.  Consider, for
instance, the range of time and space scales expected to be present in ITER (Tables~\ref{multimak1}
and~\ref{multimak2}).  In order to resolve the turbulent electron dynamics, the required
grid spacing would be on the order of $10^{-4}$ centimeters perpendicular to the magnetic field 
in space and $10^{-6}$ seconds in time.
To capture the evolution of equilibrium profiles, the simulation domain would have to 
be on the order of $200$ centimeters in space and $2-4$ seconds in time.  Assuming several grid 
points are necessary to resolve the smallest scales, this results in
$10^{6}-10^{7}$ grid points in each of two spatial dimensions (with an additional $10-100$ along 
the magnetic field) and $10^{8}-10^{9}$ grid points in time.
This is in addition to the two dimensions of velocity space, each of which requires
at least 10 grid points.  All told, this comes out to approximately $10^{25}$ grid points -- a factor of
$10^{10}$ larger than the largest fluid turbulence simulations possible on today's fastest supercomputers.

Clearly, the brute-force approach outlined above is not possible on any timescale of interest.
There is, however, a way forward.  The vast separation of time and space scales between
the fluctuations and the equilibrium can be exploited to allow for the separate evolution of
the two.  The theoretical formalism detailing the process of scale separation is described
in Chapter~\ref{chap:hierarchy}.
While these equations are coupled, the 
time and space scales addressed by each equation
are fundamentally different.  In this chapter, we detail the numerical framework of 
\verb#Trinity#~\cite{barnesAPS07}, a turbulent transport and heating code, which exploits the 
scale separation of these 
equations to greatly reduce the computational effort necessary to simulate turbulent transport 
and heating.

The moment equations~(\ref{eqn:nevo}) and~(\ref{eqn:pevo2}) describing the evolution of equilibrium
density and temperature profiles involve only slowly varying scales in space and time.  This
is by construction -- in deriving the equations, we averaged over the intermediate time and spatial
scales defined by Eqs.~(\ref{eqn:timeaverage}) and~\ref{eqn:spaceavg}.  Consequently, numerical
solution of these equations can be achieved using a course space-time grid, with spatial variation
only in the radial direction.

In order to evaluate the equilibrium evolution equations, one must specify values for the
time- and space-averaged turbulent fluxes and heating.  Whle the averaged quantities evolve on
equilibrium scales, the fluxes and heating themselves evolve on much faster fluctuation scales.
We must therefore conduct turbulence simulations on the fluctuation scales and space-time
average.  Since we only need the fluxes and heating on a coarse space-time grid for the
equilibrium evolution equations, there is no need to simulate the turbulence everywhere in the device,
nor to simulate it over the entire length of the discharge.  Instead, we can simulate the turbulence
in small regions of space and time (Fig.~\ref{fig:msgrid}) and couple these regions together using 
the equilibrium evolution equations.  
The only constraint is that the space-time domain for the turbulence
must be sufficiently large to include the longest wavelengths of the turbulence and to reach a steady-state.
This clearly provides a significant savings in both spatial and temporal resolution.  Returning to ITER
as our example, we see in Table~\ref{multimak1} that the longest turbulent timescale is on the order of 
$10^{-4}$ seconds.  Since the equilibrium evolution scale is approximately $1$ second, we obtain a factor
of $10^{2}$ savings by coarse-graining in time.  The longest turbulent wavelengths perpendicular to the field 
are on the order of $10$ centimeters.
Assuming several turbulence regions are necessary to sample the device volume, we find little savings for
a device the size of ITER in the radial direction.  There is a way to obtain considerable savings in the poloidal
direction however, to which we now turn our attention.

\section{Coupled flux tube approach}
\label{sec:flxtube}

The local, or flux tube, model~\cite{beerPoP95} introduced in Chapter 1 allows for the simulation of 
microturbulence in a thin tube, several turbulence decorrelation lengths in each dimension,
enclosing a single magnetic field line (depicted in Figs.~\ref{fig:flxtube} and~\ref{fig:masttube}).  
Because the typical spatial scale of fluctuations is much shorter across field lines
than along field lines, the flux tube is highly elongated along the field line.  The critical assumption
that allows for the reduction of the simulation domain to a single flux tube is statistical
periodicity.  That is, we assume the variation of equilibrium quantities occurs on 
such a large scale (relative to the flux tube) that the turbulence is homogeneous within the flux tube.  We stress
that this does not disallow large-scale gradients in the problem -- only variation
of these gradients over the small width of a flux tube.
The flux tube approximation is thus valid as long as the turbulent spatial scale is well
separated from all other spatial scales in the problem.  This means that any effects arising
from intermediate spatial scales (such as magnetic islands, if present) are not included in the flux
tube model.  The validity of the flux tube approximation in the limit of 
$\rho_{*}\equiv \rho/a \sim (k_{\perp}L)^{-1} \ll 1$
has been verified numerically~\cite{candyPoP04}.  This is illustrated in Fig.~\ref{fig:local},
where we see that flux tube simulations agree very well with global simulations 
(which do not assume statistical periodicity) when the turbulent scale length is much smaller
than the equilibrium scale length.

\begin{figure}
\centering
\includegraphics[height=5.0in]{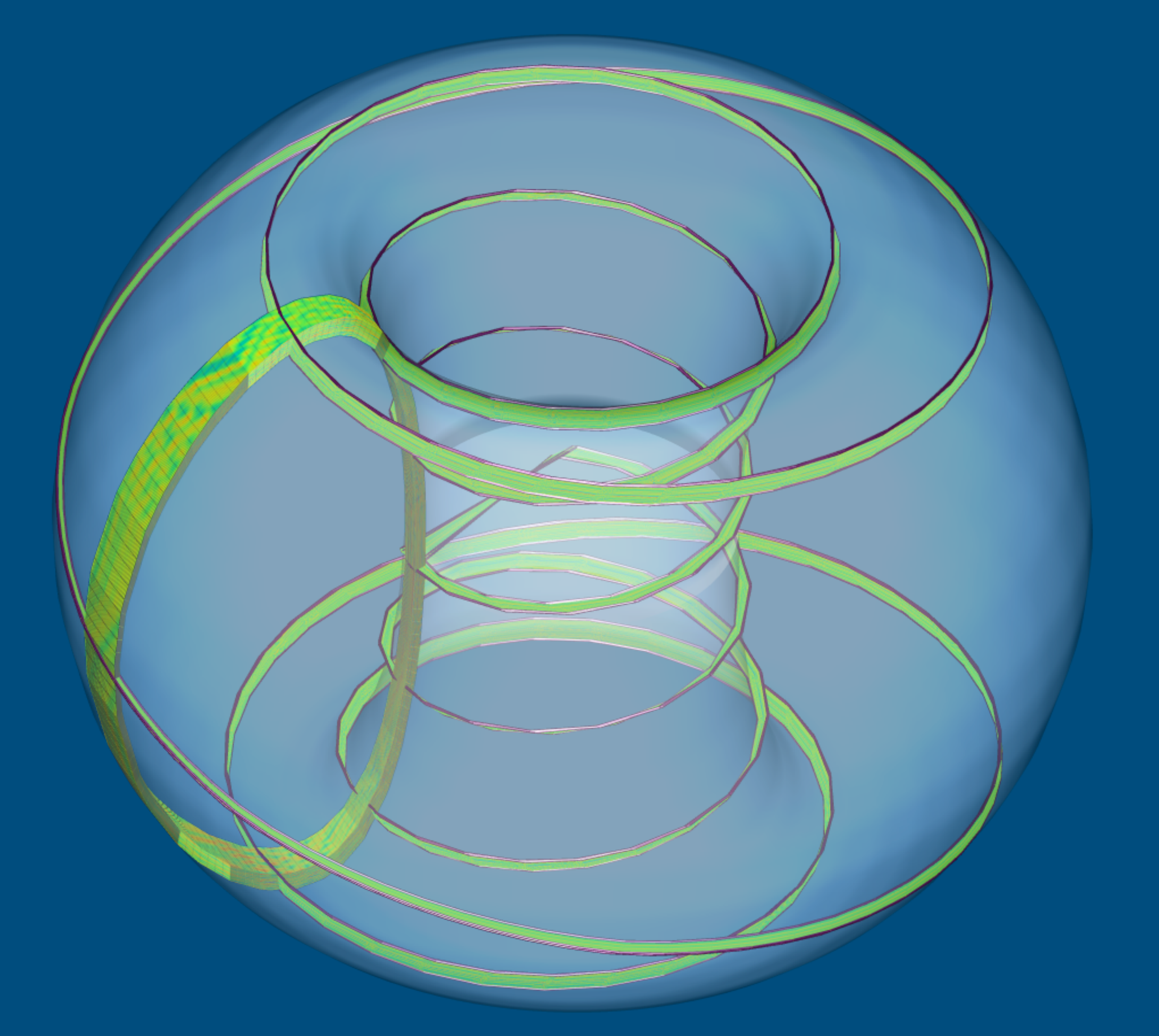}
\caption{Flux tube from \texttt{GS2} simulation of the spherical tokamak, MAST.  The flux tube
simulation domain wraps multiple times around the toroidal circumference, but covers
only a fraction of the anular flux surface it is used to map out (shown in light blue).  Graphic
courtesy of G. Stantchev.}
\label{fig:masttube}
\end{figure}

In axisymmetric magnetic field configurations, a single flux tube, which comprises only a
fraction of a flux surface, can be used to map out the entire flux surface (due to the notion
of statistical periodicity of the turbulence).  This constitutes a significant savings in 
simulation volume, which depends on the toroidal mode numbers of interest in the experiment
(in particular, the longest significant wavelength present along the field line).  We can
estimate the savings as follows: in the direction perpendicular to the magnetic field line, but
contained within the flux surface, the flux tube must be large enough to resolve the longest
turbulent wavelength.  Taking $k_{\perp}\sim n_{\phi}q/r$, where 
$n_{\phi}$ is toroidal mode number, $q$ is the safety factor, and $r$ is the distance
from the magnetic axis to the flux surface of interest, we have
\begin{equation}
L_{\perp} \sim \frac{2\pi}{k_{\perp}} \sim \frac{2\pi r}{n_{\phi}q}\sim\frac{L_{\theta}}{n_{\phi}q},
\end{equation}
where $L_{\theta}$ is the approximate circumference of the tokamak in the poloidal 
direction.  Therefore, the simulation domain of a flux tube covers a fraction of approximately 
$1/n_{\phi}q$ of a magnetic flux surface.  For ITER-like fusion devices, the longest 
perpendicular wavelength expected
to be important is approximately $k_{\perp}\rho_{i}\sim 0.1$, corresponding to a
toroidal mode number of $n_{\phi}\sim 100$.  Combined with a safety factor $q>1$,
this translates into a savings of factor greater than $100$ in simulation volume to simulate a 
single flux surface.

A single flux surface is not sufficient, however, when evolving radial equilibrium profiles.
In that case, a flux surface represents a single radial grid point on our coarse spatial grid.
This is illustrated in Figs.~\ref{fig:fsctok} and~\ref{fig:fsctok3}.  
In Fig.~\ref{fig:fsctok}, we show
a poloidal cut of a tokamak and indicate a series of flux surfaces for a representative magnetic field configuration.
We see that the area of the cross-section accounted for by several flux tubes is a small fraction of the
total area (as long as the device has a sufficiently large radial extent).  The actual simulation 
domain for a single flux tube is a rectangular box, which is 
deformed in physical space as it follows along the sheared magnetic field line (Fig.~\ref{fig:fsctok2}).  
Depending
on the parallel decorrelation length of the turbulence, the flux tube may traverse the toroidal
circumference multiple times, sampling a poloidal cut of the flux surfae at multiple locations. 
In Fig.~\ref{fig:fsctok3}, we 
plot a respresentative radial temperature profile and indicate the coarse grid composed
of several flux tubes.  In each case, notice that the flux tubes have a finite radial extent.
This is necessary in order to contain several turbulence decorrelation lengths.  In the limit
of an infinitely large device (or, equivalently, infinitely small turbulence decorrelation 
lengths), the radial extent of the flux tubes would shrink to a single radial point.  There
is an intermediate regime in which the flux tubes have finite size and, depending on the
number of flux tubes and the size of the device, may overlap radially.  This is not 
necessarily cause for alarm.  While each flux tube has finite radial extent, it is representative
of a single radial point: equilibrium profiles and gradients are taken to be constant across
each flux tube.  Consequently, overlapping flux tubes do not actually sample the same
physical space and do not lead to double-counting of the turbulence.

\begin{figure}
\centering
\includegraphics[height=4.5in]{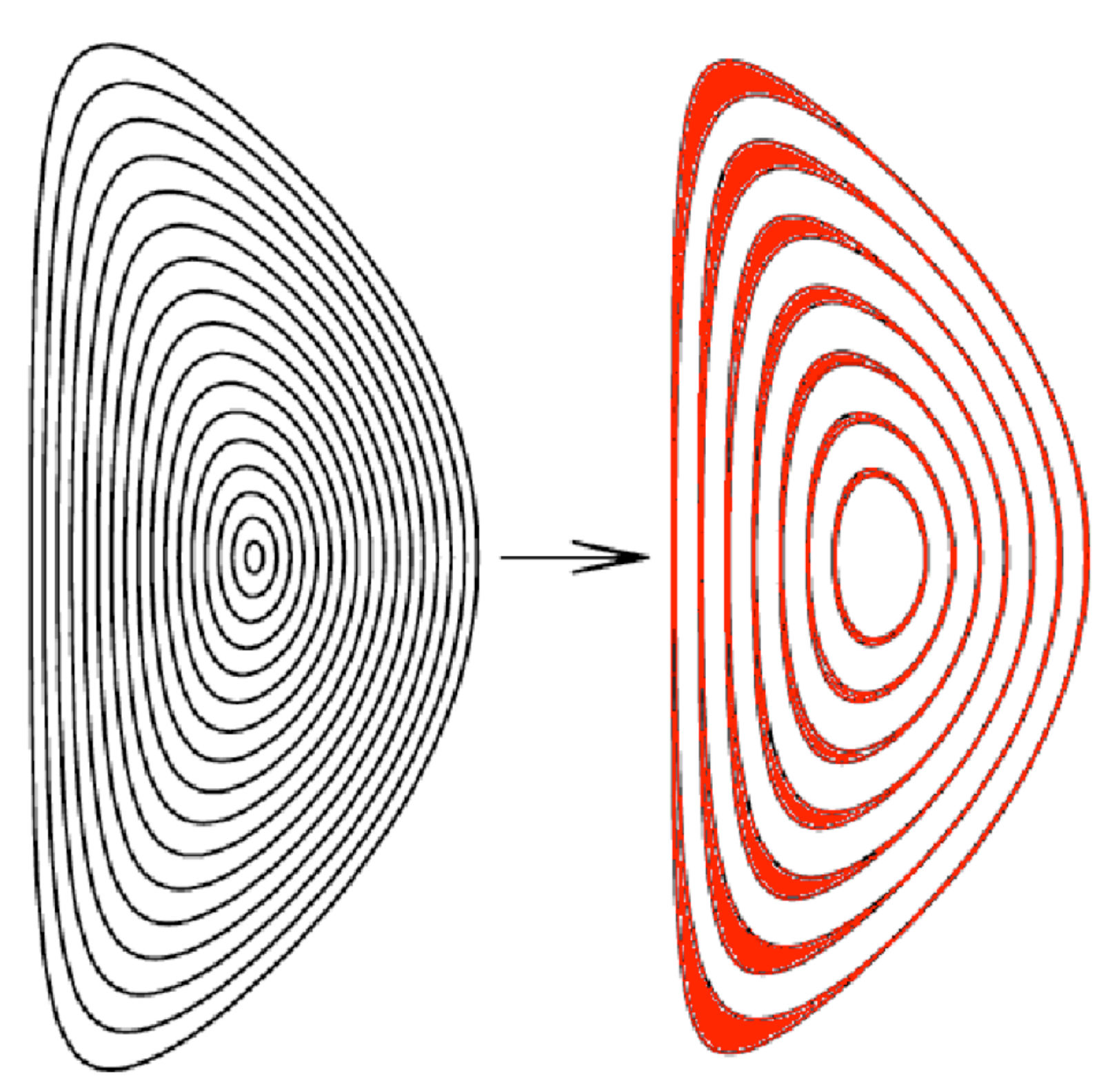}
\caption{(left): poloidal cross section of a typical tokamak.  solid lines indicate the shape
of magnetic flux surfaces and colored regions indicate a typical portion of the tokamak 
represented by the coupled flux tube approach.  (right): cartoon illustrating the simulation
domain (illustrated in blue) in a poloidal cut at the outboard midplane for a single flux tube (representing
a radial point, or flux surface)}
\label{fig:fsctok}
\end{figure}

\begin{figure}
\centering
\includegraphics[height=4.5in]{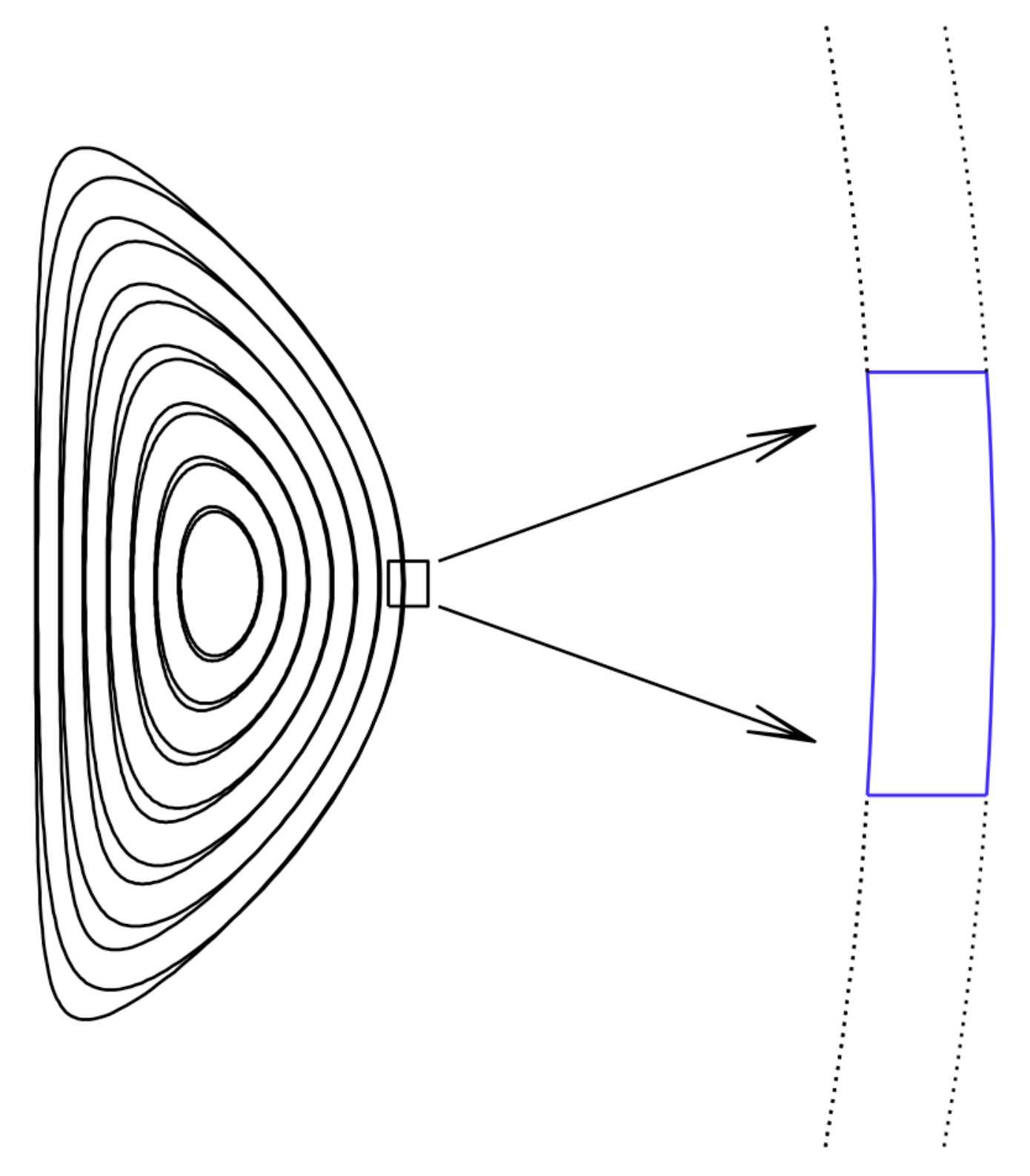}
\caption{Cartoon illustrating the flux tube simulation domain (illustrated in blue)
for a poloidal cut at the outboard midplane.  This flux tube represents the entire flux
surface, which serves as a radial grid point in our transport equations.}
\label{fig:fsctok2}
\end{figure}

\begin{figure}
\centering
\includegraphics[height=5.0in]{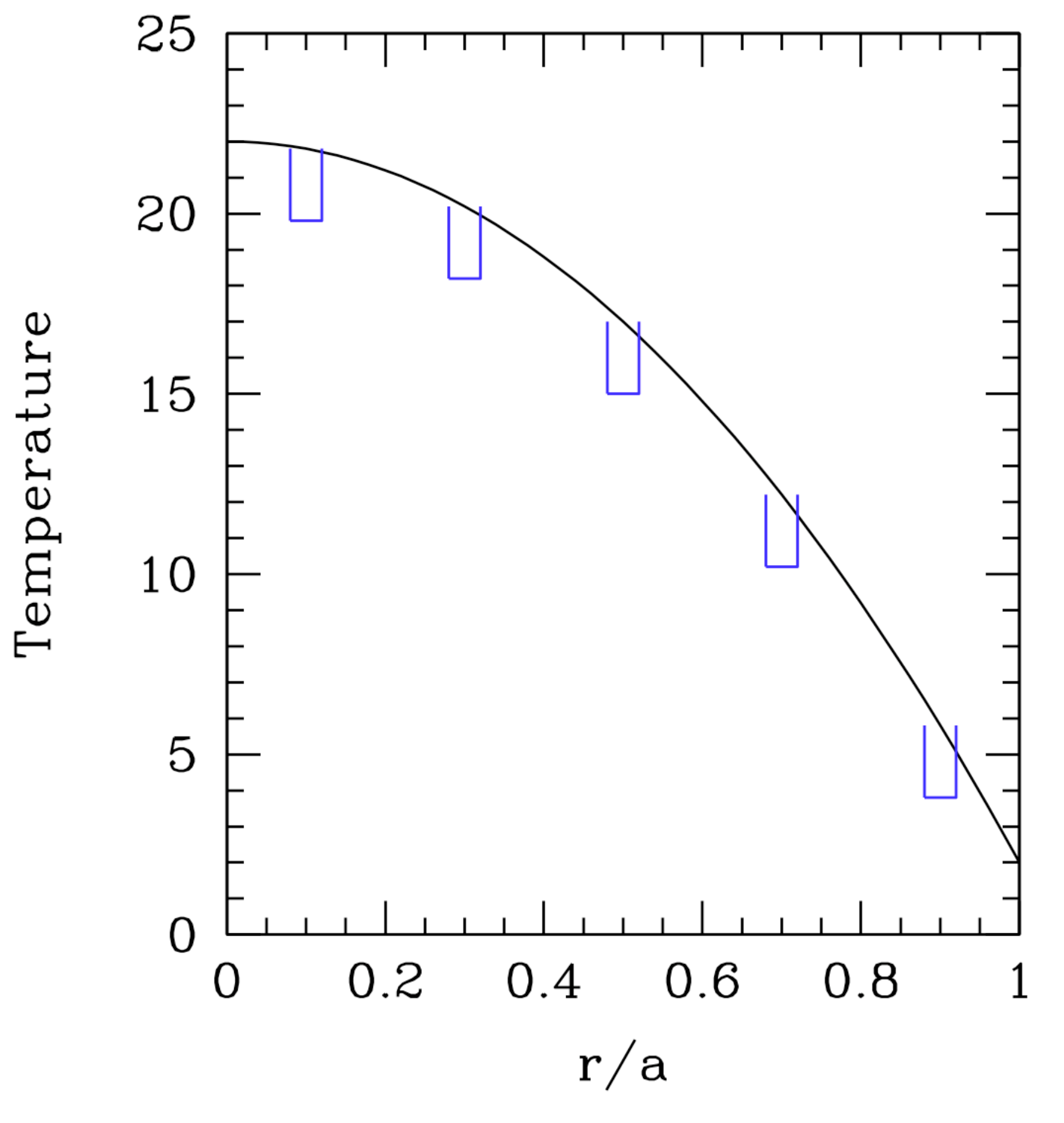}
\caption{Cartoon illustrating the portion of the radial temperature profile sampled
by the use of coupled flux tubes.  Each of the blue 'U' shapes represent a flux tube.
Although each flux tube has finite radial extent, it represents a single radial point at 
the center of its domain.}
\label{fig:fsctok3}
\end{figure}

%

\section{Normalization of the transport equations}
\label{sec:teqnorm}

We start from the gryokinetic transport equations~(\ref{eqn:nevo}) and~(\ref{eqn:pevo})
derived in Chapter~\ref{chap:hierarchy}.  For simplicity, we neglect classical effects
and time variation of the equilibrium magnetic field (which formally evolves on the 
equilibrium time scale, but in practice evolves on the slower, resistive time scale).  As mentioned in Chapter~\ref{chap:hierarchy},
we assert that external sources enter our hierarchical equations via the transport equations.
Consequently, we include general sources in our treatment in this chapter.  We also choose not to calculate
$h_{nc}$, and therefore drop the neoclassical terms appearing in our equilibrium evolution equations.
However, we do not neglect neoclassical effects completely; we include them by using
an analytic estimate for the neoclassical ion heat flux from Ref.~\cite{changPoF82}, which
we add to the turbulent ion heat flux appearing in our equations.
Since we will not be employing $h_{nc}$, we henceforth drop the $t$ subscript on the
turbulent part of $h$, using $h=h_{t}$.  Also, since we are dealing exclusively with equilibrium
densities, temperatures, and pressures, we drop the nought subscript on these quantities.
The resulting equations are:
\begin{eqnarray}
\left<\left< \pd{n_{s}}{t} \right>\right> &=& \pd{\psi}{V}\pd{}{\psi}\left[\pd{V}{\psi}
\left<\left< \int\vvol \left(R^{2}\nabla\phi \cdot
\nabla \chi\right) h_{s} \right>\right> \right]\\
\frac{3}{2}\bigg<\bigg<\pd{p_{s}}{t}\bigg>\bigg> &=&
\pd{\psi}{V}\pd{}{\psi}\left[\pd{V}{\psi} \lsavg\int\vvol \frac{mv^{2}}{2}\left(R\unit{\phi}\cdot\nabla\chi\right)h_{s}\rsavg \right] \\
&-&\lsavg\int\vvol\frac{h_{s}T_{s}}{F_{M,s}}\left(R^{2}\nabla\phi\cdot\nabla\chi\right)\pd{F_{M,s}}{\psi}\rsavg \\
&-&\lsavg\int\vvol\frac{h_{s}T_{s}}{F_{0,s}}\left< C[h_{s}] \right>_{\mbf{R}} \rsavg
+ \sum_{u}n_{s} \nu_{\epsilon}^{su}\left(T_{u}-T_{s}\right),
\end{eqnarray}
where $n_{s}$ is the equilibrium density, $T_{s}$ is the equilibrium temperature, $h_{s}$ is
the non-Boltzmann part of the perturbed gyrokinetic distribution function, $\chi$
is the generalized potential defined in Eq.~(\ref{eqn:chi}), $R$ is the major radius, 
$\phi$ is the toroidal angle, $\psi=\psi_{p}/2\pi$ is a measure of the poloidal flux, 
$\nu_{\epsilon}^{su}$ is the energy exchange rate between species $s$ and species $u$ 
given in Eq.~(\ref{eqn:nueq}), $V$ is the flux-tube volume defined in Eq.~(\ref{eqn:vol}),
and the flux surface average of $\mathcal{F}$, $\lbb \mathcal{F} \rbb$, is defined
in Eq.~(\ref{eqn:flxavg}).

We would like to rewrite the term $(R^{2}\nabla\phi\cdot \nabla \chi)$ in a more enlightening form.  
To begin this process, we need the following identity,
derived in Eq.~(\ref{eqn:vdotpsi2p5}) by assuming an axisymmetric equilibrium magnetic field ($\mbf{B}_{0}=\nabla\psi \times 
\nabla \phi + RB_{T}\nabla \phi$):
\begin{equation}
\mbf{v}\cdot\nabla \psi
= - R^{2}\nabla\phi \cdot \left(\mbf{v}\times\mbf{B}_{0}\right).
\end{equation}
Using the definition of $\mbf{v_{\chi}}$ from Eq.~(\ref{eqn:vexb}), we have
\begin{eqnarray}
\mbf{v_{\chi}}\times\mbf{B}_{0} &=& c\left(\unit{b}\times\nabla\chi\right)
\times \unit{b}\\
&=&c\left[\nabla \chi - \unit{b}\left(\unit{b}\cdot\nabla\chi\right)\right]\\
\Rightarrow R^{2}\nabla\phi\cdot\mbf{v_{\chi}}\times\mbf{B}_{0} &=& cR\left[R\nabla\phi
\cdot \nabla \chi - \frac{B_{T}}{B_{0}}\left(\unit{b}\cdot\nabla\chi\right)\right]
\label{eqn:rphichi1}
\end{eqnarray}
From the gyrokinetic ordering, $\nabla_{\parl}\chi \ll \nabla_{\perp}\chi$.  Consequently, 
the second term on the right-hand side of Eq.~(\ref{eqn:rphichi1}) can be neglected, leaving
\begin{eqnarray}
cR^{2} \nabla\phi \cdot \nabla \chi
&=& R^{2}\nabla\phi\cdot\mbf{v_E}\times\mbf{B}_{0}\\
&=& -\mbf{v_{\chi}}\cdot\nabla \psi
\label{rphichi2}
\end{eqnarray}
Using the result of Eq.~(\ref{rphichi2}), and taking 
advantage of the fact that $n_{s}$ and $T_{s}$ are constant on flux surfaces, we have
\begin{eqnarray}
\pd{n_{s}}{t} &=& -\pd{\psi}{V}\pd{}{\psi}\left[\pd{V}{\psi}
\lbb \int\vvol \left(\mbf{v_{\chi}}\cdot\nabla \psi\right) h_{s}\rbb\right]\\
\frac{3}{2}\pd{p_{s}}{t} &=&
-\pd{\psi}{V}\pd{}{\psi}\left[\pd{V}{\psi} \lbb\int\vvol 
\frac{mv^{2}}{2}\left(\mbf{v_{\chi}}\cdot\nabla \psi\right)h_{s}\rbb \right]\\
&+&\lbb\int\vvol\frac{h_{s}T_{s}}{F_{M,s}}\left(\mbf{v_{\chi}}\cdot\nabla \psi\right)\pd{F_{M,s}}{\psi}\rbb \\
&-&\lbb\int\vvol\frac{h_{s}T_{s}}{F_{0,s}}\left< C[h_{s}] \right>_{\mbf{R}}\rbb
+ \sum_{u}n_{s} \nu_{\epsilon}^{su}\left(T_{u}-T_{s}\right).
\end{eqnarray}
Defining the nonlinear fluxes
\begin{eqnarray}
\mbf{\Gamma}_{s} &=& \int\vvol \left(\mbf{v_{\chi}} h_{s}\right)\\
\mbf{Q}_{s}&=&\int\vvol \left(\frac{mv^{2}}{2}\mbf{v_{\chi}} h_{s}\right)
\end{eqnarray}
and noting
\begin{equation}
\pd{\ln F_{M}}{\psi} = \pd{\ln n}{\psi} + \left(\frac{mv^{2}}{2T}-\frac{3}{2}\right)
\pd{\ln T}{\psi}
\end{equation}
we get
\begin{equation}
\pd{n_{s}}{t} = -\pd{\psi}{V}\pd{}{\psi}\left[\lbb\left|\nabla\psi\right|\rbb\pd{V}{\psi}
\frac{\lbb  \mbf{\Gamma}_{s} \cdot \nabla \psi \rbb}{\lbb \left| \nabla\psi\right|\rbb} \right] 
+ \mathcal{S}_{n}
\end{equation}
and
\begin{equation}
\begin{split}
\frac{3}{2}\pd{p_{s}}{t} &=
-\pd{\psi}{V}\pd{}{\psi}\left[\lbb\la\nabla\psi\ra\rbb \pd{V}{\psi} 
\frac{\lbb \mbf{Q}_{s}\cdot\nabla\psi\rbb}{\lbb\left|\nabla\psi\right|\rbb}\right]\\
&+ \lbb\la \nabla \psi \ra\rbb \left[T_{s} \left(\pd{\ln n_{s}}{\psi}
-\frac{3}{2}\pd{\ln T_{s}}{\psi}\right)\frac{\lbb\mbf{\Gamma}_{s}\cdot\nabla\psi\rbb}{\lbb\la\nabla\psi\ra\rbb}
 + \pd{\ln T_{s}}{\psi}\frac{\lbb\mbf{Q}_{s}\cdot\nabla\psi\rbb}{\lbb\la\nabla\psi\ra\rbb}\right]\\
&-\lbb\int\vvol\frac{h_{s}T_{s}}{F_{0,s}}\left< C[h_{s}] \right>_{\mbf{R}}\rbb
+ \sum_{u}n_{s} \nu_{\epsilon}^{su}\left(T_{u}-T_{s}\right) + \frac{3}{2}\mathcal{S}_{p},
\end{split}
\end{equation}
where we have added general external sources, $\mathcal{S}_{n}$ and $\mathcal{S}_{p}$,
to the equations.

The normalized, flux-surface-averaged, nonlinear radial fluxes and turbulent collisional heating calculated in \verb#Trinity# are
\begin{eqnarray}
\Gamma_{s}^{N} &\equiv& \frac{\lbb\mbf{\tilde{\Gamma}}_{s}\cdot\nabla\psi\rbb}{\lbb\la\nabla\psi\ra\rbb}\\
Q_{s}^{N} &\equiv& \frac{\lbb\mbf{\tilde{Q}}_{s}\cdot\nabla\psi\rbb}{\lbb\la\nabla\psi\ra\rbb}\\
\mathcal{H}_{s}^{N} &\equiv& \lbb \int d^{3}v \frac{h_{s}}{F_{0,s}}\frac{T_{s}}{T_{r}}\frac{a}{v_{th,r}}\frac{\left< C[h_{s}] \right>_{\mbf{R}}}{n_{r}} \frac{a^{2}}{\rho_{r}^{2}}\rbb,
\end{eqnarray}
with
\begin{eqnarray}
\mbf{\tilde{\Gamma}}_{s} &=& \frac{\mbf{\Gamma}_{s}}{\left(n v_{th}\right)_{r}}\frac{a^{2}}{\rho_{r}^{2}}\\
\mbf{\tilde{Q}}_{s} &=& \frac{\mbf{Q}_{s}}{\left(nTv_{th}\right)_{r}}\frac{a^{2}}{\rho_{r}^{2}},
\end{eqnarray}
where the subscript $r$ denotes the reference species in the \verb#Trinity# calculation, and $a$ is
a user-specified normalization length.  For definiteness, we choose $a$ to be half the diameter of the
last closed flux surface (LCFS) at the elevation of the magnetic axis.  We note 
that $\rho_{r}$ is defined so that it is a flux
surface quantity.  Specifically, it is given by $\rho_{r}=v_{th,r}/\Omega_{a}$, where
$\Omega_{a}=\la e\ra B_{a}/m_{r}c$ and $B_{a}$ is the toroidal field on the flux
surface $R_{a}$, the average of the minimum and maximum values of $R$ for the flux
surface of interest.  We further define the normalized quantities
\begin{eqnarray}
\tilde{n}_{s} \equiv \frac{n_{s}}{n_{0,r}} && \tilde{T}_{s} \equiv \frac{T_{s}}{T_{0,r}}\\
\tilde{\nabla} \equiv a \nabla && \tau \equiv \frac{v_{th0,r}}{a}\frac{\rho_{r,0}^{2}}{a^{2}}t \\
\tilde{\mathcal{S}}_{n} \equiv \frac{a}{v_{th0,r}} \frac{\mathcal{S}_{n}}{n_{0,r}}\frac{a^{2}}{\rho_{r,0}^{2}} && \tilde{\mathcal{S}}_{p} \equiv  \frac{a}{v_{th0,r}}\frac{\mathcal{S}_{p}}{n_{0,r}T_{0,r}}\frac{a^{2}}{\rho_{r,0}^{2}}\\
\tilde{\nu}_{\epsilon}^{su} \equiv \frac{\nu_{\epsilon}^{su}a}{v_{th0,r}}\frac{a^{2}}{\rho_{r,0}^{2}} && \tilde{h}_{s} \equiv \frac{h_{s}}{F_{0,s}}\frac{a}{\rho_{r}},
\end{eqnarray}
where $n_{0,r}$ and $T_{0,r}$ are chosen to be $10^{20} \ m^{-3}$ and $1 \ keV$, 
respectively, and 
\begin{equation}
\rho_{0,r}\equiv\frac{v_{th0,r}}{\Omega_{r}} = \sqrt{\frac{2T_{0,r}}{m_{r}}}\frac{m_{r}c}{eB_{r}}\approx 4.57\times10^{-3} \ m,
\end{equation}
with $B_{r}$ chosen to be $1 \ T$.  For later convenience, we list here the values
of some of the normalizing quantities:
\begin{gather}
v_{th0,r} \approx 4.38 \times 10^{5} \sqrt{\frac{m_{p}}{m_{r}}} \ m/s\\
t \approx \tau \frac{\tld{a}^{3}}{9.14}\sqrt{\frac{m_{p}}{m_{r}}} \ s
\end{gather}
where $m_{p}$ is the proton mass and $\tld{a}$ is the minor radius given in meters.

Rewriting the transport equations in terms of these normalized 
quantities gives
\begin{equation}
\pd{\tld{n}_{s}}{\tau} = -\pd{\psi}{V}\pd{}{\psi}
\left[\lbb\left|\tld{\nabla}\psi\right|\rbb\pd{V}{\psi}
\Gamma_{s}^{N} \tld{n}_{r}\tld{T}_{r}^{1/2}\frac{\rho_{r}^{2}}{\rho_{r,0}^{2}}\right] + \mathcal{\tilde{S}}_{n}
\end{equation}
and
\begin{equation}
\begin{split}
\frac{3}{2}\pd{\tld{p}_{s}}{\tau} &=
-\pd{\psi}{V}\pd{}{\psi}\left[\lbb\la\tld{\nabla}\psi\ra\rbb \pd{V}{\psi} 
Q_{s}^{N} \tld{n}_{r}\tld{T}_{r}^{3/2}\frac{\rho_{r}^{2}}{\rho_{r,0}^{2}}\right]\\
&+\lbb\la \tld{\nabla} \psi \ra\rbb \tld{n}_{r}\tld{T}_{r}^{3/2}\frac{\rho_{r}^{2}}{\rho_{r,0}^{2}}
\left[\frac{T_{s}}{T_{r}}\left(\pd{\ln \tld{n}_{s}}{\psi}
-\frac{3}{2}\pd{\ln \tld{T}_{s}}{\psi}\right)\Gamma_{s}^{N}
 + \pd{\ln \tld{T}_{s}}{\psi}Q_{s}^{N}\right]\\
&-\frac{\rho_{r}^{2}}{\rho_{r,0}^{2}} \mathcal{H}_{s}^{N}\tld{n}_{r}\tld{T}_{r}^{3/2}
+ \tld{n}_{s} \tld{\nu}_{\epsilon}^{su}\tld{T}_{s}\left(\frac{T_{u}}{T_{s}}-1\right) + \frac{3}{2}\mathcal{\tld{S}}_{p},
\end{split}
\end{equation}

To proceed, we need some more definitions from \verb#Trinity#:
\begin{eqnarray}
A^{N} &\equiv& \int J_{N} d\phi d\theta \la \tld{\nabla} \rho \ra\\
\psi_{N} &\equiv& a^{2}B_{a}\psi\\
\Rightarrow J_{N}&=&J a B_{a}(\psi).
\end{eqnarray}
Noting that $\frac{dV}{d\psi}=\int J d\phi d\theta$ and $\lbb \la \nabla \psi\ra\rbb
\pd{V}{\psi}= \lbb\la\nabla\rho\ra\rbb \pd{V}{\rho}$, we have
\begin{equation}
\lbb \la \nabla \psi\ra\rbb\pd{V}{\psi} \equiv A = a^{2}A^{N}.
\end{equation}
Using the above result, we get
\begin{equation}
\pd{\tld{n}_{s}}{\tau} = -\frac{\lbb\la\tld{\nabla}\rho\ra\rbb}{A^{N}}\pd{}{\rho}
\left[A^{N}\Gamma_{s}^{N} \tld{n}_{r}\tld{T}_{r}^{1/2}\frac{\rho_{r}^{2}}{\rho_{r,0}^{2}}\right] + \mathcal{\tilde{S}}_{n}
\end{equation}
and
\begin{equation}
\begin{split}
\frac{3}{2}\pd{\tld{p}_{s}}{\tau} &=
-\frac{\lbb\la\tld{\nabla}\rho\ra\rbb}{A^{N}}\pd{}{\rho}\left[A^{N}
Q_{s}^{N} \tld{n}_{r}\tld{T}_{r}^{3/2}\frac{\rho_{r}^{2}}{\rho_{r,0}^{2}}\right]\\
&+ \lbb\la \tld{\nabla} \rho \ra\rbb \tld{n}_{r}\tld{T}_{r}^{3/2}\frac{\rho_{r}^{2}}{\rho_{r,0}^{2}}
\left[\frac{T_{s}}{T_{r}}\left(\pd{\ln \tld{n}_{s}}{\rho}
-\frac{3}{2}\pd{\ln \tld{T}_{s}}{\rho}\right)\Gamma_{s}^{N}
+ \pd{\ln \tld{T}_{s}}{\rho}Q_{s}^{N}\right]\\
&-\frac{\rho_{r}^{2}}{\rho_{r,0}^{2}} \mathcal{H}_{s}^{N}\tld{n}_{r}\tld{T}_{r}^{3/2}
+ \tld{n}_{s} \tld{\nu}_{\epsilon}^{su}\tld{T}_{s}\left(\frac{T_{u}}{T_{s}}-1\right) 
+ \frac{3}{2}\mathcal{\tld{S}}_{p},
\end{split}
\end{equation}
The normalized form of $\rho_{r}$ is given by
\begin{equation}
\frac{\rho_{r}}{\rho_{r0}}=\tld{T_{r}}^{1/2}\frac{B_{r}}{B_{a}}\equiv\frac{\tld{T}_{r}^{1/2}}{\tld{B}_{a}},
\end{equation}
leading to the following form for the transport equations:
\begin{equation}
\pd{\tld{n}_{s}}{\tau} = -\frac{\lbb\la\tld{\nabla}\rho\ra\rbb}{A^{N}}\pd{}{\rho}
\left[\frac{A^{N}}{\tld{B}_{a}^{2}}
\Gamma_{s}^{N} \tld{n}_{r}\tld{T}_{r}^{3/2}\right] + \mathcal{\tilde{S}}_{n}
\label{eqn:ndndt}
\end{equation}
\begin{equation}
\begin{split}
\frac{3}{2}\pd{\tld{p}_{s}}{\tau} &=
-\frac{\lbb\la\tld{\nabla}\rho\ra\rbb}{A^{N}}\pd{}{\rho}\left[\frac{A^{N}}{\tld{B}_{a}^{2}}
Q_{s}^{N} \tld{n}_{r}\tld{T}_{r}^{5/2}\right]\\
&+ \frac{\lbb\la \tld{\nabla} \rho \ra\rbb}{\tld{B}_{a}^{2}} \tld{n}_{r}\tld{T}_{r}^{5/2}
\left[\frac{T_{s}}{T_{r}}\left(\pd{\ln \tld{n}_{s}}{\rho}
-\frac{3}{2}\pd{\ln \tld{T}_{s}}{\rho}\right)\Gamma_{s}^{N}
 + \pd{\ln \tld{T}_{s}}{\rho}Q_{s}^{N}\right]\\
&-\frac{1}{\tld{B}_{a}^{2}} \mathcal{H}_{s}^{N}\tld{n}_{r}\tld{T}_{r}^{5/2}
+ \tld{n}_{s} \tld{\nu}_{\epsilon}^{su}\tld{T}_{s}\left(\frac{T_{u}}{T_{s}}-1\right) 
+ \frac{3}{2}\mathcal{\tld{S}}_{p}
\label{eqn:ndTdt}
\end{split}
\end{equation}

As an aside, we note that in \verb#Trinity# we choose to keep the density and temperature for the 
reference species equal to unity in each flux tube.  This means that the physical $\Delta t$ 
and $k_{\perp}\rho$ range vary from flux tube to flux tube (since these quantities are 
normalized by $v_{th,r}$ and $\rho_{r}$ factors, respectively, which have radial 
dependence).

\section{Discretization of the transport equations}
\label{sec:discret}


Now that we have a set of normalized equilibrium evolution equations, we proceed to
discretize them.  In doing so, we wish to maximize computational efficiency.  Primarily,
this is achieved by developing an implicit scheme (based on Newton's method). 
While the implicit scheme requires considerably more
computational effort at each time step than an explicit scheme, it allows for much larger
time steps.  Since the calculation of the steady-state turbulent fluxes and heating at each
equilibrium time step is by far more expensive than the advancement of the equilibrium,
the time step size is much more important than the time spent in calculation during each step.

For simplicity, we assume our system consists of electrons and a single ion species
(which we take to be the reference species), and 
we use quasineutrality to relate the electron and ion densities.  Further, we restrict ourselves 
to the use of a three-point spatial stencil.  Also, in the interest of notational convenience, we drop
the tilde on all normalized quantities.

\subsection{Particle transport}

We begin by writing the normalized particle transport equation (\ref{eqn:ndndt}) for
the reference (ion) species in the convenient form
\begin{equation}
\pd{n}{\tau} = -\frac{\rhoavg}{A}\pd{F}{\rho}+\mathcal{S}_{n},
\label{eqn:ddndt}
\end{equation}
where we have defined
\begin{equation}
F \equiv \frac{A}{\tld{B}_{a}^{2}}\Gamma_{i} \frac{p_{i}^{3/2}}{n^{1/2}}.
\end{equation}
Note that we have switched variables from $T$ to $p=nT$ to reduce the heat transport 
equation (\ref{eqn:ndTdt}) to an equation involving the evolution of a single variable.
A general time discretization for Eq.~(\ref{eqn:ddndt}) takes the form
\begin{equation}
\frac{n^{m+1}-n^{m}}{\Delta \tau} = \alpha\left[-\frac{\rhoavg}{A}\pd{F}{\rho}+\mathcal{S}_{n}\right]^{m+1}+\left(1-\alpha\right)\left[-\frac{\rhoavg}{A}\pd{F}{\rho}+\mathcal{S}_{n}\right],
\label{eqn:timediff}
\end{equation}
where $\Delta\tau$ is the transport time step size, and $\alpha \in [0,1]$, with 
$\alpha=0$ corresponding to a fully explicit scheme
and $\alpha=1$ corresponding to a fully implicit scheme.  The superscripts $m$ and $m+1$
represent the time step.  From now on we will drop the superscript $m$ wherever it appears;
whenever a time superscript is absent from a time-dependent quantity, it is understood to be 
evaluated at time step $m$.

Discretizing the spatial derivative using centered differences, we obtain
\begin{equation}
\frac{n_{j}^{m+1}-n_{j}^{m}}{\Delta \tau} = \alpha\left[-\frac{\rhoavg_{j}}{A_{j}}\frac{F_{+}-F_{-}}{\Delta\rho}+\mathcal{S}_{n}\right]^{m+1}+\left(1-\alpha\right)\left[-\frac{\rhoavg_{j}}{A_{j}}\frac{F_{+}-F_{-}}{\Delta\rho}+\mathcal{S}_{n}\right],
\label{eqn:ntranspdiff}
\end{equation}
where $\Delta\rho=x_{j+1}-x_{j}$ is the spatial grid spacing, and the subscripts $\pm$ 
indicate evaluation at the spatial locations
$x_{j\pm1/2}=(x_{j}+x_{j\pm 1})/2$, with the subscript $j$ denoting the spatial grid index.

We see that Eq.~(\ref{eqn:ntranspdiff}) is a nonlinear partial differential equation.  We
would like to treat it implicitly in order to take large transport time steps.  This requires
linearization of the problem.  We accomplish this by employing Newton's method, in 
which we expand the $m+1$ time level nonlinear term $F$ about its value at the 
time step $m$.  Keeping terms in this Taylor expansion through linear order, we have
\begin{equation}
F_{\pm}^{m+1}\approx F_{\pm}^{m} + \left(\mbf{y}-\mbf{y}_{0}\right)\left[\pd{F_{\pm}}{\mbf{y}}\right]_{\mbf{y}=\mbf{y}_{0}},
\label{eqn:fpm0}
\end{equation}
where $\mbf{y}\equiv \{\{n_{k}\},\{p_{i_{k}}\},\{p_{e_{k}}\}\}$ is a vector 
containing the values for density and electron/ion pressure
at each of the spatial grid points, and $\mbf{y}_{0}$ is the vector $\mbf{y}$ evaluated
at time step $m$.  For convenience, we will henceforth drop the $\mbf{y}=\mbf{y}_{0}$
specifier on the term $\partial F_{\pm} / \partial \mbf{y}$.

Explicitly writing the second term in Eq.~(\ref{eqn:fpm0}), we have
\begin{equation}
\left(\mbf{y}-\mbf{y}_{0}\right)\pd{F_{\pm}}{\mbf{y}} 
= \sum_{k}\left[\left(n_{k}^{m+1}-n_{k}\right)\pd{F_{\pm}}{n_{k}}
+\left(p_{i_{k}}^{m+1}-p_{i_{k}}\right)\pd{F_{\pm}}{p_{i_{k}}}
+\left(p_{e_{k}}^{m+1}-p_{e_{k}}\right)\pd{F_{\pm}}{p_{e_{k}}}\right].
\label{eqn:linf}
\end{equation}
Evaluating the partial derivatives of $F_{\pm}$ in this expression yields
\begin{gather}
\label{eqn:pdfp1}
\pd{F_{\pm}}{n_{k}} = F_{\pm}\left[\pd{\ln \Gamma_{\pm}}{n_{k}}-\frac{1}{4}\frac{\delta_{j,k}+\delta_{j\pm1,k}}{n_{\pm}}\right]\\
\label{eqn:pdfp2}
\pd{F_{\pm}}{p_{i_{k}}} = F_{\pm}\left[\pd{\ln \Gamma_{\pm}}{p_{i_{k}}}+\frac{3}{4}\frac{\delta_{j,k}+\delta_{j\pm1,k}}{p_{i_{\pm}}}\right]\\
\label{eqn:pdfp3}
\pd{F_{\pm}}{p_{e_{k}}} = F_{\pm}\pd{\ln \Gamma_{\pm}}{p_{e_{k}}},
\end{gather}
where $\delta_{j,k}$ is the Dirac delta function.
Substituting Eqs.~(\ref{eqn:linf})-(\ref{eqn:pdfp3}) in Eq.~(\ref{eqn:fpm0}), we obtain 
the following expression for $F_{\pm}^{m+1}$:
\begin{equation}
\begin{split}
F_{\pm}^{m+1}&\approx F_{\pm}\left[\frac{3}{4}\left(\frac{p_{i_{j}}^{m+1}}{p_{i_{\pm}}}+\frac{p_{i_{j\pm1}}^{m+1}}{p_{i_{\pm}}}\right)-\frac{1}{4}\left(\frac{n_{j}^{m+1}}{n_{\pm}}+\frac{n_{j\pm1}^{m+1}}{n_{\pm}}\right)\right]\\
&+F_{\pm}\sum_{k}\left[\dn\pd{\ln \Gamma_{\pm}}{n_{k}} + \dpi\pd{\ln \Gamma_{\pm}}{p_{i_{k}}}+\dpe\pd{\ln \Gamma_{\pm}}{p_{e_{k}}}\right].
\label{eqn:fpj}
\end{split}
\end{equation}

The above equation involves derivatives of the particle flux, $\Gamma$, with respect to 
the equilibrium density and pressure at each of the grid locations, $n_{k}$ and $p_{k}$.  
Unfortunately, this information is not readily available and would be prohibitively expensive
to compute directly.  In order to make calculation of these derivatives feasible, we
make the assumption that $\Gamma$ depends on the $\{n_{k}\}$ and $\{p_{k}\}$
only through the gradient scale lengths $R/L_{n}$ and $R_{L_{p}}$.  This assumption
is motivated by empirical results from both numerical simulation and experiment.

With this assumption, the derivatives of the particle flux, $\Gamma$, can be written
\begin{gather}
\pd{\Gamma_{\pm}}{n_{k}} \approx \pd{\Gamma_{\pm}}{(R/L_{n})_{\pm}}\frac{d(R/L_{n})_{\pm}}{dn_{k}}\\
\pd{\Gamma_{\pm}}{p_{k}} \approx \pd{\Gamma_{\pm}}{(R/L_{p})_{\pm}}\frac{d(R/L_{p})_{\pm}}{dp_{k}}.
\end{gather} 
In order to evaluate these expressions, we need discretized forms for $R/L_{n}$ and 
$R/{L_{p}}$, as well as estimates for the derivatives of the flux with respect to these 
equilibrium gradients.  We defer discussion of the latter issue until later.  For the 
discretization of $\left(R/{L_{n}}\right)_{\pm}$ we use
\begin{equation}
\left(\frac{R}{L_{n}}\right)_{\pm} = -\frac{R}{a}\left(\pd{\ln n}{\rho}\right)_{\pm}
\approx \mp\frac{R}{a\Delta\rho}\frac{n_{j\pm1}-n_{j}}{n_{\pm}}\approx\mp\frac{2R}{a\Delta\rho}\frac{n_{j\pm1}-n_{j}}{n_{j\pm1}+n_{j}}.
\label{eqn:rlpm}
\end{equation}
This derivative approximation is accurate to $\mathcal{O}[(\Delta\rho)^{2}]$, and
the same discretization scheme is used for $\left(R/L_{p}\right)_{\pm}$.

We next compute the discrete derivatives of the equilibrium gradients with respect to
the equilibrium density and pressure:
\begin{equation}
\sum_{k}\dn\pd{(R/L_{n})_{\pm}}{n_{k}}=\mp\frac{R}{a}\frac{1}{\Delta\rho}
\left[\frac{n_{j\pm1}^{m+1}}{n_{\pm}}\frac{n_{j}}{n_{\pm}}-\frac{n_{j}^{m+1}}{n_{\pm}}\frac{n_{j\pm1}}{n_{\pm}}\right],
\end{equation}
with a similar expression for the derivative of $R/L_{p}$ with respect to species pressure.
Using this expression in Eq.~(\ref{eqn:fpj}), we arrive at the following:
\begin{equation}
\begin{split}
&F_{\pm}^{m+1}\approx 
F_{\pm}\left[\frac{3}{4}\left(\frac{p_{i_{j}}^{m+1}}{p_{i_{\pm}}}+\frac{p_{i_{j\pm1}}^{m+1}}{p_{i_{\pm}}}\right)-\frac{1}{4}\left(\frac{n_{j}^{m+1}}{n_{\pm}}+\frac{n_{j\pm1}^{m+1}}{n_{\pm}}\right)\right]\\
&\pm\frac{RF_{\pm}}{a\Delta\rho}\left[\left(\frac{n_{j}^{m+1}}{n_{\pm}}\frac{n_{j\pm1}}{n_{\pm}}-\frac{n_{j\pm1}^{m+1}}{n_{\pm}}\frac{n_{j}}{n_{\pm}}\right)\pd{\ln \Gamma_{\pm}}{(R/L_{n})_{\pm}}\right.\\
&\left.+\left(\frac{p_{i_{j}}^{m+1}}{p_{i_{\pm}}}\frac{p_{i_{j\pm1}}}{p_{i_{\pm}}}-\frac{p_{i_{j\pm1}}^{m+1}}{p_{i_{\pm}}}\frac{p_{i_{j}}}{p_{i_{\pm}}}\right)\pd{\ln \Gamma_{\pm}}{(R/L_{p_{i}})_{\pm}}+\left(\frac{p_{e_{j}}^{m+1}}{p_{e_{\pm}}}\frac{p_{e_{j\pm1}}}{p_{e_{\pm}}}-\frac{p_{e_{j\pm1}}^{m+1}}{p_{e_{\pm}}}\frac{p_{e_{j}}}{p_{e_{\pm}}}\right)\pd{\ln \Gamma_{\pm}}{(R/L_{p_{e}})_{\pm}}\right].
\label{eqn:fpj2}
\end{split}
\end{equation}

With Eq.~(\ref{eqn:fpj2}), we can now compute $\partial F/\partial \rho$:
\begin{equation}
\begin{split}
&\left(\pd{F^{m+1}}{\rho}\right)_{j}\approx \frac{F_{+}^{m+1}-F_{-}^{m+1}}{\Delta\rho}\approx \frac{1}{\Delta\rho}\left(\frac{n_{j-1}^{m+1}}{n_{-}}\left(\mathcal{A}_{-}[n]+\frac{F_{-}}{4}\right)\right.\\
&\left.+\frac{n_{j}^{m+1}}{n_{+}}\left[\frac{1}{4}\left(F_{-}\frac{n_{+}}{n_{-}}-F_{+}\right)+\mathcal{B}_{j}[n]\right]+\frac{n_{j+1}^{m+1}}{n_{+}}\left[\mathcal{A}_{+}[n]-\frac{1}{4}F_{+}\right]\right.\\
&\left.+\frac{p_{i_{j-1}}^{m+1}}{p_{i_{-}}}\left(\mathcal{A}_{-}[p_{i}]-\frac{3}{4}F_{-}\right)+\frac{p_{i_{j}}^{m+1}}{p_{i_{+}}}\left[-\frac{3}{4}\left(F_{-}\frac{p_{i_{+}}}{p_{i_{-}}}-F_{+}\right)+\mathcal{B}_{j}[p_{i}]\right]\right.\\
&\left.+\frac{p_{i_{j+1}}^{m+1}}{p_{i_{+}}}\left[\mathcal{A}_{+}[p_{i}]+\frac{3}{4}F_{+}\right]+\frac{p_{e_{j-1}}^{m+1}}{p_{e_{-}}}\mathcal{A}_{-}[p_{e}]+\frac{p_{e_{j}}^{m+1}}{p_{e_{+}}}\mathcal{B}_{j}[p_{e}]+\frac{p_{e_{j+1}}^{m+1}}{p_{e_{+}}}\mathcal{A}_{+}[p_{e}]\right),
\label{eqn:fpj3}
\end{split}
\end{equation}
where we define
\begin{gather}
\mathcal{A}_{\pm}[w]\equiv -\frac{RF_{\pm}}{a\Delta\rho}\frac{w_{j}}{w_{\pm}^{2}}\pd{\ln \Gamma_{\pm}}{(R/L_{w})_{\pm}}\\
\mathcal{B}_{j}[w]\equiv \frac{R}{a\Delta\rho}\left(F_{+}\frac{w_{j+1}}{w_{+}^{2}}\pd{\ln \Gamma_{+}}{(R/L_{w})_{+}}+F_{-}\frac{w_{j-1}}{w_{-}^{2}}\pd{\ln \Gamma_{-}}{(R/L_{w})_{-}}\right).
\end{gather}

Plugging Eq.~(\ref{eqn:fpj3}) back into Eq.~(\ref{eqn:ntranspdiff}), we
arrive at the final form for our discretized particle transport equation:
\begin{equation}
\begin{split}
&n_{j}^{m+1}-\sum_{k=-1}^{1} \Delta \tau\alpha\left(n_{j+k}^{m+1}\psi_{n,k}+p_{i_{j+k}}^{m+1}\psi_{p_{i},k}+p_{e_{j+k}}^{m+1}\psi_{p_{e},k}\right)\\
&=n_{j}+\Delta \tau\left(1-\alpha\right)\left[-\frac{\rhoavg}{A}\pd{F}{\rho}\right]+\Delta\tau S_{n}
\end{split}
\end{equation}
where
\begin{eqnarray}
\label{eqn:ncoefs}
\psi_{n,-1}&\equiv& -\frac{\rhoavg}{A\Delta\rho}\left[\mathcal{A}_{-}[n]+\frac{1}{4}\frac{F_{-}}{n_{-}}\right]\\
\psi_{n,0}&\equiv& -\frac{\rhoavg}{A\Delta\rho}\left[\mathcal{B}_{j}[n]+\frac{1}{4}\left(\frac{F_{-}}{n_{-}}-\frac{F_{+}}{n_{+}}\right)\right]\\
\psi_{n,1}&\equiv& -\frac{\rhoavg}{A\Delta\rho}\left(\mathcal{A}_{+}[n]-\frac{1}{4}\frac{F_{+}}{n_{+}}\right)\\
\psi_{p_{i},-1}&\equiv& -\frac{\rhoavg}{A\Delta\rho}\left(\mathcal{A}_{-}[p_{i}]-\frac{3}{4}\frac{F_{-}}{p_{i_{-}}}\right)\\
\psi_{p_{i},0}&\equiv& -\frac{\rhoavg}{A\Delta\rho}\left[\mathcal{B}_{j}[p_{i}]-\frac{3}{4}\left(\frac{F_{-}}{p_{i_{-}}}-\frac{F_{+}}{p_{i_{+}}}\right)\right]\\
\psi_{p_{i},1}&\equiv& -\frac{\rhoavg}{A\Delta\rho}\left(\mathcal{A}_{+}[p_{i}]+\frac{3}{4}\frac{F_{+}}{p_{i_{+}}}\right)\\
\psi_{p_{e},-1}&\equiv& - \frac{\rhoavg}{A\Delta\rho}\mathcal{A}_{-}[p_{e}]\\
\psi_{p_{e},0}&\equiv& -\frac{\rhoavg}{A\Delta\rho}\mathcal{B}_{j}[p_{e}]\\
\psi_{p_{e},1}&\equiv& -\frac{\rhoavg}{A\Delta\rho}\mathcal{A}_{+}[p_{e}].
\label{eqn:pecoefs}
\end{eqnarray}

\subsection{Heat transport}

We begin by recasting the normalized heat transport equation (\ref{eqn:ndTdt}) in terms
of species pressure $p=nT$:
\begin{equation}
\begin{split}
\frac{3}{2}\pd{p_{s}}{\tau} &= -\frac{\left<\la\nabla\rho\ra\right>}{A}\pd{}{\rho}\left(\frac{A}{\tld{B}_{a}^{2}}Q_{s}\frac{p_{r}^{5/2}}{n_{r}^{3/2}}\right)
+\frac{\rhoavg}{\tld{B}_{a}^{2}} \frac{p_{r}^{5/2}}{n_{r}^{3/2}}\frac{a}{R}\left[\frac{p_{s}}{p_{r}}\frac{n_{r}}{n_{s}}\left(\frac{3}{2}\rlps - \frac{5}{2}\rln\right)\Gamma_{s}\right.\\
&\left.+\left(\rln-\rlps\right)Q_{s}\right]+\tld{\nu}_{\epsilon}^{su}p_{s}\left(\frac{p_{u}}{p_{s}}\frac{n_{s}}{n_{u}}-1\right)-\frac{\mathcal{H}_{s}}{\tld{B}_{a}^{2}}\frac{p_{r}^{5/2}}{n_{r}^{3/2}} + \frac{3}{2}S_{p}
\label{eqn:ptransp0}
\end{split}
\end{equation}

To keep notation compact, we define
\begin{gather}
E_{s}\equiv \tld{\nu}_{\epsilon}^{su}p_{s}\left(\frac{p_{u}}{p_{s}}\frac{n_{s}}{n_{u}}-1\right)\\
F_{s}\equiv \frac{A}{\tld{B}_{a}^{2}}Q_{s}\frac{p_{i}^{5/2}}{n^{3/2}}\\
G_{s}\equiv \frac{\rhoavg}{\tld{B}_{a}^{2}}\frac{Z_{s}}{Z_{i}}\ar\frac{p_{i}^{3/2}p_{s}}{n^{3/2}}\kappa_{s}\Gamma_{s}\\
H_{s}\equiv \frac{\rhoavg}{\tld{B}_{a}^{2}}\ar\frac{p_{i}^{5/2}}{n^{3/2}}\tld{\kappa}_{s}Q_{s}\\
K_{s}\equiv -\frac{\mathcal{H}_{s}}{\tld{B}_{a}^{2}}\frac{p_{r}^{5/2}}{n_{r}^{3/2}}\\
\kappa\equiv \frac{3}{2}\rlps - \frac{5}{2}\rln\\
\tld{\kappa}\equiv \rln - \rlps.
\end{gather}
With these definitions, Eq.~(\ref{eqn:ptransp0}) becomes
\begin{equation}
\frac{3}{2}\pd{p_{s}}{\tau} = -\frac{\rhoavg}{A}\pd{F_{s}}{\rho} + G_{s}+H_{s} + E_{s}+K_{s}+\frac{3}{2}S_{p}.
\end{equation}
Henceforth, we drop the subscript $s$; whenever a species subscript is not present, the
subscript $s$ is assumed.

Our treatment of heat transport in this subsection follows closely our the treatment of
particle transport in the previous subsection.  Our time discretization is of the form
\begin{equation}
\begin{split}
\frac{3}{2}&\left(\frac{p^{m+1}-p^{m}}{\Delta \tau}\right) = \alpha \left[-\frac{\rhoavg}{A}\pd{F}{\rho}
+G+H+E+K + \frac{3}{2}S_{p}\right]^{m+1}\\
&+\left(1-\alpha\right)\left[-\frac{\rhoavg}{A}\pd{F}{\rho}
+G+H+E+K + \frac{3}{2}S_{p}\right],
\label{eqn:ptransptdiff}
\end{split}
\end{equation}
where $\alpha$ was defined following Eq.~(\ref{eqn:timediff}).  As before, any
time-dependend quantity without a time superscript is understood to be evaluated at time 
step $m$.

Linearizing the nonlinear terms via Taylor expansion about density and pressure at time
step $m$, we again obtain expressions of the form
\begin{equation}
G_{j}^{m+1} \approx G_{j}^{m} + \left(\mbf{y}-\mbf{y}_{0}\right)\left[\pd{G_{j}}{\mbf{y}}\right]_{\mbf{y}=\mbf{y}_{0}}.
\label{eqn:g0}
\end{equation}
Explicitly writing the second term in Eq.~(\ref{eqn:g0}), we have
\begin{equation}
\left(\mbf{y}-\mbf{y}_{0}\right)\pd{G_{j}}{\mbf{y}} 
= \sum_{k}\left[\left(n_{k}^{m+1}-n_{k}\right)\pd{G_{j}}{n_{k}}
+\left(p_{i_{k}}^{m+1}-p_{i_{k}}\right)\pd{G_{j}}{p_{i_{k}}}
+\left(p_{e_{k}}^{m+1}-p_{e_{k}}\right)\pd{G_{j}}{p_{e_{k}}}\right].
\label{eqn:ling}
\end{equation}
Evaluating each of the partial derivatives in this expression, we obtain
\begin{gather}
\label{eqn:pdg1}
\pd{G_{j}}{n_{k}} = G_{j}\left[\pd{\ln \Gamma_{j}}{n_{k}}-\frac{3}{2n_{j}}\delta_{jk}
-\frac{5}{2\kappa_{j}}\pd{(R/L_{n})_{j}}{n_{k}}\right]\\
\label{eqn:pdg2}
\pd{G_{j}}{p_{i_{k}}} = G_{j}\left[\pd{\ln \Gamma_{j}}{p_{i_{k}}}+\left(\frac{3}{2p_{i_{j}}}+\pd{\ln p_{j}}{p_{i_{k}}}\right)\delta_{jk}
+\frac{3}{2\kappa_{j}}\pd{(R/L_{p})_{j}}{p_{i_{k}}}\right] \\
\label{eqn:pdg3}
\pd{G_{j}}{p_{e_{k}}} = G_{j}\left[\pd{\ln \Gamma_{j}}{p_{e_{k}}}+\pd{\ln p_{j}}{p_{e_{k}}}\delta_{jk}
+\frac{3}{2\kappa_{j}}\pd{(R/L_{p})_{j}}{p_{e_{k}}}\right].
\end{gather}
Substituting Eqs.~(\ref{eqn:ling})-(\ref{eqn:pdg3}) into Eq.~(\ref{eqn:g0}) gives
\begin{equation}
\begin{split}
&G_{j}^{m+1}= G_{j}\left[\left(\frac{3}{2}+\delta_{si}\right)\frac{p_{i_{j}}^{m+1}}{p_{i_{j}}}-\frac{3}{2}\frac{n_{j}^{m+1}}{n_{j}}+\delta_{se}\frac{p_{e_{j}}^{m+1}}{p_{e_{j}}}\right]\\
&+G_{j} \sum_{k}\left[\dn\left(\pd{\ln \Gamma_{j}}{n_{k}}-\frac{5}{2\kappa_{j}}\pd{(R/L_{n})_{j}}{n_{k}}\right)\right.\\
&\left.+\dpi\left(\pd{\ln \Gamma_{j}}{p_{i_{k}}}+\frac{3}{2\kappa_{j}}\pd{(R/L_{p})_{j}}{p_{i_{k}}}\right)\right.\\
&\left.+\dpe\left(\pd{\ln \Gamma_{j}}{p_{e_{k}}}+\frac{3}{2\kappa_{j}}\pd{(R/L_{p})_{j}}{p_{e_{k}}}\right) \right].
\label{eqn:gj}
\end{split}
\end{equation}

The expressions for the other nonlinear terms are derived in a similar manner.  The analogs to
Eqs.~(\ref{eqn:pdg1})-(\ref{eqn:pdg3}) for $H$, $K$, and $E$ are
\begin{gather}
\label{eqn:pdh1}
\pd{H_{j}}{n_{k}} = H_{j}\left[\pd{\ln Q_{j}}{n_{k}}-\frac{3}{2n_{j}}\delta_{jk}
+\frac{1}{\tld{\kappa}_{j}}\pd{(R/L_{n})_{j}}{n_{k}}\right]\\
\label{eqn:pdh2}
\pd{H_{j}}{p_{i_{k}}} = H_{j}\left[\pd{\ln Q_{j}}{p_{i_{k}}}+\frac{5}{2p_{i_{j}}}\delta_{jk}
-\frac{1}{\tld{\kappa}_{j}}\pd{(R/L_{p})_{j}}{p_{i_{k}}}\right] \\
\label{eqn:pdh3}
\pd{H_{j}}{p_{e_{k}}} = H_{j}\left[\pd{\ln Q_{j}}{p_{e_{k}}}
-\frac{1}{\tld{\kappa}_{j}}\pd{(R/L_{p})_{j}}{p_{e_{k}}}\right]\\
\label{eqn:pdk1}
\pd{K{j}}{n_{k}} = K_{j}\left[\pd{\ln \mathcal{H}_{j}}{n_{k}}-\frac{3}{2n_{j}}\delta_{jk}\right]\\
\label{eqn:pdk2}
\pd{K_{j}}{p_{i_{k}}} = K_{j}\left[\pd{\ln \mathcal{H}_{j}}{p_{i_{k}}}+\frac{5}{2p_{i_{j}}}\delta_{jk}\right]\\
\label{eqn:pdk3}
\pd{K_{j}}{p_{e_{k}}} = K_{j}\pd{\ln \mathcal{H}_{j}}{p_{e_{k}}}\\
\pd{E_{j}}{n_{k}} = \frac{5}{2n_{j}}E_{j}\delta_{jk}\\
\pd{E_{j}}{p_{i_{k}}} = \left[\frac{\delta_{se}Z_{u}-\delta_{si}Z_{s}}{p_{u_{j}}Z_{u}-p_{s_{j}}Z_{s}}-\frac{3}{2}\frac{\delta_{se}m_{s}Z_{u}+\delta_{si}m_{u}Z_{s}}{m_{s}Z_{u}p_{u_{j}}+m_{u}Z_{s}p_{s_{j}}}\right]E_{j}\delta_{jk}\\
\pd{E_{j}}{p_{e_{k}}} = \left[\frac{\delta_{si}Z_{u}-\delta_{se}Z_{s}}{p_{u_{j}}Z_{u}-p_{s_{j}}Z_{s}}-\frac{3}{2}\frac{\delta_{si}m_{s}Z_{u}+\delta_{se}m_{u}Z_{s}}{m_{s}Z_{u}p_{u_{j}}+m_{u}Z_{s}p_{s_{j}}}\right]E_{j}\delta_{jk},
\end{gather}
giving
\begin{equation}
\begin{split}
&H_{j}^{m+1}\approx H_{j}\left(\frac{5}{2}\frac{p_{i_{j}}^{m+1}}{p_{i_{j}}}-\frac{3}{2}\frac{n_{j}^{m+1}}{n_{j}}\right)
+H_{j}\sum_{k}\left[\dn\left(\pd{\ln Q_{j}}{n_{k}}+\frac{1}{\tld{\kappa}_{j}}\pd{(R/L_{n})_{j}}{n_{k}}\right)\right.\\
&\left. + \dpi\left(\pd{\ln Q_{j}}{p_{i_{k}}}-\frac{1}{\tld{\kappa}_{j}}\pd{(R/L_{p})_{j}}{p_{i_{k}}}\right)
+\dpe\left(\pd{\ln Q_{j}}{p_{e_{k}}}-\frac{1}{\tld{\kappa}_{j}}\pd{(R/L_{p})_{j}}{p_{e_{k}}}\right)\right],
\label{eqn:hj}
\end{split}
\end{equation}
\begin{equation}
\begin{split}
K_{j}^{m+1}&\approx K_{j}\left(\frac{5}{2}\frac{p_{i_{j}}^{m+1}}{p_{i_{j}}}-\frac{3}{2}\frac{n_{j}^{m+1}}{n_{j}}\right)
+K_{j}\sum_{k}\left[\dn\pd{\ln \mathcal{H}_{j}}{n_{k}}\right.\\
&\left. + \dpi\pd{\ln \mathcal{H}_{j}}{p_{i_{k}}}+\dpe\pd{\ln \mathcal{H}_{j}}{p_{e_{k}}}\right],
\label{eqn:kj}
\end{split}
\end{equation}
and
\begin{equation}
\begin{split}
E_{j}^{m+1} &= E_{j}\left(\frac{5}{2}\frac{n_{j}^{m+1}}{n_{j}}+\frac{p_{i_{j}}^{m+1}}{p_{i_{j}}}\left[p_{i_{j}}\left(\frac{\delta_{se}Z_{u}-\delta_{si}Z_{s}}{p_{u_{j}}Z_{u}-p_{s_{j}}Z_{s}}-\frac{3}{2}\frac{\delta_{se}m_{s}Z_{u}+\delta_{si}m_{u}Z_{s}}{m_{s}Z_{u}p_{u_{j}}+m_{u}Z_{s}p_{s_{j}}}\right)\right]\right.\\
&\left.+\frac{p_{e_{j}}^{m+1}}{p_{e_{j}}}\left[p_{e_{j}}\left(\frac{\delta_{si}Z_{u}-\delta_{se}Z_{s}}{p_{u_{j}}Z_{u}-p_{s_{j}}Z_{s}}-\frac{3}{2}\frac{\delta_{si}m_{s}Z_{u}+\delta_{se}m_{u}Z_{s}}{m_{s}Z_{u}p_{u_{j}}+m_{u}Z_{s}p_{s_{j}}}\right)\right]-1\right).
\label{eqn:ej}
\end{split}
\end{equation}

Finally, we consider $F_{\pm}^{m+1}$.  We have
\begin{gather}
\label{eqn:pdf1}
\pd{F_{\pm}}{n_{k}} = F_{\pm}\left[\pd{\ln Q_{\pm}}{n_{k}}-\frac{3}{4}\frac{\delta_{j,k}+\delta_{j\pm1,k}}{n_{\pm}}\right]\\
\label{eqn:pdf2}
\pd{F_{\pm}}{p_{i_{k}}} = F_{\pm}\left[\pd{\ln Q_{\pm}}{p_{i_{k}}}+\frac{5}{4}\frac{\delta_{j,k}+\delta_{j\pm1,k}}{p_{i_{\pm}}}\right]\\
\label{eqn:pdf3}
\pd{F_{\pm}}{p_{e_{k}}} = F_{\pm}\pd{\ln Q_{\pm}}{p_{e_{k}}},
\end{gather}
which gives us
\begin{equation}
\begin{split}
F_{\pm}^{m+1}&\approx F_{\pm}\left(\frac{5}{2}\frac{p_{i_{\pm}}^{m+1}}{p_{i_{\pm}}}-\frac{3}{2}\frac{n_{\pm}^{m+1}}{n_{\pm}}\right)
+F_{\pm}\sum_{k}\left[\dn\pd{\ln Q_{\pm}}{n_{k}}\right.\\
&\left. + \dpi\pd{\ln Q_{\pm}}{p_{i_{k}}}+\dpe\pd{\ln Q_{\pm}}{p_{e_{k}}}\right].
\label{eqn:fj}
\end{split}
\end{equation}
At this point in the calculation, we once again make the assumption that $\Gamma$, $Q$, 
and $\mathcal{H}$ depend on the $\{n_{k}\}$ and $\{p_{k}\}$
only through the gradient scale lengths $R/L_{n}$ and $R_{L_{p}}$.

With this assumption, the derivatives of the fluxes at grid locations can be written
\begin{gather}
\pd{\Gamma_{j}}{n_{k}} \approx \pd{\Gamma_{j}}{(R/L_{n})_{j}}\frac{d(R/L_{n})_{j}}{dn_{k}}\\
\pd{\Gamma_{j}}{p_{k}} \approx \pd{\Gamma_{j}}{(R/L_{p})_{j}}\frac{d(R/L_{p})_{j}}{dp_{k}},
\end{gather}
with similar expressions for derivatives of the heat flux, $Q$, and the heating, $\mathcal{H}$.  
We have already given discrete forms for $R/L_{n}$ and $R/L_{p}$ at the off-grid
locations $x_{j\pm1}$ in Eq.~(\ref{eqn:rlpm}).  We now give an expression for evaluation
at grid locations:
\begin{equation}
\left(\frac{R}{L_{n}}\right)_{j} = -\frac{R}{a}\left(\pd{\ln n}{\rho}\right)_{j}
\approx -\frac{R}{a}\frac{n_{j+1}-n_{j-1}}{2n_{j}\Delta\rho},
\label{eqn:rlj}
\end{equation}
This derivative approximation is accurate to $\mathcal{O}[(\Delta\rho)^{2}]$, 
and the same discretization scheme is used for $R/L_{p}$.
We next compute the discrete derivative of the equilibrium gradients with respect to
the equilibrium density and pressure:
\begin{equation}
\sum_{k}\dn\pd{(R/L_{n})_{j}}{n_{k}}=-\frac{R}{a}\frac{1}{2\Delta\rho}
\left[\frac{n_{j+1}^{m+1}}{n_{j}}-\frac{n_{j-1}^{m+1}}{n_{j}}-\frac{n_{j}^{m+1}}{n_{j}}\frac{n_{j+1}-n_{j-1}}{n_{j}}\right],
\label{eqn:drlj}
\end{equation}
with similar expressions for the derivatives of $R/L_{p}$ with respect to species pressure.

Substituting Eqs~(\ref{eqn:rlj}) and~(\ref{eqn:drlj}) into Eqs.~(\ref{eqn:gj}),~(\ref{eqn:hj}),~(\ref{eqn:kj})
and~(\ref{eqn:ej}) and combining $E$, $G$, $H$, and $K$ results in the following:
\begin{equation}
\begin{split}
&G_{j}^{m+1}+H_{j}^{m+1}+K_{j}^{m+1}+E_{j}^{m+1} = 
\frac{n_{j-1}^{m+1}}{n_{j}}\mu_{1,j}+\frac{n_{j}^{m+1}}{n_{j}}\left[\frac{5}{2}E_{j}-\frac{3}{2}\left(G_{j}+H_{j}+K_{j}\right)\right.\\
&\left.+\mu_{1,j}\left(\frac{n_{j+1}-n_{j-1}}{n_{j}}\right)\right]-\frac{n_{j+1}^{m+1}}{n_{j}}\mu_{1,j}+\frac{p_{i_{j-1}}^{m+1}}{p_{i_{j}}}\mu_{2,j}+\frac{p_{i_{j}}^{m+1}}{p_{i_{j}}}\left[G_{j}\left(\frac{3}{2}+\delta_{si}\right)\right.\\
&+\left.\frac{5}{2}\left(H_{j}+K_{j}\right)+E_{j}p_{i_{j}}\left(\frac{\delta_{se}Z_{u}-\delta_{si}Z_{s}}{p_{u_{j}}Z_{u}-p_{s_{j}}Z_{s}}-\frac{3}{2}\frac{\delta_{se}m_{s}Z_{u}+\delta_{si}m_{u}Z_{s}}{m_{s}Z_{u}p_{u_{j}}+m_{u}Z_{s}p_{s_{j}}}\right)\right.\\
&\left.+\mu_{2,j}\left(\frac{p_{i_{j+1}}-p_{i_{j-1}}}{p_{i_{j}}}\right)\right]-\frac{p_{i_{j+1}}^{m+1}}{p_{i_{j}}}\mu_{2,j}
+\frac{p_{e_{j-1}}^{m+1}}{p_{e_{j}}}\mu_{3,j}+\frac{p_{e_{j}}^{m+1}}{p_{e_{j}}}\left[G_{j}\delta_{se}\right.\\
&\left.+E_{j}p_{e_{j}}\left(\frac{\delta_{si}Z_{u}-\delta_{se}Z_{s}}{p_{u_{j}}Z_{u}-p_{s_{j}}Z_{s}}-\frac{3}{2}\frac{\delta_{si}m_{s}Z_{u}+\delta_{se}m_{u}Z_{s}}{m_{s}Z_{u}p_{u_{j}}+m_{u}Z_{s}p_{s_{j}}}\right)+\mu_{3,j}\left(\frac{p_{e_{j+1}}-p_{e_{j-1}}}{p_{e_{j}}}\right)\right]\\
&-\frac{p_{e_{j+1}}^{m+1}}{p_{e_{j}}}\mu_{3,j}-E_{j},
\end{split}
\end{equation}
where
\begin{gather}
\mu_{1,j}\equiv \frac{R}{2a\Delta\rho}\left[G_{j}\left(\pd{\ln \Gamma_{j}}{(R/L_{n})_{j}}-\frac{5}{2\kappa_{j}}\right)
+H_{j}\left(\pd{\ln Q_{j}}{(R/L_{n})_{j}} +\frac{1}{\tld{\kappa}_{j}}\right)
+K_{j}\pd{\ln \mathcal{H}_{j}}{(R/L_{n})_{j}}\right]\\
\mu_{2,j}\equiv \frac{R}{2a\Delta\rho}\left[G_{j}\left(\pd{\ln \Gamma_{j}}{(R/L_{p_{i}})_{j}}+\frac{3\delta_{si}}{2\kappa_{j}}\right)
+H_{j}\left(\pd{\ln Q_{j}}{(R/L_{p_{i}})_{j}} -\frac{\delta_{si}}{\tld{\kappa}_{j}}\right)
+K_{j}\pd{\ln \mathcal{H}_{j}}{(R/L_{p_{i}})_{j}}\right]\\
\mu_{3,j}\equiv \frac{R}{2a\Delta\rho}\left[G_{j}\left(\pd{\ln \Gamma_{j}}{(R/L_{p_{e}})_{j}}+\frac{3\delta_{se}}{2\kappa_{j}}\right)
+H_{j}\left(\pd{\ln Q_{j}}{(R/L_{p_{e}})_{j}} -\frac{\delta_{se}}{\tld{\kappa}_{j}}\right)
+K_{j}\pd{\ln \mathcal{H}_{j}}{(R/L_{p_{e}})_{j}}\right].
\end{gather}
Next we compute $\partial F/\partial \rho$:
\begin{equation}
\begin{split}
&\left(\pd{F^{m+1}}{\rho}\right)_{j}\approx \frac{F_{+}^{m+1}-F_{-}^{m+1}}{\Delta\rho}\approx \frac{1}{\Delta\rho}\left[\frac{n_{j-1}^{m+1}}{n_{-}}\left(\mathcal{A}_{-}[n]+\frac{3}{4}F_{-}\right)\right.\\
&\left.+\frac{n_{j}^{m+1}}{n_{+}}\left[\frac{3}{4}\left(F_{-}\frac{n_{+}}{n_{-}}-F_{+}\right)+\mathcal{B}_{j}[n]\right]+\frac{n_{j+1}^{m+1}}{n_{+}}\left[\mathcal{A}_{+}[n]-\frac{3}{4}F_{+}\right]\right.\\
&\left.+\frac{p_{i_{j-1}}^{m+1}}{p_{i_{-}}}\left(\mathcal{A}_{-}[p_{i}]-\frac{5}{4}F_{-}\right)+\frac{p_{i_{j}}^{m+1}}{p_{i_{+}}}\left[-\frac{5}{4}\left(F_{-}\frac{p_{i_{+}}}{p_{i_{-}}}-F_{+}\right)+\mathcal{B}_{j}[p_{i}]\right]\right.\\
&\left.+\frac{p_{i_{j+1}}^{m+1}}{p_{i_{+}}}\left[\mathcal{A}_{+}[p_{i}]+\frac{5}{4}F_{+}\right]+\frac{p_{e_{j-1}}^{m+1}}{p_{e_{-}}}\mathcal{A}_{-}[p_{e}]+\frac{p_{e_{j}}^{m+1}}{p_{e_{+}}}\mathcal{B}_{j}[p_{e}]+\frac{p_{e_{j+1}}^{m+1}}{p_{e_{+}}}\mathcal{A}_{+}[p_{e}]\right],
\end{split}
\end{equation}
where we define
\begin{gather}
\mathcal{A}_{\pm}[w]\equiv -\frac{RF_{\pm}}{a\Delta\rho}\frac{w_{j}}{w_{\pm}^{2}}\pd{\ln Q_{\pm}}{(R/L_{w})_{\pm}}\\
\mathcal{B}_{j}[w]\equiv \frac{R}{a\Delta\rho}\left(F_{+}\frac{w_{j+1}}{w_{+}^{2}}\pd{\ln Q_{+}}{(R/L_{w})_{+}}+F_{-}\frac{w_{j-1}}{w_{-}^{2}}\pd{\ln Q_{-}}{(R/L_{w})_{-}}\right).
\end{gather}

Finally, we add all of these terms up to get
\begin{equation}
\begin{split}
&p_{s_{j}}^{m+1}-\sum_{k=-1}^{1} \frac{2\Delta \tau\alpha}{3}\left(n_{j+k}^{m+1}\psi_{n,k}+p_{i_{j+k}}^{m+1}\psi_{p_{i},k}+p_{e_{j+k}}^{m+1}\psi_{p_{e},k}\right)\\
&=p_{s_{j}}+\frac{2\Delta \tau}{3}\left(1-\alpha\right)\left[-\frac{\rhoavg}{A}\pd{F}{\rho}+G+H+E+K\right]+\Delta\tau S_{p}
\end{split}
\end{equation}
where
\begin{eqnarray}
\label{eqn:ncoefs2}
\psi_{n,-1}&\equiv& \frac{\mu_{1,j}}{n_j} - \frac{\rhoavg}{A\Delta\rho}\left(\mathcal{A}_{-}[n]+\frac{3}{4}\frac{F_{-}}{n_{-}}\right)\\
\psi_{n,0}&\equiv& \frac{\mu_{1,j}}{n_j}\left(\frac{n_{j+1}-n_{j-1}}{n_{j}}\right)+\frac{5}{2n_{j}}E_{j} - \frac{3}{2n_{j}}\left(G_{j}+H_{j}+K_{j}\right)\\
 &&- \frac{\rhoavg}{A\Delta\rho}\left[\mathcal{B}_{j}[n]+\frac{3}{4}\left(\frac{F_{-}}{n_{-}}-\frac{F_{+}}{n_{+}}\right)\right]\\
\psi_{n,1}&\equiv& -\frac{\mu_{1,j}}{n_j} - \frac{\rhoavg}{A\Delta\rho}\left(\mathcal{A}_{+}[n]-\frac{3}{4}\frac{F_{+}}{n_{+}}\right)\\
\psi_{p_{i},-1}&\equiv& \frac{\mu_{2,j}}{p_{i_{j}}} - \frac{\rhoavg}{A\Delta\rho}\left(\mathcal{A}_{-}[p_{i}]-\frac{5}{4}\frac{F_{-}}{p_{i_{-}}}\right)\\
\psi_{p_{i},0}&\equiv& \frac{G_{j}}{p_{i_{j}}}\left(\frac{3}{2}+\delta_{si}\right) + \frac{5}{2p_{i_{j}}}\left(H_{j}+K_{j}\right)+\frac{\mu_{2,j}}{p_{i_{j}}}\left(\frac{p_{i_{j+1}}-p_{i_{j-1}}}{p_{i_{j}}}\right)\\
&&+E_{j}\left(\frac{\delta_{se}Z_{u}-\delta_{si}Z_{s}}{p_{u_{j}}Z_{u}-p_{s_{j}}Z_{s}}-\frac{3}{2}\frac{\delta_{se}m_{s}Z_{u}+\delta_{si}m_{u}Z_{s}}{m_{s}Z_{u}p_{u_{j}}+m_{u}Z_{s}p_{s_{j}}}\right)\\
&&-\frac{\rhoavg}{A\Delta\rho}\left[\mathcal{B}_{j}[p_{i}]-\frac{5}{4}\left(\frac{F_{-}}{p_{i_{-}}}-\frac{F_{+}}{p_{i_{+}}}\right)\right]\\
\psi_{p_{i},1}&\equiv& -\frac{\mu_{2,j}}{p_{i_{j}}} - \frac{\rhoavg}{A\Delta\rho}\left(\mathcal{A}_{+}[p_{i}]+\frac{5}{4}\frac{F_{+}}{p_{i_{+}}}\right)\\
\psi_{p_{e},-1}&\equiv& \frac{\mu_{3,j}}{p_{e_{j}}} - \frac{\rhoavg}{A\Delta\rho}\mathcal{A}_{-}[p_{e}]\\
\psi_{p_{e},0}&\equiv& \frac{G_{j}}{p_{e_{j}}}\delta_{se}+E_{j}\left(\frac{\delta_{si}Z_{u}-\delta_{se}Z_{s}}{p_{u_{j}}Z_{u}-p_{s_{j}}Z_{s}}-\frac{3}{2}\frac{\delta_{si}m_{s}Z_{u}+\delta_{se}m_{u}Z_{s}}{m_{s}Z_{u}p_{u_{j}}+m_{u}Z_{s}p_{s_{j}}}\right)\\
&&+ \frac{\mu_{3,j}}{p_{e_{j}}}\left(\frac{p_{e_{j+1}}-p_{e_{j-1}}}{p_{e_{j}}}\right)-\frac{\rhoavg}{A\Delta\rho}\mathcal{B}_{j}[p_{e}]\\
\psi_{p_{e},1}&\equiv& -\frac{\mu_{3,j}}{p_{e_{j}}} - \frac{\rhoavg}{A\Delta\rho}\mathcal{A}_{+}[p_{e}].
\label{eqn:pecoefs2}
\end{eqnarray}

\subsection{Boundary conditions}

To complete our numerical prescription, we must supply boundary conditions at the
innermost and outermost radial grid locations.  At the outer boundary, which corresponds 
to a location in the fusion device just inside the edge pedestal region, we are free to
specify density and pressure.  The purpose of the simulation is then to determine
the core temperature as a function of the pedestal density and pressure and of the external
heat source strength.  

At the internal boundary, which corresponds to the magnetic axis,
we use the physical constraint that the product of the flux surface area and the flux 
surface-averaged fluxes is zero:
\begin{equation}
\lim_{\rho\rightarrow0} \ A \begin{pmatrix}
\Gamma \\
Q_{i}\\
Q_{e}
\end{pmatrix}
= 0.
\end{equation}
We take the magnetic axis to correspond to the spatial index $j=1/2$.  Consequently,
the terms in Eqs.~(\ref{eqn:timediff}) and~(\ref{eqn:ptransptdiff}) involving the radial derivative of the 
fluxes reduce to
\begin{equation}
\left(\pd{F}{\rho}\right)_{j=1} \approx \frac{F_{3/2}-F_{1/2}}{\Delta\rho} = \frac{F_{3/2}}{\Delta\rho}.
\end{equation}
All other inner boundary terms only involve evaluation at $j=1$.  However, the linearization
of the nonlinear terms introduces quantities like $\partial (R/L_{n})_{j}/\partial n_{k}$.
Since we have used a three-point, centered stencil for $R/L_{n}$ and $R/L_{p}$, we would 
need to evaluate the density and pressure inside the inner boundary.  To avoid this, we must 
employ an alternate discretization
for $R/L_{n}$ and $R/L_{p}$ at $j=1$.  We choose to use a shifted three-point stencil to 
retain second order accuracy of derivatives:
\begin{equation}
\left(\frac{R}{L_{n}}\right)_{j} = -\frac{R}{a}\left(\pd{\ln n}{\rho}\right)_{j}
\approx -\frac{R}{a}\frac{1}{n_{j}\Delta\rho}\left(-\frac{3}{2}n_{j}+2n_{j+1}-\frac{1}{2}n_{j+2}\right).
\end{equation}
Computing the derivative of this expression with respect to the density at grid points, we have
\begin{equation}
\sum_{k}\dn\pd{(R/L_{n})_{j}}{n_{k}}=\frac{R}{a\Delta\rho}
\left[\frac{n_{j+2}^{m+1}}{2n_{j}}-2\frac{n_{j+1}^{m+1}}{n_{j}}+\frac{n_{j}^{m+1}}{n_{j}}\frac{4n_{j+1}-n_{j+2}}{2n_{j}}\right],
\end{equation}
with similar expressions for $p_{i}$ and $p_{e}$.

A consequence of this inner boundary condition is that the coefficients given in 
Eqs.~(\ref{eqn:ncoefs})-(\ref{eqn:pecoefs}) and~(\ref{eqn:ncoefs2})-(\ref{eqn:pecoefs2}) are modified by 
taking $F_{-}$ to be zero.  There are additional modifications to Eqs.~(\ref{eqn:ncoefs2})-(\ref{eqn:pecoefs2})
due to the new discretization of $R/L_{n}$ and $R/L_{p}$.  The new coefficients are
\begin{eqnarray}
\psi_{n,2}&\equiv& \frac{\mu_{1,j}}{n_j}\\
\psi_{n,0}&\equiv& \frac{\mu_{1,j}}{n_j}\left(\frac{4n_{j+1}-n_{j+2}}{n_{j}}\right)+\frac{5}{2n_{j}}E_{j} - \frac{3}{2n_{j}}\left(G_{j}+H_{j}+K_{j}\right)\\
 &&- \frac{\rhoavg}{A\Delta\rho}\left[\mathcal{B}_{j}[n]-\frac{3}{4}\frac{F_{+}}{n_{+}}\right]\\
\psi_{n,1}&\equiv& -4\frac{\mu_{1,j}}{n_j} - \frac{\rhoavg}{A\Delta\rho}\left(\mathcal{A}_{+}[n]-\frac{3}{4}\frac{F_{+}}{n_{+}}\right)\\
\psi_{p_{i},2}&\equiv& \frac{\mu_{2,j}}{p_{i_{j}}}\\
\psi_{p_{i},0}&\equiv& \frac{G_{j}}{p_{i_{j}}}\left(\frac{3}{2}+\delta_{si}\right) + \frac{5}{2p_{i_{j}}}\left(H_{j}+K_{j}\right)+\frac{\mu_{2,j}}{p_{i_{j}}}\left(\frac{4p_{i_{j+1}}-p_{i_{j+2}}}{p_{i_{j}}}\right)\\
&&+E_{j}\left(\frac{\delta_{se}Z_{u}-\delta_{si}Z_{s}}{p_{u_{j}}Z_{u}-p_{s_{j}}Z_{s}}-\frac{3}{2}\frac{\delta_{se}m_{s}Z_{u}+\delta_{si}m_{u}Z_{s}}{m_{s}Z_{u}p_{u_{j}}+m_{u}Z_{s}p_{s_{j}}}\right)\\
&&-\frac{\rhoavg}{A\Delta\rho}\left[\mathcal{B}_{j}[p_{i}]+\frac{5}{4}\frac{F_{+}}{p_{i_{+}}}\right]\\
\psi_{p_{i},1}&\equiv& -4\frac{\mu_{2,j}}{p_{i_{j}}} - \frac{\rhoavg}{A\Delta\rho}\left(\mathcal{A}_{+}[p_{i}]+\frac{5}{4}\frac{F_{+}}{p_{i_{+}}}\right)\\
\psi_{p_{e},2}&\equiv& \frac{\mu_{3,j}}{p_{e_{j}}}\\
\psi_{p_{e},0}&\equiv& \frac{G_{j}}{p_{e_{j}}}\delta_{se}+E_{j}\left(\frac{\delta_{si}Z_{u}-\delta_{se}Z_{s}}{p_{u_{j}}Z_{u}-p_{s_{j}}Z_{s}}-\frac{3}{2}\frac{\delta_{si}m_{s}Z_{u}+\delta_{se}m_{u}Z_{s}}{m_{s}Z_{u}p_{u_{j}}+m_{u}Z_{s}p_{s_{j}}}\right)\\
&&+ \frac{\mu_{3,j}}{p_{e_{j}}}\left(\frac{4p_{e_{j+1}}-p_{e_{j+2}}}{p_{e_{j}}}\right)-\frac{\rhoavg}{A\Delta\rho}\mathcal{B}_{j}[p_{e}]\\
\psi_{p_{e},1}&\equiv& -4\frac{\mu_{3,j}}{p_{e_{j}}} - \frac{\rhoavg}{A\Delta\rho}\mathcal{A}_{+}[p_{e}].
\end{eqnarray}

\section{Time averaging}
\label{sec:tavg}

The steady-state turbulent fluxes used in the transport equations are time-averaged values.
We would like to minimize simulation time by running each set of
turbulence calculations just long enough to obtain good statistics on the converged fluxes.  
Consequently, we have developed an adaptive time averaging procedure for the turbulent 
fluxes and collisional heating that automatically detects when the fluxes have converged.

For each flux tube simulation, we keep track of the instantaneous time average of the fluxes
and heating:
\begin{equation}
\overline{\Gamma}_{m}= \sum_{i=1}^{m} \Gamma_{i} (\Delta t)_{i},
\end{equation}
where $m$ denotes the turbulent time step and $\Delta t$ is the size of the turbulent time step.
Analogous expressions are used for the heat fluxes and heating.  While this quantity
is sufficient for use as the time-averaged flux in our transport calculation, we have the 
additional burden of determining when this flux has converged so that we may end the 
turbulence calculation.  In order to accomplish this, we store each of the $\overline{\Gamma}_{i}$
for $i=1\rightarrow m$.  At each time step we then compute a measure of the rms deviation
of each of the $\overline{\Gamma}^{i}$ (for $i=j\rightarrow m-1$, where $j$ is by default $m/2$, 
but can be specified by the user) from the current value:
\begin{equation}
\epsilon_{\Gamma}\equiv\sqrt{\frac{1}{m-j}\sum_{i=j}^{m-1}\left(\overline{\Gamma}_{m}^{2}-\overline{\Gamma}_{i}^{2}\right)}
\end{equation}
When $\epsilon_{\Gamma}$ is less than a user-specified tolerance, the flux is 
determined to be converged, and the turbulence calculation terminates.

\section{Quasilinear fluxes}
\label{sec:qlflx}

Since turbulent flux calculations are computationally expensive, we find it convenient
at times to use quasilinear estimates for the fluxes and heating.  We do not claim
that these estimates are quantitatively correct; we merely use them as a computationally
inexpensive tool to test our numerical scheme and to gain quick, qualitative insight
into transport and heating processes.

To obtain quasilinear estimates for the fluxes and collisional heating, we normalize the 
fluxes and heating from linear \verb#Trinity#
simulations by $\la\Phi\ra^{2}$ and multiply by a factor derived from mixing length
theory.  The argument goes as follows: one expects saturation to occur when nonlinear
effects become dominant, i.e. when $\partial h / \partial t \sim\mbf{v_{\chi}}\cdot\nabla h$.
From this balance, we obtain an approximation for the growth rate in terms of $\Phi$:
\begin{equation}
\gamma \sim k_{\perp}^{2} \rho v_{th} \frac{e\Phi}{T}.
\end{equation}
Defining
$\tilde{\Phi}\equiv (a/\rho)(e\Phi/T)$, $\tilde{k}\equiv
k_{\perp}\rho$, and $\gamma\equiv (a/v_{th})\gamma$, we have
\begin{eqnarray}
\la\tld{\Phi}\ra&\sim&\frac{\tld{\gamma}}{\tld{k}^{2}}\\
\Rightarrow \Gamma^{N}_{QL}&\sim& \frac{\tld{\gamma}^{2}}{\tld{k}^{4}}
\frac{\Gamma^{N}}{\la\tld{\Phi}\ra^{2}},
\end{eqnarray}
with analagous expressions for the heat fluxes and collisional heating.

\section{Heat source}
\label{sec:source}

We now relate the power input to the normalized source $\tilde{\mathcal{S}}_{p}$.  The power input, $\mathcal{P}$, is defined as
\begin{equation}
\mathcal{P} \equiv \int dV S_{p}.
\end{equation}
Taking into account normalizations, we have
\begin{eqnarray}
\mathcal{P} &=& \int dV \frac{n_{0,r}T_{0,r}^{3/2}}{a\sqrt{m_{r}}}\rho_{*0}^{2}\tld{S}_{p}\\
&=& a^{2} \int d\phi d\theta d\rho \left(\nabla \rho \times \nabla \theta \cdot \nabla \phi\right)^{-1} \frac{n_{0,r}T_{0,r}^{3/2}}{a\sqrt{m_{r}}}\rho_{*0}^{2}\tld{S}_{p}
\end{eqnarray}
For circular flux surfaces, we have
\begin{equation}
\mathcal{P} = 4\pi^{2}a R_{0} \frac{n_{0,r}T_{0,r}^{3/2}}{\sqrt{m_{r}}}\rho_{*0}^{2}\int_{0}^{1} d\rho \ \rho \tld{S}_{p}(\rho)
\label{eqn:pcircle}
\end{equation}
We are currently using an analytic source of the form 
\begin{equation}
\tld{S}_{p}=\frac{\tld{A}}{\sigma}\exp\left[-\frac{\rho^{2}}{2\sigma^{2}}\right],
\end{equation}
for which
\begin{equation}
\int_{0}^{1} d\rho \ \rho \tld{S}_{p}(\rho)=\tld{A}\left(1-\exp\left[-\frac{1}{2\sigma^{2}}\right]\right)\sigma.
\end{equation}
Using this result, we can
rearrange Eq.~(\ref{eqn:pcircle}) to solve for the normalized source amplitude $\tld{A}$
in terms of the input power and the width of the Gaussian power deposition profile $\sigma$:
\begin{equation}
\begin{split}
\tld{A} &= \left[4\pi^{2}aR_{0}\rho_{*0}^{2}\frac{n_{0,r}T_{0,r}^{3/2}}{m_{r}^{1/2}}\sigma\left(1-\exp\left[-\frac{1}{2\sigma^{2}}\right]\right)\right]^{-1}\mathcal{P}\\
&=\frac{R}{a}\frac{5.11\times10^{-6}\tld{m}_{i}^{1/2}\tld{\mathcal{P}}}{\tld{R}^{2}\rho_{*0}^{2}\sigma\left(1-\exp\left[-\frac{1}{2\sigma^{2}}\right]\right)},
\end{split}
\end{equation}
where $\tld{\mathcal{P}}$ is given in MegaWatts and $\tld{m}_{i}$ is given in units
of the proton mass.

\section{\texttt{Trinity} simulations}
\label{sec:trinsim}

In this section we present simple tests showing that our implicit transport solver is 
well behaved and that it provides significant computational savings over an explicit
solver.  Furthermore, we present preliminary results from \verb#Trinity# simulations
using gyrokinetic, turbulent fluxes and heating.  These simulations, which are the first
of their kind, demonstrate that the coupled flux tube approach can routinely be used to
obtain steady-state equilibrium profiles of density and pressure, as well as the corresponding
turbulent fluxes and heating.

\subsection{Tests}

The first test we present is intended to demonstrate that the heat transport and turbulent heating
terms from Eq.~(\ref{eqn:pevo2}) have been properly implemented and that the implicit transport
solver algorithm developed in this chapter is well behaved, even for multi-channel transport (here we
are evolving density and ion and electron pressures).  We artificially set the temperature equilibration
term to zero and use the following analytic model for our normalized fluxes and turbulent heating:
\begin{gather}
\Gamma_{s} = \frac{n^{1/2}}{p_{i}^{3/2}}\\
Q_{s} = \frac{p_{s}}{p_{i}^{5/2}}\\
\mathcal{H}_{s} = -\frac{3}{p_{i}^{5/2}}\pd{p_{s}}{\rho},
\end{gather}
where all quantities are understood to be the normalized versions defined in \secref{sec:teqnorm}.  We
note that these fluxes do not mesh well with the approximation employed in our transport solver algorithm
that the fluxes and heating depend primarily on $R/L_{n}$, $R/L_{T_{i}}$, and $R/L_{T_{e}}$.
Consequently, we are also testing the resiliency of our scheme.
The resultant transport equations are
\begin{gather}
\label{eqn:tst1}
\pd{n}{\tau} = 0\\
\pd{p_{s}}{\tau} = \pd{p_{s}}{\rho},
\label{eqn:tst2}
\end{gather}
which has the solution $p_{s}=F[\tau+\rho]$, with $F$ an arbitrary functional.  For our initial conditions,
we take $n(\tau=0)=1$ and $p_{s}(\tau=0)=\exp[-\rho]$.  Our boundary conditions are $n(\rho_{\pm},\tau)=1$
and $p_{s}(\rho_{\pm},\tau)=\exp[-(\rho_{\pm}+\tau)]$, were $\rho_{\pm}$ represents the the inner and
outer radii in the simulation.  The solution to this system is $n(\rho,\tau)=1$ and $p_{s}(\rho,\tau)=\exp[-(\rho+\tau)]$.
In Fig.~\ref{fig:test1}, we show the numerical solution for this system, which is in excellent agreement
with our analytic prediction.

\begin{figure}
\centering
\includegraphics[height=5.0in]{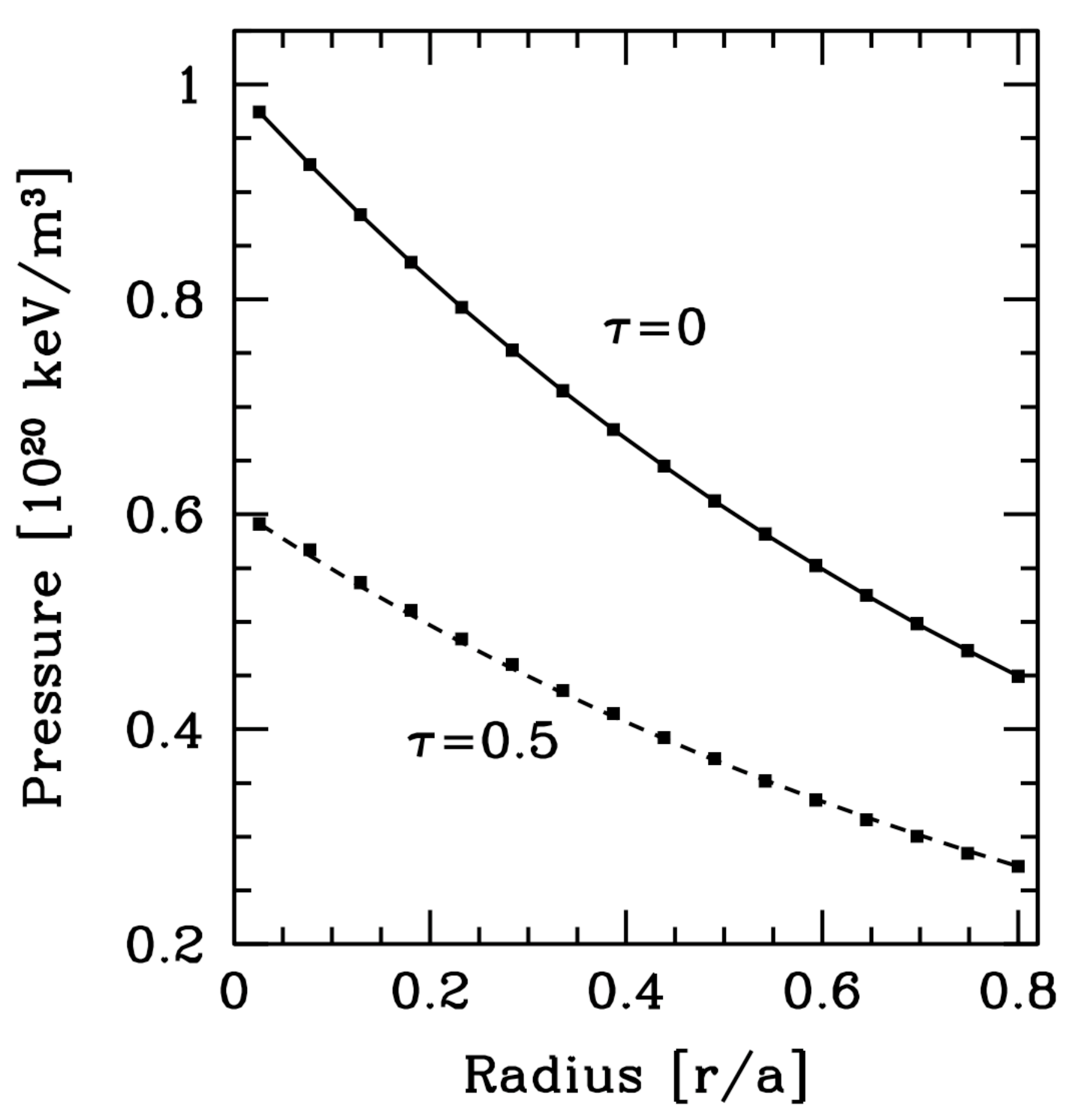}
\caption{Comparison of the analytic and numerical solutions to the system defined by
Eqs.~(\ref{eqn:tst1}) and~(\ref{eqn:tst2}) at 
$\tau=0$ and $\tau=2$.  Lines represent analytic solution and dots represent numerical solution from
\texttt{Trinity}.
Here we are showing only the ion pressure, but the solution for the electron pressure is identical (and
the density remains approximately constant in time).
Simulation conducted with $\Delta\tau=0.02$ and 16 equally spaced radial grid points (flux tubes).}
\label{fig:test1}
\end{figure}

Our second test illustrates the superiority of our implicit implentation to an explicit scheme when
considering fluxes that lead to diffusive behavior.  We artificially set the turbulent heating and temperature
equilibration terms to zero and use the following form for the turbulent heat flux (we do not evolve 
the density in this case, so there is no need to define the particle flux):
\beq
Q_{s} = \frac{3D}{2}\frac{a}{L_{p_{s}}}\frac{p_{s}}{p_{i}^{5/2}},
\eeq
where $D$ is a constant diffusion coefficient.  The resultant transport equations are
\beq
\pd{p_{s}}{\tau} = D \pd{^{2}p_{s}}{\rho^{2}},
\label{eqn:diffuse}
\eeq
which is simply the diffusion equation.  Taking the initial condition of the form 
$p_{s}(\rho,\tau=0)=\exp[-\rho^{2}/4D]$ and boundary conditions of the form
$p_{s}(\rho_{\pm},\tau)=\sqrt{1/\tau}\exp[-\rho_{\pm}^{2}/4D\tau]$, the solution
is $p_{s}(\rho,\tau)=\sqrt{1/\tau}\exp[-\rho^{2}/4D\tau]$.  In Fig.~\ref{fig:test2}, we
show the numerical solution using \verb#Trinity#'s implicit transport solver.  After conducting
a number of both explicit ($\alpha=0$) and implicit ($\alpha=1/2$) numerical simulations,
we find that the implicit scheme gives good results (relative error less than $10\%$) for at least
$\Delta\tau=2.0$, whereas the explicit scheme is numerically unstable for approximately
$\Delta\tau > 0.02$.
After taking into account the fact that an additional set of flux tube simulations must be run for each
transport channel at each transport time step when running implicitly, we find that the implicit scheme 
provides a savings of a factor of $\sim25-50$ over the explicit scheme, depending on the number of transport 
channels used (from $1-3$ currently).

\begin{figure}
\centering
\includegraphics[height=5.0in]{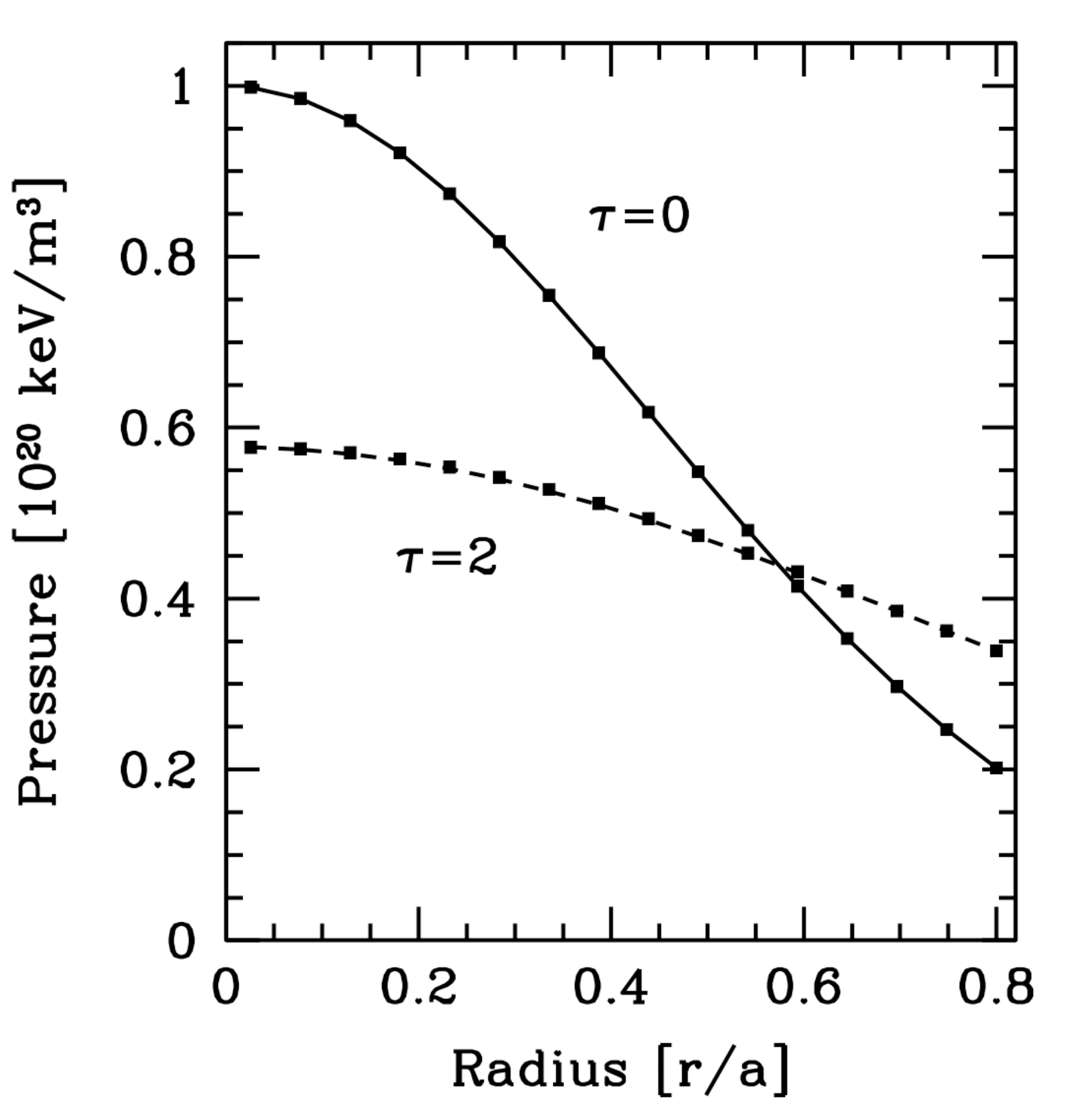}
\caption{Comparison of the analytic solution to the $D=0.1$ diffusion equation (\ref{eqn:diffuse}) at 
$\tau=0$
(solid line) and $\tau=2$ (dashed line) to the numerical solution from \texttt{Trinity} (square dots).
Here we are showing only the ion pressure, but the solution for the electron pressure is identical.
Simulation conducted with $\Delta\tau=0.1$ and 16 equally spaced radial grid points (flux tubes).}
\label{fig:test2}
\end{figure}

\subsection{Preliminary results}

The simulation results presented in this section are taken from relatively low resolution simulations with 
reduced physics models.  They are meant to be demonstrations that the coupled flux tube approach 
detailed in this chapter can be routinely used to obtain predictions for steady-state equilibrium profiles
and turbulent fluxes.  All simulations considered here were run with a hyperviscous dissipation 
model~\cite{hammettbelli} that allows us to obtain
reasonable, converged turbulent fluxes with a relatively coarse spatial grid for the turbulence.
For each simulation we used a $16\times16$ grid in the spatial plane perpendicular to the magnetic field,
$26$ grid points along the magnetic field line, $12$ energies, $20$ untrapped pitch angles, and a variable
number of trapped pitch angles (the number of trapped pitch angles in \verb#Trinity# depends on 
location along the equilibrium magnetic field line.  See Chapter 4 for more details.).  

We consider a three different systems.  All of them have: a single kinetic, hydrogenic ion species; 
electrostatic fluctuations;
major radius of $6.2$ meters; aspect ratio of $3.1$; local (Miller) geometry with concentric, circular flux
surfaces; fixed edge temperature of $4$ keV; and external heat input to the ions (via a Gaussian 
deposition profile with $\sigma=0.2\rho$).  The first two systems we consider both have adiabatic electrons
and $60$ MW deposited in the ions from an external source, but they have different magnetic field strengths.
Evolving only the ion pressure gradient, the simulations were run with eight radial 
grid points and $\Delta\tau=0.02$ ($\Delta t \approx 0.018$ seconds) for $25$ time steps.  
The simulations took approximately $20$ minutes each on $2048$ processors.  The steady state ion 
temperature profiles for the two different toroidal magnetic field strengths are shown in 
Fig.~\ref{fig:ti2Bs}.  As 
expected, the case with the stronger magnetic field leads to higher core temperatures.  In 
Fig.~\ref{fig:neocomp} we compare the $B_{a}=5.3$ T result for the ion temperature profile with the
same result calculated using only neoclassical fluxes (obtained using the analytic expression
for the ion heat flux from Ref.~\cite{changPoF82}).  We see that in the absence of microturbulence,
the core ion temperature is well in excess of what is required to ignite a burning plasma.

\begin{figure}
\centering
\includegraphics[height=5.0in]{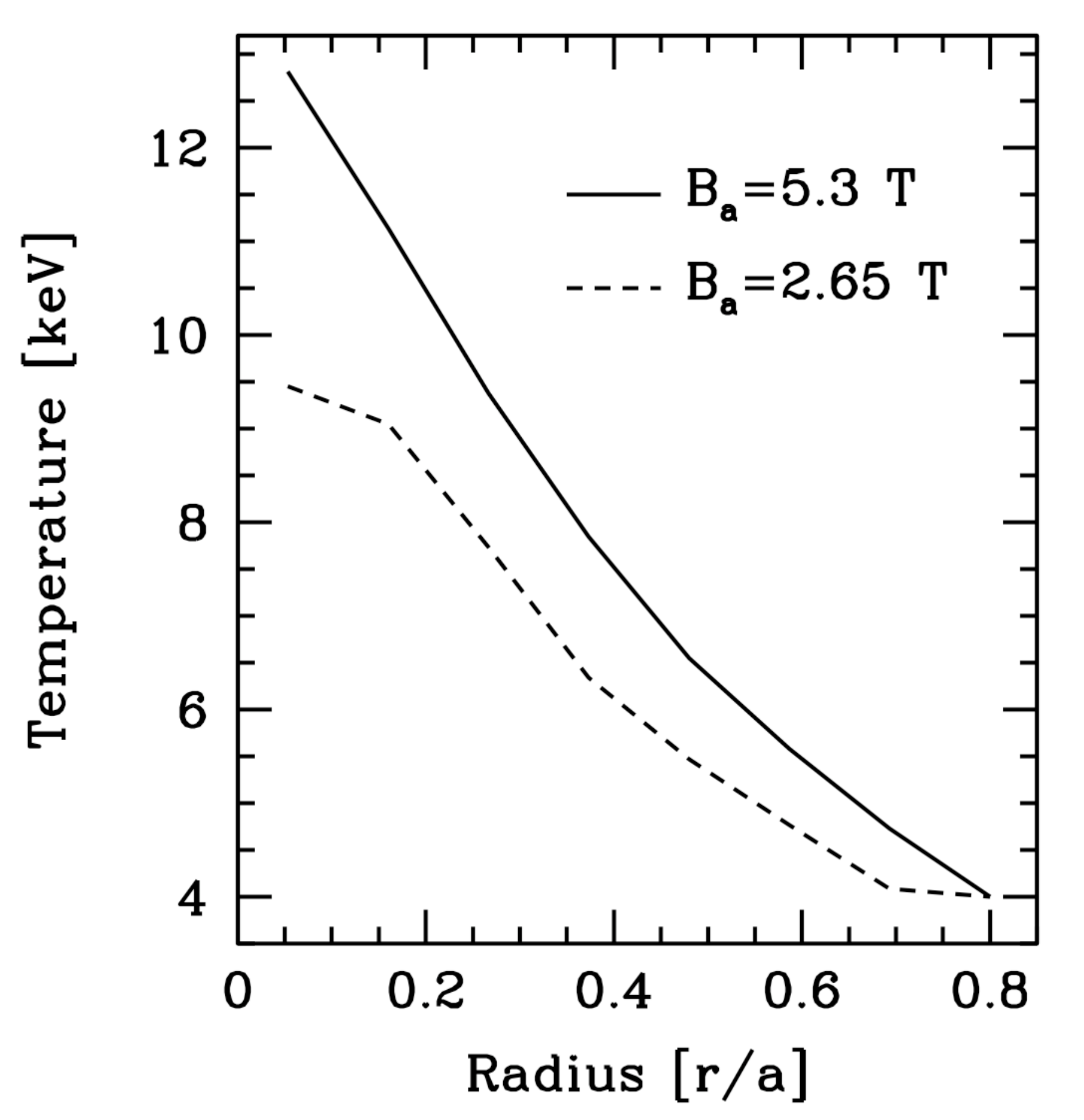}
\caption{Steady-state ion temperature profile for two different values of $B_{a}$,
the magnetic field magnitude at the center of the LCFS.  As expected, an increase in $B_{a}$
leads to an increase in the core temperature.}
\label{fig:ti2Bs}
\end{figure}

\begin{figure}
\centering
\includegraphics[height=5.0in]{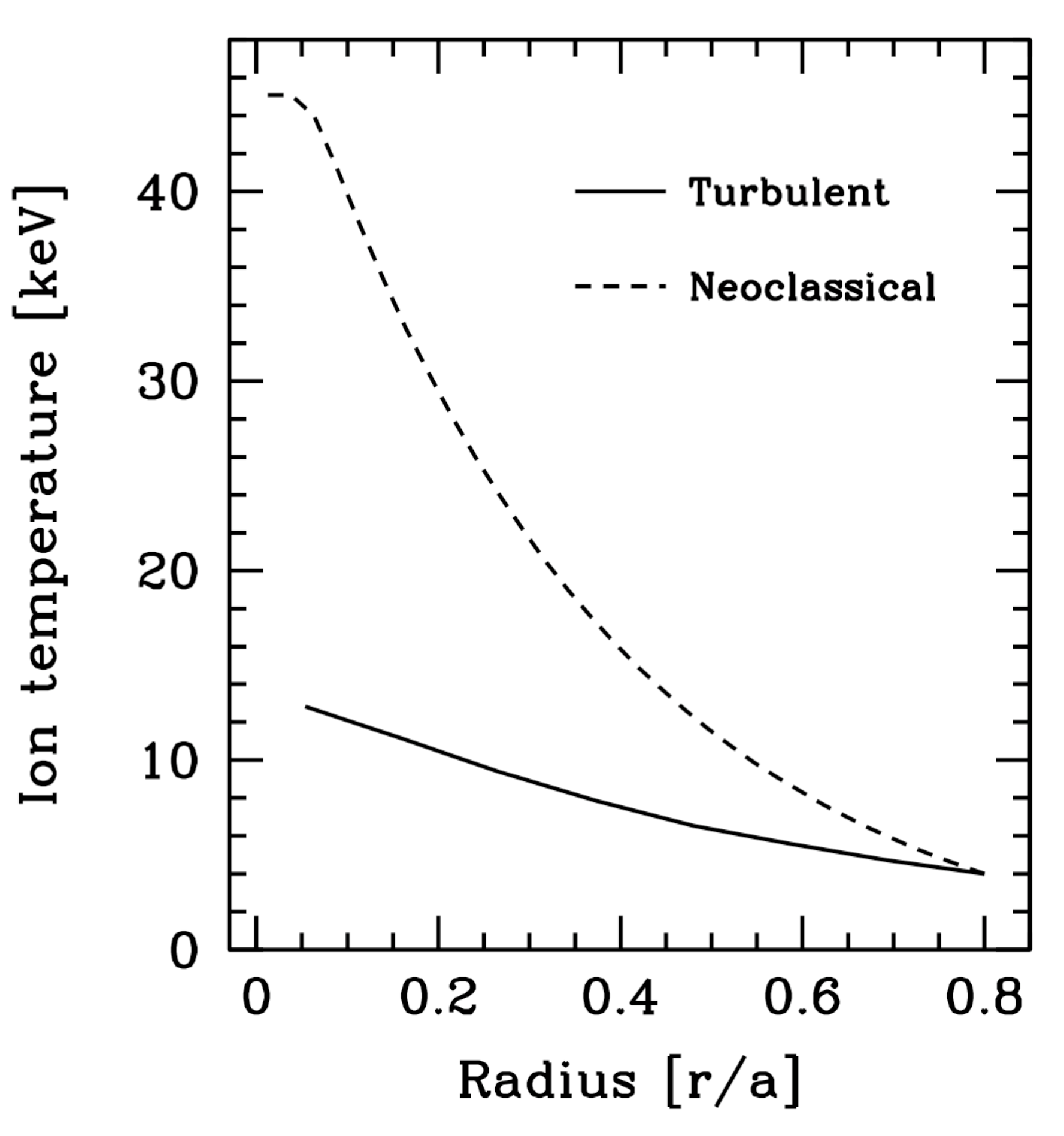}
\caption{Comparison of steady-state ion temperature profiles for simulations using
turbulent fluxes (solid line) and only neoclassical fluxes (dashed line).  Without the fluxes 
arising from microturbulence, core plasma temperatures would easily be sufficient to ignite 
the plasma.}
\label{fig:neocomp}
\end{figure}

The final case we consider has kinetic electrons, a magnetic field strength of $B_{a}=5.3$ T, and $120$ MW
external heat source, with $30\%$ going into the ion channel.  Again using $8$ radial grid points,
we evolved both the ion and electron equilibrium pressure profiles.  For our time step, we used
$\Delta\tau=0.005$ ($\Delta t \approx 0.004$ seconds) and evolved for 25 time steps.  The simulation
took approximately one hour on $4096$ processors.  The results, shown in Fig.~\ref{fig:tgrads} indicate
that the use of kinetic electrons (instead of the adiabatic electron model used to obtain Fig.~\ref{fig:ti2Bs})
leads to a significant (approximately $65\%$) reduction in the core ion temperature.  The fact that the
core ion temperature is reduced upon taking into account kinetic electron effects is not suprising since
the trapped electron mode (TEM) is enabled when considering kinetic electrons.  However, the large size of 
the reduction may be misleading, since we are employing a relatively coarse grid in phase space.

\begin{figure}
\centering
\includegraphics[height=5.0in]{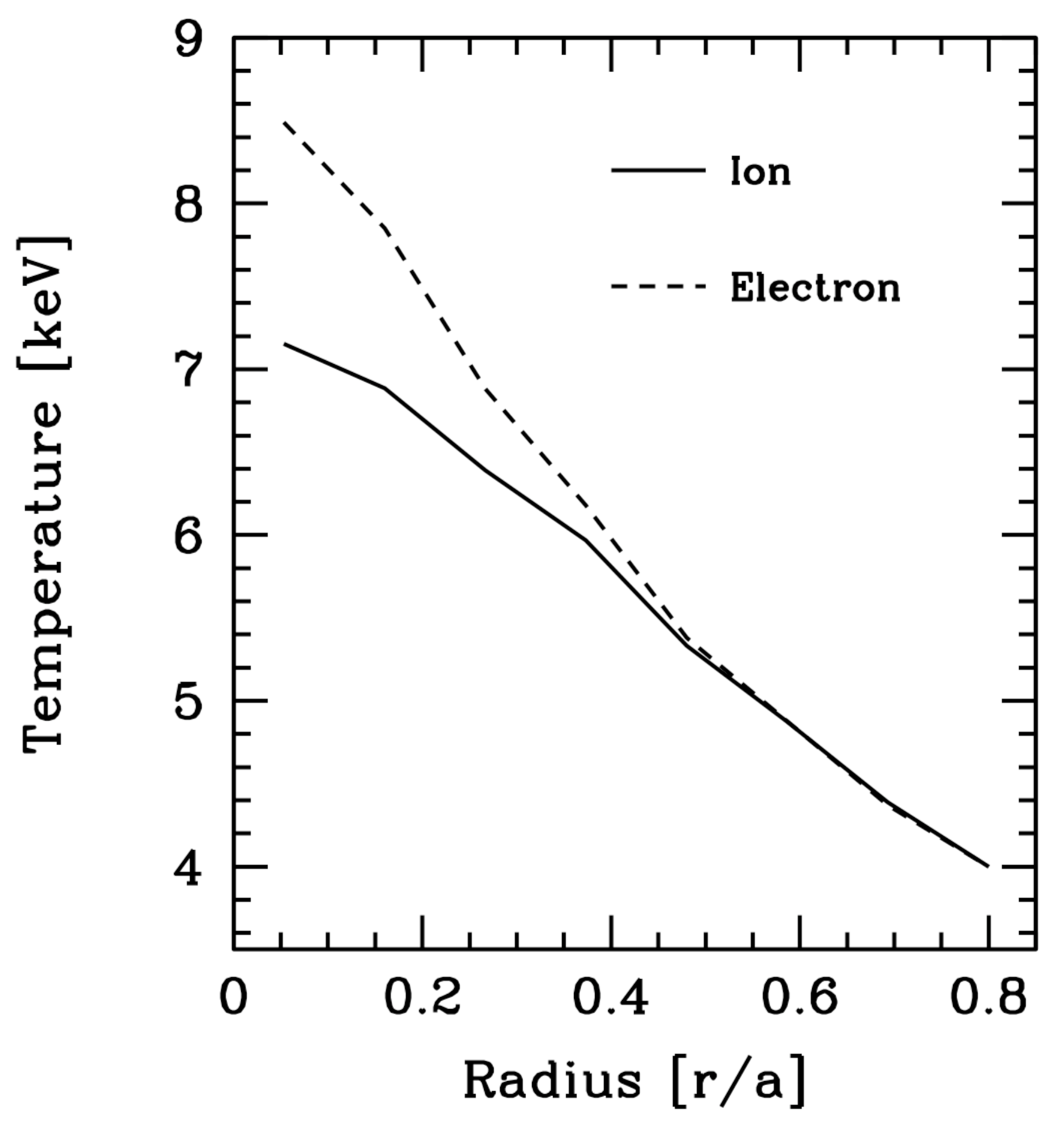}
\caption{Steady-state ion and electron temperature profiles for the same system used to obtain
the $B_{a}=5.3$ T plot in Fig.~\ref{fig:ti2Bs}, with the exception that here we retain kinetic electron
effects.  Temperature equilibration is strong enough near the edge (due to low electron temperature, 
moderate collisionality, and weak local external heating) to keep the ions and electrons in
thermal equilibrium, but this is not true as we approach the core.  Comparing with Fig.~\ref{fig:ti2Bs},
we see that the core ion temperature is significantly decreased by retaining kinetic electron effects.}
\label{fig:tgrads}
\end{figure}

\section{Summary}

In this chapter, we have detailed a numerical framework for efficiently simulating turbulent transport
and heating in magnetic confinement fusion devices.  In \secref{sec:flxtube} we introduced the
local approximation, which allows for the use of a turbulence simulation domain consisting of a thin tube 
encompassing a single magnetic field line.  Each flux
tube is used to map out an entire flux surface, constituting a significant saving in simulation volume.  
These flux surfaces are then used as radial grid points
in a coarse spatial grid when solving the equilibrium evolution equations (\ref{eqn:nevo}) and 
(\ref{eqn:pevo2}) derived in Chapter 3.
The steady-state turbulent fluxes and heating calculated in each of these flux tube simulations are then 
time-averaged, representing a single step in a coarse, equilibrium timescale grid.  

In Secs.~\ref{sec:teqnorm} and~\ref{sec:discret}, we normalized and discretized the equilibrium evolution 
equations.  An important consideration in our time discretization was the stiffness of the equations, which
led us to develop a fully implicit scheme.  This was accomplished using Newton's method, in which we expanded
the nonlinear terms about their values at the previous timestep and kept only terms through linear order.
As a result, we are forced to evaluate derivatives of the averaged turbulent fluxes and heating with respect to 
the density and pressures at each of the radial grid locations.  Since this is computationally very expensive,
we made the approximation that the fluxes depend primarily on gradient scale lengths; the dependence on the
local density and pressures, as well as the dependence on higher order derivatives, is considered to be weak
enough so that it can be neglected in taking the flux derivatives.

In \secref{sec:tavg}, we detailed the method by which we obtain numerical time averages of the
turbulent fluxes and heating.  By comparing the running time average to the history of accumulated 
time averages, we defined a criterion that is used to determine when the time averaged turbulent fluxes and 
heating have converged to their steady state value.  Once they have converged, the flux tube calculation
is terminated and the time averaged fluxes and heating are passed to the transport solver.

We described our simple quasilinear flux model in \secref{sec:qlflx} and our external heating source
in \secref{sec:source}.  Finally, we presented the results of \verb#Trinity# simulations in \secref{sec:trinsim}.
These results included simple tests showing that the implicit, multi-channel transport solver employed
in \verb#Trinity# is well behaved and computationally efficient.  Additionally, we presented preliminary
results from full-volume \verb#Trinity# simulations of the entire discharge of ITER-like plasmas.  These
simulations, which calculated steady state equilibrium profiles and corresponding gyrokinetic, turbulent 
fluxes, constitute the first such simulations ever conducted.  Each simulation took less than an hour on
no more than $4096$ processors, making it possible to routinely run such simulations in the future.


\renewcommand{\thechapter}{8}

\chapter{\textbf{Summary and discussion}}
\label{chap:sum}
\vspace{+90pt}

In this thesis, we have presented a complete theoretical (Chapter 3) and numerical (Chapter 7)
prescription for studies of the self-consistent interaction between turbulence and equilibrium 
profiles.  In order to 
make such numerical studies feasible and to ensure that the relevant physics processes are 
accurately modeled, we developed and implemented velocity space resolution diagnostics 
(Chapter 4) and a model physical collision operator for gyrokinetics (Chapters 5 and 6).
Combining all of these elements, we have developed a new code, \verb#Trinity#, with 
which we have produced the first ever nonlinear, gyrokinetic simulations of coupled 
turbulence, transport, and heating over a full fusion device volume and discharge time 
(Chapter 7).

Thus far, the physical systems we have considered have been somewhat simplified.
However, the capability currently exists to do more physically realistic simulations, including
multiple species, electromagnetic effects, electron scales, and general geometry.
Consequently, \verb#Trinity# can immediately be used to explore a variety of interesting,
experimentally relevant problems.  It can be used to conduct both qualitative and quantitative 
studies of possible novel effects of the turbulence-equilibrium interaction, such as the 
formation of internal transport barriers and the effect of turbulent heating on the electron-ion 
temperature ratio.  The fact that \verb#Trinity# simulation runtimes are relatively short
also allows us to routinely carry out parameter scans to study things such as: shaping effects; scaling of 
performance with device size, aspect ratio, and magnetic field strength; and dependence
of core temperature profiles on edge temperature.

There are still numerous improvements which could be made to the numerical algorithms
employed by \verb#Trinity#.  Consideration should be given to how one can quickly
determine whether or not a given nonlinear simulation is stable (below the critical gradient
threshold) so that runtime is not wasted simulating decaying turbulence.  Runtime could
also be saved by developing a scheme to minimize the number of, or utilize use the resources 
from, idle processors (which appear because some flux tube calculations converge faster than
others).  A preconditioner
for the profiles calculated using quasilinear or gyrofluid estimates for the fluxes could be
employed to provide initial profiles to \verb#Trinity# that will quickly converge to
steady state.  One could explore whether multiple iterations of the Newton solver (instead
of the single iteration method developed in Chapter 7) allows for the use of larger transport
time steps and more rapid convergence to steady state.  This could also lead to the development
of an adaptive transport time step.  The spatial stencil used for finite differences could be
widened at essentially no additional computational cost.  Finally, it may be possible
to develop a scheme for evaluating terms such as $\partial Q / \partial(R/L_{p})$
that does not require additional nonlinear flux calculations (through the use of 
quasilinear flux estimates or something similar).

Possible future directions for improvement to the physics model in \verb#Trinity# include:
treatment of the large scale radial electric field, equilibrium flows, and momentum transport; 
inclusion of additional plasma parameter dependencies (such as the electron-ion temperature 
ratio) when approximating the fluxes at the new time step in our transport solver algorithm; 
a more sophisticated (numerical) calculation of neoclassical transport and heating effects; 
treatment of fast particles; more realistic particle, momentum, and heat sources; and 
treatment of the slow evolution of magnetic flux surfaces.  Equilibrium shear flows
associated with the large scale radial electric field profile are believed to play a critical
role in the reduction of turbulence and formation of the edge pedestal.  The development
and implementation of momentum transport equations would thus allow for quantitative studies
of transport barrier formation.  Inclusion of electron-ion temperature ratio dependence
in our transport solver would allow us to calculate heating in the homogeneous plasmas
present in astrophysical systems.  In particular, it would allow us to determine the ratio
of the turbulent energy deposited in ions to the turbulent energy deposited in electrons,
giving the steady state electron-ion temperature ratio.

By itself, the first-principles turbulent transport code
presented here is not sufficient to provide self-contained, comprehensive predictions for the 
performance of fusion devices such as ITER.  There are a number of critical physics 
phenomena not currently present in our model, such as equilibrium flows, edge physics, and MHD 
processes.  However, it can provide first-principles predictions for core profile evolution over a wide range 
of experimental configurations and plasma parameter sets.  Furthermore, a code such
as \verb#Trinity# is a necessary component in full-physics, predictive simulations
of tokamak discharges.  Full-physics, predictive simulations are a critical component for the 
fusion program as we develop ITER and look beyond to the next generation of fusion devices.

\appendix
\renewcommand{\thechapter}{A}

\chapter{Geometry}
\label{app:geo}

\section{General geometry}

Our development closely follows that of Ref.~\cite{beerPoP95}.
Since the divergence of the magnetic field is zero, one may
use a Clebsch formulation~\cite{kruskalPoF58}:
\beq
\label{cB}
{\bf B} = \grad \alpha \times \grad \psi.
\eeq
To represent an equilibrium magnetic field composed of closed surfaces, it is 
sufficient to define~\cite{kruskalPoF58}:
$
\alpha = \phi - q(\psi) \theta - \nu(\psi, \theta, \phi) \qquad {\rm and}
\qquad \psi = \Psi.
$
Here, $\theta$ and $\phi$ are the physical poloidal and toroidal angles,
respectively, $\Psi = (2 \pi)^{-2} \int_V d\!\tau {\bf B}\cdot \grad
\theta$ is the poloidal flux, $q(\Psi) = d \Psi_T/d \Psi$, $\Psi_T = (2
\pi)^{-2} \int_V d\!\tau {\bf B}\cdot \grad \phi$ is the toroidal flux, and
$d\tau$ is the volume element.  The quantity $\nu$ should be periodic in
$\theta$ and $\phi$. 

It is convenient to define a new angle $\zeta = \phi - \nu$.  With these
definitions, Eq.~\ref{cB} becomes 
$$
{\bf B_0} = \grad \Psi \times \grad (q \theta - \zeta),
$$
where the subscript on ${\bf B}$ is included to emphasize that we are
concerned with the equilibrium, unperturbed magnetic field.  The field
lines are straight in the $(\zeta, \theta)$ plane, and are labeled by
$\alpha$.  Useful coordinates are therefore $(\rho, \alpha, \theta)$, where
$\rho(\Psi)$ determines the flux surface, $\alpha$ chooses a field line in
that surface, and $\theta$ measures the distance along that field line.

In an axisymmetric system, one may also represent the magnetic field as
\beq
\label{axiB}
{\bf B_0} = I(\Psi) \grad \phi + \grad \Psi \times \grad \phi, 
\eeq 
where $I(\Psi) = R B_T$.  We will find it useful to take advantage of this
representation, although not necessary.  

In the ballooning or field-line following limit, we assume that the
perturbed quantities vary as  
$$
A = \hat{A}(\theta) \exp{(i S)}
$$
where $\bhat \cdot \grad S = 0$.  This takes into account the fact that
the perturbations tend to be slowly varying along the field line, and
allows for rapid variation across the field line~\cite{antonPoF80}.

The latter condition implies
$$
(\grad \alpha \times \grad \Psi) \cdot \grad S = 0
$$
which, in turn, implies $S = S(\alpha, \Psi)$.  To make contact with the
ballooning approximation and with field-line following coordinates, one may
choose $S = n_0 \left(\alpha + q \theta_0 \right)$, where $n_0$ is some (large)
integer, and $\theta_0$ is the familiar ballooning parameter which, in
field-line-following coordinates, determines $k_x$ through the relation
$k_x = - k_\theta \hat{s} \theta_0$.  Here, $\hat{s} = \rho / q (dq/d\rho)$,
and $\rho$ is an arbitrary flux surface label.  

\section{Operators and arguments}

In general, we wish to simulate the nonlinear electromagnetic gyrokinetic
equation in the ballooning, or field-line-following, limit.  We choose a
field-line-following representation~\cite{beerPoP95}, which has the advantage that
the nonlinear terms are easy to evaluate and are independent of the details
of the magnetic geometry.  Further details may found in
Ref.~(\cite{beerPoP95}).  Below, we focus on the linear terms, which may be
affected by the geometry.

Effects of the magnetic geometry in this limit enter through only a small
number of terms, regardless of whether one proceeds with a moment-based
approach~\cite{beerPoP95}, a $\delta \!f$ approach~\cite{dimitsJCP93,dentonJCP95}, or a
gyrokinetic approach~\cite{rewoldtPoF82,kotschCPC95}.  Consider, for example,
Eqs.~(23--24) of Ref.~\cite{antonPoF80}:
\beq
\label{Ant_1}
\hat{g} = \hat{h} - \frac{1}{B_0} \pd{F_0}{\mu} \left[ J_0 \left(\frac{\vpe 
|\grad S|}{\Omega}\right) \left(q \hat{\phi} - \frac{\vpa }{ c} q
\hat{\psi} \right) + q \hat{\sigma} \frac{\vpe |\grad S| }{ c} J_1\left(
\frac{\vpe |\grad S| }{ \Omega} \right) \right], 
\eeq
and
$$
-i\left( \omega - \omega_d + i \vpa \bhat \cdot \grad \right) \hat{h} =
\int_{-\pi}^\pi \frac{d\xi }{ 2 \pi} \exp{(-iL)} {\rm
st}\left(\hat{f}_0\right) 
$$
$$
+ i \omega \left( \pd{F_0}{\epsilon} - \frac{\bf{
B_0} \times \grad S \cdot \grad F_0 }{ B_0 m \Omega \omega} \right)
$$
\beq
\label{Ant_2}
\left[ J_0\left(\frac{\vpe |\grad S| }{ \Omega}\right) \left(q \hat{\phi} -
\frac{\vpa }{ c} q \hat{\psi} \right) + q \hat{\sigma} \frac{ |\grad S| \vpe
}{ c} J_1\left(\frac{\vpe |\grad S| }{ \Omega}\right) \right].  
\eeq
Here, $\omega_d \equiv \grad S \cdot {\bf B_0} \times \left(m \vpa^2
\bhat \cdot \grad \bhat + \mu \grad B_0 + q \grad \Phi_0\right)/(m B_0
\Omega)$.  The notation is explained in Ref.~\cite{antonPoF80}.  Note that
the unperturbed magnetic field $B_0 = B_0(\theta)$.    

These equations, together with Maxwell's equations, describe the linear
properties of a wide range of microinstabilities.  In the limit of large
toroidal mode number $n_0$, only the following components of these equations
depend on $\theta$: $\bhat \cdot \grad$, $|\grad S|^2$, ${\bf B_0} \times
\left(\bhat \cdot \grad \bhat\right) \cdot \grad S$, $\left({\bf B_0} \times
\grad B_0 \right) \cdot \grad S$, and $B_0(\theta)$.  To perform volume
integrations and flux surface averages in the nonlinear simulations, it is
also necessary to have the Jacobian $J$ and $|\grad \rho|$ as functions of
$\theta$.  We now consider the terms individually.

To make our normalizations clear, we treat the $\omega_*$ term in
detail. The $\omega_*$ term may be written as 
$$ -i \frac{{\bf B_0} \times \grad S \cdot \grad F_0 }{ B_0 m \Omega} q
 \hat{\chi} = -i n_0 \frac{c }{ B_0 } \hat{\chi} \left[ \bhat \times
 \grad \left(\alpha + q \theta_0\right) \cdot \grad F_0 \right]$$
 where $$\hat{\chi} = \left(\hat{\phi} - \frac{\vpa }{ c}
 \hat{\psi}\right) J_0 + \frac{\hat{\sigma} |\grad S| \vpe }{ c} J_1.$$
 This, in turn, is
$$
-i n_0 \frac{c }{ B_0 } \hat{\chi} \left[ \bhat \times \grad \left(\alpha + q
\theta_0 \right) \cdot \grad F_0 \right]
= -i n_0 \frac{c }{ B_0} \hat{\chi} \left( \bhat \cdot \grad \alpha \times
\grad \Psi \right) \pd{F_0}{\Psi} = -i n_0 c \hat{\chi} \pd{F_0}{\Psi},
$$
where we have assumed that $F_0 = F_0(\Psi)$.  

We now introduce normalizing quantities.  Lengths are normalized to
$a$, which we choose to be half the diameter of the last closed flux
surface (LCFS), measured at the elevation of the magnetic axis.  The
magnetic field is normalized to the toroidal field on the flux surface
at $R_a$, ($B_a = I(\Psi)/R_a$) where $R_a$ is the average of the
minimum and maximum of $R$ on the flux surface and $I(\psi)$ is as
used in Eq.~(\ref{axiB}). Time is normalized to $a/v_t$, where $v_t =
\sqrt{T/m_i}$.  Thus, for example, $\grad = (1/a) \grad_N$ and $\Psi =
a^2 B_a \Psi_N$.  Perturbed quantities are scaled up by $a/\rho_{ia}$,
where $\rho_{ia} = v_t/\Omega_a$ and $\Omega_a = |e| B_a / (m_i c)$.
The perturbed field is normalized by $T_i/|e|$, so that, for example,
$\hat{\chi}_N = (|e| \hat{\chi} /T_i) (a / \rho_{ia})$.  [Here, we
consider only the one-species problem.  The generalization to multiple
species is straightforward.]  Finally, we introduce an arbitrary flux
surface label $\rho$, normalized so that $\rho = 0$ at the magnetic
axis and $\rho = 1$ at the LCFS.  Note that the Larmor radius $\rho_i$
should not be confused with the flux surface label $\rho$.  Upon
adopting these normalizations, one finds
$$
-i n_0 c \hat{\chi} \pd{F_0}{\Psi} 
= - i \frac{n_0 }{ a^2} \frac{c T}{ e B_a} \frac{\rho_{ia} }{ a} \hat{\chi}_N
\pd{F_0}{\rho} \frac{d\rho }{ d\Psi_N} = -i k_\theta \rho_{ia} \frac{\rho_{ia}
v_t }{ a^2} \pd{F_0}{\rho} \hat{\chi}_N
$$
which serves to define $k_\theta \equiv (n_0/a) d\rho/d\Psi_N$.  In the
high aspect ratio, zero $\beta$, circular flux surface limit, $k_\theta = n_0
q / r$.  For the case in which there is a background density gradient, one
finds  
$$
 -i k_\theta \rho_{ia} \frac{\rho_{ia}
v_t }{ a^2} \pd{F_0}{\rho}  \hat{\chi}_N
= i (k_\theta \rho_{ia}) \hat{\chi}_N
\frac{ F_0 }{ (L_n)_N} \frac{\rho_{ia} v_t }{ a^2} = i (k_\theta \rho_{ia}) \hat{\chi}_N
F_0 \frac{ a }{ L_n} \left(\frac{\rho_{ia} v_t }{ a^2}\right)
$$
in which the dimensionless quantity $(L_n)_N^{-1} = -(1/n) dn/d\rho$, and may
also be written as $L_n/a$.  With the specified normalizations for time, space, and
perturbed quantities, the factor ${\rho_{ia} v_t / a^2}$ scales out of the
gyrokinetic equation.  Compare, for example, the $\omega_*$ term
with the first term in Eq.~(\ref{Ant_2}),
$$
i \omega \hat{h} = i \omega_N \hat{h}_N  \left( \frac{\rho_{ia} v_t }{ a^2}
\right).
$$
The factor in parentheses is common to all terms in the equation, and does
not appear in any other form.  It may therefore be considered to be
arbitrary.

In the $\omega_*$ term, note that $k_\theta$ is multiplied by $\rho_{ia}$,
confirming that it is natural to consider perpendicular gradients normalized
by the gyroradius $\rho_{ia}$ rather than to the minor radius $a$, as
expected in the ballooning or field-line-following limit.  

To summarize, upon adopting the above normalizations, the $\omega_*$ term
in Eq.~(\ref{Ant_2}) in field-line-following coordinates becomes 
\beq
\label{omega_*}
-i \frac{{\bf B_0} \times \grad S \cdot \grad F_0 }{ B_0 m \Omega} q
\hat{\chi} = i \omega_{*N} \hat{\chi}_N F_0 \left( \frac{\rho_{ia} v_t
  }{ a^2} \right) = - i k_{\theta} \rho_{ia} \frac{1 }{ F_0} \frac{dF_0
  }{ d\rho} \hat{\chi}_N F_0 \left( \frac{\rho_{ia} v_t }{ a^2}
\right) \eeq Note that $\omega_{*N} = - k_\theta \rho_{ia} (1/F_0)
(dF_0/d\rho)$ is dimensionless, independent of $\theta$, and related
to the dimensional $\omega_*$ by $\omega_* = \omega_{*N} v_t/a$.  

We now turn to the $\bhat \cdot \grad$ operator.  We begin by using the
{\bf B} field in the form of Eq.~(\ref{cB}) to find $\alpha$:
$$
{\bf B} \cdot \grad \phi = 
\grad \theta \times \grad \Psi \cdot \grad \phi \pd{\alpha}{\theta}
$$
which implies
\beq
\label{alpha}
\alpha = \int_0^\theta d\theta \frac{{\bf B_0} \cdot \grad \phi }{ \grad
\theta \times \grad \Psi \cdot \grad \phi}.
\eeq
For an axisymmetric {\bf B} field, this integral may be evaluated with the
use of Eq.~(\ref{axiB}).  In this case, the $\bhat \cdot \grad$ operator
may be explicitly evaluated.  It is
$$
\bhat \cdot \grad \hat{h}(\theta) = \frac{{\bf B_0} \cdot \grad \theta }{
B_0} \pd{\hat{h}}{\theta} = - \frac{I_N }{ a B_N }
\left(\pd{\alpha}{\theta} \right)^{-1} |\grad_N \phi|^2
\pd{\hat{h}}{\theta}, 
$$
which serves to define 
\beq
\label{gradpar}
\left(\bhat \cdot \grad\right)_N = - \frac{I_N }{ B_N }
\left(\pd{\alpha}{\theta} \right)^{-1} |\grad_N \phi|^2.
\eeq
In the high aspect ratio, zero $\beta$, circular flux surface limit, $\left(\bhat
\cdot \grad\right)_N = a / q R_0$, where $R_0$ is the major radius at the
center of the flux surface. 

Next, we consider the $\grad B$ part of the $\omega_d$ operator.  This term
is given by 
$$
\frac{\vpe^2 }{ 2} \frac{ \hat{h} }{ \Omega B_0^2} {\bf B_0} \times \grad B_0 \cdot \grad S =
\left(\frac{\rho_{ia} v_t }{ a^2}\right) \left(\frac{k_\theta \rho_{ia} }{ 2}\right) 
\frac{v_{\perp N}^2 }{ 2} \hat{h}_N \left[\frac{2 }{ B_N^2}  \frac{d \Psi_N }{ d \rho} 
\bhat \times \grad_N B_N \cdot \grad_N \left(\alpha + q \theta_0\right) \right].
$$
The module released here produces the factors in square brackets, {\it i.e.},
\beq
\label{gradb}
\omega_{\grad B} = \frac{2 }{ B_N^2} \frac{d \Psi_N }{ d \rho} \bhat \times
\grad_N B_N \cdot \grad_N \alpha 
\quad {\rm and} \quad 
\omega_{\grad B}^{(0)} = \frac{2 }{ B_N^2} \frac{d \Psi_N }{ d \rho} \bhat
\times \grad_N B_N \cdot \grad_N q. 
\eeq
In the high aspect ratio, zero $\beta$, circular flux surface limit,
$\omega_{\grad B} = 2 a/R_0 \left(\cos{\theta} + \hat{s} \theta
\sin{\theta}\right)$, and  $\omega_{\grad B}^{(0)} = -2 \left(a/R_0\right) \hat{s}
\sin{\theta}$.  

The curvature drift is nearly the same as the $\grad B$ drift, except that
$\vpe^2 \rightarrow 2 \vpa^2$, and the fact that there is an additional
component of the curvature drift given by
\beq
\label{gradP}
\vpa^2 \frac{4 \pi \hat{h} }{ \Omega B_0^2} \bhat \times \grad p \cdot \grad S =
\left(\frac{\rho_{ia} v_t }{ a^2}\right) \left(\frac{k_\theta \rho_{ia} }{ 2}\right) 
\hat{h}_N v_{\parl N}^2  \left[ \frac{1 }{ B_N^3} \frac{d \Psi_N }{ d\rho}  
\bhat \times \grad_N \beta_a \cdot \grad_N (\alpha +q \theta_0) \right].
\eeq
The module released here produces the factors in square brackets, {\it i.e.}, 
\beq
\label{curvature}
\omega_{\bf \kappa}       = \omega_{\grad B} + \frac{1 }{ B_N^3} \frac{d \Psi_N }{ d \rho} 
\bhat \times \grad_N \beta_a \cdot \grad_N \alpha, 
\quad 
\omega_{\bf \kappa}^{(0)} = \omega_{\grad B}^{(0)} 
\eeq
Here, $\beta_a = 8 \pi p / B_a^2$.  Note that a perpendicular gradient 
of $\beta_a$ gets no contribution from the gradient of the magnetic
field, since $B_a$ is a constant.

We do not explicitly consider the remaining component of $\omega_d$,
proportional to $\grad \Phi_0$.  To the extent that the electrostatic
potential is constant on a flux surface, this term may be specified using
the information provided by the module.

To summarize, in field-line-following coordinates, the term in
Eq.~(\ref{Ant_2}) that is proportional to $\omega_d$ is given by
$$
i \omega_d \hat{h} = i \left(\frac{k_\theta \rho_{ia} }{ 2}\right) \hat{h}_N 
\left(\frac{\rho_{ia} v_t }{ a^2}\right) \left[
\frac{v_{\perp N}^2 }{ 2} 
\left(\omega_{\grad B} + \theta_0 \, \omega_{\grad B}^{(0)}\right)
+ v_{\parl N}^2 \left( \omega_{\bf \kappa} + \theta_0 \, \omega_{\bf \kappa}^{(0)}\right)
\right]
$$ 

The form of Eqs.~(\ref{Ant_1}--\ref{Ant_2}) and of the gyrokinetic
Maxwell's equations~\cite{antonPoF80} (not shown) guarantees that $|\grad S|$
always appears as the square, $|\grad S|^2$.  In general geometry, this
term may be written 
$$
|\grad S|^2 = \frac{n_0^2 }{ a^2}  
\left|\grad_N \left(\alpha + q \theta_0 \right) \right|^2 =
k_\theta^2 \left(\frac{d \Psi_N }{ d \rho}\right)^2 
\left|\left(\grad_N \alpha + \theta_0 \grad_N q \right)
\cdot
\left(\grad_N \alpha + \theta_0 \grad_N q \right)
\right|.
$$
The module released here produces the factors $(g_1, g_2, g_3)$, where 
\beq
\label{gds2}
|\grad S|^2 = k_\theta^2 \left|
g_1 + 2 \theta_0 g_2 + \theta_0^2 g_3 \right| =
k_\theta^2 \left(\frac{d \Psi_N }{ d \rho}\right)^2 
\left| \grad_N \alpha \cdot \grad_N \alpha 
+ 2 \theta_0 \grad_N \alpha \cdot \grad_N q
+ \theta_0^2 \grad_N q \cdot \grad_N q \right|. 
\eeq
In the high aspect ratio, zero $\beta$, circular flux surface limit, $g_1 = 1
+ \hat{s}^2 \theta^2$, $g_2 = -\theta \hat{s}$, and $g_3 = \hat{s}^2$.
Note that $|\grad S|^2$ typically appears with a factor of $1/\Omega^2$,
which is not included in Eq.~(\ref{gds2}). 

The remaining quantities are straightforward.  The variation of the
unperturbed magnetic field along the field line is reported by the module as
$B_N$, with 
\beq
\label{bmag}
B_N(\theta) = B_0(\theta)/B_a.
\eeq
The quantity $|\grad_N \rho|$ is also reported by the module, and is unity in
the high aspect ratio, zero $\beta$, circular flux surface limit.

For numerical applications, it is sometimes necessary to choose the field line
coordinate so that $(\bhat \cdot \grad)_N$ is constant.  This choice allows
the straightforward evaluation of terms proportional to $|k_\parl|$ in the
transform space.  Thus, we use $(\rho, \alpha, \theta')$ coordinates, where
$\theta'$ is the equal arc periodic coordinate defined by 
\beq
\label{theta'}
\theta'(\theta) = 2 \pi L_N(\theta) / L_N(\pi) - \pi
\eeq
between $-\pi$ and $\pi$, and $L_N(\theta) = {\int_{-\pi}^\theta d\theta \, 
\left( \bhat \cdot \grad \right)_N^{-1}}$. 
In this coordinate system, the coefficient of the parallel gradient
operator of Eq.~(\ref{gradpar}) becomes  
\beq
\label{gradpar'}
\left(\bhat \cdot \grad\right)'_N =  2 \pi / L_N(\pi).
\eeq
The Jacobian is $J_N = \left(d \Psi_N /d \rho \right) \left(L_N /2 \pi
B_N\right).$  With these definitions, the flux surface average of a
quantity $\Gamma$ is defined to be 
$$
\langle \Gamma \rangle = \frac{\int \Gamma J_N \, d\theta' \, d\alpha }{ 
\int J_N \, d\theta' \, d\alpha}.
$$
The normalized area of the flux surface is $A_N = 2 \pi \langle \, |\grad_N
\rho | \, \rangle \int J d\theta'$.   

The field-line variation of the quantities 
$\omega_{\grad B}$  and $\omega_{\grad B}^{(0)}$ [Eq.~(\ref{gradb})], 
$\omega_\kappa$ and $\omega_\kappa^{(0)}$, [Eq.~(\ref{curvature})], 
$(g_1, g_2, g_3)$ [Eq.~(\ref{gds2})],
$B_N(\theta')$ [Eq.~(\ref{bmag})], and 
$\left(\bhat \cdot \grad\right)_N$ [Eq.~(\ref{gradpar'})], 
together with the quantities 
$|\grad_N \rho|$, 
$d\rho/d\Psi_N$ and 
$d\beta /d\rho$ 
are the outputs of this geometry module.   These
coefficients contain all of the geometric information necessary for
numerical calculations of high-$n$ microstability and turbulence in
axisymmetric toroidal configurations with nested magnetic surfaces.  

\section{Module details}
\label{module}

Input numerical equilibria may be specified in numerous ways, as documented
in the module.  Interfaces to direct and inverse Grad-Shafranov equilibrium
solvers are available.  These include using output from several equilibrium
codes in use in the fusion community, such as TOQ~\cite{millerNF87,zeeuwJCAM90},
EFIT~\cite{laoNF90}, VMOMS~\cite{laoPoF81}, JSOLVER~\cite{deluciaJCP80}, and
CHEASE~\cite{luetjensCPC96}, as well as the local equilibrium model of Ref.~\cite{millerPoP98}.

Here, we describe our implementation of the Miller local equilibrium
model~\cite{millerPoP98} for completeness.  This model extends the usual
zero-beta, high-aspect ratio equilibrium to arbitrary aspect ratio, cross
section and beta, and allows one to consider geometric effects on
microinstabilities in a controlled way.  

The shape of the reference flux surface and its perpendicular derivative 
are specified in the $(R,Z)$ plane by 
\beq
\label{R}
R_N(\theta) = R_{0N}(\rho) + \rho \cos{\left[\theta + \delta(\rho) \sin{\theta}\right]},
\eeq
\beq
\label{Z}
Z_N(\theta) = \kappa(\rho) \rho \sin{\theta}.
\eeq
Here, $R_N = R / a$, {\it etc.,} $R_{0N}(\rho) = R_{0N}(\rho_f) + R_{0N}' \,  d\rho$, 
$\delta(\rho) = \delta(\rho_f) + \delta' \, d\rho$, 
$\kappa(\rho) = \kappa(\rho_f) + \kappa' \, d\rho$, 
and $\rho_f$ denotes the flux surface of interest.  In the remainder of
this section, the ``N'' subscripts will be dropped, since no ambiguities
will arise.  

As emphasized in Ref.~\cite{millerPoP98}, the actual shape of
neighboring flux surfaces $(\rho \neq \rho_f)$ is not determined by
Eqs.~(\ref{R}) and (\ref{Z}).  Instead, this is determined by solving the
Grad-Shafranov equation in the neighborhood of $\rho_f$.  As noted by
Mercier and Luc~\cite{mercier}, one may find this solution provided
$R(\theta)$, $Z(\theta)$, $B_p(\theta)$, $p'(\rho_f)$, and
$I'(\rho_f)$.  One additional piece of information is required to determine
either the safety factor $q$ or $d\Psi/d\rho$.  Finally, the normalization
of the magnetic field determines $I(\rho_f)$.

In our implementation, we take $q$ to be an input parameter, and upon
noting that $\oint \alpha \, d\theta = - 2 \pi q$, use it to
define $d\Psi/d\rho$ from Eq.~(\ref{alpha}): 
\beq
\label{dpsidrho}
\frac{d\Psi }{ d \rho} = \frac{I }{ 2 \pi q}  \oint \frac{d\theta  }{ R^2} 
\left(\grad \theta \times \grad \rho \cdot \grad \phi\right)^{-1}. 
\eeq
[For numerical equilibria, $d\Psi/d\rho$ may be calculated directly, and
this expression defines the safety factor.]  The poloidal magnetic
field $B_p(\rho_f)$ is specified by 
$$
B_p = \frac{d\Psi }{ d\rho} \frac{|\grad \rho| }{ R},
$$
where $|\grad \rho|$ may be found from Eqs.~(\ref{R}) and (\ref{Z}).  

The remaining steps may be used with the Miller local equilibrium model or with
full numerical equilibria.  We allow arbitrary values of $dp/d\rho$ and
$\hat{s}$ by using the mercier expressions to find $\grad
S$~\cite{millerPoP98,mercier,bishopNF84}.  As noted in
Ref.~(\cite{millerPoP98}), the result is exactly equivalent to the
generalized $s - \alpha$ analysis of Greene and Chance~\cite{greeneNF81}.
To proceed, we define  
$$ 
A(\theta) = \int \frac{d\theta }{ \grad \theta \times \grad
\Psi \cdot \grad \phi} 
\left[\frac{1 }{ R^2} + \left(\frac{I }{ B_p R^2}\right)^2 \right], 
\qquad
B(\theta) = I \int \frac{d\theta }{ \grad \theta \times \grad
\Psi \cdot \grad \phi} \left[ \frac{1 }{ \left(B_p R\right)^{2}}\right], 
$$
$$
C(\theta) = I \int \frac{d\theta }{ \grad \theta \times \grad
\Psi \cdot \grad \phi} \left[ \frac{\sin{u} + R/R_c }{ B_p R^4}
\right],
$$
where $u(\theta)$ is the angle between the horizontal and the tangent to
the magnetic surface in the poloidal plane, and $R_c$ is the local radius
of curvature of the surface in the poloidal plane.  If we define $\bar{A} =
\oint \cdots$, {\it etc.,} it can be shown that 
\beq
\label{shat}
\hat{s} = \frac{\rho }{ q} \frac{d q }{ d \rho} = 
\frac{\rho }{ 2 \pi q} \frac{d\Psi }{ d \rho} \left(
\bar{A} I' + \bar{B} p' + 2 \bar{C}
\right)
\eeq
where the primes denote derivatives with respect to $\Psi$.  Thus, one may
specify any two of $p'$, $I'$, and $\hat{s}$.  This freedom is a direct
consequence of the two free functions in the Grad-Shafranov equation. 

It can also be shown~\cite{bishopNF84,millerPoP98} that the perpendicular component
of the gradient of $\alpha$ is given by
$$
{\grad \alpha \cdot \hat{e}_\Psi} = |\grad \Psi| \left(A I' + B p' + 2
C\right). 
$$
The parallel component of the gradient of $\alpha$ may be easily found from
Eq.~(\ref{alpha}).  With $\grad \alpha$ in hand, the remainder of the
calculation is straightforward.  We note that $\grad B$ may also be
calculated using the mercier formulas; our treatment is the same as can
be found in Refs.~\cite{millerPoP98} and~\cite{bishopNF84}.  To
wit, the perpendicular component is 
$$
{\grad B \cdot \hat{e}_\Psi} = \frac{B_p }{ B_0} \left(\frac{B_p }{ R_c} + R p'
- \frac{I^2 \sin{u} }{ R^3 B_p}\right), 
$$
and since $B(\rho_f)$ does not depend on $p'$ or $I'$, the component of
$\grad B$ along the field line depends on neither quantity.  

The expressions for $\hat{s}$ and the gradients of $\alpha$ and $B$ make it
clear that once the safety factor, the shape of the flux surface, and $B_p$
are determined (either from a numerical equilibrium or from the local
equilibrium), one may vary $p'$ and $\hat{s}$ independently to find a
family of solutions, all of which satisfy the Grad-Shafranov equation.
This flexibility allows one to carry out the Greene-Chance kind of analysis
for microinstabilities.  Such an analysis simplifies the interpretation of
the numerical calculations, since all other parameters can easily be held
fixed.

Within the context of the local equilibrium model~\cite{millerPoP98}, one may
also vary individual shape parameters one at a time, to explore the
dependences in a controlled fashion.

The eleven dimensionless parameters that determine the local MHD
equiliibrium in this implementation of the Miller model are summarized in
Table I.

\begin{center}
\begin{tabular}{||l|c||}\hline
*Minor radius & $\rho_f$ \\ \hline 
*Safety factor & $q$  \\ \hline
Magnetic shear & $\hat{s} = (\rho/q) dq/d\rho$ \\ \hline
Elongation & $\kappa$ \\ \hline
$d\kappa/d\rho$ & $\kappa'$ \\ \hline
Triangularity & $\delta$ \\ \hline
$d\delta/d\rho$ & $\delta'$ \\ \hline
Center of LCFS & $R_{\rm geo N}$ \\ \hline
Center of flux surface  & $R_{0N}$ \\ \hline
$dR_0/d\rho$    & $R_{0N}'$ \\ \hline
*$d\beta /d\rho$ & $\beta'$ \\ \hline
\end{tabular}
\end{center}

In addition to these eleven parameters, there are two normalizing
dimensional parameters, $a$ and $B_a$.  In all, there are two more
parameters than are found in Ref.~\cite{millerPoP98}. We include
the additional parameters to allow a somewhat more natural correspondence
between {\it reported} equilibria and the input variables.  We emphasize
that there is nothing ``extra'' in our implementation of the model as
result; it is only modestly easier to use for some applications.  For
example, our choice of the normalization of the magnetic field $(B_a)$ is
the vacuum magnetic field at $R_{\rm geo}$, the center of the LCFS.  This
quantity is the most commonly reported magnetic field value.  By allowing
$R_0$ to be specified separately, we also make it conceptually easier to
separate the effects of Shafranov shift from the derivative of the
Shafranov shift.  The inclusion of the normalized minor radius as a
separate variable is a natural choice as soon as one allows for separate
specification of $R_0$ and $R_{\rm geo}$.

The starred quantities ($\rho_f$, $\hat{s}$, and $d\beta/d\rho$) may be
specified when reading in numerical equilibria.  Values of the latter two
quantities that are different from the actual equilibrium values are
incorporated by using Eq.~(\ref{shat}) to define $I'$.

Finally, when using numerically generated equilibria, the module allows one
to choose from the most common definitions of $\rho$, such as the
normalized poloidal or toroidal fluxes, the horizontal minor radius, {\it
etc.}  The user may also provide his or her own definition of $\rho$ by
supplying a simple function.

\appendix
\renewcommand{\thechapter}{B}

\chapter{Landau damping of the ion acoustic wave}
\label{app:landamp}

We consider the collisionless ion acoustic wave in slab geometry with adiabatic electrons.
The gyrokinetic equation for this system has the particularly simple form
\begin{equation}
\pd{h}{t} + v_{z}\pd{h}{z} = \frac{qF_{0}}{T}\pd{\left< \Phi \right>}{t}.
\label{eqn:iaw}
\end{equation}
Changing variables from $h$ to $g\equiv\left< f_{1} \right>$ and assuming solutions of the form
\begin{equation}
g = \tilde{g}(\mbf{v})e^{i\left(k_{\parl}z - i \omega t\right)},
\end{equation}
we obtain
\begin{equation}
\left( \omega - kv\right) g = kv \frac{e \left< \Phi \right>_{\mbf{R}}}{T_{i}} F_{M},
\end{equation}
where I am using $v=v_{\parl}$ and $k=k_{\parl}$ for convenience. 
Neglecting FLR effects and assuming quasineutrality gives
\begin{equation}
\left(\omega - kv\right)g = kv \tau \frac{F_{M}}{n_{0}}\int d^{3}v' g(v').
\label{depart}
\end{equation}
Defining
\begin{equation}
\overline{g}(v) = 2\pi \int_{0}^{\infty}v_{\perp} dv_{\perp} g(\mbf{v})
\end{equation}
and integrating over the perpendicular velocities in the gyrokinetic equation yields
\begin{equation}
\left(\omega-kv\right) \overline{g}(v) = k F(v),
\label{start}
\end{equation}
where
\begin{eqnarray}
F(v) &=& v \tau \frac{n_{1}}{\sqrt{2\pi}v_{t}}e^{-\frac{v^{2}}{2v_{t}^{2}}},\\
n_{1} &=& \int dv' \overline{g}(v').
\end{eqnarray}
Following the analysis of Case and van Kampen, we see that this equation has 
solutions of the form
\begin{eqnarray}
\overline{g}(v) &=& F(v)\left[\mathcal{P}\frac{1}{u-v}+\lambda(k,u) \delta(u-v)\right], \label{g}
\end{eqnarray}
with $u=\frac{\omega}{k}$, provided that $\lambda$ is chosen to satisfy the condition
\begin{equation}
n_{1} = \int dv' \overline{g}(v') = \mathcal{P}\int dv' \frac{F(v')}{u-v'} + \lambda(k,u) F(u).
\label{lamcon}
\end{equation}
A general solution is given in the form
\begin{equation}
\overline{f}(z,v,t) = \int_{-\infty}^{\infty} \int_{-\infty}^{\infty} C(k,u) \overline{g}_{k,u}(v) e^{ik\left(z-ut\right)}dk \ du,
\label{gensol}
\end{equation}
where $C(k,u)$ is determined by the initial condtion
\begin{equation}
\overline{f}(z,v,0) = \int \int C(k,u) \overline{g}_{k,u}(v)e^{ikz}dk \ du.
\end{equation}

Taking the inverse Fourier transform of the above expression gives
\begin{equation}
\mathcal{F}(k,v) = \int C(k,u) \overline{g}_{k,u}(v) du,
\label{ic}
\end{equation}
where
\begin{equation}
\mathcal{F}(k,v) = \frac{1}{2\pi}\int \overline{f}(z,v,0) e^{-ikz} dz.
\end{equation}
Plugging the expression (\ref{g}) for $\overline{g}$ into the initial condition (\ref{ic}) yields
\begin{equation}
\frac{\mathcal{F}(k,v)}{F(v)} = \mathcal{P}\int\frac{C(k,u)}{u-v}du + \lambda(k,v) C(v).
\label{ccon}
\end{equation}

We now have two equations, (\ref{lamcon}) and (\ref{ccon}), for two unknowns ($\lambda$ and $C$).  In order to solve
this linear system, it is convenient to define some new notation.  Any square integrable function $H$ can be written
\begin{equation}
H(q) = \int_{-\infty}^{\infty} K(p) e^{ipq} dp. 
\end{equation} 
We define the positive and negative frequency parts of $H$ as
\begin{eqnarray}
H_{\pm}(q) &=& \pm \int_{0}^{\pm\infty} K(p) e^{ipq} dp,
\end{eqnarray}
so that $H = H_{+} + H_{-}$. Further we define the function $H_{*}=H_{+}-H_{-}$.  It 
can be shown that $H_{*}$ has the alternate form
\begin{equation}
H_{*}(v) =  \mathcal{P} \frac{1}{\pi i}\int_{-\infty}^{\infty} \frac{H(v')}{v'-v}dv'.
\end{equation}
With these definitions in hand, we rewrite eqns (\ref{lamcon}) and (\ref{ccon}) as
\begin{eqnarray}
n_{1} &=& -\pi i F_{*}(u) + \lambda F(u),\\
\frac{\mathcal{F}(k,v)}{F(v)} &=& \left(\lambda + \pi i\right)C_{+}(v) + \left(\lambda - \pi i\right)C_{-}(v).
\end{eqnarray}
Eliminating $\lambda$ gives an expression involving $C_{+}$ and $C_{-}$:
\begin{equation}
\mathcal{F}(k,u) = \left(n_{1} + 2\pi i F_{+}(u)\right) C_{+}(k,u) + \left(n_{1} - 2\pi i F_{-}(u)\right)C_{-}(k,u). 
\end{equation}
The transform $\mathcal{F}$ can also be broken down into negative and positive frequency parts to give two separate equations.
\begin{eqnarray}
\mathcal{F}_{\pm}(k,u) &=& \left(n_{1} \pm 2\pi i F_{\pm}(u)\right)C_{\pm}(u)
\end{eqnarray}
These can then be used to construct $C(k,u)$:
\begin{equation}
C(k,u) = \frac{\mathcal{F}_{+}(k,u)}{n_{1}+2\pi i F_{+}(u)} + \frac{\mathcal{F}_{-}(k,u)}{n_{1}-2\pi i F_{-}(u)}.
\label{c}
\end{equation}

Substituting the expressions $(\ref{g})$ and $(\ref{c})$ for $\overline{g}$ and $C$ into the equation (\ref{gensol}) for $\overline{f}$
gives
\begin{equation}
\begin{split}
\overline{f}(z,v,t)&=\int \int \left[\frac{\mathcal{F}_{+}(k,u)}{n_{1}+2\pi i F_{+}(u)}+\frac{\mathcal{F}_{-}(k,u)}{n_{1}-2\pi i F_{-}(u)}\right]F(v)\\
&\times\left[\mathcal{P}\frac{1}{u-v}+ \lambda(k,u)\delta(u-v)\right]e^{ik\left(z-ut\right)}dk \ du.
\label{f1}
\end{split}
\end{equation}
We can use the identity
\begin{equation}
\mathcal{F}_{\pm}(k,u) = \frac{1}{2\pi}\int_{-\infty}^{\infty} e^{-ikz'} dz' \int_{-\infty}^{\infty} \delta_{\pm}(u-v') \overline{f}(z',v',0) dv'
\end{equation}
to rewrite eqn (\ref{f1}) in the more convenient form
\begin{equation}
\begin{split}
\overline{f}(z,v,t) &= \int \left[\frac{\delta_{+}(u-v')}{n_{1}+2\pi i F_{+}(u)}+\frac{\delta_{-}(u-v')}{n_{1}-2\pi i F_{-}(u)}\right]\frac{\overline{f}(z',v',0)}{2\pi}F(v)\\
&\times\left[\mathcal{P}\frac{1}{u-v} + \lambda(k,u)\delta(u-v)\right]e^{ik\left(z-z'-ut\right)}dz' dv' dk \ du.
\end{split}
\end{equation}
Now we pick an initial condition of the form
\begin{eqnarray}
\overline{f}(z,v,0) &=&\tilde{f}(v,0) e^{ik_{0}z},
\end{eqnarray}
which gives
\begin{equation}
\begin{split}
\overline{f}(z,v,t) &= e^{ik_{0}\left(z-vt\right)}\left(n_{1}+\pi i F_{*}(v)\right)\left(\frac{\tilde{f}_{+}(v,0)}{n_{1}+2\pi i F_{+}(v)}+\frac{\tilde{f}_{-}(v,0)}{n_{1}-2\pi i F_{-}(v)}\right)\\
&+ \mathcal{P}\int \frac{F(v)}{u-v}\left(\frac{\tilde{f}_{+}(u,0)}{n_{1}+2\pi i F_{+}(u)}+\frac{\tilde{f}_{-}(u,0)}{n_{1}-2\pi i F_{-}(u)}\right)e^{ik_{0}\left(z-ut\right)} du.
\end{split}
\end{equation}
\appendix
\renewcommand{\thechapter}{C}

\chapter{Proof of the $H$-Theorem for our model collision operator}
\label{app:hthm}

In the case of the expansion $f=F_0 + \delta f$ about a Maxwellian the entropy generation 
by like particle collisions takes the form
\begin{eqnarray}
\nonumber
\frac{d S}{dt} &=& - \frac{d}{dt} \iint f \ln f \,d\bm{v} \,d\bm{r}\\
&=& - \iint \hat{f} {C}[\hat{f} F_0] \,d\bm{v}\,d\bm{r}\ge0,
\label{entropy}
\end{eqnarray}
where we use the compact notation $\hat{f} = \delta f / F_0$.
The statement of the $H$-theorem is that the right-hand side of \eqref{entropy} is nonnegative 
and that it is exactly zero when $\delta f$ is a perturbed Maxwellian. 

We represent $\hat{f}$ as a Cartesian tensor expansion (or equialently spherical harmonic 
expansion) in velocity space:
\begin{eqnarray}
\hat{f}(\bm{r},\bm{v}) = \hat{f}_{0}(\bm{r},v) + \bm{v} \cdot \bm{\hat{f}}_{1}(\bm{r},v) + {R}[\hat{f}](\bm{r},\bm{v}),
\label{expansion}
\end{eqnarray}
where $R[\hat{f}]$ comprises the higher order terms. 
It is then possible to recast the statement of the $H$-Theorem in terms of this expansion 
using linearity of the model collision operator $C$ [\eqref{model}], orthogonality of the 
expansion and the fact that spherical harmonics are eigenfunctions of the Lorentz operator ${L}$. 
By construction, $R[\hat{f}]$ satisfies 
$\int R[\hat{f}]F_0 {d}\bm{v} = 0$ and 
$\int \bm{v} R[\hat{f}] F_0{d}\bm{v} = \bm{0}$, from which it follows that
$R[\hat{f}]$ does not contribute to the field-particle parts of the model operator:
$\bm{U}[ R [ \hat{f} ]F_0 ] = \bm{0}$ and $Q[ R [ \hat{f} ] F_0] = 0$. 
Substituting \eqref{expansion} into the right-hand side of \eqref{entropy}, where the 
operator $C$ is given by \eqref{model}, and integrating by parts those terms involving 
derivatives of $R[\hat{f}]$, we find that they all give nonnegative contributions, so we 
have 
\begin{eqnarray}
 - \int\hat{f} {C}[\hat{f}F_0] d\bm{v} \ge \sigma_0 + \sigma_1,
\end{eqnarray}
where 
\begin{align}
\label{sigma0_def}
\sigma_0 &= -\int \hat{f}_0 C[\hat{f}_0 F_0]d\bm{v},\\
\sigma_1 &= -\int \bm{v} \cdot \bm{\hat{f}}_1 {C}[\bm{v}\cdot\bm{\hat{f}}_1 F_0] d\bm{v}.
\label{sigma1_def}
\end{align}
In order to prove the $H$-theorem, 
it is now sufficient to show that $\sigma_0 \ge 0$ and $\sigma_1 \ge 0$. 

Starting with $\sigma_0$ and using \eqref{model}, we integrate over angles and 
use the identity 
$v^{4}\nu_E F_{0} = -(\partial/\partial v)(v^5\nu_{\parallel} F_{0})$ to express 
the term containing $Q$: 
\begin{equation}
\sigma_0 = -2\pi \int \hat{f}_0\pd{}{v}\left[ v^4 \nu_{\parallel}F_{0} \pd{}{v}
\left( \hat{f_0} - \frac{v^2}{\vth^2}\,Q[\hat{f}_0 F_0] \right) \right] dv,
\end{equation}
Using the aforementioned identity again in the expression for $Q$ [\eqref{def_Q}] and 
integrating by parts where opportune, we get
\begin{equation}
\begin{split}
\sigma_0 &= 4\pi\left[\frac{1}{2}\int\left(\pd{\hat{f}_0}{v}\right)^2 v^4\nu_\parallel F_0dv 
\right.\\
&- \left.\left.\left(\int \pd{\hat{f}_0}{v} v^5\nu_\parallel F_0 dv\right)^2
\right/ \int v^6\nu_E F_0 dv.
\right]
\end{split}
\label{sigma0}
\end{equation}
It is easy to see from the Cauchy-Schwarz inequality that
\begin{equation}
\label{sigma0_CS}
\left(\int{ \pd{\hat{f}_{0}}{v} v^5 \nu_\parallel F_{0} dv } \right)^2 \le 
\int \left(\pd{\hat{f}_{0}}{v} \right)^2 v^4 \nu_\parallel F_{0} dv 
\int v^6 \nu_\parallel F_{0} dv.
\end{equation}
Using this in the second term of \eqref{sigma0}, we infer
\begin{equation}
\begin{split}
\sigma_0 &\ge 4\pi \int \left(\pd{\hat{f}_{0}}{v} \right)^2 v^4 \nu_\parallel F_{0} dv\\
&\times\left(\frac{1}{2}-\left.\int v^6\nu_\parallel F_0 dv\right/ \int v^6\nu_E F_0 dv\right)
= 0, 
\end{split}
\label{sigma0_ineq}
\end{equation}
where to prove that the right-hand side vanishes, we  
again used the identity  $v^{4}\nu_E F_{0} = -(\partial/\partial v)(v^5\nu_{\parallel} F_{0})$ 
and integrated by parts. 
Thus, we have proved that $\sigma_0 \ge 0$. 

Turning now to $\sigma_1$ [\eqref{sigma1_def}], using \eqref{model}, and integrating 
by parts where opportune, we get
\begin{equation}
\begin{split}
\sigma_1 &= \int ( \bm{v}\cdot\bm{\hat{f}}_1 )^2 \nu_D F_{0} d\bm{v} 
- \frac{1}{2} \int \left(\pd{}{v} \bm{v}\cdot\bm{\hat{f}}_1 \right)^2 v^2 \nu_\parallel F_{0} d\bm{v}\\ 
&\quad- 3 \vth^2 \left.\left|\int \bm{x} \bm{x}\cdot\bm{\hat{f}}_1 \nu_s F_{0} d\bm{v} \right|^2 \right/ \int x^2 \nu_s F_{0} d\bm{v}, 
\end{split}
\end{equation}
where we have used the standard notation that $\bm{x} = \bm{v} / \vth$ and $x = v/\vth$. 
Integrating over angles and using the simple identity 
$\bm{a} \cdot \int \hat{\bm{v}}\hat{\bm{v}} d\Omega = (4\pi/3)\bm{a}$, 
where $\hat{\bm{v}}=\bm{v}/v$ and $\bm{a}$ is an arbitrary vector, we have
\begin{equation}
\begin{split}
\sigma_1 &= \frac{4\pi\vth^5}{3} \Biggl( \int\! |\bm{\hat{f}}_1|^2 x^4 \nu_D F_{0} dx
+ \frac{1}{2} \int \left| \pd{}{x}\,x\bm{\hat{f}}_1\right|^2\!\! x^4 \nu_\parallel F_{0} dx \Biggr. \\  
&\quad - \left.\left.\left| \int \bm{\hat{f}}_1 x^4 \nu_s F_{0} dx\right|^2\right/ \int x^4 \nu_s F_{0} dx  
\right).
\end{split}
\label{sigma1}
\end{equation}
Once again applying the Cauchy-Schwarz inequality, we find that
\begin{equation}
\left| \int \bm{\hat{f}}_1 x^4\nu_s F_{0} dx \right|^2 \le 
\int \left| \bm{\hat{f}}_1 \right |^2 x^4 \nu_s F_{0} dx
\int x^4 \nu_s F_{0} dx.
\end{equation}
Using this in the last term in \eqref{sigma1}, we get 
\begin{equation}
\begin{split}
\sigma_1   \ge  \frac{4\pi}{3}\vth^5
&\left( \int\left|\bm{\hat{f}}_1\right|^2 x^4 \Delta\nu F_{0} {d}x \right.\\
&+ \left.\frac{1}{2} \int \left|\pd{}{x}\,x\bm{\hat{f}}_1\right|^2 
x^4\nu_\parallel F_{0} dx\right),
\end{split}
\end{equation}
where $\Delta\nu = \nu_D-\nu_s$.  
Upon using the identity 
$2x^3\Delta\nu F_{0} = ({\partial/\partial x}) (x^4\nu_\parallel F_{0})$ 
in the first term of the above expression and integrating the resulting expression by parts, 
we finally obtain
\begin{eqnarray}
\sigma_1 \ge \frac{4\pi}{3}\vth^5
\int \left|\pd{\bm{\hat{f}}_1}{x} \right|^2 x^6\nu_\parallel F_{0} {d}x\ge0.
\end{eqnarray}

We now consider when these inequalities becomes equalities, i.e., 
when the right-hand side of \eqref{entropy} is zero. 
Firstly, this requires $\partial R[\hat{f}]/\partial \xi = 0$ and thus $R[\hat{f}] = 0$, 
so $\hat{f}$ contains no 2nd or higher-order spherical harmonics. 
Secondly, $\sigma_0 = 0$ if either $\hat{f}_0$ is independent of $v$ or we have equality 
in the invocation of the Cauchy-Schwarz inequality [\eqref{sigma0_CS}], which occurs 
if $\hat{f}_0 \propto v^2$. Similarly $\sigma_1 = 0$ iff $\bm{\hat{f}}_1$ is 
independent of $v$.  Thus, the right-hand side of \eqref{entropy} 
vanishes iff $\hat{f} \propto 1, \bm{v}, v^2$, i.e., $\delta f= \hat{f} F_0$ 
is a perturbed Maxwellian.   

This completes the proof of the $H$-theorem for our model operator.
\appendix
\renewcommand{\thechapter}{D}

\chapter{Gyroaveraging our model collision operator}
\label{app:gyroC}

To transform the derivatives in \eqref{model} from the original phase-space coordinates 
$(\bm{r},v,\xi,\vartheta)$ to the new coordinates $(\bm{R},v,\xi,\vartheta)$, 
we require the following formulae:
\begin{align}
\left(\pd{}{{v}}\right)_{\bm{r}} &= \left(\pd{}{{v}}\right)_{\bm{R}} 
- \frac{1}{v}\,\bm{\rho}\cdot\left(\pd{}{\bm{R}}\right)_{\bm{v}},\\
\left(\pd{}{\xi}\right)_{\bm{r}} &= \left(\pd{}{\xi}\right)_{\bm{R}} 
+ \frac{\xi}{1-\xi^2}\,\bm{\rho} \cdot \left(\pd{}{\bm{R}}\right)_{\bm{v}}, \\
\left(\pd{}{\vartheta}\right)_{\bm{r}} &= \left(\pd{}{\vartheta}\right)_{\bm{R}} 
+ \frac{\bm{v}_\perp}{\Omega}\cdot \left(\pd{}{\bm{R}}\right)_{\bm{v}},
\end{align}
where $\bm{\rho}=\unit{b}\times\bm{v}_\perp/\Omega$. 
In Fourier-transformed perpendicular guiding center variables, we can replace 
in the above formulae $(\partial/\partial\bm{R})_{\bm{v}}\to i\bm{k}$, where 
$\bm{k}\equiv\bm{k}_\perp$. It is convenient to align (without loss of generality) 
the $\vartheta=0$ axis with $\bm{k}$, so we have 
$\bm{v}_\perp\cdot\bm{k} = k_\perp v\sqrt{1-\xi^2}\cos\vartheta$ 
and $\bm{\rho}\cdot\bm{k} = -k_\perp v\sqrt{1-\xi^2}\sin\vartheta$. 
Using the above formulae, we gyroaverage the Lorentz operator in \eqref{model}:
\begin{equation}
\label{avg_L}
\left<L[\hk]\right> = {1\over2}\pd{}{\xi}(1-\xi^2)\pd{\hk}{\xi}
- \frac{v^2(1+\xi^2)}{4\Omega^2}k_\perp^2\hk,
\end{equation}
where we have used 
$\gyroR{\bm{\rho}\bm{\rho}} = \gyroR{\bm{v}_\perp\bm{v}_\perp}=(1/2)\bm{I}$. 
Note that both the terms containing $\xi$ and $\vartheta$ derivatives in the 
original operator \eqref{def_L} produce non-zero gyrodiffusive contributions 
[the second term in \eqref{avg_L}]. Another such gyrodiffusive term, equal to 
$-\nu_\parallel \left[v^2(1-\xi^2)/4\Omega\right]k_\perp^2\hk$, arises from the 
energy-diffusion part of the test-particle operator in \eqref{model}. 
Collecting these terms together and defining the thermal Larmor 
radius $\rho=\vth/\Omega$, we arrive at the gyrodiffusion term in 
\eqref{gyroav}. 

It remains to gyroaverage the field-particle terms. 
For the energy-conservation term [\eqref{def_Q}] we have
\begin{equation}
\left<e^{i\bm{k}\cdot\bm{\rho}}Q[e^{-i\bm{k}\cdot\bm{\rho}}\hk]\right> = 
\left<e^{i\bm{k}\cdot\bm{\rho}}\right>Q[e^{-i\bm{k}\cdot\bm{\rho}}\hk], 
\label{avgQ}
\end{equation}
where 
\begin{equation}
\begin{split}
&Q[e^{-i\bm{k}\cdot\bm{\rho}}\hk] = \\
&\qquad\int v^2 \nu_E \left<e^{-i\bm{k}\cdot\bm{\rho}}\right>\hk d\bm{v}
\left/ \int v^2 \left(v/\vth\right)^2 \nu_E F_{0} d\bm{v}. \right.
\end{split}
\label{Qhk_ap}
\end{equation}
Note that the $\vartheta$ integration in $Q$ only affected 
$e^{-i\bm{k}\cdot\bm{\rho}}$, hence the above expression. 
Using the standard Bessel function identity\cite{watson1966ttb}
$\int\limits_0^{2\pi} e^{i a \sin \vartheta } \,d\vartheta = 2\pi J_0(a)$,
we find $\left<{e^{-i\bm{k}\cdot\bm{\rho}}}\right> = J_0(a)$, 
where $a=k_\perp v_\perp/\Omega$.
Substituting this into \eqsref{avgQ}{Qhk_ap}, we arrive at the energy-conservation term 
in \eqref{gyroav}, where expression in the right-hand side of \eqref{Qhk_ap} is denoted 
$Q[\hk]$ [\eqref{Qhk}].  

The momentum-conserving terms are handled in an analogous way: details can be found in 
Appendix\ B of Ref.\ \cite{schekAPJ07}, where a simpler model operator was gyroaveraged.

\appendix
\renewcommand{\thechapter}{E}

\chapter{Comparison of our collision operator with previous model operators}

\label{app:ccomp}
In order to compare and contrast with previously suggested operators that do include energy 
diffusion, we first rewrite in our notation the operator derived by Catto and 
Tsang\cite{cattoPoF76} [their Eqs.~(14) and (16)],
\begin{equation}
\begin{split}
{C}_\mathrm{CT}\left[\delta f\right] &= \nu_D {L}[ \delta f ] 
+ \frac{1}{v^2} \pd{}{v}\left(\frac{1}{2}\, v^4 \nu_\parallel F_{0} \pd{}{v}\frac{\delta f}{F_0} \right)
+ \frac{2 F_0}{n_{0} \vth^2 }\,\bm{v} \cdot \int \bm{v} \nu_s\delta f d\bm{v}\\
&+ \frac{2 F_0}{3n_{0}} \left( \frac{v^2}{\vth^2}-\frac{3}{2}\right)
\int \frac{v^2}{\vth^2}\, \nu_E\delta f d\bm{v}. 
\label{CT_op}
\end{split}
\end{equation}
This operator, whilst it conserves particle number, momentum and energy, neither obeys the 
$H$-Theorem nor vanishes on a perturbed Maxwellian. 

The latter point can be demonstrated most easily by letting $\delta f = x^2F_0$, 
where $x=v/\vth$. This $\delta f$ is proportional to a perturbed Maxwellian with 
non-zero $\delta n$ and $\delta T$. We can then evaluate the test-particle and 
field-particle parts of the operator to find
\begin{equation}
\frac{1}{v^2} \pd{}{v}\left( \frac{1}{2} v^4 \nu_\parallel F_{0} \pd{}{v}\frac{\delta f}{F_0} \right) 
= \frac{1}{x^2} \pd{}{x} \left( x^5 \nu_\parallel F_{0} \right) = - x^2 \nu_E F_{0}
\end{equation}
and
\begin{equation}
\int x^2 \nu_E \delta f d\bm{v} 
= \frac{4 n_{0}}{\sqrt{\pi}} \int\limits_0^{\infty} x^6 \nu_E e^{-x^2} {d}x 
= \sqrt\frac{2}{\pi} \,n_0\nu.
\end{equation}
Substituting into \eqref{CT_op}, we get
\begin{equation}
{C}_\mathrm{CT}[\delta f] = -x^2\nu_E F_{0} + \sqrt\frac{2}{\pi} 
\left(x^2-\frac{3}{2}\right) \nu F_{0},
\end{equation}
which is non-zero despite $\delta f$ being a perturbed Maxwellian. 

In order to show that the $H$-theorem can be violated by the operator \eqref{CT_op}, 
let us consider a perturbed distribution function of the form $\delta f = x^3 F_0$. Then
\begin{equation}
C_{\mathrm{CT}}[\delta f] = \left[ \frac{3}{2} \left(\nu_\parallel -\nu_E\right) + \left(x^2 - \frac{3}{2}\right) \nu \right] F_0,
\end{equation}
so the entropy generation is,
\begin{equation}
\frac{dS}{dt} = - \iint \frac{\delta f}{F_0}\,{C}_\mathrm{CT}\left[\delta f\right] d\bm{v}d\bm{r}
=-\frac{3}{64}(32-21\sqrt{2})\,\nu V < 0, 
\end{equation}
where $V$ is the volume of the system. The above expression is negative, which breaks the 
$H$-theorem and produces unphysical plasma cooling for the particular form of the perturbed 
distribution function that we have examined. 

The second case we examine here is the sequence of operators derived by Hirshman and 
Sigmar.\cite{hirshmanPoF76} The general operator proposed by these authors is given in 
their Eq.~(25). In our notation, we rewrite here the $N=0$, $l=0,1$ restriction of the 
like-particle form of their operator with $\Delta \nu$ set to $0$ for simplicity (this does 
not affect the discussion that follows): 
\begin{equation}
\begin{split}
C_\mathrm{HS}[\delta f] 
&=
\nu_D {L}[\delta f] + \frac{1}{v^2} \pd{}{v}
\left[
\frac{1}{2} v^4 \nu_{\parallel} F_{0} \pd{}{v} 
\left(
\frac{1}{4\pi}\int\limits_{-1}^{1} {d} \xi \int\limits_{0}^{2\pi} {d} \vartheta \frac{\delta f}{F_0} - \frac{v^2}{\vth^2}\,Q[\delta f]
\right)
\right]\\
&+ \nu_D\, \frac{2\bm{v} \cdot \bm{U}}{\vth^2} F_0, 
\end{split}
\end{equation}
where $\bm{U}$ and $Q$ are defined by \eqsref{def_U}{def_Q}. 
The primary concern here comes from the angle averaging operation in the energy-diffusion 
part of the operator. Firstly, the energy diffusion only acts on the spherically symmteric 
(in velocity space) part of the perturbed distribution function. However, there is no reason 
why there cannot arise perturbations that have very large energy derivatives but 
angle-average to zero (for example, $\delta f\propto\xi$). Clearly, such perturbations 
will not damped correctly. Secondly, upon conversion to gyrokinetic coordinates and 
gyroaveraging (see \secref{sec_gk} and Appendix D), the operator becomes
\begin{equation}
\begin{split}
C_\mathrm{HS,GK}[\hk] &=
\nu_D(v) \left[ \frac{1}{2} \frac{\partial}{\partial \xi} \left(1-\xi^2\right) \pd{\hk}{\xi} 
- \frac{1}{4}(1+\xi^2)\frac{v^2}{\vth^2}\,k_{\perp}^2\rho^2 \hk \right] \\
&+ \frac{J_{0}(a)}{v^2}\pd{}{v}\left[ \frac{1}{2}\, v^4 \nu_\parallel F_{0}
\pd{}{v}\int\limits_{{-1}}^{1} {d\xi} \frac{J_0(a) \hk}{2\pi F_{0}} \right]\\
&\qquad + 2\nu_D\frac{v_\perp J_1(a) U_\perp \left[ \hk \right] +  v_\parallel J_0(a) U_\parallel \left[ \hk \right] }{ \vth^2} F_{0} + \nu_E\, \frac{v^2}{\vth^2}\,J_0(a)Q[\hk] F_{0},
\end{split}
\end{equation}
where the conservation functionals $U_\perp$, $U_\parallel$ and $Q$ are the same as 
defined in \eqsdash{Uperp}{Qhk}. The immediatly obvious problem is that the angle 
averaging has introduced two new Bessel functions into the energy diffusion term. The 
energy diffusion is therefore supressed it by one power of $k_\perp \rho$ in the limit 
$k_\perp \rho\gg1$, while it is precisely in this limit that we expect the small-scale 
structure in the velocity space to be particularly important.\cite{schekAPJ07,schekPPCF08} This 
means that the energy cutoff in phase space is artificially pushed to smaller scales and one 
might encounter all the problems associated with insufficient energy 
diffusion.\cite{barnesPoP08}

While, for the reasons outlined above, we expect the Hirshman--Sigmar operator not to be a suitable 
model for collisions, we would like to note that for many purposes the Hirshman-Sigmar 
operators are superior to the model operator we presented in \secref{sec_newop}. Taken as 
a sequence, they provide a rigorous way of obtaining classical and neoclassical transport 
coefficients to any desired degree of accuracy, and it is relatively easy to solve the Spitzer 
problem for them, while the Spitzer functions for our operator are hard to find analytically.

\appendix
\renewcommand{\thechapter}{F}

\chapter{Sherman-Morrison formulation}
\label{app:sm}

The repeated application of the Sherman-Morrison formula considered here
is an extension of the scheme presented in Tatsuno and Dorland.~\cite{tatsunoAN08} 
Throughout this calculation, we adopt general notation applicable to both
Eqns.~(\ref{eqn:Lsplit}) and (\ref{eqn:DMEsplit}) and provide specific variable definitions in 
Table~\ref{tab:sm}.  
Both Eqns.~(\ref{eqn:Lsplit}) and (\ref{eqn:DMEsplit}) can be written in the form
\begin{equation}
A\mbf{x}=\mbf{b}.
\end{equation}  
\begin{table}
\begin{center}
\begin{tabular} {| c || c | c |}
\hline
variable & $L$ & $D$ \\
\hline \hline
$A$ & $\ 1-\Delta t\left(L+U_L\right) \ $ & $\ 1-\Delta t\left(D+U_D+E\right) \ $\\
\hline
$\mbf{x}$ & $h^{**}$ & $h^{n+1}$ \\
\hline
$\mbf{b}$ & $h^{*}$ & $h^{**}$\\
\hline
$A_{0}$ & $1-\Delta t L$ & $1-\Delta t D$\\
\hline
$\mbf{v}_{0}$ & $\nu_{D}v_{\perp}J_{1}$ & $-\Delta \nu v_{\perp}J_{1}$ \\
\hline
$\mbf{v}_{1}$ & $\nu_{D}v_{\parl}J_{0}$ & $-\Delta \nu v_{\parl}J_{0}$ \\
\hline
$\mbf{v}_{2}$ & $v_{\parl}$ (electrons) & $\nu_{E}v^{2}J_{0}$ \\
& $0$ (ions) &\\
\hline
$\mbf{u}_{0}$ & $-\Delta t\mbf{v}_0/d_{u}$ & $-\Delta t\mbf{v}_0/d_{u}$ \\
\hline
$\mbf{u}_{1}$ & $-\Delta t\mbf{v}_1/d_{u}$ & $-\Delta t\mbf{v}_1/d_{u}$ \\
\hline
$\mbf{u}_{2}$ & $-\Delta t\nu_{D}^{ei}v_{\parl}/d_{q}$ (electrons) & $-\Delta t\mbf{v}_2/d_{q}$ \\
& $0$ (ions) &\\
\hline
$d_{u}$ & $\int d^{3}v \ \nu_{D} v_{\parl}^{2} F_{0}$ & $\int d^{3}v \ \Delta\nu v_{\parl}^{2} F_{0}$ \\
\hline
$d_{q}$ & $v_{th,e}^{2}/2$ & $\int d^{3}v \ \nu_{E} v^{4} F_{0}$ \\
\hline
\end{tabular}
\end{center}
\caption{Sherman-Morrison variable definitions for Lorentz and energy diffusion
operator equations}
\label{tab:sm}
\end{table}
Because $\left< \mathcal{U} \right>$ and $\left< \mathcal{E} \right>$ are integral operators, we can write
them as tensor products so that
\begin{equation}
A\equiv A_{0}+\mbf{u}_{0}\otimes\mbf{v}_{0}+\mbf{u}_{1}\otimes\mbf{v}_{1}
+\mbf{u}_{2}\otimes\mbf{v}_{2}.
\end{equation}
We now define 
\begin{equation}
A_{1} = A_{0}+\mbf{u}_{0}\otimes\mbf{v}_{0}, \ \ A_{2}=A_{1}+\mbf{u}_{1}\otimes\mbf{v}_{1},
\end{equation}
so that
\begin{equation}
\left(A_{2}+\mbf{u}_{2}\otimes\mbf{v}_{2}\right)\mbf{x} = \mbf{b}.
\end{equation}
Applying the Sherman-Morrison formula to this equation, we find
\begin{equation}
\mbf{x} = \mbf{y}_{2} - \frac{\mbf{v}_{2}\cdot\mbf{y}_{2}}{1+\mbf{v}_{2}\cdot\mbf{z}_{2}}\mbf{z}_{2},
\end{equation}
where $A_{2}\mbf{y}_{2}=\mbf{b}$ and $A_{2}\mbf{z}_{2}=\mbf{u}_{2}$.

Applying the Sherman-Morrison formula to each of these equations gives
\begin{gather}
\mbf{y}_{2} = \mbf{y}_{1} - \frac{\mbf{v}_{1}\cdot\mbf{y}_{1}}{1+\mbf{v}_{1}\cdot\mbf{w}_{1}}\mbf{w}_{1}\\
\mbf{z}_{2} = \mbf{z}_{1} - \frac{\mbf{v}_{1}\cdot\mbf{z}_{1}}{1+\mbf{v}_{1}\cdot\mbf{w}_{1}}\mbf{w}_{1},
\end{gather}
where $A_{1}\mbf{y}_{1}=\mbf{b}$, $A_{1}\mbf{w}_{1}=\mbf{u}_{1}$, and 
$A_{1}\mbf{z}_{1}=\mbf{u}_{2}$.  A final application of Sherman-Morrison to 
these three equations yields
\begin{gather}
\mbf{y}_{1} = \mbf{y}_{0} - \frac{\mbf{v}_{0}\cdot\mbf{y}_{0}}{1+\mbf{v}_{0}\cdot\mbf{s}_{0}}\mbf{s}_{0}\\
\mbf{w}_{1} = \mbf{w}_{0} - \frac{\mbf{v}_{0}\cdot\mbf{w}_{0}}{1+\mbf{v}_{0}\cdot\mbf{s}_{0}}\mbf{s}_{0}\\
\mbf{z}_{1} = \mbf{z}_{0} - \frac{\mbf{v}_{0}\cdot\mbf{z}_{0}}{1+\mbf{v}_{0}\cdot\mbf{s}_{0}}\mbf{s}_{0},
\end{gather}
where $A_{0}\mbf{y}_{0}=\mbf{b}$, $A_{0}\mbf{s}_{0}=\mbf{u}_{0}$,
$A_{0}\mbf{w}_{0}=\mbf{u}_{1}$, and $A_{0}\mbf{z}_{0}=\mbf{u}_{2}$.

We can simplify our expressions by noting that $\mbf{v}_{0,1,2}$ and $\mbf{u}_{0,1,2}$ have
definite parity in $v_{\parl}$.  A number of inner products then vanish by symmetry, leaving
the general expressions
\begin{gather}
\mbf{y}_{2} = \mbf{y}_{0} - \left[\frac{\mbf{v}_{0}\cdot\mbf{y}_{0}}{1+\mbf{v}_{0}\cdot\mbf{s}_{0}}\right]\mbf{s}_{0} - \left[\frac{\mbf{v}_{1}\cdot\mbf{y}_{0}}{1+\mbf{v}_{1}\cdot\mbf{w}_{0}}\right]\mbf{w}_{0}\\
\mbf{z}_{2} = \mbf{z}_{0} - \left[\frac{\mbf{v}_{0}\cdot\mbf{z}_{0}}{1+\mbf{v}_{0}\cdot\mbf{s}_{0}}\right]\mbf{s}_{0}- \left[\frac{\mbf{v}_{1}\cdot\mbf{z}_{0}}{1+\mbf{v}_{1}\cdot\mbf{w}_{0}}\right]\mbf{w}_{0}.
\end{gather}
\appendix
\renewcommand{\thechapter}{G}

\chapter{Compact differencing the test-particle operator}
\label{app:cmpctdiff}

In this appendix, we derive a second order accurate compact difference scheme for pitch-angle 
scattering and energy diffusion on an unequally spaced grid.  The higher order of accuracy
of this scheme is desirable, but it does not possess discrete versions of the Fundamental
Theorem of Calculus and integration by parts when used with Gauss-Legendre quadrature
(or any other integration scheme with grid spacings unequal to integration weights).
Consequently, one should utilize this scheme only if integration weights and grid spacings 
are equal, or if exact satisfaction of conservation properties is not considered important.

For convenience, we begin by noting that Eqns.~(\ref{eqn:Lsplit}) and 
(\ref{eqn:DMEsplit}) can both be written in the general form
\begin{equation}
\left(\pd{h}{t}\right)_{C} = H\left(Gh'\right)' + S=H\left(G'h'+Gh''\right)+S,
\label{eqn:cmpctgk}
\end{equation}
where: for the Lorentz operator equation (\ref{eqn:Lsplit}) we identify $H=1$, 
$G=\nu_{D}\left(1-\xi^{2}\right)/2$,
$S=U_{L}[h]-k^{2}v^{2}\nu_{D}\left(1+\xi^{2}\right)h/4\Omega_{0}^{2}$,
and the prime denotes differentiation with respect to $\xi$; and for the energy diffusion
operator equation (\ref{eqn:DMEsplit}), we identify $H=1/2v^{2}F_{0}$,
$G=\nu_{\parl}v^{4}F_{0}$, $S=U_{D}[h]+E[h]-k^{2}v^{2}\nu_{\parl}\left(1-\xi^{2}\right)h/4\Omega_{0}^{2}$,
and the prime denotes differentiation with respect to $v$.  Here, the $h$ we are using
is actually normalized by $F_{0}$.

Employing Taylor Series expansions of $h$, we obtain the expressions
\begin{equation}
h_{i}' = \frac{\delta_{-}^{2}\left(h_{i+1}-h_{i}\right) + \delta_{+}^{2}\left(h_{i}-h_{i-1}\right)}{\delta_{+}\delta_{-}\left(\delta_{+}+\delta_{-}\right)} + \mathcal{O}[\delta^{2}],
\end{equation}
and
\begin{equation}
h_{i}'' = 2\frac{\delta_{-}\left(h_{i+1}-h_{i}\right) - \delta_{+}\left(h_{i}-h_{i-1}\right)}{\delta_{+}\delta_{-}\left(\delta_{+}+\delta_{-}\right)}+\frac{\delta_{-}-\delta_{+}}{3}h_{i}''' + \mathcal{O}[\delta^{2}],
\label{eqn:hpp}
\end{equation}
where $i$ denotes evaluation at the velocity space gridpoint $x_{i}$, and $\delta_{\pm}\equiv \left|x_{i\pm 1}-x_{i}\right|$ (here $x$ is a dummy
variable representing either $\xi$ or $v$).  In order for the $h_{i}''$ expression to be second order accurate, we must obtain
a first order accurate expression for $h_{i}'''$ in terms of $h_{i}$, $h_{i}'$, and $h_{i}''$.  We accomplish this by differentiating Eqn.~(\ref{eqn:cmpctgk}) with 
respect to $x$:
\begin{equation}
\left(\pd{h'}{t}\right)_{C} = H'\left(Gh'\right)' + H\left(Gh'\right)''+ S'
\label{eqn:hp}
\end{equation}
and rearranging terms to find
\begin{equation}
\begin{split}
h_{i}''' &= \frac{1}{H_{i}G_{i}}\Big[\left(\pd{h_{i}}{t}\right)_{C} - H_{i}'\left(G_{i}'h_{i}' + G_{i}h_{i}''\right)\\
&-H_{i}\left(G_{i}''h_{i}+2G_{i}'h_{i}''\right)-S_{i}'\Big] + \mathcal{O}[\delta],
\end{split}
\end{equation}
where, unless denoted otherwise, all quantities are taken at the $n+1$ time level.
Plugging this result into Eqn.~\ref{eqn:hpp} and grouping terms, we have
\begin{equation}
\begin{split}
\mu_{i}h_{i}''& =  \frac{\delta_{-}-\delta_{+}}{3H_{i}G_{i}} \left[h_{i}'\left(\frac{1}{\Delta t}-H_{i}'G_{i}'-H_{i}G_{i}''\right) - \frac{\left(h_{i}^{n}\right)'}{\Delta t}-S_{i}'\right] \\
&+ 2\frac{\delta_{-}\left(h_{i+1}-h_{i}\right)-\delta_{+}\left(h_{i}-h_{i-1}\right)}{\delta_{+}\delta_{-}\left(\delta_{+}+\delta_{-}\right)}+ \mathcal{O}[\delta^{2}],
\end{split}
\end{equation}
where $\mu_{i}=1+\left(\delta_{+}-\delta_{-}\right)\left(H_{i}'G_{i}+2H_{i}G_{i}'\right)/3H_{i}G_{i}$,
and we have taken $(\partial h/\partial t)_{C} = (h^{n+1}-h^{n})/\Delta t$.  Using Eqn.~(\ref{eqn:hp})
and the above result in Eqn.~(\ref{eqn:cmpctgk}), we find
\begin{equation}
\begin{split}
\left(\pd{h}{t}\right)_{C} &= h_{i}'\left(H_{i}G_{i}' - \frac{\delta_{+}+\delta_{-}}{3\mu_{i}}\left[\frac{1}{\Delta t} - H_{i}'G_{i}' - H_{i}G_{i}''\right]\right)\\
&+\frac{H_{i}G_{i}}{\mu_{i}}\left(2\frac{\delta_{-}\left(h_{i+1}-h_{i}\right) - \delta_{+}\left(h_{i}-h_{i-1}\right)}{\delta_{+}\delta_{-}\left(\delta_{+}+\delta_{-}\right)}\right.\\
&+\left.\frac{\delta_{+}+\delta_{-}}{3H_{i}G_{i}}\left[\frac{\left(h_{i}^{n}\right)'}{\Delta t} + S_{i}'\right]\right) + S_{i}+\mathcal{O}[\delta^{2}].
\label{eqn:cmpctgen}
\end{split}
\end{equation}This is the general compact differenced form to be used when solving Eqns.~(\ref{eqn:Lsplit}) and (\ref{eqn:DMEsplit}).

In order to illustrate how compact differencing affects the implicit solution using Sherman-Morrison,
we present the result of using the particular form of $S$ for energy diffusion in Eqn.~(\ref{eqn:cmpctgen}):
\begin{equation}
\begin{split}
\frac{h_{i}^{n+1}-h_{i}^{n}}{\Delta t}& = \frac{h_{i+1}}{\delta_{+}\left(\delta_{+}+\delta_{-}\right)}\left(\frac{2H_{i}G_{i}}{\mu_{i}}+\delta_{-}\zeta_{i}\right)\\
&\frac{h_{i-1}}{\delta_{-}\left(\delta_{+}+\delta_{-}\right)}\left(\frac{2H_{i}G_{i}}{\mu_{i}}-\delta_{+}\zeta_{i}\right)\\
&\frac{h_{i}}{\delta_{-}\delta_{+}}\left(-\frac{2H_{i}G_{i}}{\mu_{i}}+\left(\delta_{+}-\delta_{-}\right)\zeta_{i}-\delta_{+}\delta_{-}K\tilde{\nu}_{s}\right)\\
&\frac{\sigma_{i}}{\Delta t}\Bigg(h_{i+1}^{n}\frac{\delta_{-}}{\delta_{+}\left(\delta_{+}+\delta_{-}\right)} - h_{i-1}^{n}\frac{\delta_{+}}{\delta_{-}\left(\delta_{+}+\delta_{-}\right)}\\
&+h_{i}^{n}\frac{\delta_{+}-\delta_{-}}{\delta_{+}\delta_{-}}\Bigg) + \tilde{U}_{\parl}V_{\parl} + \tilde{U}_{\perp}V_{\perp} + \tilde{U}_{q}q + \mathcal{O}[\delta^{2}],
\label{eqn:cmpctEdiff}
\end{split}
\end{equation}
where 
\begin{gather*}
\sigma_{i} = \left(\delta_{+}-\delta_{-}\right)/3\mu_{i}\\
K=k^{2}v_{th}^{2}\left(1-\xi^{2}\right)/8\Omega_{0}^{2}\\
\zeta_{i}=H_{i}G_{i}'-\sigma_{i} \left(1/\Delta t - H_{i}'G_{i}' -H_{i}G_{i}'' + K\nu_{s}\right)\\
\nu_{s}=\nu_{\parl}v_{th}^{2}/2v^{2}\\
\tilde{A}=A+\sigma A'\\
U_{\perp,\parl,q}=\mbf{u}_{0,1,2}\\
V_{\perp,\parl,q}=\mbf{v}_{0,1,2}
\end{gather*}
with $\mbf{u}$ and $\mbf{v}$ given in Table~\ref{tab:sm}.

We see
that the only significant effects of compact differencing on numerical implementation are: 
modification of $h^{**}$ in Eqn.~(\ref{eqn:DMEsplit}) to reflect the $h^{n}$ terms
on the right-hand side of Eqn.~(\ref{eqn:cmpctEdiff}); and modification
of the $U_{\parl}$, $U_{\perp}$, and $U_{q}$ terms that appear in Sherman-Morrison 
($\mbf{u}_{0,1,2}$ from Appendix A) to include an additional $\sigma U'$ term.
\appendix
\renewcommand{\thechapter}{H}

\chapter{Sample \texttt{Trinity} input file}
\label{app:trinin}
\verb#! case with 2 species, nonlinear fluxes, 4 keV edge temperature#\\
\verb#&nt_params#\\
\\
\verb#! geometry#\\
\verb# rad_out = 0.8  ! outer rad. bound (normalized by minor radius, a)#\\
\verb# rmaj = 6.2     ! major radius, R (in meters)#\\
\verb# aspr = 3.1     ! aspect ratio (R/a)#\\
\verb# bmag = 5.3     ! B-field amplitude at center of LCFS (in Tesla)#\\
 \\
\verb#! species parameters#\\
\verb# ntspec = 2  ! number of species to evolve in transport equations#\\
\verb# qi = 1      ! ion charge (in units of proton charge)#\\
\verb# mi = 1      ! ion mass (in units of proton mass)#\\
 \\
\verb#! time advancement#\\
\verb# ntstep = 25         ! number of transport time steps#\\
\verb# ntdelt = 0.005      ! transport time step size#\\
\verb# subcycle = .false.  ! set to T to subcycle temperature equilibration#\\
\verb# nsub = 1            ! number of subcycles per time step#\\
\verb# impfac = 0.5        ! time-centering (0=explicit, 1=implicit)#\\
 \\
\verb#! fluxes#\\
\verb# grad_option = "tgrads"  ! assume fluxes depend on ion and electron#\\
\verb#                         ! pressure grads#\\
\verb# ql_flag = .false.       ! T for quasilinear estimate for fluxes#\\
\verb# model_flux = .false.    ! T for offset linear estimate for fluxes#\\
\verb# include_neo = .true.    ! T to include neoclassical ion heat flux#\\
\verb# dfprim = 1.0            ! amount by which to perturb dens grad when#\\
\verb#                         ! obtaining estimate for flux derivative#\\
\verb# dtprim = 1.0            ! amount by which to perturb temp grad#\\
\verb# pflx_min =1.0e-5        ! minimum particle flux to use if GS2#\\
\verb#                         ! calculation does not converge#\\
\verb# qflx_min = 5.0e-4       ! minimum heat flux#\\
 \\
\verb# temp_equil = .true.    ! F to neglect temperature equilibration#\\
\verb#                        ! between species#\\
 \\
\verb#! initialization#\\
\verb# rln = 1.0       ! initial R/Ln for density profile#\\
\verb# rlti = 5.0      ! initial R/LTi for ion temperature profile#\\
\verb# rlte = 5.0     ! initial R/LTe for electron temperature profile#\\
\verb# nedge = 1.0    ! fixed dens at outer rad. bound (in 10^{20}/m^{3})#\\
\verb# tiedge = 4.0   ! fixed ion temp at outer rad. bound (in keV)#\\
\verb# teedge = 4.0   ! fixed electron temp at outer rad. bound (in keV)#\\
 \\
\verb#! sources#\\
\verb# densin = 0.0     ! amplitude of particle source#\\
\verb# powerin = 120.0  ! heat source power (in MW)#\\
\verb# src_ratio = 0.5  ! fraction of powerin going into ions#\\
\verb# psig = 0.2       ! width of Gaussian heat source profile#\\
\verb# nsig = 0.2       ! width of Gaussian particle source profile#\\
 \\
\verb# /#

\end{document}